\documentclass[11pt]{article}
\pdfoutput=1

\usepackage[T1]{fontenc}
\usepackage{epsfig,amsmath,amssymb}
\usepackage{graphicx}
\usepackage{xspace}
\usepackage{listings}
\usepackage{ulem}
\usepackage{slashed}
\usepackage{cite}
\usepackage[colorlinks,citecolor=blue,urlcolor=blue,linkcolor=blue]{hyperref}
\usepackage{xcolor}
\usepackage{rotating}
\usepackage{multirow}
\usepackage{url} 

\usepackage{feynmp}
\graphicspath{{./Diagrams/}}

\def\hc{\text{h.c.}}

\newcommand\FlavorKit{{\tt FlavorKit}\xspace}
\newcommand\SPheno{{\tt SPheno}\xspace}
\newcommand\SARAH{{\tt SARAH}\xspace}

\newcommand{\VeezL}{E^L}
\newcommand{\VeezR}{E^R}
\newcommand{\VeeaL}{A^L}
\newcommand{\VeeaR}{A^R}
\newcommand{\VeehL}{H^L}
\newcommand{\VeehR}{H^R}
\newcommand{\VeeZ}{\tilde{E}}
\newcommand{\Vaee}{\tilde{A}}
\newcommand{\VeeH}{\tilde{H}}

\newcommand{\VddzL}{D^L}
\newcommand{\VddzR}{D^R}
\newcommand{\VddaL}{A^{d,L}}
\newcommand{\VddaR}{A^{d,R}}
\newcommand{\VddhL}{H^{d,L}}
\newcommand{\VddhR}{H^{d,R}}

\newcommand{\VuuzL}{U^L}
\newcommand{\VuuzR}{U^R}
\newcommand{\VuuaL}{A^{u,L}}
\newcommand{\VuuaR}{A^{u,R}}
\newcommand{\VuuhL}{H^{u,L}}
\newcommand{\VuuhR}{H^{u,R}}

\newcommand{\VneeL}{N^L}
\newcommand{\VneeR}{N^R}
\newcommand{\VeneL}{\bar{N}^L}
\newcommand{\VeneR}{\bar{N}^R}

\newcommand{\VnddL}{N^{d,L}}
\newcommand{\VnddR}{N^{d,R}}
\newcommand{\VdndL}{\bar{N}^{d,L}}
\newcommand{\VdndR}{\bar{N}^{d,R}}

\newcommand{\VnuuL}{N^{u,L}}
\newcommand{\VnuuR}{N^{u,R}}
\newcommand{\VunuL}{\bar{N}^{u,L}}
\newcommand{\VunuR}{\bar{N}^{u,R}}

\newcommand{\VnnzL}{M^L}
\newcommand{\VnnzR}{M^R}
\newcommand{\VnnhL}{S^L}
\newcommand{\VnnhR}{S^R}
\newcommand{\VnnaL}{P^L}
\newcommand{\VnnaR}{P^R}

\newcommand{\VcduL}{W^{d,L}}
\newcommand{\VcduR}{W^{d,R}}

\newcommand{\VdcuL}{\bar{W}^{d,L}}
\newcommand{\VdcuR}{\bar{W}^{d,R}}

\newcommand{\VcudL}{W^{u,L}}
\newcommand{\VcudR}{W^{u,R}}

\newcommand{\VcudLc}{\bar{W}^{u,L}}
\newcommand{\VcudRc}{\bar{W}^{u,R}}

\newcommand{\VcevL}{W^L}
\newcommand{\VcevR}{W^R}
\newcommand{\VecviL}{\bar{X}^L}
\newcommand{\VecviR}{\bar{X}^R}
\newcommand{\VecvrL}{\hat{\bar{X}}^L}
\newcommand{\VecvrR}{\hat{\bar{X}}^R}
\newcommand{\VceviL}{{X}^L}
\newcommand{\VceviR}{{X}^R}
\newcommand{\VcevrL}{\hat{{X}}^L}
\newcommand{\VcevrR}{\hat{{X}}^R}
\newcommand{\VcczL}{C^L}
\newcommand{\VcczR}{C^R}
\newcommand{\VcchL}{S^{c,L}}
\newcommand{\VcchR}{S^{c,R}}
\newcommand{\VccaL}{P^{c,L}}
\newcommand{\VccaR}{P^{c,R}}

\newcommand{\VvivrZ}{\tilde{V}}
\newcommand{\Vhvivi}{\tilde{H}^{ii}}
\newcommand{\Vhvrvr}{\tilde{H}^{rr}}
\newcommand{\Vhvivr}{\tilde{H}^{ir}}
\newcommand{\Vavivi}{\tilde{A}^{ii}}
\newcommand{\Vavrvr}{\tilde{A}^{rr}}
\newcommand{\Vavivr}{\tilde{A}^{ir}}

\newcommand{\VvehpL}{V^{+,L}}
\newcommand{\VvehpR}{V^{+,R}}
\newcommand{\VevhmL}{\bar{V}^{+,L}}
\newcommand{\VevhmR}{\bar{V}^{+,R}}
\newcommand{\VvewpL}{\hat{V}^{+,L}}
\newcommand{\VvewpR}{\hat{V}^{+,R}}
\newcommand{\VevwmL}{\hat{\bar{V}}^{+,L}}
\newcommand{\VevwmR}{\hat{\bar{V}}^{+,R}}

\newcommand{\VudhpL}{V^{u,L}}
\newcommand{\VudhpR}{V^{u,R}}
\newcommand{\VduhmL}{\bar{V}^{d,L}}
\newcommand{\VduhmR}{\bar{V}^{d,R}}
\newcommand{\VudwpL}{\hat{V}^{u,L}}
\newcommand{\VudwpR}{\hat{V}^{u,R}}
\newcommand{\VduwmL}{\hat{\bar{V}}^{d,L}}
\newcommand{\VduwmR}{\hat{\bar{V}}^{d,R}}

\newcommand{\Vhmhpz}{Z^{hh}}
\newcommand{\Vhmwpz}{Z^{hw}}
\newcommand{\Vhpwmz}{\bar{Z}^{hw}}
\newcommand{\Vwpwmz}{Z^{ww}}
\newcommand{\Vhhmhp}{H^{hh}}
\newcommand{\Vhhmwp}{H^{hw}}
\newcommand{\Vhhpwm}{\bar{H}^{hw}}
\newcommand{\Vhwmwp}{H^{ww}}
\newcommand{\Vahmhp}{A^{hh}}
\newcommand{\Vahmwp}{A^{hw}}
\newcommand{\Vahpwm}{\bar{A}^{hw}}

\newcommand{\VvvzL}{V^L}
\newcommand{\VvvzR}{V^R}
\newcommand{\VvvhL}{H^{\nu,L}}
\newcommand{\VvvhR}{H^{\nu,R}}
\newcommand{\VvvaL}{A^{\nu,L}}
\newcommand{\VvvaR}{A^{\nu,R}}

\newcommand{\VLL}{V^{LL}}
\newcommand{\VLR}{V^{LR}}
\newcommand{\SLL}{S^{LL}}
\newcommand{\SLR}{S^{LR}}
\newcommand{\TLL}{T^{LL}}
\newcommand{\TLR}{T^{LR}}

\newcommand{\SLRa}[1]{\SLR_{{A^0_p},#1}}
\newcommand{\SLLa}[1]{\SLL_{{A^0_p},#1}}
\newcommand{\SLRh}[1]{\SLR_{{h_p},#1}}
\newcommand{\SLLh}[1]{\SLL_{{h_p},#1}}
\newcommand{\VLRZ}[1]{\VLR_{Z,#1}}
\newcommand{\VLLZ}[1]{\VLL_{Z,#1}}

\newcommand{\XLRb}[1]{X^{LR}_{#1}}
\newcommand{\XLLb}[1]{X^{LL}_{#1}}

\newcommand{\SLRb}[1]{\SLR_{#1}}
\newcommand{\SLLb}[1]{\SLL_{#1}}
\newcommand{\TLRb}[1]{\TLR_{#1}}
\newcommand{\TLLb}[1]{\TLL_{#1}}
\newcommand{\VLRb}[1]{\VLR_{#1}}
\newcommand{\VLLb}[1]{\VLL_{#1}}

\newcommand{\KmonoL}[1]{A^{1L}_{#1}}
\newcommand{\KdiL}[1]{A^{2L}_{#1}}

\newcommand{\VhmhpP}{F^{h}}
\newcommand{\Veep}{F^{\tilde{e}}}
\newcommand{\VhmwpP}{F^{hw}}
\newcommand{\VhpPwm}{\bar{F}^{hw}}
\newcommand{\VwpPwm}{F^{w}}
\newcommand{\VccpL}{F^{c,L}}
\newcommand{\VccpR}{F^{c,R}}

\newcommand{\eq}[1]{Eq.~(\ref{#1})}

\newcommand{\fig}[1]{Fig.~\ref{#1}}

\definecolor{mkgreen}{rgb}{0.2,.70,.3}

\def\lsim{\raise0.3ex\hbox{$\;<$\kern-0.75em\raise-1.1ex\hbox{$\sim\;$}}}

\begin{document}
 
\vspace*{-2cm}
\begin{flushright}
LPT-Orsay-14-43 \\
BONN-TH-14-11 \\
IFT-UAM/CSIC-14-061 \\
FTUAM-14-25 \\
\vspace*{2mm}
\today
\end{flushright}

\begin{center}
\vspace*{15mm}

\vspace{1cm}
{\Large\bf 
Lepton flavor violation in low-scale seesaw models: SUSY and non-SUSY contributions
} \\
\vspace{1cm}

{\bf A. Abada$^{a}$, M. E. Krauss$^{b}$, W. Porod$^{b}$, F. Staub$^{c}$, A. Vicente$^{d}$ and C. Weiland$^{e}$}

 \vspace*{.5cm} 
$^{a}$ Laboratoire de Physique Th\'eorique, CNRS -- UMR 8627, \\
Universit\'e de Paris-Sud 11, F-91405 Orsay Cedex, France
\vspace*{.2cm} 

$^{b}$ Institut f\"ur Theoretische Physik und Astronomie, Universit\"at W\"urzburg\\
97074 W\"urzburg,Germany
\vspace*{.2cm} 

$^{c}$ Physikalisches Institut der Universit\"at Bonn, 53115 Bonn, Germany
\vspace*{.2cm} 

$^{d}$ IFPA, Dep. AGO, Universit\'e de Li\`ege, \\
Bat B5, Sart-Tilman B-4000 Li\`ege 1, Belgium
\vspace*{.2cm} 

$^{e}$ Departamento de F\'{\i}sica Te\'orica and Instituto de F\'{\i}sica Te\'orica, IFT-UAM/CSIC,\\
Universidad Aut\'onoma de Madrid, Cantoblanco, 28049 Madrid, Spain

\end{center}

\begin{abstract}
Taking the supersymmetric inverse seesaw mechanism as the explanation
for neutrino oscillation data, we investigate charged lepton flavor
violation in radiative and 3-body lepton decays as well as in
neutrinoless $\mu-e$ conversion in muonic atoms. In contrast to former
studies, we take into account all possible contributions:
supersymmetric as well as non-supersymmetric. We take CMSSM-like
boundary conditions for the soft supersymmetry breaking parameters.
We find several regions where cancellations between various
contributions exist, reducing the lepton flavor violating rates by an
order of magnitude compared to the case where only the dominant
contribution is taken into account. This is in particular important
for the correct interpretation of existing data as well as for
estimating the reach of near future experiments where the sensitivity
will be improved by one to two orders of magnitude.  Moreover, we
demonstrate that ratios like BR($\tau\to 3 \mu$)/BR($\tau\to \mu e^+
e^-$) can be used to determine whether the supersymmetric
contributions dominate over the $W^\pm$ and $H^\pm$ contributions or vice
versa.
\end{abstract}

\newpage

\tableofcontents

\vspace*{3mm}
\newpage

\section{Introduction}

The recent discovery of a bosonic state at the Large Hadron Collider
(LHC) \cite{Aad:2012tfa,Chatrchyan:2012ufa} stands as a major
breakthrough in particle physics. Although further confirmation is
required, all data are compatible with the long-awaited Higgs boson,
thus completing the Standard Model (SM) particle content. Furthermore, the
properties and decay modes of this scalar are in good agreement with
the SM expectations, making the SM picture more motivated than ever.

In this context, it is crucial to keep in mind that the SM cannot be
the ultimate theory. In fact, and besides theoretical arguments such
as the hierarchy problem, there are very good experimental reasons
to go beyond the SM (BSM). The best of these motivations is the
existence of non-zero neutrino masses and mixing angles, now firmly
established by neutrino oscillation experiments
\cite{Tortola:2012te,GonzalezGarcia:2012sz,Capozzi:2013csa}. Since the
SM lepton sector does not include them, one has to go beyond the SM.

A generic prediction in most of these neutrino mass models is lepton
flavor violation (LFV), not only in the neutrino sector but also for
the charged leptons. Depending of the exact realization of the
neutrino mass model, the rates for the LFV processes can be very
different. For instance, high-scale models typically predict small
branching ratios, thus making LFV hard (if not impossible) to be
discovered. In contrast, one expects measurable LFV rates if the scale
of new physics is not far from the electroweak (EW) scale. These
low-scale mechanisms generating neutrino masses are thus more
attractive from a phenomenological point of view, since they offer a
window to new physics thanks to their LFV promising
perspectives. Moreover, they can be directly tested at the LHC via the
production of new particles if these are light enough.

On the experimental side, the field of LFV physics will live an era of
unprecedented developments in the near future, with dedicated
experiments in different fronts\footnote{See \cite{Mihara:2013zna} for
  a recent review.}. In the case of the muon radiative decay $\mu \to
e \gamma$, the MEG collaboration has announced plans for future
upgrades. These will allow for an improvement of the current bound,
$\text{BR}(\mu \to e \gamma) < 5.7 \cdot 10^{-13}$
\cite{Adam:2013mnn}, reaching a sensitivity of about $6 \cdot
10^{-14}$ after 3 years of acquisition time
\cite{Baldini:2013ke}. Limits on $\tau$ radiative decays are less
stringent, but they are expected to be improved at Belle II
\cite{Aushev:2010bq}.  These will also search for lepton flavor
violating $B$-meson decays. Moreover, the perspectives for the 3-body
decays $\ell_\alpha \to 3 \, \ell_\beta$ are good as well. The decay
$\mu \to 3 \, e$ was searched for long ago by the SINDRUM experiment
\cite{Bellgardt:1987du}, setting the limit $\text{Br}(\mu \to 3 \, e)
< 1.0 \cdot 10^{-12}$. The future Mu3e experiment announces a
sensitivity of $\sim 10^{-16}$ \cite{Blondel:2013ia}, which would
imply a 4 orders of magnitude improvement. In the case of $\tau$
decays to three charged leptons, Belle II will again be the facility
where improvements are expected \cite{Bevan:2014iga}, although
recently the LHCb collaboration has reported first bounds on $\tau\to
3 \, \mu$ \cite{Aaij:2013fia}.  The LFV process where the best
developments are expected in the next few years is neutrinoless
$\mu-e$ conversion in muonic atoms. In the near future, many different
experiments will search for a positive signal. These include Mu2e
\cite{Carey:2008zz,Glenzinski:2010zz,Abrams:2012er}, DeeMe
\cite{Aoki:2010zz}, COMET \cite{Cui:2009zz,Kuno:2013mha} and
PRISM/PRIME \cite{PRIME}. The expected sensitivities for the
conversion rate range from a modest $10^{-14}$ in the near future to
an impressive $10^{-18}$.  Finally, one can also search for LFV in
high-energy experiments, such as the LHC. A popular process in this
case is the Higgs boson LFV decay to a pair of charged leptons, $h \to
\ell_\alpha \ell_\beta$, with $\alpha \ne \beta$
\cite{Pilaftsis:1992st,DiazCruz:1999xe}, which has recently received
some attention
\cite{Blankenburg:2012ex,Harnik:2012pb,Davidson:2012ds,Arhrib:2012mg,%
  Arhrib:2012ax,BhupalDev:2012zg,%
  Arana-Catania:2013xma,Falkowski:2013jya,Arganda:2014dta,Bressler:2014jta,Kopp:2014rva,Sierra:2014nqa}.
First bounds on $h \to \mu \tau$ have been reported by the CMS
collaboration \cite{CMS:2014hha}~\footnote{The CMS collaboration also
  reports an intriguing $2.5 \, \sigma$ excess in $h \to \mu \tau$
  leading to BR$(h \to \tau \mu) \sim 1 \%$.}.  For other
possibilities to search for LFV at high-energy colliders, see
\cite{Porod:2002zy,Bartl:2005yy,Hirsch:2008dy,Hirsch:2008gh,Kaneko:2008re,%
  delAguila:2008cj,Atre:2009rg,Abada:2010kj,%
  Esteves:2010si,Esteves:2010ff,Abada:2011mg,Abada:2012re,Bandyopadhyay:2012px,%
  Mondal:2012jv,Das:2012ze,Teixeira:2014jza,BhupalDev:2012ru}.  In
table~\ref{tab:sensi} we collect present bounds and expected
near-future sensitivities for the most popular low-energy LFV
observables.

\begin{table}[tb!]
\centering
\begin{tabular}{|c|c|c|}
\hline
LFV Process & Present Bound & Future Sensitivity  \\
\hline
    $\mu \rightarrow  e \gamma$ & $5.7\times 10^{-13}$~\cite{Adam:2013mnn}  & $6\times 10^{-14}$~\cite{Baldini:2013ke} \\
    $\tau \to e \gamma$ & $3.3 \times 10^{-8}$~\cite{Aubert:2009ag}& $ \sim3\times10^{-9}$~\cite{Aushev:2010bq}\\
    $\tau \to \mu \gamma$ & $4.4 \times 10^{-8}$~\cite{Aubert:2009ag}& $ \sim3\times10^{-9}$~\cite{Aushev:2010bq} \\
    $\mu \rightarrow e e e$ &  $1.0 \times 10^{-12}$~\cite{Bellgardt:1987du} &  $\sim10^{-16}$~\cite{Blondel:2013ia} \\
    $\tau \rightarrow \mu \mu \mu$ & $2.1\times10^{-8}$~\cite{Hayasaka:2010np} & $\sim 10^{-9}$~\cite{Aushev:2010bq} \\
    $\tau^- \rightarrow e^- \mu^+ \mu^-$ &  $2.7\times10^{-8}$~\cite{Hayasaka:2010np} & $\sim 10^{-9}$~\cite{Aushev:2010bq} \\
    $\tau^- \rightarrow \mu^- e^+ e^-$ &  $1.8\times10^{-8}$~\cite{Hayasaka:2010np} & $\sim 10^{-9}$~\cite{Aushev:2010bq} \\
    $\tau \rightarrow e e e$ & $2.7\times10^{-8}$~\cite{Hayasaka:2010np} &  $\sim 10^{-9}$~\cite{Aushev:2010bq} \\
    $\mu^-, \mathrm{Ti} \rightarrow e^-, \mathrm{Ti}$ &  $4.3\times 10^{-12}$~\cite{Dohmen:1993mp} & $\sim10^{-18}$~\cite{PRIME} \\
    $\mu^-, \mathrm{Au} \rightarrow e^-, \mathrm{Au}$ & $7\times 10^{-13}$~\cite{Bertl:2006up} & \\
    $\mu^-, \mathrm{Al} \rightarrow e^-, \mathrm{Al}$ &  & $10^{-15}-10^{-18}$ \\
    $\mu^-, \mathrm{SiC} \rightarrow e^-, \mathrm{SiC}$ &  & $10^{-14}$~\cite{Aoki:2012zza} \\
\hline
\end{tabular}
\caption{Current experimental bounds and future sensitivities for some
  low-energy LFV observables.}
\label{tab:sensi}
\end{table}

With such a large variety of processes, a proper theoretical
understanding of potential hierarchies or correlations in a given
model becomes necessary. This goal requires detailed analytical and
numerical studies of the different contributions to the LFV processes,
in order to determine the dominant ones and to get a proper
interpretation
of the LFV bounds. Furthermore, the understanding of the LFV
  anatomy of several models allows one to discriminate among them by
using combinations of observables which have definite predictions
\cite{Buras:2010cp}.

In this work we are interested in LFV in supersymmetric and
non-supersymmetric variants of the inverse seesaw model (ISS)
\cite{Mohapatra:1986bd}. This low-scale neutrino mass model
constitutes a very interesting alternative to the usual seesaw
mechanism.  The suppression mechanism that guarantees the smallness of
neutrino masses is the introduction of a slight breaking of lepton
number in the singlet sector, in the form of a small (compared to the
EW scale) Majorana mass for the $X$ singlets. This allows for large
Yukawa couplings compatible with a low (TeV or even lower) mass for
the seesaw mediators.  With this combination, one expects a very rich
phenomenology, including sizable LFV rates and additional
contributions to the radiative corrections to the Higgs mass
\cite{Elsayed:2011de,Hirsch:2011hg,Chun:2014tfa}.  In the
supersymmetric (SUSY) version of the ISS, the new singlet fermions are
promoted to singlet superfields. The appealing features of the ISS
mechanism are kept also in the SUSY version.

LFV in models with light right-handed (RH) neutrinos has already been
studied in great detail. Early studies
\cite{Bernabeu:1987gr,Ilakovac:1994kj,Deppisch:2004fa,Deppisch:2005zm}
already pointed out the existence of large enhancements in the LFV
rates with respect to those found in high-scale models. More recently,
there has been a revived interest due to the expected experimental
improvements in the near future. Interestingly, dominant contributions
have been found in (non-SUSY) box diagrams induced by RH
neutrinos. This was first shown in \cite{Ilakovac:2009jf} and later
confirmed in \cite{Alonso:2012ji,Dinh:2012bp,Ilakovac:2012sh}. In this
case, the future $\mu- e$ conversion experiments will play a major
role in constraining light right-handed neutrino scenarios. The usual
photon penguin contributions get also enhanced in the presence of
light RH neutrinos, see for example \cite{Dev:2013oxa}. Regarding the
SUSY contributions, several studies have recently addressed the role
of the $Z$-penguins. A large enhancement with respect to the usual
dipole contribution was reported in \cite{Hirsch:2012ax}. Later, this
result was (qualitatively) confirmed in \cite{Ilakovac:2012sh} and
further exploited in several phenomenological studies
\cite{Dreiner:2012mx,Hirsch:2012kv,Abada:2012cq,Krauss:2013jva}.
However, in \cite{Krauss:2013gya} it was shown that the results in
\cite{Hirsch:2012ax} (and the subsequent studies
\cite{Dreiner:2012mx,Hirsch:2012kv,Abada:2012cq,Krauss:2013jva}) are
incorrect, due to an inconsistency in \cite{Arganda:2005ji}. 
While this has a negligible impact in the case of high-scale seesaw
models, this is not the case for low-scale seesaw models like the
supersymmetric version of the ISS.

Given that recent studies pointed out important but partial results
and the upcoming experimental improvements, we aim in this work for a
complete calculation of the various LFV observables taking into
account all contributions at the same time. One of our results will be
that there exist several regions in parameter space where
cancellations between various contributions occur, changing the
interpretation of existing and future experimental results. In order
to do so we have made use of {\tt FlavorKit} \cite{Porod:2014xia}, a
tool that combines the analytical power of {\tt SARAH}
\cite{Staub:2008uz,Staub:2009bi,Staub:2010jh,Staub:2012pb,Staub:2013tta}
with the numerical routines of {\tt SPheno}
\cite{Porod:2003um,Porod:2011nf} to obtain predictions for flavor
observables in a wide range of models. This setup makes use of {\tt
  FeynArts}/{\tt FormCalc}
\cite{Hahn:1998yk,Hahn:2000kx,Hahn:2000jm,Hahn:2004rf,Hahn:2005vh,Nejad:2013ina}
to compute generic predictions for the form factors of the relevant
operators and thus provides an automatic computation of the flavor
observables. We use this setup to compute for the first time the Higgs
penguin contributions to LFV in the inverse seesaw~\footnote{The Higgs
  penguin contributions to LFV processes were first considered in the
  context of the inverse seesaw in \cite{Abada:2011hm}. However, our
  paper goes beyond this reference in two ways: by doing the
  computation in the mass basis and by taking into account all
  contributions to the Higgs penguins.}.  In addition, we improve
previous studies in others aspects as well: (i) we make use of the
full 2-loop renormalization group equations (RGEs) including all
flavor effects in the SM and SUSY sectors to obtain the parameters
entering the calculation, and (ii) we include for the first time the
decays $\tau^- \to \mu^+ e^- e^-$ and $\tau^- \to e^+ \mu^- \mu^-$.

The paper is organized as follows: in Section~\ref{sec:model} we
present the ISS model and its supersymmetric extension. The LFV
observables induced by the extended particle content and the dominant
contributions are discussed in Section~\ref{sec:observables}, and in
Section~\ref{sec:numerics} we present our numerical results. In
Section~\ref{sec:conclusions} we draw our conclusions.  In the
appendices we first introduce the formulae for the mass matrices and
our convention for the loop integrals before presenting the additional
contributions to the 1- and 2-loop RGEs compared to the MSSM case.
More importantly, they contain the complete set of contributions to
the LFV observables discussed in this paper.

\section{Inverse seesaw model and its supersymmetric extension}
\label{sec:model}

In the inverse seesaw, the Standard Model field content is extended by
$n_R$ generations of right-handed neutrinos $\nu_R$ and $n_X$
generations of singlet fermions $X$ (such that $n_R+n_X = N_s$), both
with lepton number
$L=+1$~\cite{Mohapatra:1986aw,Mohapatra:1986bd,Bernabeu:1987gr}.  The
corresponding Lagrangian before EWSB has the form
\begin{equation}
 \mathcal{L}_\mathrm{ISS} = \mathcal{L}_\mathrm{SM} - Y^{ij}_\nu \overline{\nu_{Ri}} \widetilde{H}^\dagger L_{j}-M_R^{ij}\overline{\nu_{Ri}} X_j-\frac{1}{2}\mu_{X}^{ij}\overline{X_{i}^C} X_{j}+ h.c.\,,
\label{ISSlagrangian}
\end{equation}
where a sum over $i, j = 1, 2, 3$ is assumed\footnote{The ISS requires
  the introduction of at least two right-handed neutrinos in order to
  account for the active neutrino masses and mixings. The most minimal
  ISS realization~\cite{Malinsky:2009df,Gavela:2009cd,Abada:2014vea}
  consists in the addition of two right-handed and two sterile
  neutrinos to the SM content. However, its minimal SUSY
  realization~\cite{Hirsch:2009ra} requires only one pair of fermionic
  singlets.}.  $\mathcal{L}_\mathrm{SM}$ is the SM Lagrangian, $Y_\nu$
are the neutrino Yukawa couplings and $M_R$ is a complex mass matrix
that generates a lepton number conserving mass term for the fermion
singlets. The complex symmetric mass matrix $\mu_X$ violates lepton
number by two units and is naturally small, in the sense of 't
Hooft~\cite{'tHooft:1979bh}, since in the limit $\mu_X \to 0$ lepton
number is restored. This Majorana mass term also leads to a small mass
splitting in the heavy neutrino sector, which then become quasi-Dirac
neutrinos.

After EWSB, in the basis $(\nu_L\,,\;\nu_R^C\,,\;X)$, the $9\times 9$ neutrino mass matrix is given by
\begin{equation}
 M_{\mathrm{ISS}}=\left(\begin{array}{c c c} 0 & m_D^T 
 & 0 \\ m_D & 0 & M_R \\ 
 0 & M_R^T & \mu_X \end{array}\right)\,.\label{eq:ISSmatrix}
\end{equation}
 where $m_D = \frac{1}{\sqrt{2}} Y_\nu v$ and $v/\sqrt{2}$ is the
 vacuum expectation value (vev) of the Higgs boson.

Under the assumption $\mu_X \ll m_D \ll M_R$, the mass matrix
$M_{\mathrm{ISS}}$ can be block-diagonalized to give an effective mass
matrix for the light neutrinos~\cite{GonzalezGarcia:1988rw}
\begin{equation}
 M_{\mathrm{light}}\simeq m_D^T {M_R^T}^{-1} \mu_X M_R^{-1} m_D\,,
\label{LightMatrix}
\end{equation}
whereas the heavy quasi-Dirac neutrinos have masses corresponding
approximately to the entries of $M_R$.

As usual, one can easily obtain a supersymmetric version of the model
by promoting the corresponding fields to superfields
$\widehat{\nu}^C_i$ and $\widehat{X}_i$ ($i=1,2,3$) and including the
corresponding interactions in the superpotential. This reads
\begin{equation}
 W=  W_\mathrm{MSSM} + \varepsilon_{ab} Y^{ij}_\nu \widehat{\nu}^C_i \widehat{L}^a_j \widehat{H}_u^b+M_{R_{ij}}\widehat{\nu}^C_i\widehat{X}_j+
\frac{1}{2}\mu_{X_{ij}}\widehat{X}_i\widehat{X}_j\,.
\label{eq:SuperPotMSSMISS}
\end{equation}
$W_\mathrm{MSSM}$ is the superpotential of the MSSM given by
\begin{equation}
W_\mathrm{MSSM} = \epsilon_{ab}   Y^{ij}_u \widehat{U}_i^{C } \widehat{Q}_j^{ a} \widehat{H}^b_u 
- \epsilon_{ab}  Y^{ij}_d  \widehat{D}_i^{C } \widehat{Q}_j^{ a} \widehat{H}^b_d - 
 \epsilon_{ab} Y^{ij}_e  \widehat{E}^C_i \widehat{L}_j^a \widehat{H}^b_d + \epsilon_{ab} \mu  \widehat{H}^a_u \widehat{H}^b_d\ ,
\end{equation}
where we skipped the color indices. The corresponding soft SUSY
breaking Lagrangian is given by
\begin{align}
-\mathcal{L}^\mathrm{soft}&=-\mathcal{L}_\mathrm{MSSM}^\mathrm{soft} 
         +   \widetilde\nu^{C}_i m^2_{\widetilde \nu^C_{ij}}\widetilde\nu^{C*}_j
         + \widetilde X^{*}_i m^2_{X_{ij}} \widetilde X_j 
         \nonumber\\
      &
     + (T_{\nu}^{ij}  \varepsilon_{ab}
                 \widetilde\nu^C_i \widetilde L^a_j H_u^b +
                B_{M_R}^{ij}  \widetilde\nu^C_i \widetilde X_j 
                +\frac{1}{2}B_{\mu_X}^{ij}  \widetilde X_i \widetilde X_j
      + \widetilde X^{*}_i m^2_{X \nu^C_{ij}} \widetilde \nu^{C}_j
      +\mathrm{h.c.}) \, ,
\label{eq:softSUSY}
\end{align}
where $B_{M_R}^{ij}$ and $B_{\mu_X}^{ij}$ are the new parameters
involving the scalar partners of the sterile neutrino states. Notice
that while the former conserves lepton number, the latter violates
lepton number by two units.  Finally, ${\mathcal
  L}_\mathrm{MSSM}^\mathrm{ soft}$ collects the soft SUSY breaking
terms of the MSSM.
\begin{align}
- \mathcal{L}_\mathrm{MSSM}^\mathrm{soft} = \,\, 
& \Big(\epsilon_{ab}  T^{ij}_u \tilde{U}^{C }_i \tilde{Q}^{ a}_j H^b_u 
- \epsilon_{ab} T^{ij}_d  \tilde{D}^{C }_i \tilde{Q}^{ a}_j H^b_d - 
\epsilon_{ab} T^{ij}_e  \tilde{E}^C_i \tilde{L}^a_j H^b_d + \epsilon_{ab} B_\mu  H^a_u H^b_d + \mathrm{h.c.} \Big)  \nonumber \\
& + \frac{1}{2}\Big(M_1 \lambda_B \lambda_B + \delta_{ab} M_2 \lambda^a_W \lambda^b_W + M_3 \lambda^\alpha_G \lambda^\beta_G + \mathrm{h.c.} \Big) \nonumber \\
& + \left(\delta_{ab} (\tilde{Q}^{ a}_i)^* m_{q,ij}^2 \tilde{Q}^{ b}_j 
+ (\tilde{D}_i^{C})^* m_{d,ij}^2 \tilde{D}^{C}_j + (\tilde{U}_i^{C})^* m_{u,ij}^2 \tilde{U}^{C }_j\right) \nonumber \\
& +\tilde{E}_i^{C *} m_{e,ij}^2 \tilde{E}^C_j + \delta_{ab} (\tilde{L}_i^{a})^* m_{l,ij}^2 \tilde{L}^b_j  + m_{H_d}^2 |H_d|^2 + m_{H_u}^2 |H_u|^2 \, .
\end{align}
The neutrino mass matrix has the same form as in \eq{eq:ISSmatrix}, just
replacing $v$ by $v_u$, the vev of the up-type Higgs boson. The
mass matrices of this model are the same as in the MSSM apart from the
sneutrino sector. Neglecting for the moment being the soft-breaking
terms which lead to a splitting between the scalar and pseudoscalar
parts, the corresponding mass matrix reads
\begin{equation} 
m^2_{\tilde \nu^i} = \left( 
\begin{array}{ccc}
%m^2_{LL}
m_L^2 + \frac{1}{2} v_{u}^{2} {Y_{\nu}^{T}  Y_\nu^*} + D_L
 & -\frac{1}{\sqrt{2}} \Big(v_d \mu Y_\nu^T   - v_u T_\nu^\dagger  \Big)  &
 \frac{1}{\sqrt{2}} v_u {\Re\Big({Y_{\nu}^{T}  M_R^*}\Big)} \\ 
-\frac{1}{\sqrt{2}} \Big(v_d \mu Y_\nu^*   - v_u T_\nu  \Big) 
& m^2_{\widetilde \nu^C} + M_R  M_R^{\dagger} +  \frac{1}{2}  v_{u}^{2} Y_\nu  Y_{\nu}^{\dagger} &  - M_R  \mu_S^* \\ 
\frac{1}{\sqrt{2}} v_u {M_R^{T}  Y_\nu^*}  & - \mu_S  M_R^{\dagger}  & 
M_R^{T}  M_R^*  + m_S^2  + \mu_S  \mu_S^* \end{array} 
\right) 
 \end{equation} 
with 
\begin{align} 
D_{L} &=  - m^2_Z \cos^2 \theta_W \cos 2 \beta \, {\bf 1} \, .
\end{align} 
The complete mass matrices including the $B$-parameters as well as all
other mass matrices can be found in Appendix~\ref{app:masses}.

\section{Low energy observables}
\label{sec:observables}

The fact that the LHC has not yet seen any supersymmetric particles
\cite{Chatrchyan:2014lfa,Aad:2014wea} implies, at least in the
specific SUSY model we consider in this work, that squarks and gluinos
must be heavy.  However, it could well be that sleptons, charginos and
neutralinos are relatively light, thus having large contributions to
LFV decays.  Here we will consider the processes $\ell_\alpha \to
\ell_\beta \gamma$, $\ell_\alpha \to \ell_\beta \ell_\gamma
\ell_\delta$ and $\mu-e$ conversion in nuclei. In this section we will
present the effective low-energy lagrangian and the basic formulae for
the observables. This will also serve to fix our notation (we stay
close to the conventions of Ref.~\cite{Porod:2014xia}). The details
for the calculations of the corresponding form factors can be found in
appendices \ref{app:loopintegrals}--\ref{app:operators}.

\subsection{Effective lagrangian}

The interaction lagrangian relevant for LFV can be written as
\begin{equation} \label{eq:L-LFV}
{\cal L}_{\text{LFV}} = {\cal L}_{\ell \ell \gamma}
+ {\cal L}_{4 \ell} + {\cal L}_{2 \ell 2d}  + {\cal L}_{2 \ell 2u} \, .
\end{equation}
with
\begin{eqnarray} 
{\cal L}_{\ell \ell \gamma} &=& e \, \bar \ell_\beta \left[ \gamma^\mu \left(K_1^L P_L + K_1^R P_R \right) + i m_{\ell_\alpha} \sigma^{\mu \nu} q_\nu \left(K_2^L P_L + K_2^R P_R \right) \right] \ell_\alpha A_\mu + \hc \label{eq:L-llg} \\
{\cal L}_{4 \ell} &=& \sum_{\substack{I=S,V,T\\X,Y=L,R}} A_{XY}^I \bar \ell_\beta \Gamma_I P_X \ell_\alpha \bar \ell_\delta \Gamma_I P_Y \ell_\gamma + \hc \label{eq:L-4L} \\
{\cal L}_{2 \ell 2d} &=& \sum_{\substack{I=S,V,T\\X,Y=L,R}} B_{XY}^I \bar \ell_\beta \Gamma_I P_X \ell_\alpha \bar d_\gamma \Gamma_I P_Y d_\gamma + \hc \label{eq:L-2L2D} \\
{\cal L}_{2 \ell 2u} &=& \left. {\cal L}_{2 \ell 2d} \right|_{d \to u, \, B \to C} \label{eq:L-2L2U} \, .
\end{eqnarray}
Here $e$ is the electric charge, $q$ the 4-momenta of the photon,
$P_{L,R} =\frac{1}{2} (1 \mp \gamma_5)$ are the usual chirality
projectors and $\ell_{\alpha}$ and $d_{\alpha}$ denote the lepton and
d-quark flavors, respectively. Furthermore, we have defined $\Gamma_S
= 1$, $\Gamma_V = \gamma_\mu$ and $\Gamma_T = \sigma_{\mu \nu}$. We
omit flavor indices in the form factors for the sake of simplicity.
The underlying Feynman diagrams as well as the complete analytic
results are given in appendices \ref{app:photon}--\ref{app:operators}.

Whenever possible, we have compared the explicit analytical formulae
for the form factors with results already available in the literature.
The supersymmetric contributions to boxes, Higgs penguins, photon
penguins were found to perfectly agree
with~\cite{Arganda:2005ji,Arganda:2007jw}, while the supersymmetric
Z-penguins only differ from~\cite{Arganda:2005ji} via a constant term
as pointed out in~\cite{Krauss:2013gya}. This constant term does not
impact the result of~\cite{Arganda:2005ji} where a high-scale seesaw
mechanism is considered but it can lead to non-physical results in
low-scale seesaw models.  We have also cross-checked our calculation
of the non-SUSY boxes with~\cite{Alonso:2012ji}, confirming their
results. To the knowledge of the authors, this is the first
calculation of the non-SUSY Higgs penguins in a two Higgs doublet
Model, thus no comparison was possible.

\subsection{$\ell_\alpha \to \ell_\beta  \gamma$}

In case of the radiative decay $\ell_\alpha \to \ell_\beta \gamma$, the
corresponding decay width is given by~\cite{Hisano:1995cp}
\begin{equation}
\Gamma \left( \ell_\alpha \to \ell_\beta \gamma \right) = \frac{\alpha_{\text{em}} m_{\ell_\alpha}^5}{4} \left( |K_2^L|^2 + |K_2^R|^2 \right) \, ,
\end{equation}
where the dipole form factors $K_2^{L,R}$ are defined in
\eq{eq:L-llg}, $\alpha_{\text{em}}$ being the fine structure constant.

\subsection{$\ell_\alpha^- \to \ell_\beta^- \ell_\beta^- \ell_\beta^+$}

Next, we consider the $\ell_\alpha^-(p) \to \ell_\beta^-(p_1)
\ell_\beta^-(p_2) \ell_\beta^+(p_3)$ 3-body decays. Using the
operators in our LFV lagrangian, the decay width is given by
\begin{eqnarray}
\Gamma \left( \ell_\alpha \to 3 \, \ell_\beta \right) &=& \frac{m_{\ell_\alpha}^5}{512 \pi^3} \left[ e^4 \, \left( \left| K_2^L \right|^2 + \left| K_2^R \right|^2 \right) \left( \frac{16}{3} \log{\frac{m_{\ell_\alpha}}{m_{\ell_\beta}}} - \frac{22}{3} \right) \right. \label{L3Lwidth} \\
&+& \frac{1}{24} \left( \left| A_{LL}^S \right|^2 + \left| A_{RR}^S \right|^2 \right) + \frac{1}{12} \left( \left| A_{LR}^S \right|^2 + \left| A_{RL}^S \right|^2 \right) \nonumber \\
&+& \frac{2}{3} \left( \left| \hat A_{LL}^V \right|^2 + \left| \hat A_{RR}^V \right|^2 \right) + \frac{1}{3} \left( \left| \hat A_{LR}^V \right|^2 + \left| \hat A_{RL}^V \right|^2 \right) + 6 \left( \left| A_{LL}^T \right|^2 + \left| A_{RR}^T \right|^2 \right) \nonumber \\
&+& \frac{e^2}{3} \left( K_2^L A_{RL}^{S \ast} + K_2^R A_{LR}^{S \ast} + c.c. \right) - \frac{2 e^2}{3} \left( K_2^L \hat A_{RL}^{V \ast} + K_2^R \hat A_{LR}^{V \ast} + c.c. \right) \nonumber \\
&-& \frac{4 e^2}{3} \left( K_2^L \hat A_{RR}^{V \ast} + K_2^R \hat A_{LL}^{V \ast} + c.c. \right) \nonumber \\
&-& \left. \frac{1}{2} \left( A_{LL}^S A_{LL}^{T \ast} + A_{RR}^S A_{RR}^{T \ast} + c.c. \right) - \frac{1}{6} \left( A_{LR}^S \hat A_{LR}^{V \ast} + A_{RL}^S \hat A_{RL}^{V \ast} + c.c. \right) \right]  \nonumber \, .
\end{eqnarray}
Here we have defined
\begin{equation} \label{eq:defAV}
\hat A_{XY}^V = A_{XY}^V + e^2 K_1^X \qquad \left( X,Y = L,R \right) \, .
\end{equation}
The mass of the leptons in the final state has been neglected in this
formula, with the exception of the numerical factors that multiply the
$K_2^{L,R}$ contribution~\footnote{We note that the correct form for
  the terms proportional to $K_2^{L,R}$ was first obtained in
  Ref. \cite{Ilakovac:1994kj}.}. \eq{L3Lwidth} agrees with the one in
ref.~\cite{Arganda:2005ji}, but includes in addition $A_{LR}^S$ and
$A_{RL}^S$. In \cite{Arganda:2005ji}, these contributions were
absorbed in the corresponding vector form factors, $A_{LR}^V$ and
$A_{RL}^V$, by means of a Fierz
transformation~\cite{Arganda:private}. In contrast, $A_{LR}^S$ and
$A_{RL}^S$ were explicitly added to the set of contributing form
factors in \cite{Ilakovac:2012sh}. The relation between our
coefficients and the ones of \cite{Arganda:2005ji} is given in table
\ref{tab:relcoeffs}.
\begin{table}
\centering
{
\renewcommand{\arraystretch}{1.55}
\begin{tabular}{cc}
\hline
This paper & \cite{Arganda:2005ji} \\
\hline
$\displaystyle K_{1,2}^{L,R}$ & $\displaystyle A_{1,2}^{L,R}$ \\
$\displaystyle A_{LL}^S$ & $\displaystyle e^2 \hat B_3^L$ \\
$\displaystyle A_{RR}^S$ & $\displaystyle e^2 \hat B_3^R$ \\
$\displaystyle A_{LL}^V$ & $\displaystyle e^2 \left( \frac{1}{2} B_1^L + F_{LL}\right)$ \\
$\displaystyle A_{RR}^V$ & $\displaystyle e^2 \left( \frac{1}{2} B_1^R + F_{RR} \right)$ \\
$\displaystyle A_{LR}^V - \frac{1}{2} A_{LR}^S$ & $\displaystyle e^2 \left( \hat B_2^L + F_{LR} \right)$ \\
$\displaystyle A_{RL}^V - \frac{1}{2} A_{RL}^S$ & $\displaystyle e^2 \left( \hat B_2^R + F_{RL} \right)$ \\
$\displaystyle A_{LL}^T$ & $\displaystyle e^2 B_4^L$ \\
$\displaystyle A_{RR}^T$ & $\displaystyle e^2 B_4^R$ \\
\hline
\end{tabular}
}
\caption{Relation between the form factors defined in this
  paper and the ones in \cite{Arganda:2005ji}. Here $\hat{B}_2^{L,R} =
  B_2^{L,R} + B_{2, \rm Higgs}^{L,R}$ and $\hat{B}_3^{L,R} = B_3^{L,R}
  + B_{3, \rm Higgs}^{L,R}$, and $F_{XY} = \frac{F_X E^Y}{e^2 \,
    m_Z^2}$, with $E^L$ and $E^R$ the tree-level $Z$-boson couplings
  to a pair of charged leptons (see appendix \ref{app:vertices}).}
\label{tab:relcoeffs}
\end{table}

\subsection{$\ell_\alpha^- \to \ell_\beta^- \ell_\gamma^- \ell_\gamma^+$}

We consider the $\ell_\alpha^-(p) \to \ell_\beta^-(p_1)
\ell_\gamma^-(p_2) \ell_\gamma^+(p_3)$ 3-body decays, with $\beta \ne
\gamma$. The decay width is given by
\begin{eqnarray}
\Gamma \left( \ell_\alpha^- \to \ell_\beta^- \ell_\gamma^- \ell_\gamma^+ \right) &=& \frac{m_{\ell_\alpha}^5}{512 \pi^3} \left[ e^4 \, \left( \left| K_2^L \right|^2 + \left| K_2^R \right|^2 \right) \left( \frac{16}{3} \log{\frac{m_{\ell_\alpha}}{m_{\ell_\gamma}}} - 8 \right) \right. \label{L3Lwidth2} \\
&+& \frac{1}{12} \left( \left| A_{LL}^S \right|^2 + \left| A_{RR}^S \right|^2 \right) + \frac{1}{12} \left( \left| A_{LR}^S \right|^2 + \left| A_{RL}^S \right|^2 \right) \nonumber \\
&+& \frac{1}{3} \left( \left| \hat A_{LL}^V \right|^2 + \left| \hat A_{RR}^V \right|^2 \right) + \frac{1}{3} \left( \left| \hat A_{LR}^V \right|^2 + \left| \hat A_{RL}^V \right|^2 \right) + 4 \left( \left| A_{LL}^T \right|^2 + \left| A_{RR}^T \right|^2 \right) \nonumber \\
&-& \left. \frac{2 e^2}{3} \left( K_2^L \hat A_{RL}^{V \ast} + K_2^R \hat A_{LR}^{V \ast} + K_2^L \hat A_{RR}^{V \ast} + K_2^R \hat A_{LL}^{V \ast} + c.c. \right) \right]  \nonumber \, .
\end{eqnarray}
Here we have used the same definition as in
\eq{eq:defAV}. Furthermore, as for $\ell_\alpha \to 3 \, \ell_\beta$,
the mass of the leptons in the final state has been neglected in the
decay width formula, with the exception of the dipole terms
$K_2^{L,R}$.

Finally, we note that Eqs. \eqref{L3Lwidth} and \eqref{L3Lwidth2} are
in perfect agreement with the expressions given in
Ref. \cite{Ilakovac:2012sh}.

\subsection{$\ell_\alpha^- \to \ell_\beta^+ \ell_\gamma^- \ell_\gamma^-$}

Finally, we consider the $\ell_\alpha^-(p) \to \ell_\beta^+(p_1)
\ell_\gamma^-(p_2) \ell_\gamma^-(p_3)$ 3-body decays, with $\beta \ne
\gamma$. The decay width is given by
\begin{eqnarray}
\Gamma \left( \ell_\alpha^- \to \ell_\beta^+ \ell_\gamma^- \ell_\gamma^- \right) &=& \frac{m_{\ell_\alpha}^5}{512 \pi^3} \left[ \frac{1}{24} \left( \left| A_{LL}^S \right|^2 + \left| A_{RR}^S \right|^2 \right) + \frac{1}{12} \left( \left| A_{LR}^S \right|^2 + \left| A_{RL}^S \right|^2 \right)  \right. \label{L3Lwidth3} \\
&+& \frac{2}{3} \left( \left| \hat A_{LL}^V \right|^2 + \left| \hat A_{RR}^V \right|^2 \right) + \frac{1}{3} \left( \left| \hat A_{LR}^V \right|^2 + \left| \hat A_{RL}^V \right|^2 \right) + 6 \left( \left| A_{LL}^T \right|^2 + \left| A_{RR}^T \right|^2 \right) \nonumber \\
&-& \left. \frac{1}{2} \left( A_{LL}^S A_{LL}^{T \ast} + A_{RR}^S A_{RR}^{T \ast} + c.c. \right) - \frac{1}{6} \left( A_{LR}^S \hat A_{LR}^{V \ast} + A_{RL}^S \hat A_{RL}^{V \ast} + c.c. \right) \right]  \nonumber \, .
\end{eqnarray}
The same definitions and conventions as in the previous two
observables have been used. Notice that this process does not receive
contributions from penguin diagrams, but only from boxes.

\subsection{Coherent $\mu-e$ conversion in nuclei}

We now turn to the discussion of $\mu-e$ conversion in nuclei, which
will follow the conventions and approximations described in
Ref.~\cite{Kuno:1999jp,Arganda:2007jw} (see also
\cite{Vergados:1985pq,Bernabeu:1993ta,Faessler:1999jf} for detailed
works regarding the effective lagrangian at the nucleon level,
\cite{Kitano:2002mt,Alonso:2012ji} for a calculation including the
effects of the atomic electric field and \cite{Crivellin:2014cta} for
recent improvements on the hadronic uncertainties). The conversion
rate, relative to the the muon capture rate, can be expressed as
\begin{align}
{\rm CR} (\mu- e, {\rm Nucleus}) &= 
\frac{p_e \, E_e \, m_\mu^3 \, G_F^2 \, \alpha_{\text{em}}^3 
\, Z_{\rm eff}^4 \, F_p^2}{8 \, \pi^2 \, Z}  \nonumber \\
&\times \left\{ \left| (Z + N) \left( g_{LV}^{(0)} + g_{LS}^{(0)} \right) + 
(Z - N) \left( g_{LV}^{(1)} + g_{LS}^{(1)} \right) \right|^2 + 
\right. \nonumber \\
& \ \ \ 
 \ \left. \,\, \left| (Z + N) \left( g_{RV}^{(0)} + g_{RS}^{(0)} \right) + 
(Z - N) \left( g_{RV}^{(1)} + g_{RS}^{(1)} \right) \right|^2 \right\} 
\frac{1}{\Gamma_{\rm capt}}\,.
\end{align}   
$Z$ and $N$ are the number of protons and neutrons in the nucleus and
$Z_{\rm eff}$ is the effective atomic charge~\cite{Chiang:1993xz}.
Similarly, $G_F$ is the Fermi constant, $F_p$ is the nuclear matrix
element and $\Gamma_{\rm capt}$ represents the total muon capture
rate. $p_e$ and $E_e$ ($\simeq m_\mu$ in our numerical evaluation) are
the momentum and energy of the electron and $m_\mu$ is the muon mass.
In the above, $g_{XK}^{(0)}$ and $g_{XK}^{(1)}$ (with $X = L, R$ and
$K = S, V$) can be written in terms of effective couplings at the
quark level as
\begin{align}
g_{XK}^{(0)} &= \frac{1}{2} \sum_{q = u,d,s} \left( g_{XK(q)} G_K^{(q,p)} +
g_{XK(q)} G_K^{(q,n)} \right)\,, \nonumber \\
g_{XK}^{(1)} &= \frac{1}{2} \sum_{q = u,d,s} \left( g_{XK(q)} G_K^{(q,p)} - 
g_{XK(q)} G_K^{(q,n)} \right)\,.
\end{align}
For coherent $\mu-e$ conversion in nuclei, only scalar ($S$) and
vector ($V$) couplings contribute~\cite{Kuno:1999jp}. Furthermore,
sizable contributions are expected only from the $u,d,s$ quark
flavors.  The numerical values of the relevant $G_K$ factors
are~\cite{Kuno:1999jp,Kosmas:2001mv}
\begin{align}
&
G_V^{(u, p)}\, =\, G_V^{(d, n)\,} =\, 2 \,;\, \ \ \ \ 
G_V^{(d, p)}\, =\, G_V^{(u, n)}\, = 1\,; \nonumber \\
&
G_S^{(u, p)}\, =\, G_S^{(d, n)}\, =\, 5.1\,;\, \ \ 
G_S^{(d, p)}\, =\, G_S^{(u, n)}\, = \,4.3 \,;\, \nonumber \\
&
G_S^{(s, p)}\,=\, G_S^{(s, n)}\, = \,2.5\,.
\end{align}
Finally, the $g_{XK(q)}$ coefficients can be written in terms of the
form factors in Eqs.\eqref{eq:L-llg}, \eqref{eq:L-2L2D} and
\eqref{eq:L-2L2U} as
\begin{eqnarray}
g_{LV(q)} &=& \frac{\sqrt{2}}{G_F} \left[ e^2 Q_q \left( K_1^L - K_2^R \right)- \frac{1}{2} \left( C_{\ell\ell qq}^{VLL} + C_{\ell\ell qq}^{VLR} \right) \right] \\
g_{RV(q)} &=& \left. g_{LV(q)} \right|_{L \to R} \\ 
g_{LS(q)} &=& - \frac{\sqrt{2}}{G_F} \frac{1}{2} \left( C_{\ell\ell qq}^{SLL} + C_{\ell\ell qq}^{SLR} \right) \\
g_{RS(q)} &=& \left. g_{LS(q)} \right|_{L \to R} \, .
\end{eqnarray}
Here $Q_q$ is the quark electric charge ($Q_d = -1/3$, $Q_u = 2/3$)
and $C_{\ell\ell qq}^{IXK} = B_{XY}^K \, \left( C_{XY}^K \right)$ for
d-quarks (u-quarks), with $X = L, R$ and $K = S, V$.

\section{Results}
\label{sec:numerics}

\subsection{Numerical setup}

For the numerical examples we have implemented the model in the
Mathematica package \SARAH
\cite{Staub:2008uz,Staub:2009bi,Staub:2010jh,Staub:2012pb,Staub:2013tta},
which creates the required modules for \SPheno
\cite{Porod:2003um,Porod:2011nf} to calculate the masses and mixing
matrices including the complete 1-loop corrections. In the Higgs
sector we include in addition the known 2-loop corrections to the
Higgs mass from the MSSM
\cite{Brignole:2001jy,Degrassi:2001yf,Brignole:2002bz,Dedes:2002dy,Dedes:2003km,Allanach:2004rh}.
However, this does not include 2-loop corrections stemming from the
extended neutrino and sneutrino sectors, where we can have sizable
Yukawa couplings. Moreover, \SARAH calculates also the full 2-loop
RGEs including the entire flavor structure for the model, which we
have summarized in Appendix~\ref{app:rges}.  This will be of great
importance in our numerical studies, as we use CMSSM-like boundary
conditions, see below for their definition.  In the flavor observables
we include all possible contributions. These are calculated using the
\FlavorKit interface \cite{Porod:2014xia}.  In the context of this
project we have extended the lists of observables implemented in
\FlavorKit by $\ell^-_\alpha \to \ell_\beta^- \ell_\beta^+
\ell^-_\gamma$ and $\ell^-_\alpha \to \ell_\beta^- \ell_\beta^-
\ell^+_\gamma$.

\begin{table}[hbt]
\centering
\begin{tabular}{|c|c||c|c|} \hline
 $\alpha_{em}^{-1}$ & 127.92783 & $G_\mu$ & $1.11639 \cdot 10^{-5} \text{GeV}^{-2}$ \\
 $\alpha_S$ & 0.11720 & $M_Z$ & 91.18760 GeV \\
 $m_b(m_b)$ & 4.2 GeV & $m_t$ & 172.9 GeV \\
 $m_\tau $ & 1.777 GeV &  &\\
 \hline 
\end{tabular}
\caption{Input values for the SM parameters taken at $M_Z$ unless
  otherwise specified.}
\label{tab:SMinput}
\end{table}

The numerical evaluation of each parameter point is performed as
follows: the $Y_\nu$ Yukawa couplings are
calculated using a modified Casas-Ibarra parameterization
\cite{Casas:2001sr}, properly adapted for the inverse seesaw
\cite{Basso:2012ew,Abada:2012mc} (and fixing $M_R=2$~TeV,
$\mu_X=10^{-5}$~GeV and the lightest neutrino mass
$m_{\nu_1}=10^{-4}$~eV):
\begin{equation}
Y_\nu = \frac{\sqrt{2}}{v_u} V^\dagger D_{\sqrt{X}} R D_{\sqrt{m_\nu}} U_{\rm{PMNS}}^\dagger\,.
\label{eq:casas_ibarra}
\end{equation}
   Here
$D_{\sqrt{m_\nu}} = {\rm{diag}}(\sqrt{m_{\nu_i}})$, $D_{\sqrt{X}} =
   {\rm{diag}} (\sqrt{\hat X_i})$, $\hat X_i$ being the eigenvalues of
   $X = M_R \mu_X^{-1} M_R^T$, and $V$ is the matrix that diagonalizes
   $X$ as $V X V^T = \hat X$. Furthermore, we parameterize the complex
   orthogonal $R$ matrix as
\begin{align}
R &=  \begin{pmatrix}
1 & 0 & 0 \\
0 & \cos \theta^R_{23} & \sin \theta^R_{23} \\
0 & - \sin \theta^R_{23} & \cos \theta^R_{23}
\end{pmatrix} \begin{pmatrix}
\cos \theta^R_{13} & 0 & \sin \theta^R_{13} \\
0 & 1 & 0 \\
- \sin \theta^R_{13} & 0 & \cos \theta^R_{13}
\end{pmatrix} \begin{pmatrix}
\cos \theta^R_{12} & \sin \theta^R_{12} & 0 \\
- \sin \theta^R_{12} & \cos \theta^R_{12} & 0 \\
0 & 0 & 1
\end{pmatrix}.
\label{eq:Rmatrix}
\end{align}
Below we will set $R$ to the unit matrix except when stated
otherwise. We use the best-fit values for the neutrino oscillation
parameters as given in \cite{Forero:2014bxa}:
\begin{align}
\Delta m^2_{21} &= 7.60 \cdot 10^{-5}~\rm{eV}^2\,,~
\Delta m^2_{31} = 2.48 \cdot 10^{-3}~\rm{eV}^2\,, \nonumber \\
\sin^2\theta_{12} &= 0.323\,,~
\sin^2\theta_{23} = 0.467\,,~
\sin^2\theta_{13} = 0.0234\,.
\end{align}
which are close to the ones obtained in
\cite{Tortola:2012te,GonzalezGarcia:2012sz,Capozzi:2013csa}.  We make
use in our scans of the values
\begin{equation}
Y_\nu =  f \cdot 10^{-2} \cdot \begin{pmatrix}
0.0956 & -0.0589 & 0.0348 \\
0.616 & 0.594 & -0.687 \\
0.404 & 1.78 & 1.91
\end{pmatrix}
\label{eq:Ynu_ref}
\end{equation}
fixed with $f=1$ even if we vary $M_R$. This is because one can always
adjust $\mu_X$ to fulfill neutrino oscillation data without affecting
any of our observables.

\SPheno derives the SM gauge and Yukawa couplings at $M_Z$ where we
take the masses and couplings given in table~\ref{tab:SMinput} as
input.  2-loop RGEs for the dimensionless parameters are then used to
evaluate these couplings at $M_{GUT}$, defined by the requirement $g_1
= g_2$, where $g_1$ and $g_2$ are the couplings for the $U(1)_Y$ and
$SU(2)_L$ gauge groups, respectively. At $M_{GUT}$ the CMSSM boundary
conditions are applied
\begin{eqnarray*}
& m_{\nu^C}^2 = m^2_X = m_l^2 = m_e^2 = m_q^2  = m_d^2 = m_u^2 \equiv m_0^2 \, {\bf 1} & \\
& m_{H_d}^2 = m_{H_u}^2 \equiv m_0^2 & \\
& M_1 = M_2 = M_3  \equiv M_{1/2} & \\
& T_i \equiv A_0 Y_i  \hspace{1cm} i=e,d,u,\nu
\end{eqnarray*}
The mixed soft-term $m_{X \nu^C}^2$ is set to zero at the GUT scale
and is not generated via RGE effects.  Moreover, the phase of $\mu$,
which is an RGE invariant, is given as input. The ratio of the Higgs
vevs, $\tan\beta =\frac{v_u}{v_d}$, completes the list of input
parameters.  Then 2-loop RGEs are used to evolve these parameters to
$Q_{\rm{EWSB}} = \sqrt{\tilde t_1 \tilde t_2}$.  The numerical values
for superpotential terms $M_R$ and $\mu_X$, as well as for their
corresponding soft terms $B_{M_R}$ and $B_{\mu_X}$, are used as input
at the SUSY scale. $B_\mu$ and $|\mu|$ are obtained as usual from the
minimization conditions of the vacuum\footnote{In principle one could
  require that all $B$-parameters are proportional to each other,
  e.g.\ $B_\mu : B_{M_R} : B_{\mu_X} = \mu : M_R : \mu_X$. However, as
  their actual value does not have any significant impact as long as
  this ratio is fulfilled up to a factor 2-3 we fix for simplicity
  $B_{M_R}$ and $B_{\mu_X}$.}.
At $M_{SUSY}$ the 1-loop corrected masses are calculated before the
RGEs run down to $M_Z$ to re-calculate gauge and Yukawa couplings
using the new SUSY corrections.  These steps are iterated until the
mass spectrum has converged with a numerical precision of
$10^{-4}$. Afterwards, \SPheno runs the RGEs to $Q=160$~GeV for the
calculation of the operators which contribute to quark flavor
violating observables and to $Q=M_Z$ for the calculation of the
operators needed for lepton flavor violating observables. These
operators are then combined to compute the different observables using
$\alpha(0)$, which includes to a large extent the effects from running
the operators between $M_Z$ and the energy scale where the LFV
processes take place (usually given by the mass of the decaying
particle).

\subsection{Numerical results}
\begin{figure}[t]
\centering
\includegraphics[width=0.49 \linewidth]{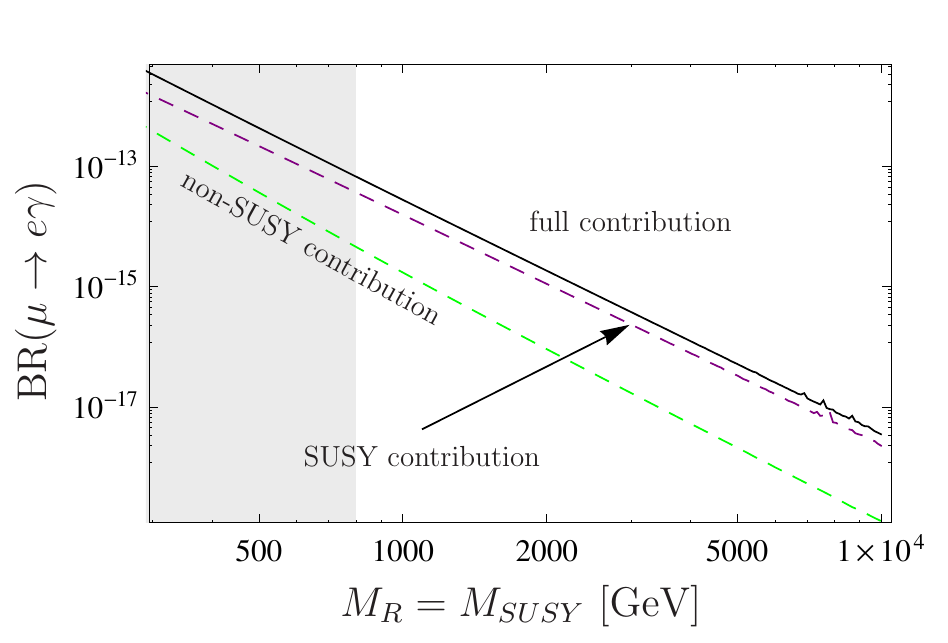}
\includegraphics[width=0.49 \linewidth]{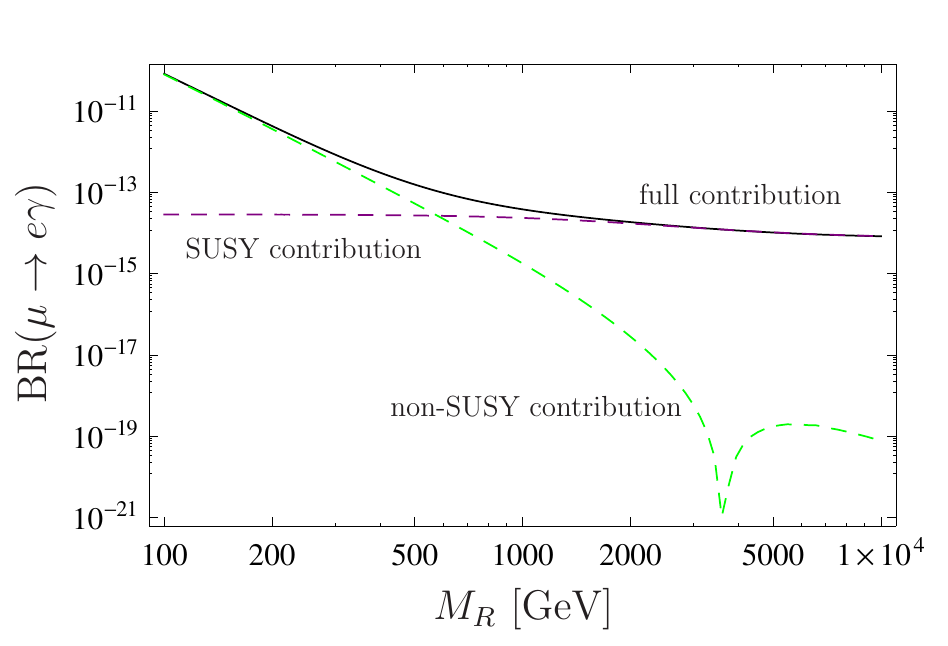} \\
\includegraphics[width=0.49 \linewidth]{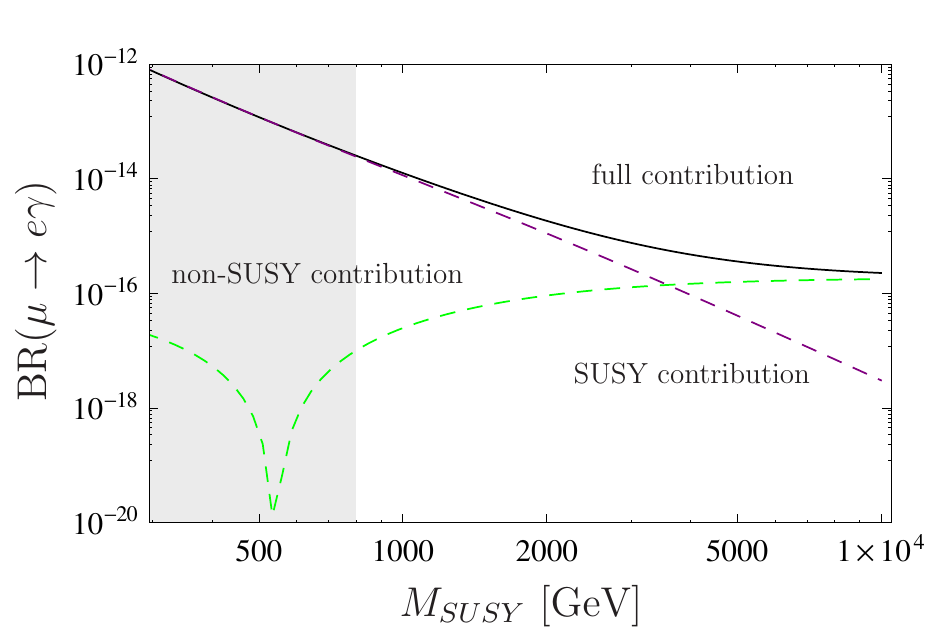}
\caption{BR($\mu\to e \gamma$) as a function of $M_{SUSY}$ and $M_R$.
  The other parameters are given in the text. The gray area roughly
  corresponds to the parameter space excluded by the LHC experiments.}
\label{fig:MuEgamma}
\end{figure}
\begin{table}[b]
\begin{center}
\begin{tabular}{|c|c||c|c|} \hline
$m_0$ & 1 TeV & $M_{1/2}$ & 1 TeV \\
$A_0$ & -1.5 TeV & $M_R$ & 2 TeV \\
$B_{\mu_X}$ & $100 \, \mu_X$ & $B_{M_R}$ & $100 \, M_R$ \\
$\tan \beta$ &   10 & sign$(\mu)$ & + \\ \hline
\end{tabular}
\end{center}
\caption{Standard values for the various parameters. $M_R$ and $\mu_X$
  are taken proportional to the unit matrix.}
\label{tab:SUSYinput} 
\end{table}
We will use the parameter values given in table~\ref{tab:SUSYinput} as
starting point for our numerical computations unless stated otherwise.
A variation of the soft SUSY parameters is denoted by a variation of
$M_{SUSY}$, which actually implies a variation of three parameters at
the same time $M_{SUSY} = m_0 = M_{1/2} = -A_0$. For completeness, we
note that fixing the ratio $m_0/A_0$ usually gives a Higgs boson mass,
$m_h$, that does not agree with the ATLAS and CMS measurements.
Nevertheless, we emphasize that (1) our results depend only weakly on
the value of $A_0$, and (2) contributions mediated by $h$ itself are
subdominant. Therefore, the actual Higgs boson mass is of little
importance for our investigations here.

We start with the discussion of $\mu$ decays as the bounds are
strongest in this case.  In \fig{fig:MuEgamma} we show the dependence
of BR$(\mu\to e \gamma)$ on $M_R$ and $M_{SUSY}$ as well as the
individual dependence of the SUSY and non-SUSY contributions. The
latter consist of $\nu$-$W^\pm$ and the $\nu$-$H^\pm$
contributions. There are two particular features: (i) if $M_R
=M_{SUSY}$ the SUSY contributions are more important than the non-SUSY
ones and the relative importance of the SUSY contributions increases
with the scale. The reason for the latter is that the mixing between
light and heavy neutrinos decreases like $\sim m_D / M_R$, whereas the
mixing in the sneutrino sector decreases only logarithmically with the
scale.  (ii) The non-SUSY contributions can flip its sign. This is due
to a sign-difference between the $\nu$-$H^\pm$ and the $\nu$-$W^\pm$
contributions to the coefficients $K_2^{L,R}$. This is in contrast to
the analogous decay in the quark sector, $b\to s\gamma$, where the
$W^\pm$- and $H^\pm$-contributions have always the same sign. The
reason for this difference can be found in
Eqs.~(\ref{eq:Hphoton1})--(\ref{eq:Wphoton2}), presented in
appendix~\ref{app:photon}, where the light neutrino masses appear
instead of the mass of the heavy $t$-quark.  We have checked
explicitly, both numerically and analytically, that we recover the
$b\to s \gamma$ result if we replace the corresponding masses and
Yukawa couplings.  Finally, we stress that the scalar masses are
functions of $M_{SUSY}$, which explains why also the non-SUSY
contribution actually depends on the SUSY scale.
With our specific structure of the $Y_\nu$ matrices we find that $M_R$
has to be larger than $M_{SUSY}$ for the sign flip to occur, which is
also the reason why we do not observe it in case of $M_R=M_{SUSY}$.
The grey area corresponds to the part of the parameter space which is
excluded in the CMSSM by the most recent ATLAS results
\cite{Aad:2014wea}. However, we want to stress that even though the
squark and gluino masses are essentially the same in our model as in
the CMSSM, the cascade decays can be quite different due to (i) the
enlarged sneutrino sector with additional light states and (ii) the
different slepton masses. Thus, this is a rather conservative bound.

\begin{figure}[t]
\includegraphics[width=0.49 \linewidth]{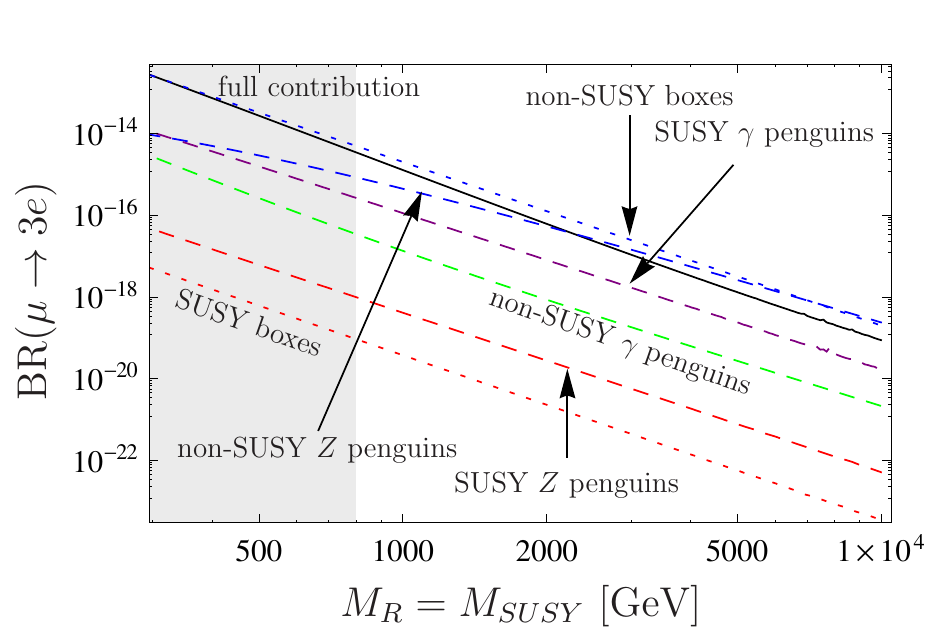}
\includegraphics[width=0.49 \linewidth]{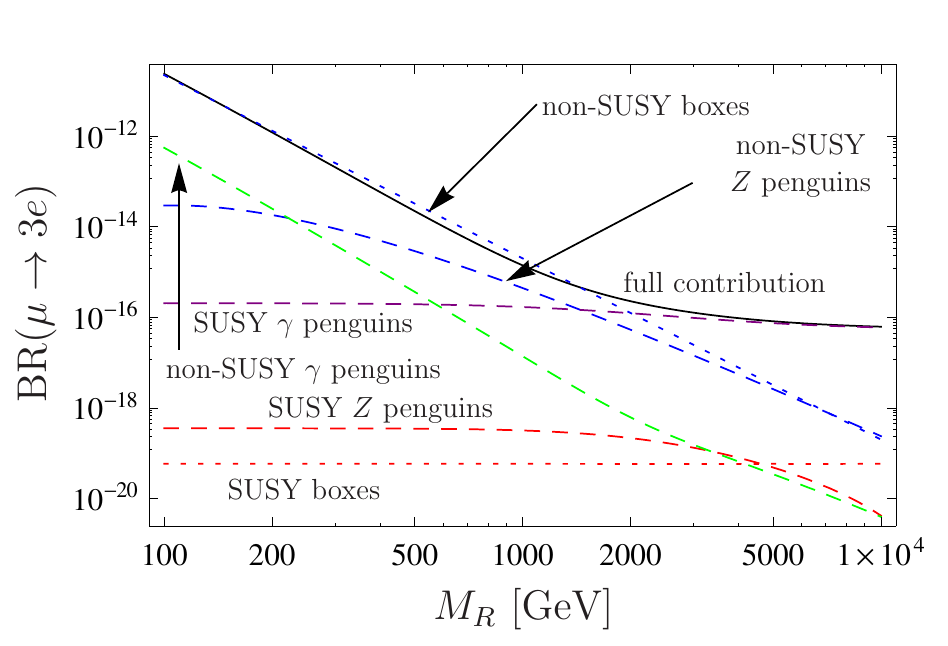} \\
\includegraphics[width=0.49 \linewidth]{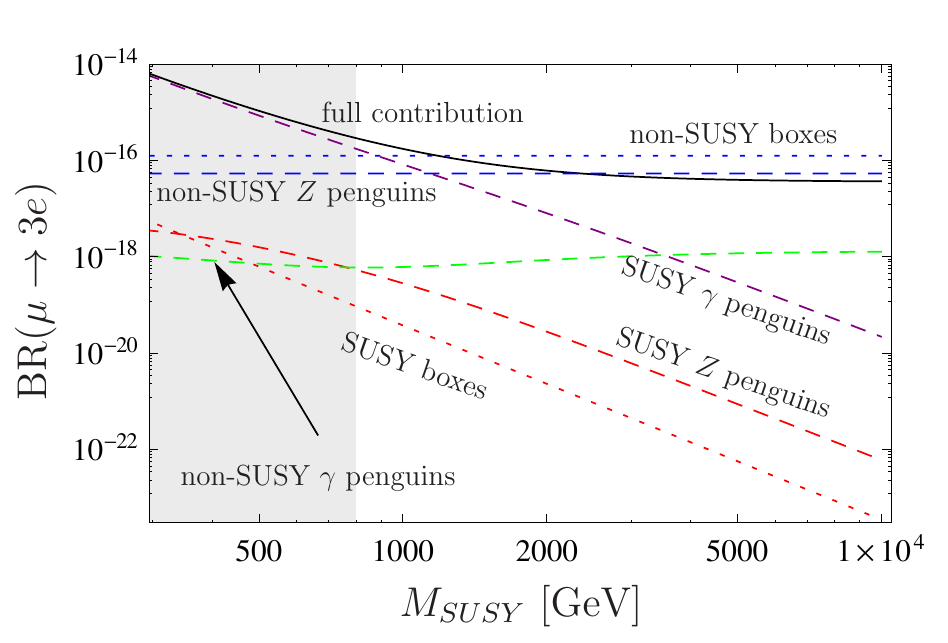}
\includegraphics[width=0.49 \linewidth]{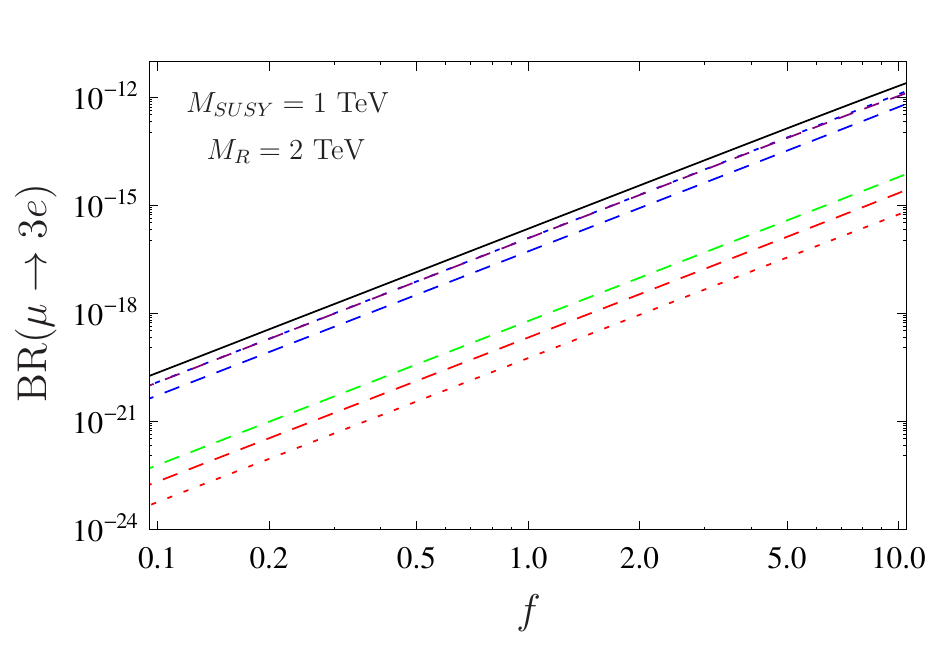}
\caption{BR($\mu\to 3 e$) as a function of $M_{SUSY}$, $M_R$ and an
  overall scaling parameter $f$ for $Y_\nu$.  The other parameters are
  given in the text. The gray area roughly corresponds to the
  parameter space excluded by the LHC experiments.}
\label{fig:Mu3E}
\end{figure}

In \fig{fig:Mu3E} we display our results for the branching ratio
BR($\mu \to 3 \, e$) as well as the various contributions to this
decay. Here we find that for the case $M_R=M_{SUSY}$ the non-SUSY
boxes dominate. This fact was first noted in \cite{Ilakovac:2009jf}
and later confirmed by
\cite{Alonso:2012ji,Dinh:2012bp,Ilakovac:2012sh}. 
Note that this does not depend on
the overall strength of the $Y_\nu$ couplings, which we rescale as
$Y_\nu \to f \, Y_\nu$. This can be seen from the lower right plot:
all contributions scale in the same way.  However, the situation can
change in principle if one allows for additional flavor violation in
the soft SUSY breaking parameters. Note that the sign-flip induced by
the $H^\pm$ contributions is not as pronounced as in the case of $\mu\to
e \gamma$, where it led to a change of the overall sign, as the
different contributions to the off-shell photon appear with different
weights. However, it is the reason for the observed kink in the
non-SUSY $\gamma$-penguin.  We also observe that we have negative
interference between non-SUSY $Z$-penguins and the corresponding box
contributions.  In particular, for larger values of $M_R$ this can
reduce BR($\mu \to 3 e$) by up to an order of magnitude.  Since this
is precisely the region which will be probed by future experiments,
the possible appearance of these cancellations has to be taken into
account in order to interpret the experimental results properly.

\begin{figure}[t]
\includegraphics[width=0.49 \linewidth]{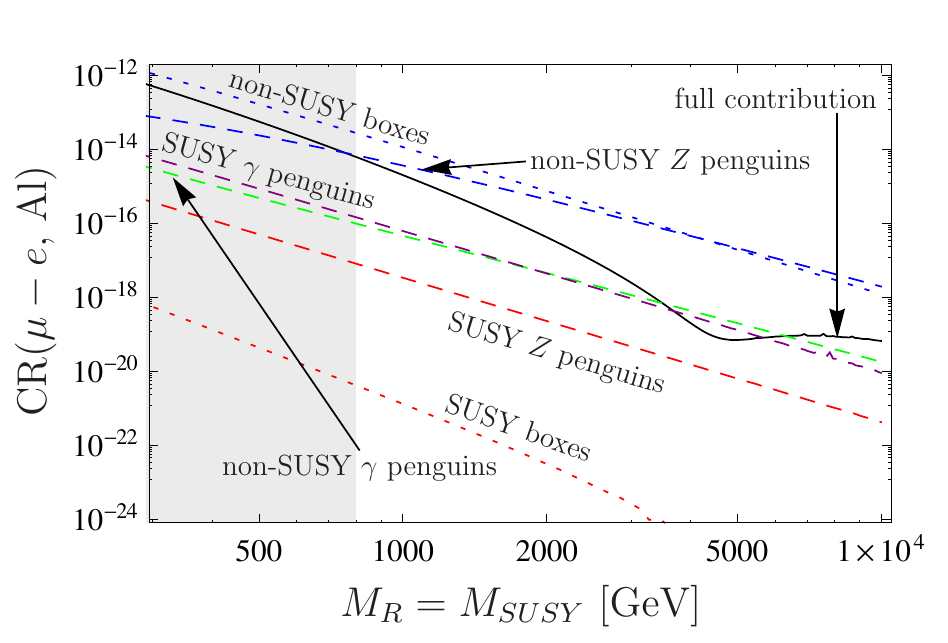}
\includegraphics[width=0.49 \linewidth]{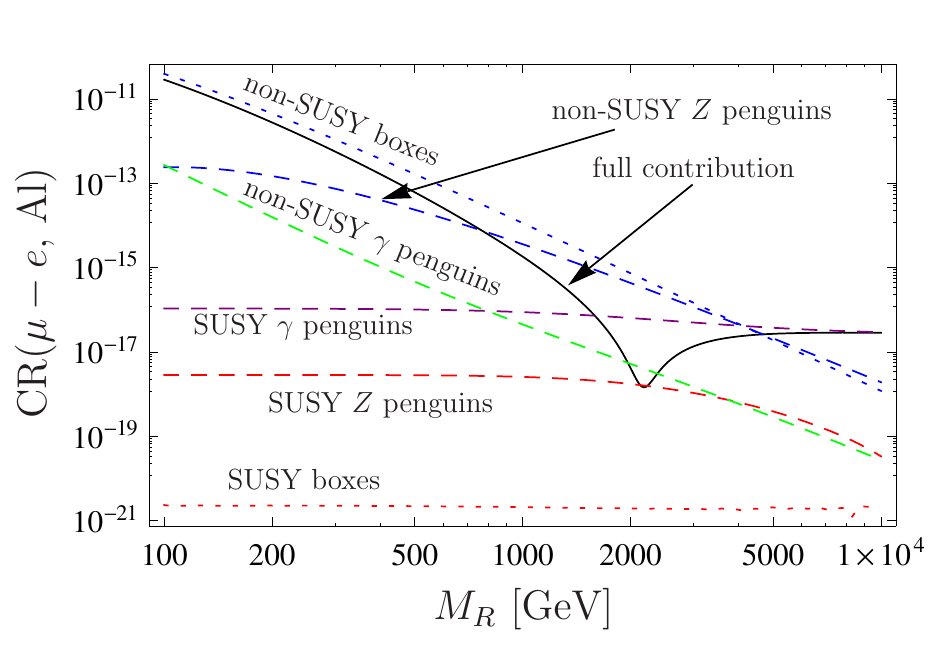} \\
\includegraphics[width=0.49 \linewidth]{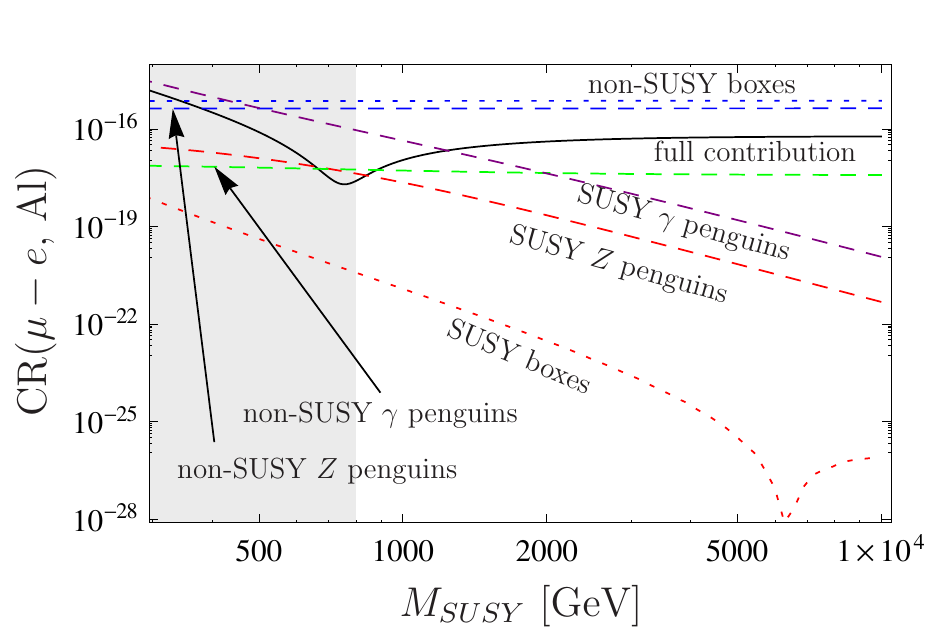}
\includegraphics[width=0.49 \linewidth]{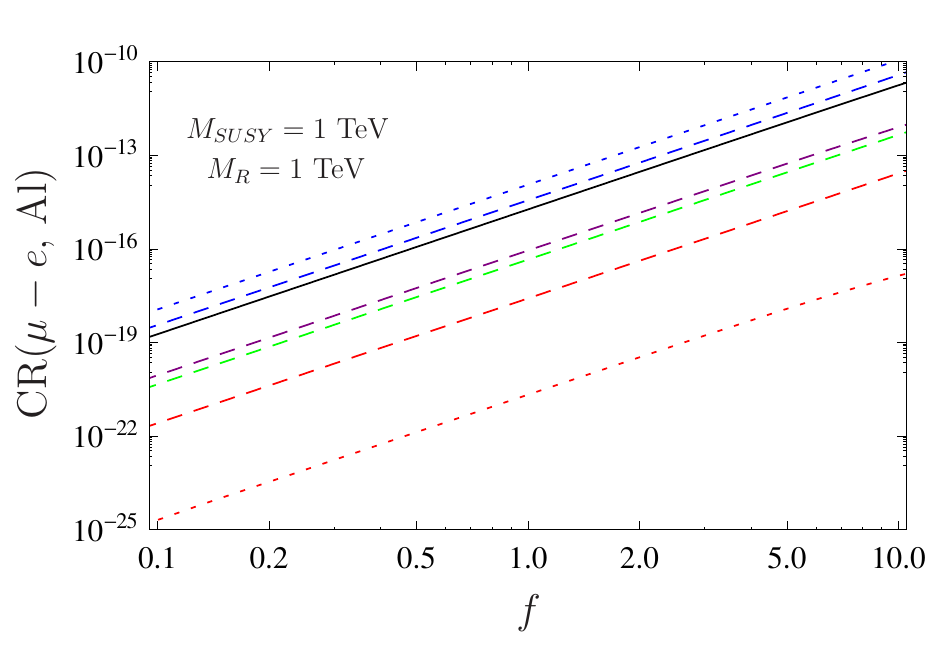}
\caption{$\mu-e$ conversion on $\mathrm{Al}$ as a function of
  $M_{SUSY}$, $M_R$ and the scaling parameter $f$ for $Y_\nu$. The
  gray area roughly corresponds to the parameter space excluded by the
  LHC experiments.}
\label{fig:MuEconversion}
\end{figure}

\begin{figure}[t]
\centering
\includegraphics[width=0.49 \linewidth]{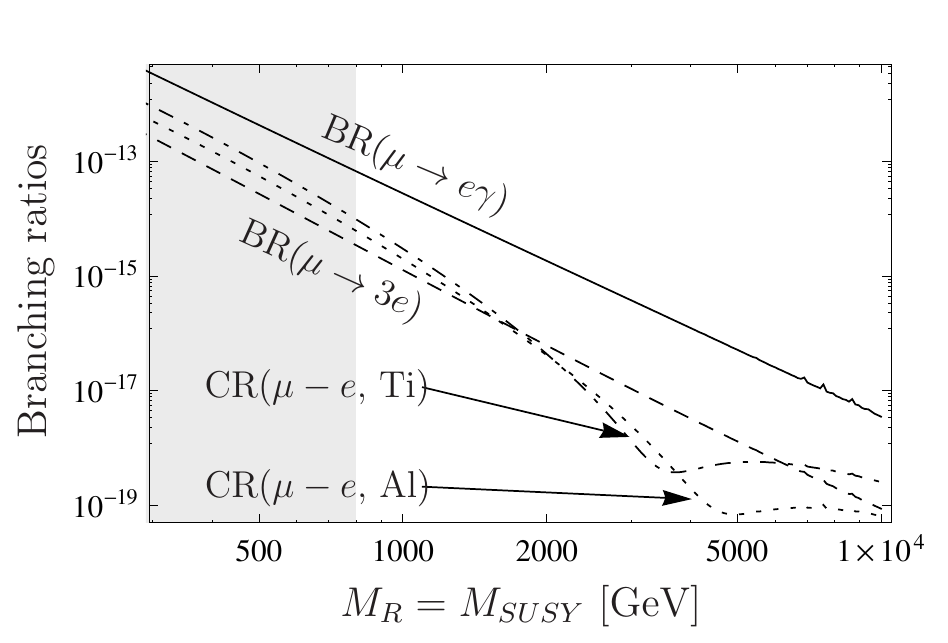}
\includegraphics[width=0.49 \linewidth]{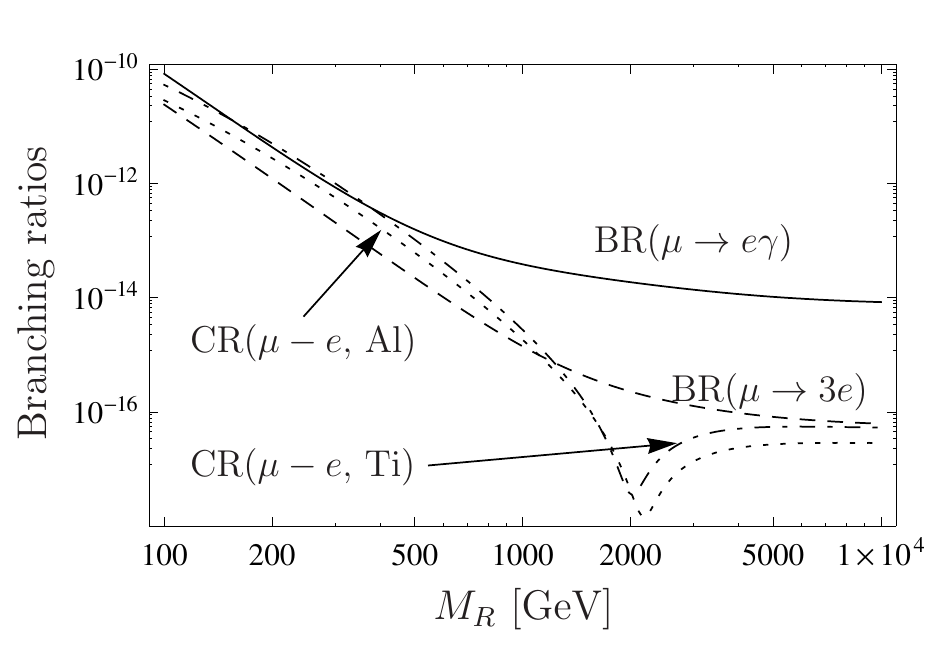} \\
\includegraphics[width=0.49 \linewidth]{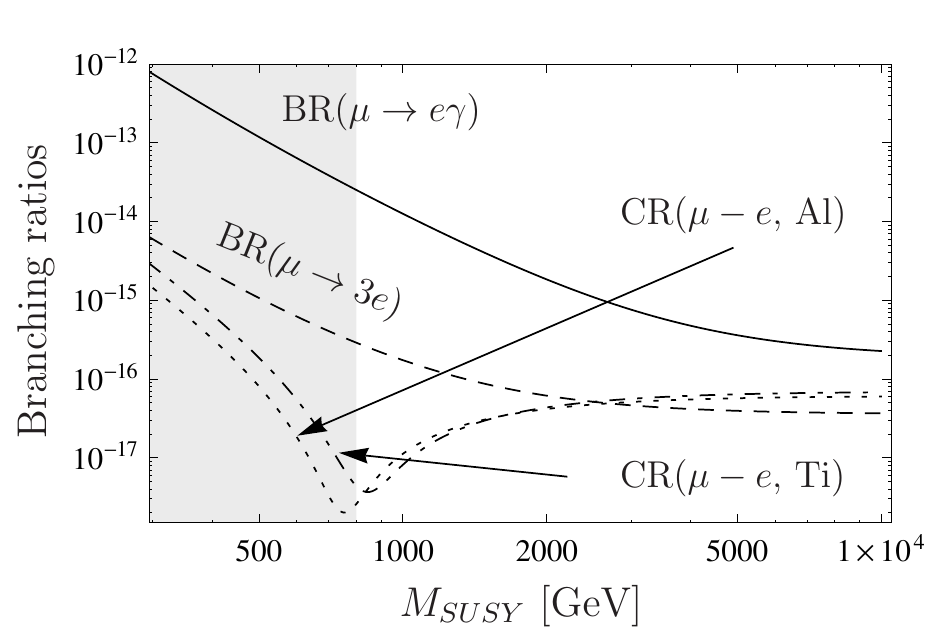}
\caption{BR($\mu\to e \gamma$), BR($\mu\to 3 e$), $\mu-e$ conversion
  in $\mathrm{Ti}$ and $\mathrm{Al}$ as a function of $M_R$ and $M_{SUSY}$. The gray
  area roughly corresponds to the parameter space excluded by the LHC
  experiments.}
\label{fig:various}
\end{figure}

Similar features appear in case of $\mu-e$ conversion in nuclei, as
exemplified for the case of an aluminium ($\mathrm{Al}$) nucleus in
\fig{fig:MuEconversion}. The main difference is that there is a large
part of parameter space where a pronounced negative interference
between the non-SUSY $Z$-penguin and the corresponding box
contributions can occur.  Note that with the expected sensitivity of
$10^{-18}$ one can probe $Y_\nu$ couplings down to a few $\times
10^{-6}$ for $M_R=M_{SUSY}=1$~TeV or, equivalently, to a mass scale of
about $5$ TeV in case of $Y_\nu$ as given in
Eq.~(\ref{eq:Ynu_ref}). As we found for the 3-body decays, for higher
mass scales the non-SUSY $Z$-penguins can be as important as the
corresponding box-diagrams. The overall features are essentially
element independent as can be seen in \fig{fig:various} where we show
all three observables discussed so far together and include also
$\mu-e$ conversion in titanium ($\mathrm{Ti}$).  In case $M_R \simeq M_{SUSY}$,
we find that $\mu-e$ conversion in nuclei is the most stringent LFV
observable in our model.

\begin{figure}[t]
\includegraphics[width=0.49 \linewidth]{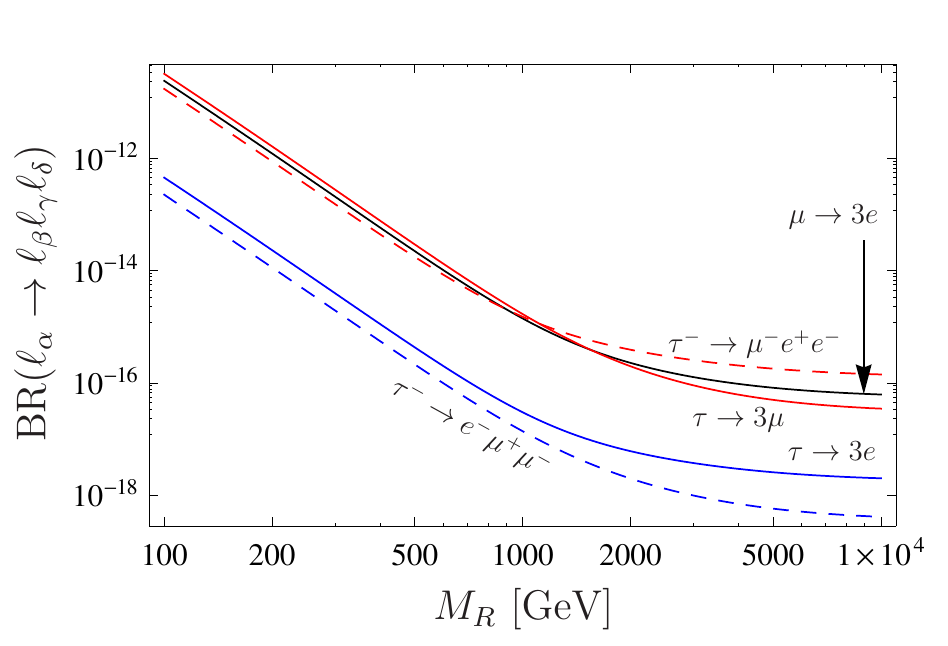}
\includegraphics[width=0.49 \linewidth]{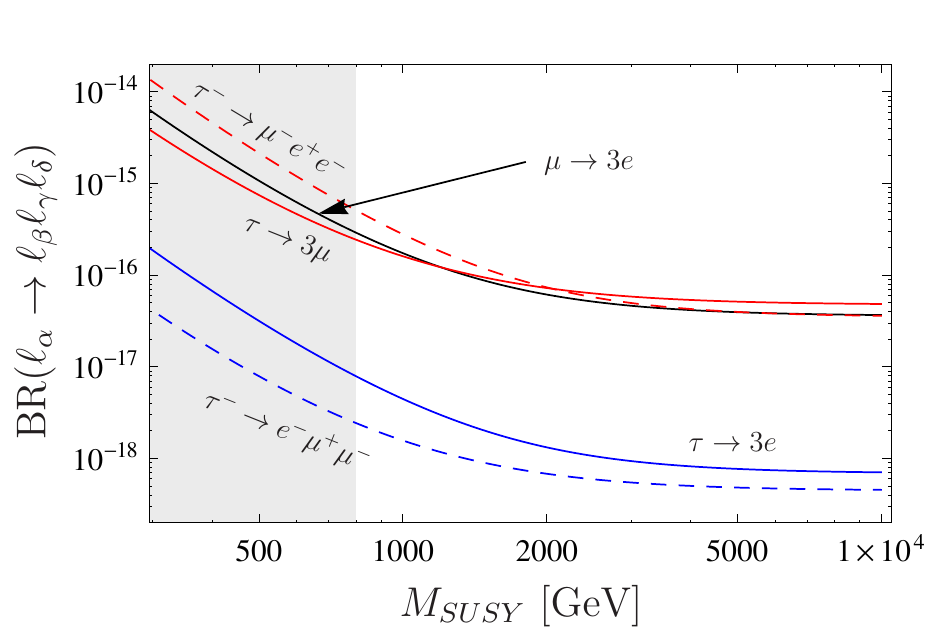} \\
\includegraphics[width=0.49 \linewidth]{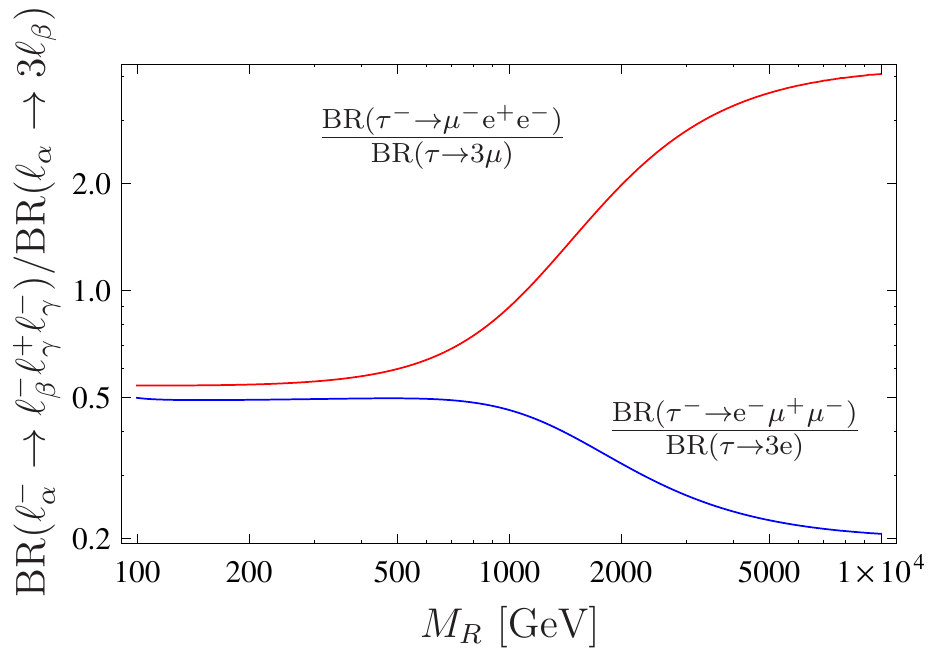}
\includegraphics[width=0.49 \linewidth]{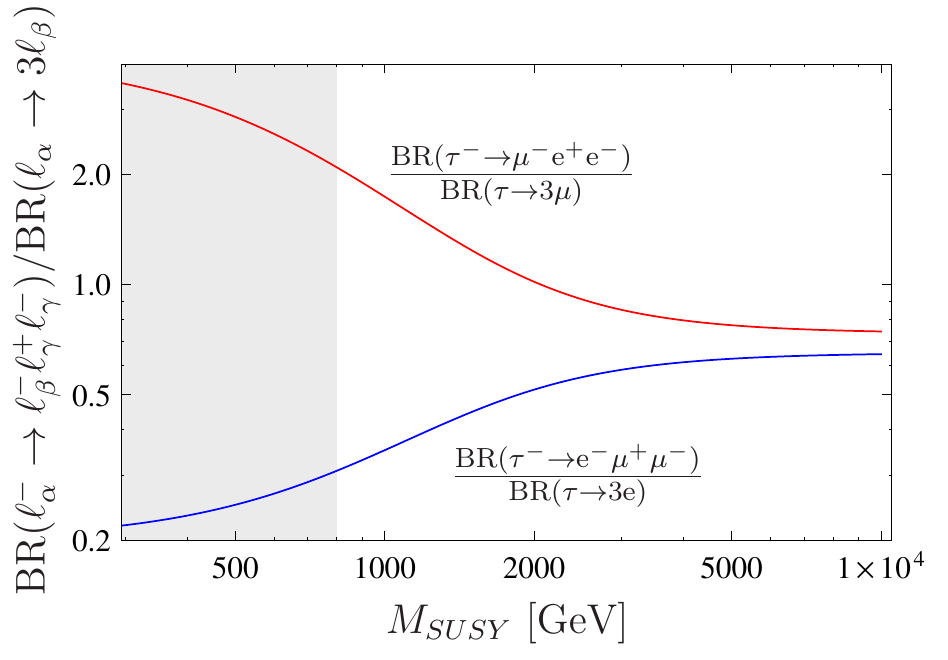}
\caption{Branching ratios for $\tau$ decays as a function of $M_R$ and
  $M_{SUSY}$.  In the upper two plots the lines correspond to
  BR($\mu\to 3 \, e$) (black solid), BR($\tau\to 3e$) (blue solid),
  BR($\tau\to 3 \, \mu$) (red solid), BR($\tau^- \to e^- \mu^+ \mu^-$)
  (blue dashed) and BR($\tau^- \to \mu^- e^+ e^-$) (red dashed). The
  gray area roughly corresponds to the parameter space excluded by the
  LHC experiments.  }
\label{fig:tauDecays}
\end{figure}

Turning now to the LFV $\tau$ decays, we show in \fig{fig:tauDecays}
several branching ratios for the scenario defined
above. Unfortunately, they are too small to be observed in the near
future. Below we will show alternative scenarios (in which the $R$
matrix is not assumed to be the unit matrix) where this is not the
case. Nevertheless, they show an interesting feature which is quite
generic in this model: BR$(\tau \to \mu e^+ e^-) \simeq$ BR($\tau \to
3 \, \mu)$ and BR$(\tau \to e \mu^+ \mu^-) \simeq$ BR($\tau \to 3 \,
e)$. Particularly interesting is that these branching ratios are
sensitive to the relative size of the non-SUSY contributions compared
to the SUSY ones.  We also stress that the various contributions
contribute similarly as in case of $\mu \to 3 \, e$.  For completeness
we note that BR$(\tau \to e \mu^+ e^-)$ and BR$(\tau \to \mu e^+
\mu^-)$ are strongly suppressed, at least a factor of $10^{-6}$ with
respect to the other 3-body decays, as they require at least one
additional flavor violating vertex in the dominant contributions.

\begin{figure}[t]
\includegraphics[width=0.49 \linewidth]{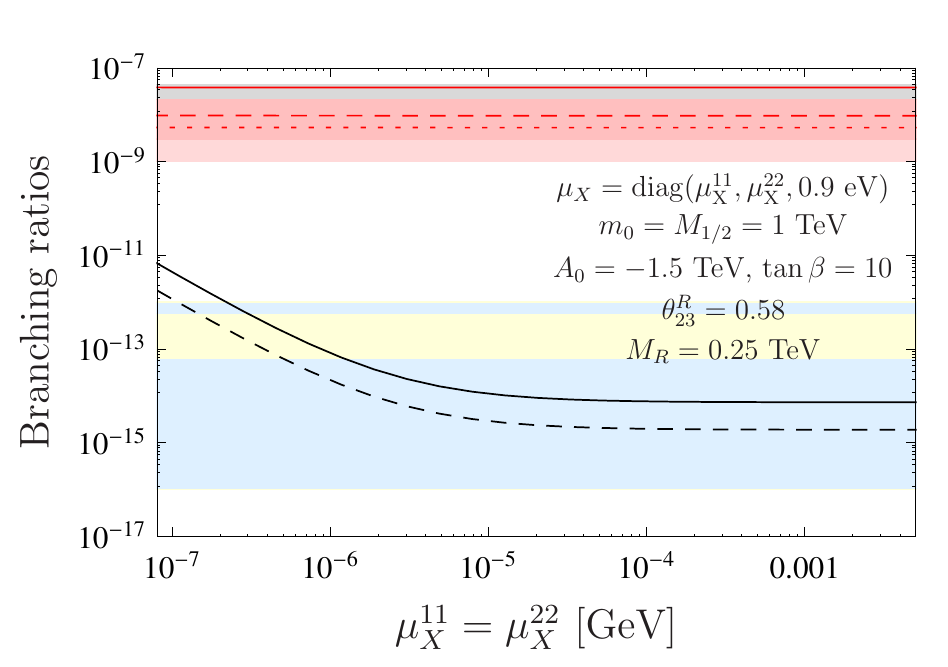}
\includegraphics[width=0.49 \linewidth]{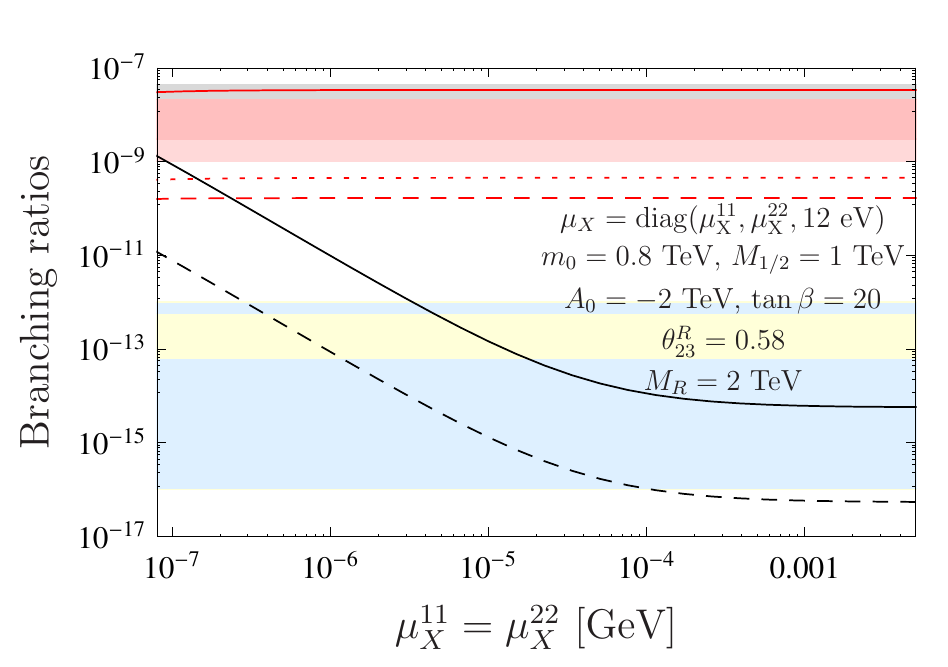}
\caption{$\mu$- and $\tau$-observables as a function of $\mu_X$. The
  underlying parameters are given in the plots.  The lines correspond
  to BR$(\tau \to \mu \gamma)$ (full red), BR$(\tau \to 3 \, \mu)$
  (dashed red), BR($\tau^- \to \mu^- e^+ e^-$) (dotted red), BR($\mu
  \to e \gamma$) (full black) and BR($\mu \to 3 \, e$) (dashed black).
  The light gray, red, yellow and blue bands show the expected future
  reach of the dedicated experiments to $\tau \to \mu \gamma$, $\tau
  \to 3 \, \mu$, $\mu \to e \gamma$ and $\mu \to 3 \, e $ as given in
  table~\ref{tab:sensi}.}
\label{fig:specialCases}
\end{figure}
It is worth stressing that the fact that the $\mu$ observables are more
constraining than the $\tau$ decays is correct in large parts of the
parameter space.  However, there is also a substantial part where the
opposite is true, as exemplified in \fig{fig:specialCases} where we
tune the parameters such that both, $\mu$- and $\tau$-observables can
be discovered in the next generation of experiments. For this we have
adjusted the diagonal entries of $\mu_X$ as well as $\theta_{23}^R$
and calculated $Y_\nu$ using \eq{eq:casas_ibarra}. Clearly, this part
of the parameter space requires quite some hierarchy in $\mu_X$ to explain
neutrino data correctly.
Note that even in this part of parameter space the ratios BR$(\tau \to
\mu e^+ e^-) \simeq$ BR($\tau \to 3 \, \mu)$ and BR$(\tau \to e \mu^+
\mu^-) \simeq$ BR($\tau \to 3 \, e)$ show the same dependence on the
ratio $M_R/M_{SUSY}$ as in the previous case.

\begin{figure}[t]
\includegraphics[width=0.49 \linewidth]{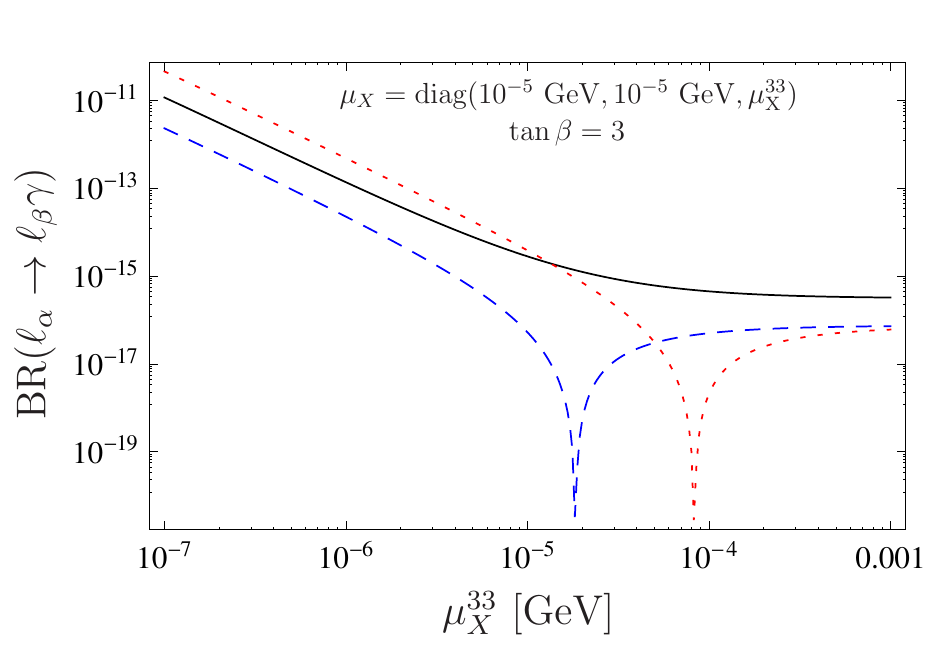}
\includegraphics[width=0.49 \linewidth]{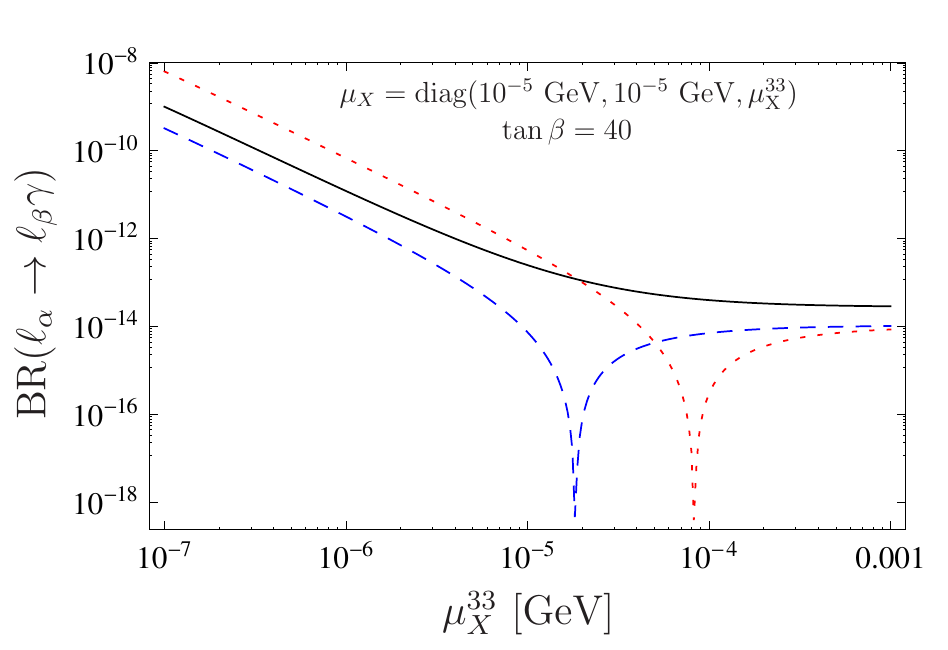} \\
\includegraphics[width=0.49 \linewidth]{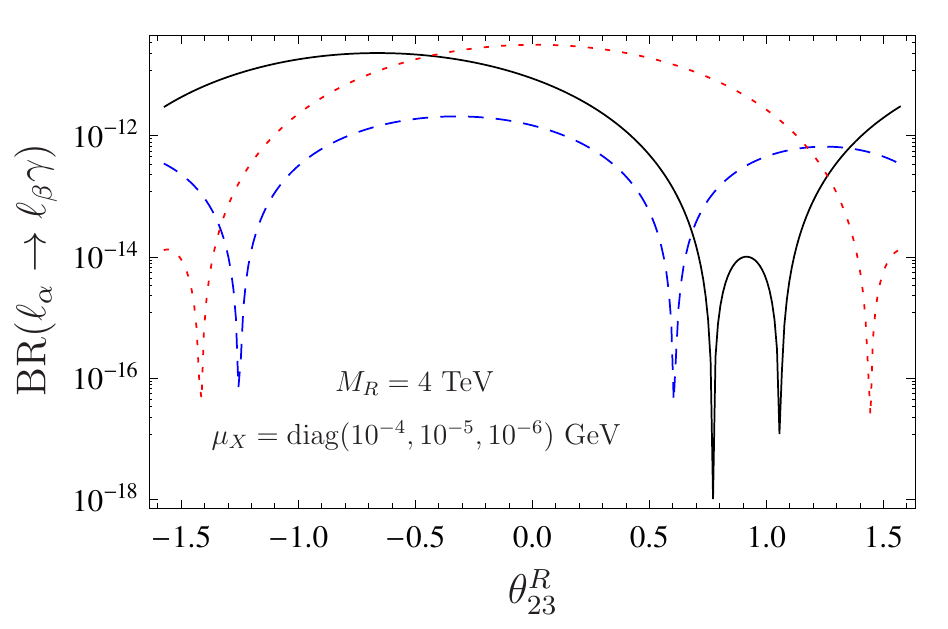}
\includegraphics[width=0.49 \linewidth]{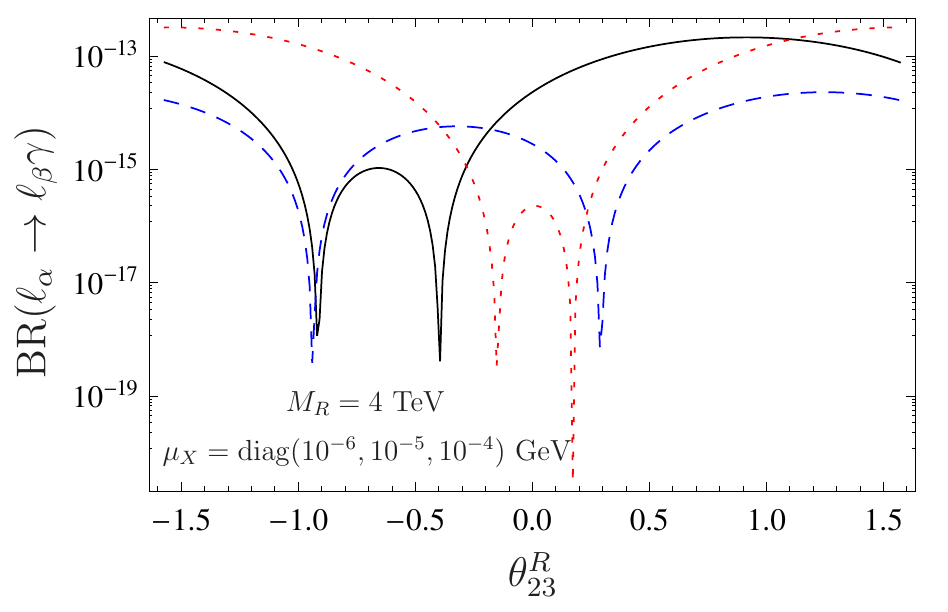} 
\caption{Dependence of $\ell_\alpha \to \ell_\beta \gamma$ on  $\mu_X^{33}$ and
$\theta_{23}^R$. The lines correspond to BR($\mu\to e \gamma$) (black),
BR($\tau\to e \gamma$) (blue dashed) and BR($\tau\to \mu \gamma$) (red dotted).
}
\label{fig:hierarchies_ltolgamma}
\end{figure}
The impact of the $R$ matrix and the hierarchy in the
$\mu_X$ entries is further illustrated for the decays $\ell_\alpha \to
\ell_\beta \gamma$ in \fig{fig:hierarchies_ltolgamma}.  Again, we have
calculated $Y_\nu$ via \eq{eq:casas_ibarra}, such that the results
from neutrino oscillation experiments are explained correctly. One
finds that, depending on the region in the parameter space, either the
$\mu$ decay or the $\tau$ decays are more important. As in case of the
3-body decays, one finds fine-tuned combinations of the parameters
where all decays can be observed in future experiments. Note that for
fixed $\theta_{23}^R$ the branching ratios scale like $f^2_R/f_X$
where $f_R$ and $f_X$ denote an overall scaling of $M_R$ and $\mu_X$,
respectively. Moreover, the branching ratios scale like $\tan^2\beta$
if the SUSY contributions dominate. In case the non-SUSY contributions
dominate we find only a slight $\tan\beta$ dependence for very large
$\tan\beta$ values.

Finally, let us comment on the Higgs penguin contributions to the
different LFV observables. In all our numerical scans they have been
found to be completely negligible and that is why we have decided not
to include them in our figures. In principle, one could look for
sizable Higgs penguin contributions by going to regions in parameter
space with large $\tan \beta$ and low pseudoscalar masses
\cite{Babu:2002et,Abada:2011hm}. This, however, would require
dedicated parameter scans in order to overcome the constraints from
flavor data, as these regions are already in strong tension after the
LHCb measurement of the $B_s \to \mu^+ \mu^-$ branching ratio
\cite{Aaij:2012nna}. For this reason, we have not pursued this goal
any further. Nevertheless, we have checked that the Higgs penguins
contributions to $\ell_\alpha \to \ell_\beta \ell_\gamma \ell_\delta$
and $\mu$-$e$ conversion in nuclei have the expected decoupling
behavior for large $M_R$ and/or $M_{SUSY}$ scales.

\clearpage

\section{Conclusions} \label{sec:conclusions}

This paper represents the first complete computation of selected LFV
observables in scenarios with light right-handed neutrinos. These
include the radiative decays $\ell_\alpha \to \ell_\beta \gamma$, the
3-body decays $\ell_\alpha \to \ell_\beta \ell_\gamma \ell_\delta$ (in
several variants) and neutrinoless $\mu-e$ conversion in nuclei. Our
results are valid in the inverse seesaw and should also hold in
low-scale type-I seesaw models with nearly conserved lepton number,
the inverse seesaw being a specific realization of these
models. Compared to previous studies, we have also included
Higgs-penguins and considered non-supersymmetric as well as
supersymmetric contributions to the corresponding LFV amplitudes
simultaneously.

For the numerical examples we took a CMSSM inspired scenario where we
also considered the limiting cases with either $M_R\gg M_{SUSY}$ and
$M_{SUSY}\gg M_R$.  Our main conclusions can be summarized as follows:

\begin{itemize}

\item The SUSY contributions dominate the induced photon penguins if
  both, $M_R$ and $M_{SUSY}$, are about the same size.  For $M_R \lsim
  M_{SUSY}/2$ the non-SUSY contributions start to dominate the
  radiative decays $\ell_\alpha \to \ell_\beta\gamma$.

\item For low $M_R$ scales the LFV phenomenology is dominated by
  non-SUSY contributions. This holds in particular for the 3-body
  decays and $\mu$-$e$ conversion in nuclei. These are mainly given by
  boxes and $Z$-penguin diagrams containing right-handed neutrinos in
  the loop. In contrast to the usual high-scale seesaw models, in
  which their contributions to LFV processes are tiny, the
  right-handed neutrinos can play a major role in low-scale seesaw
  scenarios. In what concerns the non-SUSY box contributions, our
  results confirm previous claims in the literature
  \cite{Ilakovac:2009jf,Alonso:2012ji,Dinh:2012bp,Ilakovac:2012sh}. Furthermore,
  we have highlighted the relevance of the non-SUSY $Z$-penguins,
  previously regarded as subdominant in most
  studies~\footnote{Non-SUSY $Z$-penguins were also included in
    Ref. \cite{Ilakovac:2012sh}, where their potentially large
    contributions were also shown.}. They are particularly relevant
  for larger values of $M_R$, where we often find a negative
  interference between the $Z$-penguins and the box
  contributions. This will be particularly important when the next
  generation of experiments start to probe this mass region.
  
\item The proper decoupling of the different contributions has been
  checked explicitly, e.g.\ we have checked that the
  SUSY-contributions, the $\nu^C$-$X$ and the Higgs contributions
  decouple independently as expected.

\item Currently, the radiative decay $\mu \to e \gamma$ is the most
  constraining LFV process. However, due to the promising experimental
  prospects in the near future, the situation will change. If the
  coming experiments perform as planned, $\mu \to 3 \, e$ will be the
  most relevant LFV process in the mid term, whereas neutrinoless
  $\mu-e$ conversion in nuclei will set the strongest constraints in
  the long term.

\item Ratios of $\tau$ LFV branching ratios can provide additional
  information about the dominant contributions. In particular, when
  the non-SUSY contributions dominate, one finds $\text{BR}(\tau^- \to
  \mu^- e^+ e^-)/\text{BR}(\tau \to 3 \, \mu) \simeq \text{BR}(\tau^-
  \to e^- \mu^+ \mu^-)/\text{BR}(\tau \to 3 \, e) \simeq 0.5-0.8$, whereas
  for a SUSY dominated scenario $\text{BR}(\tau^- \to \mu^- e^+
  e^-)/\text{BR}(\tau \to 3 \, \mu) \gg \text{BR}(\tau^- \to e^- \mu^+
  \mu^-)/\text{BR}(\tau \to 3 \, e)$. This can in turn be used to get
  a hint on the hierarchy between the seesaw and SUSY scales.

\end{itemize}

\section*{Acknowledgements}

We thank Martin Hirsch and Ernesto Arganda for fruitful discussions.
A.V.\ acknowledges partial support from the EXPL/FIS-NUC/0460/2013
project financed by the Portuguese FCT.  M.E.K.\ and W.P.\ have been
supported by the DFG, project no. PO-1337/3-1 and the DFG research
training group GRK 1147. FS is supported by the BMBF PT DESY
Verbundprojekt 05H2013-THEORIE 'Vergleich von LHC-Daten mit
supersymmetrischen Modellen'. C.W. receives financial support from the
Spanish CICYT through the project FPA2012-31880 and a partial support
from the European Union FP7 ITN INVISIBLES (Marie Curie Actions,
PITN-GA-2011-289442) and the Spanish MINECO's ``Centro de Excelencia
Severo Ochoa'' Programme under grant SEV-2012-0249. A. A acknowledges
support from the European Union FP7 ITN INVISIBLES (Marie Curie
Actions, PITN-\-GA-\-2011-\-289442).

\begin{appendix}
\allowdisplaybreaks

\section{Masses and vertices}
\label{app:masses_vertices}

We give first our conventions for the mass matrices as well as for the
corresponding rotation matrices. These matrices are then used to
express in appendix \ref{app:vertices} all the vertices needed to
calculate the LFV observables.

\subsection{Mass matrices}
\label{app:masses}
\begin{itemize} 
\item {\bf Mass matrix for Neutrinos}, Basis: \( \left(\nu_L, \nu_R^C, X \right) \)\

\begin{equation} 
m_{\nu} = \left( 
\begin{array}{ccc}
0 &\frac{1}{\sqrt{2}} v_u Y_{\nu}^{T}  &0\\ 
\frac{1}{\sqrt{2}} v_u Y_\nu  &0 &M_R\\ 
0 &M_{R}^{T} &\mu_X\end{array} 
\right) 
 \end{equation} 
This matrix is diagonalized by \(U^V\): 
\begin{equation} 
U^{V,*} m_{\nu} U^{V,\dagger} = m^{dia}_{\nu} 
\end{equation} 

\item {\bf Mass matrix for CP-odd Sneutrinos}, Basis: \( \left(\sigma_L, \sigma_R, \sigma_X\right) \) 
\begin{equation} 
m^2_{\nu^i} = \left( 
\begin{array}{ccc}
m_{\sigma_L\sigma_L} &m^T_{\sigma_L\sigma_R} &\frac{1}{\sqrt{2}} v_u \, {\Re\Big({Y_{\nu}^{T}  M_R^*}\Big)} \\ 
m_{\sigma_L\sigma_R} &m_{\sigma_R\sigma_R} & \Re \Big(B_{M_R} - M_R  \mu_X^*\Big)\\ 
\frac{1}{\sqrt{2}} v_u \, {\Re\Big({M_{R}^{T}  Y_\nu^*}\Big)}  & \Re \Big(B_{M_R}^{T} - \mu_X  M_{R}^{\dagger} \Big) &m_{\sigma_X\sigma_X}\end{array} 
\right) 
 \end{equation} 
\begin{align} 
m_{\sigma_L\sigma_L} &= \frac{1}{2} v_{u}^{2} \, {\Re\Big({Y_{\nu}^{T}  Y_\nu^*}\Big)}  + {\Re\Big(m_l^2\Big)} + \frac{1}{8} \Big(g_{1}^{2} + g_{2}^{2}\Big) \Big(v_{d}^{2} - v_{u}^{2} \Big) {\bf 1} \\ 
m_{\sigma_L\sigma_R} &= - \frac{1}{\sqrt{2}} \Big(v_d \, {\Re\Big(\mu Y_\nu^* \Big)}  -v_u \, {\Re\Big(T_\nu\Big)} \Big)\\ 
m_{\sigma_R\sigma_R} &= {\Re\Big({m_\nu^2}\Big)}  + {\Re\Big({M_R  M_{R}^{\dagger}}\Big)}+ \frac{1}{2} v_{u}^{2} \, {\Re\Big({Y_\nu  Y_{\nu}^{\dagger}}\Big)}\\ 
m_{\sigma_X\sigma_X} &= {\Re\Big({M_{R}^{T}  M_R^*}\Big)} +  {\Re\Big(m_X^2\Big)} - {\Re\Big(B_{\mu_X}\Big)}  + {\Re\Big({\mu_X  \mu_X^*}\Big)}
\end{align} 
This matrix is diagonalized by \(Z^i\): 
\begin{equation} 
Z^i m^2_{\nu^i} Z^{i,\dagger} = m^{2,dia}_{\nu^i} 
\end{equation} 

\item {\bf Mass matrix for CP-even Sneutrinos}, Basis: \( \left(\phi_L, \phi_R, \phi_X\right) \) 
\begin{equation} 
m^2_{\nu^R} = \left( 
\begin{array}{ccc}
m_{\phi_L\phi_L} &m^T_{\phi_L\phi_R} &\frac{1}{\sqrt{2}} v_u \, {\Re\Big({Y_{\nu}^{T}  M_R^*}\Big)} \\ 
m_{\phi_L\phi_R} &m_{\phi_R\phi_R} &  {\Re\Big(B_{M_R} + M_R  \mu_X^*}\Big)\\ 
\frac{1}{\sqrt{2}} v_u \, {\Re\Big({M_{R}^{T}  Y_\nu^*}\Big)}  & \Re\Big(B_{M_R}^{T} + \mu_X  M_{R}^{\dagger}\Big) &m_{\phi_X\phi_X}\end{array} 
\right) 
 \end{equation} 
\begin{align} 
m_{\phi_L\phi_L} &= \frac{1}{2} v_{u}^{2} \, {\Re\Big({Y_{\nu}^{T}  Y_\nu^*}\Big)}  + {\Re\Big(m_l^2\Big)}  + \frac{1}{8} \Big(g_{1}^{2} + g_{2}^{2}\Big) \Big(v_{d}^{2} - v_{u}^{2} \Big) {\bf 1}\\ 
m_{\phi_L\phi_R} &= - \frac{1}{\sqrt{2}} \Big(v_d {\Re\Big(\mu Y_\nu^* \Big)}  - v_u {\Re\Big(T_\nu\Big)} \Big)\\ 
m_{\phi_R\phi_R} &= {\Re\Big({m_\nu^2}\Big)}  + {\Re\Big({M_R  M_{R}^{\dagger}}\Big)}  + \frac{1}{2} v_{u}^{2} \, {\Re\Big({Y_\nu  Y_{\nu}^{\dagger}}\Big)} \\ 
m_{\phi_X\phi_X} &= {\Re\Big({M_{R}^{T}  M_R^*}\Big)}  + {\Re\Big(m_X^2\Big)}  + {\Re\Big(B_{\mu_X}\Big)} + {\Re\Big({\mu_X  \mu_X^*}\Big)}
\end{align}
This matrix is diagonalized by \(Z^R\):
\begin{equation}
Z^R m^2_{\nu^R} Z^{R,\dagger} = m^{2,dia}_{\nu^R}
\end{equation}

\item {\bf Mass matrix for Down-Squarks}, Basis: \( \left(\tilde{d}_{L,{{\alpha_1}}}, \tilde{d}_{R,{{\alpha_2}}}\right)\) 
\begin{equation} 
m^2_{\tilde{d}} = \left( 
\begin{array}{cc}
m_{\tilde{d}_L\tilde{d}_L^*} &\frac{1}{\sqrt{2}} \Big(v_d T_{d}^{\dagger}  - v_u \mu Y_{d}^{\dagger} \Big)\delta_{{\alpha_1} {\beta_2}} \\ 
\frac{1}{\sqrt{2}} \delta_{{\alpha_2} {\beta_1}} \Big(v_d T_d  - v_u Y_d \mu^* \Big) &m_{\tilde{d}_R\tilde{d}_R^*}\end{array} 
\right) 
 \end{equation} 
\begin{align} 
m_{\tilde{d}_L\tilde{d}_L^*} &= -\frac{1}{24} \Big(3 g_{2}^{2}  + g_{1}^{2}\Big) \Big(v_d^2 - v_u^2\Big)\delta_{{\alpha_1} {\beta_1}}  {\bf 1} + \frac{1}{2} \delta_{{\alpha_1} {\beta_1}} \Big(2 m_q^2  + v_{d}^{2} {Y_{d}^{\dagger}  Y_d} \Big)\\ 
m_{\tilde{d}_R\tilde{d}_R^*} &= \frac{1}{12} g_{1}^{2} \Big(v_u^2 - v_d^2 \Big)\delta_{{\alpha_2} {\beta_2}}  {\bf 1} + \frac{1}{2} \delta_{{\alpha_2} {\beta_2}} \Big(2 m_d^2  + v_{d}^{2} {Y_d  Y_{d}^{\dagger}} \Big)
\end{align} 
This matrix is diagonalized by \(Z^D\): 
\begin{equation} 
Z^D m^2_{\tilde{d}} Z^{D,\dagger} = m^{2,dia}_{\tilde{d}} 
\end{equation}

\item {\bf Mass matrix for Up-Squarks}, Basis: \( \left(\tilde{u}_{L,{{\alpha_1}}}, \tilde{u}_{R,{{\alpha_2}}}\right)\) 
\begin{equation} 
m^2_{\tilde{u}} = \left( 
\begin{array}{cc}
m_{\tilde{u}_L\tilde{u}_L^*} &\frac{1}{\sqrt{2}} \Big(- v_d \mu Y_{u}^{\dagger}  + v_u T_{u}^{\dagger} \Big)\delta_{{\alpha_1} {\beta_2}} \\ 
\frac{1}{\sqrt{2}} \delta_{{\alpha_2} {\beta_1}} \Big(- v_d Y_u \mu^*  + v_u T_u \Big) &m_{\tilde{u}_R\tilde{u}_R^*}\end{array} 
\right) 
 \end{equation} 
\begin{align} 
m_{\tilde{u}_L\tilde{u}_L^*} &= -\frac{1}{24} \Big(-3 g_{2}^{2}  + g_{1}^{2}\Big) \Big(v_d^2 - v_u^2\Big)\delta_{{\alpha_1} {\beta_1}}  {\bf 1} + \frac{1}{2} \delta_{{\alpha_1} {\beta_1}} \Big(2 m_q^2  + v_{u}^{2} {Y_{u}^{\dagger}  Y_u} \Big)\\ 
m_{\tilde{u}_R\tilde{u}_R^*} &= \frac{1}{2} \delta_{{\alpha_2} {\beta_2}} \Big(2 m_u^2  + v_{u}^{2} {Y_u  Y_{u}^{\dagger}} \Big) + \frac{1}{6} g_{1}^{2}  \Big(v_d^2 - v_u^2\Big)\delta_{{\alpha_2} {\beta_2}} {\bf 1}
\end{align} 
This matrix is diagonalized by \(Z^U\): 
\begin{equation} 
Z^U m^2_{\tilde{u}} Z^{U,\dagger} = m^{2,dia}_{\tilde{u}} 
\end{equation} 

\item {\bf Mass matrix for Sleptons}, Basis: \( \left(\tilde{e}_L, \tilde{e}_R\right) \) 
\begin{equation} 
m^2_{\tilde{e}} = \left( 
\begin{array}{cc}
m_{\tilde{e}_L\tilde{e}_L^*} &\frac{1}{\sqrt{2}} \Big(v_d T_{e}^{\dagger}  - v_u \mu Y_{e}^{\dagger} \Big)\\ 
\frac{1}{\sqrt{2}} \Big(v_d T_e  - v_u Y_e \mu^* \Big) &m_{\tilde{e}_R\tilde{e}_R^*}\end{array} 
\right) 
 \end{equation} 
\begin{align} 
m_{\tilde{e}_L\tilde{e}_L^*} &= \frac{1}{2} v_{d}^{2} {Y_{e}^{\dagger}  Y_e}  + \frac{1}{8} \Big(- g_{2}^{2}  + g_{1}^{2}\Big) \Big(v_d^2 - v_u^2\Big) {\bf 1} + m_l^2\\ 
m_{\tilde{e}_R\tilde{e}_R^*} &= \frac{1}{2} v_{d}^{2} {Y_e  Y_{e}^{\dagger}}  + \frac{1}{4} g_{1}^{2}  \Big(v_u^2 - v_d^2\Big) {\bf 1} + m_e^2
\end{align} 
This matrix is diagonalized by \(Z^E\): 
\begin{equation} 
Z^E m^2_{\tilde{e}} Z^{E,\dagger} = m^{2,dia}_{\tilde{e}} 
\end{equation} 

\item {\bf Mass matrix for CP-even Higgs}, Basis: \( \left(\phi_{d}, \phi_{u}\right)\) 
\begin{equation} 
m^2_{h} = \left( 
\begin{array}{cc}
\frac{1}{8} \Big(g_{1}^{2} + g_{2}^{2}\Big)\Big(3 v_{d}^{2}  - v_{u}^{2} \Big) + m_{H_d}^2 + |\mu|^2 &-\frac{1}{4} \Big(g_{1}^{2} + g_{2}^{2}\Big)v_d v_u  - {\Re\Big(B_{\mu}\Big)} \\ 
-\frac{1}{4} \Big(g_{1}^{2} + g_{2}^{2}\Big)v_d v_u  - {\Re\Big(B_{\mu}\Big)}  &-\frac{1}{8} \Big(g_{1}^{2} + g_{2}^{2}\Big)\Big(-3 v_{u}^{2}  + v_{d}^{2}\Big) + m_{H_u}^2 + |\mu|^2\end{array} 
\right) 
 \end{equation} 
This matrix is diagonalized by \(Z^H\): 
\begin{equation} 
Z^H m^2_{h} Z^{H,\dagger} = m^{2,dia}_{h} 
\end{equation} 

\item {\bf Mass matrix for CP-odd Higgs}, Basis: \( \left(\sigma_{d}, \sigma_{u}\right) \) 
\begin{eqnarray} 
m^2_{A^0} &=& \left( 
\begin{array}{cc}
\frac{1}{8} \Big(g_{1}^{2} + g_{2}^{2}\Big)\Big(v_d^2 - v_u^2\Big) + m_{H_d}^2 + |\mu|^2 &{\Re\Big(B_{\mu}\Big)}\\ 
{\Re\Big(B_{\mu}\Big)} &-\frac{1}{8} \Big(g_{1}^{2} + g_{2}^{2}\Big)\Big(v_d^2 - v_u^2\Big) + m_{H_u}^2 + |\mu|^2\end{array} 
\right) \nonumber \\
 && \hspace{1cm} +  \xi_{Z}m^2(Z) 
 \end{eqnarray} 
Gauge fixing contributions: 
\begin{equation} 
m^2 ({Z}) = \left( 
\begin{array}{cc}
\frac{1}{4} v_{d}^{2} \Big(g_1 \sin\Theta_W   + g_2 \cos\Theta_W  \Big)^{2}  &-\frac{1}{4} v_d v_u \Big(g_1 \sin\Theta_W   + g_2 \cos\Theta_W  \Big)^{2} \\ 
-\frac{1}{4} v_d v_u \Big(g_1 \sin\Theta_W   + g_2 \cos\Theta_W  \Big)^{2}  &\frac{1}{4} v_{u}^{2} \Big(g_1 \sin\Theta_W   + g_2 \cos\Theta_W  \Big)^{2} \end{array} 
\right) 
 \end{equation} 
This matrix is diagonalized by \(Z^A\): 
\begin{equation} 
Z^A m^2_{A^0} Z^{A,\dagger} = m^{2,dia}_{A^0} 
\end{equation} 

\item {\bf Mass matrix for Charged Higgs}, Basis: \( \left(H_d^-, H_u^{+,*}\right), \left(H_d^{-,*}, H_u^+\right) \) 
\begin{equation} 
m^2_{H^-} = \left( 
\begin{array}{cc}
m_{H_d^-H_d^{-,*}} &\frac{1}{4} g_{2}^{2} v_d v_u  + B_{\mu}^*\\ 
\frac{1}{4} g_{2}^{2} v_d v_u  + B_{\mu} &m_{H_u^{+,*}H_u^+}\end{array} 
\right) +  \xi_{W^-}m^2(W^-) 
 \end{equation} 
\begin{align} 
m_{H_d^-H_d^{-,*}} &= \frac{1}{8} \Big(g_{1}^{2} \Big(v_d^2 - v_u^2\Big) + g_{2}^{2} \Big(v_{d}^{2} + v_{u}^{2}\Big)\Big) + m_{H_d}^2 + |\mu|^2\\ 
m_{H_u^{+,*}H_u^+} &= \frac{1}{8} \Big(g_{1}^{2} \Big(- v_{d}^{2}  + v_{u}^{2}\Big) + g_{2}^{2} \Big(v_{d}^{2} + v_{u}^{2}\Big)\Big) + m_{H_u}^2 + |\mu|^2
\end{align} 
Gauge fixing contributions: 
\begin{equation} 
m^2 ({W^-}) = \left( 
\begin{array}{cc}
\frac{1}{4} g_{2}^{2} v_{d}^{2}  &-\frac{1}{4} g_{2}^{2} v_d v_u \\ 
-\frac{1}{4} g_{2}^{2} v_d v_u  &\frac{1}{4} g_{2}^{2} v_{u}^{2} \end{array} 
\right) 
 \end{equation} 
This matrix is diagonalized by \(Z^+\): 
\begin{equation} 
Z^+ m^2_{H^-} Z^{+,\dagger} = m^{2,dia}_{H^-} 
\end{equation} 

\item {\bf Mass matrix for Neutralinos}, Basis: \( \left(\lambda_{\tilde{B}}, \tilde{W}^0, \tilde{H}_d^0, \tilde{H}_u^0\right) \) 
\begin{equation} 
m_{\tilde{\chi}^0} = \left( 
\begin{array}{cccc}
M_1 &0 &-\frac{1}{2} g_1 v_d  &\frac{1}{2} g_1 v_u \\ 
0 &M_2 &\frac{1}{2} g_2 v_d  &-\frac{1}{2} g_2 v_u \\ 
-\frac{1}{2} g_1 v_d  &\frac{1}{2} g_2 v_d  &0 &- \mu \\ 
\frac{1}{2} g_1 v_u  &-\frac{1}{2} g_2 v_u  &- \mu  &0\end{array} 
\right) 
 \end{equation} 
This matrix is diagonalized by \(N\): 
\begin{equation} 
N^* m_{\tilde{\chi}^0} N^{\dagger} = m^{dia}_{\tilde{\chi}^0} 
\end{equation}

\item {\bf Mass matrix for Charginos}, Basis: \( \left(\tilde{W}^-, \tilde{H}_d^-\right), \left(\tilde{W}^+, \tilde{H}_u^+\right) \) 
\begin{equation} 
m_{\tilde{\chi}^-} = \left( 
\begin{array}{cc}
M_2 &\frac{1}{\sqrt{2}} g_2 v_u \\ 
\frac{1}{\sqrt{2}} g_2 v_d  &\mu\end{array} 
\right) 
 \end{equation} 
This matrix is diagonalized by \(U\) and \(V\) 
\begin{equation} 
U^* m_{\tilde{\chi}^-} V^{\dagger} = m^{dia}_{\tilde{\chi}^-} 
\end{equation} 

\item {\bf Mass matrix for charged Leptons}, Basis: \( \left(e_L\right), \left(e_R^*\right) \) 
\begin{equation} 
m_{e} = \left( 
\begin{array}{c}
\frac{1}{\sqrt{2}} v_d Y_{e}^{T} \end{array} 
\right) 
 \end{equation} 
This matrix is diagonalized by \(U^e_L\) and \(U^e_R\) 
\begin{equation} 
U^{e,*}_L m_{e} U_{R}^{e,\dagger} = m^{dia}_{e} 
\end{equation} 

\item {\bf Mass matrix for Down-Quarks}, Basis: \( \left(d_{L,{{\alpha_1}}}\right), \left(d^*_{R,{{\beta_1}}}\right) \) 
\begin{equation} 
m_{d} = \left( 
\begin{array}{c}
\frac{1}{\sqrt{2}} v_d \delta_{{\alpha_1} {\beta_1}} Y_{d}^{T} \end{array} 
\right) 
 \end{equation} 
This matrix is diagonalized by \(U^d_L\) and \(U^d_R\) 
\begin{equation} 
U^{d,*}_L m_{d} U_{R}^{d,\dagger} = m^{dia}_{d} 
\end{equation} 

\item {\bf Mass matrix for Up-Quarks}, Basis: \( \left(u_{L,{{\alpha_1}}}\right), \left(u^*_{R,{{\beta_1}}}\right) \) 
\begin{equation} 
m_{u} = \left( 
\begin{array}{c}
\frac{1}{\sqrt{2}} v_u \delta_{{\alpha_1} {\beta_1}} Y_{u}^{T} \end{array} 
\right) 
 \end{equation} 
This matrix is diagonalized by \(U^u_L\) and \(U^u_R\) 
\begin{equation} 
U^{u,*}_L m_{u} U_{R}^{u,\dagger} = m^{dia}_{u} 
\end{equation} 
\end{itemize}

\subsection{Vertices}
\label{app:vertices}

In this appendix we list all vertices relevant for our computations.
Our conventions are as follows:
\begin{itemize}
 \item Chiral vertices are parameterized as
 \begin{eqnarray*}
  &\Gamma^L_{F_a F_b S_c} P_L + \Gamma^R_{F_a F_b S_c} P_R & \\
  &\Gamma^L_{F_a F_b V^\mu_c} \gamma_\mu P_L + \Gamma^R_{F_a F_b
V^\mu_c} \gamma_\mu P_R &
 \end{eqnarray*}
 \item The momentum flow in vector and scalar-vector vertices is
 \begin{eqnarray*}
 & \Gamma_{S_a S_b V^\mu_c}  (p^\mu_{S_b} - p^\mu_{S_a}) & \\
 & \Gamma_{V_a^\rho V_b^\sigma V_c^\mu} (g_{\rho \mu} (-
p_{V_c}^{\sigma}  + p_{V_a}^{\sigma}) + g_{\rho \sigma} (-
p_{V_a}^{\mu}  + p_{V_c}^{\mu}) + g_{\sigma \mu} (- p_{V_a}^{\rho}  +
p_{V_c}^{\rho}))&
 \end{eqnarray*}
\end{itemize}
Here we used polarization projectors $P_{L,R}$, metric $g_{\mu\nu}$
and momenta $p$ of the external fields.

\subsubsection{Fermion-Scalar vertices}
\begin{align} 
\VccaL_{ijk} = \Gamma^L_{\tilde{\chi}^+_{{i}}\tilde{\chi}^-_{{j}}A^0_{{k}}}  =  & \,- \frac{1}{\sqrt{2}} g_2 \Big(U^*_{j 1} V^*_{i 2} Z_{{k 2}}^{A}  + U^*_{j 2} V^*_{i 1} Z_{{k 1}}^{A} \Big)\\ 
\VccaR_{ijk} = \Gamma^R_{\tilde{\chi}^+_{{i}}\tilde{\chi}^-_{{j}}A^0_{{k}}}  =  & \,\frac{1}{\sqrt{2}} g_2 \Big(U_{{i 1}} V_{{j 2}} Z_{{k 2}}^{A}  + U_{{i 2}} V_{{j 1}} Z_{{k 1}}^{A} \Big) 
\\ 
\VnnaL_{ijk} = \Gamma^L_{\tilde{\chi}^0_{{i}}\tilde{\chi}^0_{{j}}A^0_{{k}}}  =  & \,\frac{1}{2} \Big(N^*_{i 3} \Big(g_1 N^*_{j 1}  - g_2 N^*_{j 2} \Big)Z_{{k 1}}^{A} - g_2 N^*_{i 2} N^*_{j 3} Z_{{k 1}}^{A} - g_1 N^*_{i 4} N^*_{j 1} Z_{{k 2}}^{A} +g_2 N^*_{i 4} N^*_{j 2} Z_{{k 2}}^{A} \nonumber \\ 
 &+g_2 N^*_{i 2} N^*_{j 4} Z_{{k 2}}^{A} +g_1 N^*_{i 1} \Big(N^*_{j 3} Z_{{k 1}}^{A}  - N^*_{j 4} Z_{{k 2}}^{A} \Big)\Big)\\ 
\VnnaR_{ijk} = \Gamma^R_{\tilde{\chi}^0_{{i}}\tilde{\chi}^0_{{j}}A^0_{{k}}}  =  & \,\frac{1}{2} \Big(Z_{{k 1}}^{A} \Big(\Big(- g_1 N_{{i 1}}  + g_2 N_{{i 2}} \Big)N_{{j 3}}  + N_{{i 3}} \Big(- g_1 N_{{j 1}}  + g_2 N_{{j 2}} \Big)\Big)\nonumber \\ 
 &+Z_{{k 2}}^{A} \Big(\Big(g_1 N_{{i 1}}  - g_2 N_{{i 2}} \Big)N_{{j 4}}  + N_{{i 4}} \Big(g_1 N_{{j 1}}  - g_2 N_{{j 2}} \Big)\Big)\Big) 
\\ 
\VddaL_{ijk} = \Gamma^L_{\bar{d}_{{i \alpha}}d_{{j \beta}}A^0_{{k}}}  =  & \, -\frac{i}{\sqrt{2}} \delta_{\alpha \beta} \sum_{b=1}^{3}U^{d,*}_{L,{j b}} \sum_{a=1}^{3}U^{d,*}_{R,{i a}} Y_{d,{a b}}   Z_{{k 1}}^{A} \\ 
\VddaR_{ijk} = \Gamma^R_{\bar{d}_{{i \alpha}}d_{{j \beta}}A^0_{{k}}}  =  & \, \frac{i}{\sqrt{2}} \delta_{\alpha \beta} \sum_{b=1}^{3}\sum_{a=1}^{3}Y^*_{d,{a b}} U_{R,{j a}}^{d}  U_{L,{i b}}^{d}  Z_{{k 1}}^{A}  
\\ 
\VeeaL_{ijk} = \Gamma^L_{\bar{\ell}_{{i}}\ell_{{j}}A^0_{{k}}}  =  & \,-\frac{i}{\sqrt{2}} \sum_{b=1}^{3}U^{e,*}_{L,{j b}} \sum_{a=1}^{3}U^{e,*}_{R,{i a}} Y_{e,{a b}}   Z_{{k 1}}^{A} \\ 
\VeeaR_{ijk} = \Gamma^R_{\bar{\ell}_{{i}}\ell_{{j}}A^0_{{k}}}  =  & \, \frac{i}{\sqrt{2}} \sum_{b=1}^{3}\sum_{a=1}^{3}Y^*_{e,{a b}} U_{R,{j a}}^{e}  U_{L,{i b}}^{e}  Z_{{k 1}}^{A}  
\\ 
\VuuaL_{ijk} = \Gamma^L_{\bar{u}_{{i \alpha}}u_{{j \beta}}A^0_{{k}}}  =  & \,-\frac{i}{\sqrt{2}} \delta_{\alpha \beta} \sum_{b=1}^{3}U^{u,*}_{L,{j b}} \sum_{a=1}^{3}U^{u,*}_{R,{i a}} Y_{u,{a b}}   Z_{{k 2}}^{A} \\ 
\VuuaR_{ijk} =  \Gamma^R_{\bar{u}_{{i \alpha}}u_{{j \beta}}A^0_{{k}}}  =  & \, \frac{i}{\sqrt{2}} \delta_{\alpha \beta} \sum_{b=1}^{3}\sum_{a=1}^{3}Y^*_{u,{a b}} U_{R,{j a}}^{u}  U_{L,{i b}}^{u}  Z_{{k 2}}^{A}  
\\ 
\VvvaL_{ijk} = \Gamma^L_{\nu_{{i}}\nu_{{j}}A^0_{{k}}}  =  & \,-\frac{i}{\sqrt{2}} \Big(\sum_{b=1}^{3}U^{V,*}_{j b} \sum_{a=1}^{3}U^{V,*}_{i 3 + a} Y_{\nu,{a b}}   + \sum_{b=1}^{3}U^{V,*}_{i b} \sum_{a=1}^{3}U^{V,*}_{j 3 + a} Y_{\nu,{a b}}  \Big)Z_{{k 2}}^{A} \\ 
\VvvaR_{ijk} = \Gamma^R_{\nu_{{i}}\nu_{{j}}A^0_{{k}}}  =  & \, \frac{i}{\sqrt{2}} \Big(\sum_{b=1}^{3}\sum_{a=1}^{3}Y^*_{\nu,{a b}} U_{{j 3 + a}}^{V}  U_{{i b}}^{V}  + \sum_{b=1}^{3}\sum_{a=1}^{3}Y^*_{\nu,{a b}} U_{{i 3 + a}}^{V}  U_{{j b}}^{V} \Big)Z_{{k 2}}^{A}  
\\ 
\VcudL_{ijk} = \Gamma^L_{\tilde{\chi}^-_{{i}}u_{{j \beta}}\tilde{d}^*_{{k \gamma}}}  =  & \,- \delta_{\beta \gamma} \Big(g_2 U^*_{i 1} \sum_{a=1}^{3}U^{u,*}_{L,{j a}} Z_{{k a}}^{D}   - U^*_{i 2} \sum_{b=1}^{3}U^{u,*}_{L,{j b}} \sum_{a=1}^{3}Y_{d,{a b}} Z_{{k 3 + a}}^{D}   \Big)\\ 
\VcudR_{ijk} = \Gamma^R_{\tilde{\chi}^-_{{i}}u_{{j \beta}}\tilde{d}^*_{{k \gamma}}}  =  & \, \delta_{\beta \gamma} \sum_{b=1}^{3}\sum_{a=1}^{3}Y^*_{u,{a b}} U_{R,{j a}}^{u}  Z_{{k b}}^{D}  V_{{i 2}}  
\\ 
\VcchL_{ijk} = \Gamma^L_{\tilde{\chi}^+_{{i}}\tilde{\chi}^-_{{j}}h_{{k}}}  =  & \,- \frac{1}{\sqrt{2}} g_2 \Big(U^*_{j 1} V^*_{i 2} Z_{{k 2}}^{H}  + U^*_{j 2} V^*_{i 1} Z_{{k 1}}^{H} \Big)\\ 
\VcchR_{ijk} = \Gamma^R_{\tilde{\chi}^+_{{i}}\tilde{\chi}^-_{{j}}h_{{k}}}  =  & \,- \frac{1}{\sqrt{2}} g_2 \Big(U_{{i 1}} V_{{j 2}} Z_{{k 2}}^{H}  + U_{{i 2}} V_{{j 1}} Z_{{k 1}}^{H} \Big) 
\\ 
\VnnhL_{ijk} = \Gamma^L_{\tilde{\chi}^0_{{i}}\tilde{\chi}^0_{{j}}h_{{k}}}  =  & \,\frac{1}{2} \Big(N^*_{i 3} \Big(g_1 N^*_{j 1}  - g_2 N^*_{j 2} \Big)Z_{{k 1}}^{H} - g_2 N^*_{i 2} N^*_{j 3} Z_{{k 1}}^{H} - g_1 N^*_{i 4} N^*_{j 1} Z_{{k 2}}^{H} +g_2 N^*_{i 4} N^*_{j 2} Z_{{k 2}}^{H} \nonumber \\ 
 &+g_2 N^*_{i 2} N^*_{j 4} Z_{{k 2}}^{H} +g_1 N^*_{i 1} \Big(N^*_{j 3} Z_{{k 1}}^{H}  - N^*_{j 4} Z_{{k 2}}^{H} \Big)\Big)\\ 
\VnnhR_{ijk} = \Gamma^R_{\tilde{\chi}^0_{{i}}\tilde{\chi}^0_{{j}}h_{{k}}}  =  & \,\frac{1}{2} \Big(Z_{{k 1}}^{H} \Big(\Big(g_1 N_{{i 1}}  - g_2 N_{{i 2}} \Big)N_{{j 3}}  + N_{{i 3}} \Big(g_1 N_{{j 1}}  - g_2 N_{{j 2}} \Big)\Big)\nonumber \\ 
 &+Z_{{k 2}}^{H} \Big(\Big(- g_1 N_{{i 1}}  + g_2 N_{{i 2}} \Big)N_{{j 4}}  + N_{{i 4}} \Big(- g_1 N_{{j 1}}  + g_2 N_{{j 2}} \Big)\Big)\Big) 
\\ 
\VnddL_{ijk} = \Gamma^L_{\tilde{\chi}^0_{{i}}d_{{j \beta}}\tilde{d}^*_{{k \gamma}}}  =  & \,\delta_{\beta \gamma} \Big( \frac{1}{\sqrt{2}} g_2 N^*_{i 2} \sum_{a=1}^{3}U^{d,*}_{L,{j a}} Z_{{k a}}^{D} - N^*_{i 3} \sum_{b=1}^{3}U^{d,*}_{L,{j b}} \sum_{a=1}^{3}Y_{d,{a b}} Z_{{k 3 + a}}^{D}    - \frac{1}{3 \sqrt{2}} \, g_1 N^*_{i 1} \sum_{a=1}^{3}U^{d,*}_{L,{j a}} Z_{{k a}}^{D}  \Big)\\ 
\VnddR_{ijk} = \Gamma^R_{\tilde{\chi}^0_{{i}}d_{{j \beta}}\tilde{d}^*_{{k \gamma}}}  =  & \,- \delta_{\beta \gamma} \Big(\sum_{b=1}^{3}\sum_{a=1}^{3}Y^*_{d,{a b}} U_{R,{j a}}^{d}  Z_{{k b}}^{D}  N_{{i 3}}  + \frac{\sqrt{2}}{3} \, g_1 \sum_{a=1}^{3}Z_{{k 3 + a}}^{D} U_{R,{j a}}^{d}  N_{{i 1}} \Big) 
\\ 
\VneeL_{ijk} = \Gamma^L_{\tilde{\chi}^0_{{i}}\ell_{{j}}\tilde{e}^*_{{k}}}  =  & \, - N^*_{i 3} \sum_{b=1}^{3}U^{e,*}_{L,{j b}} \sum_{a=1}^{3}Y_{e,{a b}} Z_{{k 3 + a}}^{E}    + \frac{1}{\sqrt{2}} \, g_1 N^*_{i 1} \sum_{a=1}^{3}U^{e,*}_{L,{j a}} Z_{{k a}}^{E}   + \frac{1}{\sqrt{2}} \, g_2 N^*_{i 2} \sum_{a=1}^{3}U^{e,*}_{L,{j a}} Z_{{k a}}^{E} \\ 
\VneeR_{ijk} =  \Gamma^R_{\tilde{\chi}^0_{{i}}\ell_{{j}}\tilde{e}^*_{{k}}}  =  & \,- \Big(\sqrt{2} g_1 \sum_{a=1}^{3}Z_{{k 3 + a}}^{E} U_{R,{j a}}^{e}  N_{{i 1}}  + \sum_{b=1}^{3}\sum_{a=1}^{3}Y^*_{e,{a b}} U_{R,{j a}}^{e}  Z_{{k b}}^{E}  N_{{i 3}} \Big) 
\\ 
\VnuuL_{ijk} = \Gamma^L_{\tilde{\chi}^0_{{i}}u_{{j \beta}}\tilde{u}^*_{{k \gamma}}}  =  & \, - \delta_{\beta \gamma} \Big(\frac{1}{\sqrt{2}} g_2 N^*_{i 2} \sum_{a=1}^{3}U^{u,*}_{L,{j a}} Z_{{k a}}^{U}   + N^*_{i 4} \sum_{b=1}^{3}U^{u,*}_{L,{j b}} \sum_{a=1}^{3}Y_{u,{a b}} Z_{{k 3 + a}}^{U}    + \frac{1}{3 \sqrt{2}} \, g_1 N^*_{i 1} \sum_{a=1}^{3}U^{u,*}_{L,{j a}} Z_{{k a}}^{U}  \Big)\\ 
\VnuuR_{ijk} = \Gamma^R_{\tilde{\chi}^0_{{i}}u_{{j \beta}}\tilde{u}^*_{{k \gamma}}}  =  & \, \delta_{\beta \gamma} \Big( \frac{2 \sqrt{2}}{3} \, g_1 \sum_{a=1}^{3}Z_{{k 3 + a}}^{U} U_{R,{j a}}^{u}  N_{{i 1}}  - \sum_{b=1}^{3}\sum_{a=1}^{3}Y^*_{u,{a b}} U_{R,{j a}}^{u}  Z_{{k b}}^{U}  N_{{i 4}} \Big) 
\\ 
\VddhL_{ijk} = \Gamma^L_{\bar{d}_{{i \alpha}}d_{{j \beta}}h_{{k}}}  =  & \,- \frac{1}{\sqrt{2}} \delta_{\alpha \beta} \sum_{b=1}^{3}U^{d,*}_{L,{j b}} \sum_{a=1}^{3}U^{d,*}_{R,{i a}} Y_{d,{a b}}   Z_{{k 1}}^{H} \\ 
\VddhR_{ijk} =  \Gamma^R_{\bar{d}_{{i \alpha}}d_{{j \beta}}h_{{k}}}  =  & \,- \frac{1}{\sqrt{2}} \delta_{\alpha \beta} \sum_{b=1}^{3}\sum_{a=1}^{3}Y^*_{d,{a b}} U_{R,{j a}}^{d}  U_{L,{i b}}^{d}  Z_{{k 1}}^{H}  
\\ 
\VcduL_{ijk} = \Gamma^L_{\tilde{\chi}^+_{{i}}d_{{j \beta}}\tilde{u}^*_{{k \gamma}}}  =  & \,- \delta_{\beta \gamma} \Big(g_2 V^*_{i 1} \sum_{a=1}^{3}U^{d,*}_{L,{j a}} Z_{{k a}}^{U}   - V^*_{i 2} \sum_{b=1}^{3}U^{d,*}_{L,{j b}} \sum_{a=1}^{3}Y_{u,{a b}} Z_{{k 3 + a}}^{U}   \Big)\\ 
\VcduR_{ijk} =  \Gamma^R_{\tilde{\chi}^+_{{i}}d_{{j \beta}}\tilde{u}^*_{{k \gamma}}}  =  & \, \delta_{\beta \gamma} \sum_{b=1}^{3}\sum_{a=1}^{3}Y^*_{d,{a b}} U_{R,{j a}}^{d}  Z_{{k b}}^{U}  U_{{i 2}}  
\\ 
\VudhpL_{ijk} = \Gamma^L_{\bar{u}_{{i \alpha}}d_{{j \beta}}H^+_{{k}}}  =  & \, \delta_{\alpha \beta} \sum_{b=1}^{3}U^{d,*}_{L,{j b}} \sum_{a=1}^{3}U^{u,*}_{R,{i a}} Y_{u,{a b}}   Z_{{k 2}}^{+} \\ 
\VudhpR_{ijk} =  \Gamma^R_{\bar{u}_{{i \alpha}}d_{{j \beta}}H^+_{{k}}}  =  & \, \delta_{\alpha \beta} \sum_{b=1}^{3}\sum_{a=1}^{3}Y^*_{d,{a b}} U_{R,{j a}}^{d}  U_{L,{i b}}^{u}  Z_{{k 1}}^{+}  
\\ 
\VvehpL_{ijk} = \Gamma^L_{\nu_{{i}}\ell_{{j}}H^+_{{k}}}  =  & \, \sum_{b=1}^{3}U^{e,*}_{L,{j b}} \sum_{a=1}^{3}U^{V,*}_{i 3 + a} Y_{\nu,{a b}}   Z_{{k 2}}^{+} \\ 
\VvehpR_{ijk} = \Gamma^R_{\nu_{{i}}\ell_{{j}}H^+_{{k}}}  =  & \, \sum_{b=1}^{3}\sum_{a=1}^{3}Y^*_{e,{a b}} U_{R,{j a}}^{e}  U_{{i b}}^{V}  Z_{{k 1}}^{+}  
\\ 
\VeehL_{ijk} = \Gamma^L_{\bar{\ell}_{{i}}\ell_{{j}}h_{{k}}}  =  & \,- \frac{1}{\sqrt{2}} \sum_{b=1}^{3}U^{e,*}_{L,{j b}} \sum_{a=1}^{3}U^{e,*}_{R,{i a}} Y_{e,{a b}}   Z_{{k 1}}^{H} \\ 
\VeehR_{ijk} = \Gamma^R_{\bar{\ell}_{{i}}\ell_{{j}}h_{{k}}}  =  & \,- \frac{1}{\sqrt{2}} \sum_{b=1}^{3}\sum_{a=1}^{3}Y^*_{e,{a b}} U_{R,{j a}}^{e}  U_{L,{i b}}^{e}  Z_{{k 1}}^{H}  
\\ 
\VceviL_{ijk} = \Gamma^L_{\tilde{\chi}^+_{{i}}\ell_{{j}}\nu^i_{{k}}}  =  & \,-\frac{i}{\sqrt{2}} \Big(- g_2 V^*_{i 1} \sum_{a=1}^{3}U^{e,*}_{L,{j a}} Z^{i,*}_{k a}   + V^*_{i 2} \sum_{b=1}^{3}U^{e,*}_{L,{j b}} \sum_{a=1}^{3}Z^{i,*}_{k 3 + a} Y_{\nu,{a b}}   \Big)\\ 
\VceviR_{ijk} = \Gamma^R_{\tilde{\chi}^+_{{i}}\ell_{{j}}\nu^i_{{k}}}  =  & \,-\frac{i}{\sqrt{2}} \sum_{b=1}^{3}Z^{i,*}_{k b} \sum_{a=1}^{3}Y^*_{e,{a b}} U_{R,{j a}}^{e}   U_{{i 2}}  
\\ 
\VcevrL_{ijk} = \Gamma^L_{\tilde{\chi}^+_{{i}}\ell_{{j}}\nu^R_{{k}}}  =  & \,- \frac{1}{\sqrt{2}} \Big(g_2 V^*_{i 1} \sum_{a=1}^{3}U^{e,*}_{L,{j a}} Z^{R,*}_{k a}   - V^*_{i 2} \sum_{b=1}^{3}U^{e,*}_{L,{j b}} \sum_{a=1}^{3}Z^{R,*}_{k 3 + a} Y_{\nu,{a b}}   \Big)\\ 
\VcevrL_{ijk} = \Gamma^R_{\tilde{\chi}^+_{{i}}\ell_{{j}}\nu^R_{{k}}}  =  & \, \frac{1}{\sqrt{2}} \sum_{b=1}^{3}Z^{R,*}_{k b} \sum_{a=1}^{3}Y^*_{e,{a b}} U_{R,{j a}}^{e}   U_{{i 2}}  
\\ 
\VuuhL_{ijk} = \Gamma^L_{\bar{u}_{{i \alpha}}u_{{j \beta}}h_{{k}}}  =  & \,- \frac{1}{\sqrt{2}} \delta_{\alpha \beta} \sum_{b=1}^{3}U^{u,*}_{L,{j b}} \sum_{a=1}^{3}U^{u,*}_{R,{i a}} Y_{u,{a b}}   Z_{{k 2}}^{H} \\ 
\VuuhR_{ijk} = \Gamma^R_{\bar{u}_{{i \alpha}}u_{{j \beta}}h_{{k}}}  =  & \,- \frac{1}{\sqrt{2}} \delta_{\alpha \beta} \sum_{b=1}^{3}\sum_{a=1}^{3}Y^*_{u,{a b}} U_{R,{j a}}^{u}  U_{L,{i b}}^{u}  Z_{{k 2}}^{H}  
\\ 
% \VduhmL_{ijk} = \Gamma^L_{\bar{d}_{{i \alpha}}u_{{j \beta}}H^-_{{k}}}  =  & \, Z^{+,*}_{k 1} \delta_{\alpha \beta} \sum_{b=1}^{3}U^{u,*}_{L,{j b}} \sum_{a=1}^{3}U^{d,*}_{R,{i a}} Y_{d,{a b}}   \\ 
% \VduhmR_{ijk} = \Gamma^R_{\bar{d}_{{i \alpha}}u_{{j \beta}}H^-_{{k}}}  =  & \, Z^{+,*}_{k 2} \delta_{\alpha \beta} \sum_{b=1}^{3}\sum_{a=1}^{3}Y^*_{u,{a b}} U_{R,{j a}}^{u}  U_{L,{i b}}^{d}   
% \\ 
\VvvhL_{ijk} = \Gamma^L_{\nu_{{i}}\nu_{{j}}h_{{k}}}  =  & \,- \frac{1}{\sqrt{2}} \Big(\sum_{b=1}^{3}U^{V,*}_{j b} \sum_{a=1}^{3}U^{V,*}_{i 3 + a} Y_{\nu,{a b}}   + \sum_{b=1}^{3}U^{V,*}_{i b} \sum_{a=1}^{3}U^{V,*}_{j 3 + a} Y_{\nu,{a b}}  \Big)Z_{{k 2}}^{H} \\ 
\VvvhR_{ijk} = \Gamma^R_{\nu_{{i}}\nu_{{j}}h_{{k}}}  =  & \,- \frac{1}{\sqrt{2}} \Big(\sum_{b=1}^{3}\sum_{a=1}^{3}Y^*_{\nu,{a b}} U_{{j 3 + a}}^{V}  U_{{i b}}^{V}  + \sum_{b=1}^{3}\sum_{a=1}^{3}Y^*_{\nu,{a b}} U_{{i 3 + a}}^{V}  U_{{j b}}^{V} \Big)Z_{{k 2}}^{H}  
\\ 
\VccpL_{ijk} = \Gamma^L_{\tilde{\chi}^+_{{i}}\nu_{{j}}\tilde{e}_{{k}}}  =  & \, V^*_{i 2} \sum_{b=1}^{3}Z^{E,*}_{k b} \sum_{a=1}^{3}U^{V,*}_{j 3 + a} Y_{\nu,{a b}}   \\ 
\VccpR_{ijk} = \Gamma^R_{\tilde{\chi}^+_{{i}}\nu_{{j}}\tilde{e}_{{k}}}  =  & \,- \Big(g_2 \sum_{a=1}^{3}Z^{E,*}_{k a} U_{{j a}}^{V}  U_{{i 1}}  - \sum_{b=1}^{3}\sum_{a=1}^{3}Y^*_{e,{a b}} Z^{E,*}_{k 3 + a}  U_{{j b}}^{V}  U_{{i 2}} \Big) 
\end{align}
In addition, we introduce 
\begin{eqnarray}
& \VdcuL_{ijk} = (\VcduR_{jik})^*  \hspace{0.5cm} \VdcuR_{ijk} = (\VcduL_{jik})^* \hspace{1cm}
 \VcudLc_{ijk} = (\VcudR_{jik})^*  \hspace{0.5cm} \VcudRc_{ijk} = (\VcudL_{jik})^* & \\
& \VecviL_{ijk} = (\VceviR_{jik})^*  \hspace{0.5cm} \VecviR_{ijk} = (\VceviL_{jik})^*  \hspace{1cm}
 \VecvrL_{ijk} = (\VcevrR_{jik})^*  \hspace{0.5cm} \VecvrR_{ijk} = (\VcevrL_{jik})^* & \\
& \VdndL_{ijk} = (\VnddR_{jik})^* \hspace{0.5cm} \VdndR_{ijk} = (\VnddL_{jik})^*  \hspace{1cm}
 \VunuL_{ijk} = (\VnuuR_{jik})^* \hspace{0.5cm} \VunuR_{ijk} = (\VnuuL_{jik})^* & \\
& \VeneL_{ijk} = (\VneeR_{jik})^*  \hspace{0.5cm} \VeneR_{ijk} = (\VneeL_{jik})^*  \hspace{1cm}
 \VduhmL_{ijk} = (\VudhpR_{jik})^* \hspace{0.5cm} \VduhmR_{ijk} = (\VudhpL_{jik})^* & \\
& \VevhmL_{ijk} = (\VvehpR_{jik})^* \hspace{0.5cm} \VevhmR_{ijk} = (\VvehpL_{jik})^* &
\end{eqnarray}

\subsubsection{Fermion-Vector vertices}
\begin{align} 
\VccpL_{ij} =\Gamma^L_{\tilde{\chi}^+_{{i}}\tilde{\chi}^-_{{j}}\gamma_{{\mu}}}  =  & \,\,\, \VccpR_{ij} = \Gamma^R_{\tilde{\chi}^+_{{i}}\tilde{\chi}^-_{{j}}\gamma_{{\mu}}}   =  \, e \delta_{ij} \\
\VcczL_{ij} =\Gamma^L_{\tilde{\chi}^+_{{i}}\tilde{\chi}^-_{{j}}Z_{{\mu}}}  =  & \, g_2 U^*_{j 1} \cos\Theta_W  U_{{i 1}}  + \frac{1}{2} \, U^*_{j 2} \Big(- g_1 \sin\Theta_W   + g_2 \cos\Theta_W  \Big)U_{{i 2}} \\ 
\VcczR_{ij} = \Gamma^R_{\tilde{\chi}^+_{{i}}\tilde{\chi}^-_{{j}}Z_{{\mu}}}  =  & \, g_2 V^*_{i 1} \cos\Theta_W  V_{{j 1}}  + \frac{1}{2} \, V^*_{i 2} \Big(- g_1 \sin\Theta_W   + g_2 \cos\Theta_W  \Big)V_{{j 2}} 
\\ 
\VnnzL_{ij} = \Gamma^L_{\tilde{\chi}^0_{{i}}\tilde{\chi}^0_{{j}}Z_{{\mu}}}  =  & \,-\frac{1}{2} \Big(g_1 \sin\Theta_W   + g_2 \cos\Theta_W  \Big)\Big(N^*_{j 3} N_{{i 3}}  - N^*_{j 4} N_{{i 4}} \Big)\\ 
\VnnzR_{ij} =  \Gamma^R_{\tilde{\chi}^0_{{i}}\tilde{\chi}^0_{{j}}Z_{{\mu}}}  =  & \,\frac{1}{2} \Big(g_1 \sin\Theta_W   + g_2 \cos\Theta_W  \Big)\Big(N^*_{i 3} N_{{j 3}}  - N^*_{i 4} N_{{j 4}} \Big) 
\\ 
\VddzL_{ij} = \Gamma^L_{\bar{d}_{{i \alpha}}d_{{j \beta}}Z_{{\mu}}}  =  & \,\frac{1}{6} \delta_{\alpha \beta} \delta_{i j} \Big(3 g_2 \cos\Theta_W   + g_1 \sin\Theta_W  \Big)\\ 
\VddzR_{ij} = \Gamma^R_{\bar{d}_{{i \alpha}}d_{{j \beta}}Z_{{\mu}}}  =  & \,-\frac{1}{3} g_1 \delta_{\alpha \beta} \delta_{i j} \sin\Theta_W   
\\ 
\VudwpL_{ij} = \Gamma^L_{\bar{u}_{{i \alpha}}d_{{j \beta}}W^+_{{\mu}}}  =  & \,- \frac{1}{\sqrt{2}} g_2 \delta_{\alpha \beta} \sum_{a=1}^{3}U^{d,*}_{L,{j a}} U_{L,{i a}}^{u}  \\ 
\VudwpR_{ij} = \Gamma^R_{\bar{u}_{{i \alpha}}d_{{j \beta}}W^+_{{\mu}}}  =  & \,0 
\\ 
\VvewpL_{ij} = \Gamma^L_{\nu_{{i}}\ell_{{j}}W^+_{{\mu}}}  =  & \,- \frac{1}{\sqrt{2}} g_2 \sum_{a=1}^{3}U^{e,*}_{L,{j a}} U_{{i a}}^{V}  \\ 
\VvewpR_{ij} =  \Gamma^R_{\nu_{{i}}\ell_{{j}}W^+_{{\mu}}}  =  & \,0 
\\ 
\VeezL_{ij} = \Gamma^L_{\bar{\ell}_{{i}}\ell_{{j}}Z_{{\mu}}}  =  & \,\frac{1}{2} \delta_{i j} \Big(- g_1 \sin\Theta_W   + g_2 \cos\Theta_W  \Big)\\ 
\VeezR_{ij} =  \Gamma^R_{\bar{\ell}_{{i}}\ell_{{j}}Z_{{\mu}}}  =  & \,- g_1 \delta_{i j} \sin\Theta_W   
\\ 
\VuuzL_{ij} = \Gamma^L_{\bar{u}_{{i \alpha}}u_{{j \beta}}Z_{{\mu}}}  =  & \,-\frac{1}{6} \delta_{\alpha \beta} \delta_{i j} \Big(3 g_2 \cos\Theta_W   - g_1 \sin\Theta_W  \Big)\\ 
\VuuzR_{ij} =  \Gamma^R_{\bar{u}_{{i \alpha}}u_{{j \beta}}Z_{{\mu}}}  =  & \,\frac{2}{3} g_1 \delta_{\alpha \beta} \delta_{i j} \sin\Theta_W   
\\ 
\VvvzL_{ij} = \Gamma^L_{\nu_{{i}}\nu_{{j}}Z_{{\mu}}}  =  & \,-\frac{1}{2} \Big(g_1 \sin\Theta_W   + g_2 \cos\Theta_W  \Big)\sum_{a=1}^{3}U^{V,*}_{j a} U_{{i a}}^{V}  \\ 
\VvvzR_{ij} = \Gamma^R_{\nu_{{i}}\nu_{{j}}Z_{{\mu}}}  =  & \,\frac{1}{2} \Big(g_1 \sin\Theta_W   + g_2 \cos\Theta_W  \Big)\sum_{a=1}^{3}U^{V,*}_{i a} U_{{j a}}^{V}   
\end{align} 
In addition, we introduce 
\begin{eqnarray}
& \VduwmL_{ij} = (\VudwpL_{ji})^*  \hspace{0.5cm} \VduwmR_{ij} = (\VudwpR_{ji})^* \hspace{1cm}
\VevwmL_{ij} = (\VvewpL_{ij})^*  \hspace{0.5cm} \VevwmR_{ij} = (\VvewpR_{ij})^* & 
\end{eqnarray}

\subsubsection{Scalar vertices}
\begin{align} 
\Vahmhp_{ijk} = \Gamma_{A^0_{{i}}H^-_{{j}}H^+_{{k}}}  = & \, -\frac{i}{4} g_{2}^{2} \Big(v_d Z_{{i 2}}^{A}  + v_u Z_{{i 1}}^{A} \Big)\Big(- Z^{+,*}_{j 1} Z_{{k 2}}^{+}  + Z^{+,*}_{j 2} Z_{{k 1}}^{+} \Big) 
\\ 
\Vaee_{ijk} = \Gamma_{A^0_{{i}}\tilde{e}_{{j}}\tilde{e}^*_{{k}}}  = & \, - \frac{i}{\sqrt{2}} \Big(\sum_{b=1}^{3}Z^{E,*}_{j b} \sum_{a=1}^{3}Z_{{k 3 + a}}^{E} T_{e,{a b}}   Z_{{i 1}}^{A} - \sum_{b=1}^{3}\sum_{a=1}^{3}Z^{E,*}_{j 3 + a} T^*_{e,{a b}}  Z_{{k b}}^{E}  Z_{{i 1}}^{A} \nonumber \\ 
 &+\Big(- \mu \sum_{b=1}^{3}\sum_{a=1}^{3}Y^*_{e,{a b}} Z^{E,*}_{j 3 + a}  Z_{{k b}}^{E}   + \mu^* \sum_{b=1}^{3}Z^{E,*}_{j b} \sum_{a=1}^{3}Y_{e,{a b}} Z_{{k 3 + a}}^{E}   \Big)Z_{{i 2}}^{A} \Big) 
\\ 
\Vavivi_{ijk} = \Gamma_{A^0_{{i}}\nu^i_{{j}}\nu^i_{{k}}}  = 0 \\
\Vavivr_{ijk} =\Gamma_{A^0_{{i}}\nu^i_{{j}}\nu^R_{{k}}}  = & \, - \frac{1}{2 \sqrt{2}} \Big(\mu \sum_{b=1}^{3}Z^{R,*}_{k b} \sum_{a=1}^{3}Y^*_{\nu,{a b}} Z^{i,*}_{j 3 + a}   Z_{{i 1}}^{A} - \mu \sum_{b=1}^{3}Z^{i,*}_{j b} \sum_{a=1}^{3}Y^*_{\nu,{a b}} Z^{R,*}_{k 3 + a}   Z_{{i 1}}^{A} \nonumber \\ 
 &+\mu^* \sum_{b=1}^{3}Z^{R,*}_{k b} \sum_{a=1}^{3}Z^{i,*}_{j 3 + a} Y_{\nu,{a b}}   Z_{{i 1}}^{A} - \mu^* \sum_{b=1}^{3}Z^{i,*}_{j b} \sum_{a=1}^{3}Z^{R,*}_{k 3 + a} Y_{\nu,{a b}}   Z_{{i 1}}^{A} \nonumber \\ 
 &+\sum_{b=1}^{3}Z^{R,*}_{k b} \sum_{a=1}^{3}Z^{i,*}_{j 3 + a} T^*_{\nu,{a b}}   Z_{{i 2}}^{A} - \sum_{b=1}^{3}Z^{i,*}_{j b} \sum_{a=1}^{3}Z^{R,*}_{k 3 + a} T^*_{\nu,{a b}}   Z_{{i 2}}^{A} \nonumber \\ 
 &+\sum_{b=1}^{3}Z^{R,*}_{k b} \sum_{a=1}^{3}Z^{i,*}_{j 3 + a} T_{\nu,{a b}}   Z_{{i 2}}^{A} - \sum_{b=1}^{3}Z^{i,*}_{j b} \sum_{a=1}^{3}Z^{R,*}_{k 3 + a} T_{\nu,{a b}}   Z_{{i 2}}^{A} \nonumber \\ 
 &+\sum_{c=1}^{3}Z^{R,*}_{k c} \sum_{b=1}^{3}Z^{i,*}_{j 6 + b} \sum_{a=1}^{3}Y^*_{\nu,{a c}} M_{R,{a b}}    Z_{{i 2}}^{A} - \sum_{c=1}^{3}Z^{i,*}_{j c} \sum_{b=1}^{3}Z^{R,*}_{k 6 + b} \sum_{a=1}^{3}Y^*_{\nu,{a c}} M_{R,{a b}}    Z_{{i 2}}^{A} \nonumber \\ 
 &- \sum_{c=1}^{3}Z^{R,*}_{k 6 + c} \sum_{b=1}^{3}Z^{i,*}_{j b} \sum_{a=1}^{3}M^*_{\nu,{a c}} Y_{\nu,{a b}}    Z_{{i 2}}^{A} +\sum_{c=1}^{3}Z^{i,*}_{j 6 + c} \sum_{b=1}^{3}Z^{R,*}_{k b} \sum_{a=1}^{3}M^*_{\nu,{a c}} Y_{\nu,{a b}}    Z_{{i 2}}^{A} \Big) 
\\ 
\Vavrvr_{ijk} =  \Gamma_{A^0_{{i}}\nu^R_{{j}}\nu^R_{{k}}}  = 0\\
\Vhhmhp_{ijk} = \Gamma_{h_{{i}}H^-_{{j}}H^+_{{k}}}  = & \, -\frac{1}{4} \Big(Z^{+,*}_{j 1} \Big(Z_{{i 1}}^{H} \Big(\Big(g_{1}^{2} + g_{2}^{2}\Big)v_d Z_{{k 1}}^{+}  + g_{2}^{2} v_u Z_{{k 2}}^{+} \Big) + Z_{{i 2}}^{H} \Big(\Big(- g_{1}^{2}  + g_{2}^{2}\Big)v_u Z_{{k 1}}^{+}  + g_{2}^{2} v_d Z_{{k 2}}^{+} \Big)\Big)\nonumber \\ 
 &+Z^{+,*}_{j 2} \Big(Z_{{i 1}}^{H} \Big(\Big(- g_{1}^{2}  + g_{2}^{2}\Big)v_d Z_{{k 2}}^{+}  + g_{2}^{2} v_u Z_{{k 1}}^{+} \Big) + Z_{{i 2}}^{H} \Big(\Big(g_{1}^{2} + g_{2}^{2}\Big)v_u Z_{{k 2}}^{+}  + g_{2}^{2} v_d Z_{{k 1}}^{+} \Big)\Big)\Big) 
\\ 
\VeeH_{ijk} = \Gamma_{h_{{i}}\tilde{e}_{{j}}\tilde{e}^*_{{k}}}  = & \, -\frac{1}{4} \Big(\Big(- g_{2}^{2}  + g_{1}^{2}\Big)\sum_{a=1}^{3}Z^{E,*}_{j a} Z_{{k a}}^{E}  \Big(v_d Z_{{i 1}}^{H}  - v_u Z_{{i 2}}^{H} \Big)\nonumber \\ 
 &+2 \Big(\sqrt{2} \sum_{b=1}^{3}Z^{E,*}_{j b} \sum_{a=1}^{3}Z_{{k 3 + a}}^{E} T_{e,{a b}}   Z_{{i 1}}^{H} +\sqrt{2} \sum_{b=1}^{3}\sum_{a=1}^{3}Z^{E,*}_{j 3 + a} T^*_{e,{a b}}  Z_{{k b}}^{E}  Z_{{i 1}}^{H} \nonumber \\ 
 &+2 v_d \sum_{c=1}^{3}Z^{E,*}_{j 3 + c} \sum_{b=1}^{3}\sum_{a=1}^{3}Y^*_{e,{c a}} Y_{e,{b a}}  Z_{{k 3 + b}}^{E}   Z_{{i 1}}^{H} +2 v_d \sum_{c=1}^{3}\sum_{b=1}^{3}Z^{E,*}_{j b} \sum_{a=1}^{3}Y^*_{e,{a c}} Y_{e,{a b}}   Z_{{k c}}^{E}  Z_{{i 1}}^{H} \nonumber \\ 
 &- \sqrt{2} \mu^* \sum_{b=1}^{3}Z^{E,*}_{j b} \sum_{a=1}^{3}Y_{e,{a b}} Z_{{k 3 + a}}^{E}   Z_{{i 2}}^{H} - \sqrt{2} \mu \sum_{b=1}^{3}\sum_{a=1}^{3}Y^*_{e,{a b}} Z^{E,*}_{j 3 + a}  Z_{{k b}}^{E}  Z_{{i 2}}^{H} \nonumber \\ 
 &+g_{1}^{2} \sum_{a=1}^{3}Z^{E,*}_{j 3 + a} Z_{{k 3 + a}}^{E}  \Big(- v_d Z_{{i 1}}^{H}  + v_u Z_{{i 2}}^{H} \Big)\Big)\Big) 
\\ 
\Vhvivi_{ijk}  = \Gamma_{h_{{i}}\nu^i_{{j}}\nu^i_{{k}}}  = & \, -\frac{1}{4} \Big(- \sqrt{2} \mu \sum_{b=1}^{3}Z^{i,*}_{k b} \sum_{a=1}^{3}Y^*_{\nu,{a b}} Z^{i,*}_{j 3 + a}   Z_{{i 1}}^{H} - \sqrt{2} \mu \sum_{b=1}^{3}Z^{i,*}_{j b} \sum_{a=1}^{3}Y^*_{\nu,{a b}} Z^{i,*}_{k 3 + a}   Z_{{i 1}}^{H} \nonumber \\ 
 &- \sqrt{2} \mu^* \sum_{b=1}^{3}Z^{i,*}_{k b} \sum_{a=1}^{3}Z^{i,*}_{j 3 + a} Y_{\nu,{a b}}   Z_{{i 1}}^{H} - \sqrt{2} \mu^* \sum_{b=1}^{3}Z^{i,*}_{j b} \sum_{a=1}^{3}Z^{i,*}_{k 3 + a} Y_{\nu,{a b}}   Z_{{i 1}}^{H} \nonumber \\ 
 &+\sqrt{2} \sum_{b=1}^{3}Z^{i,*}_{k b} \sum_{a=1}^{3}Z^{i,*}_{j 3 + a} T^*_{\nu,{a b}}   Z_{{i 2}}^{H} +\sqrt{2} \sum_{b=1}^{3}Z^{i,*}_{j b} \sum_{a=1}^{3}Z^{i,*}_{k 3 + a} T^*_{\nu,{a b}}   Z_{{i 2}}^{H} \nonumber \\ 
 &+\sqrt{2} \sum_{b=1}^{3}Z^{i,*}_{k b} \sum_{a=1}^{3}Z^{i,*}_{j 3 + a} T_{\nu,{a b}}   Z_{{i 2}}^{H} +\sqrt{2} \sum_{b=1}^{3}Z^{i,*}_{j b} \sum_{a=1}^{3}Z^{i,*}_{k 3 + a} T_{\nu,{a b}}   Z_{{i 2}}^{H} \nonumber \\ 
 &+\sqrt{2} \sum_{c=1}^{3}Z^{i,*}_{k c} \sum_{b=1}^{3}Z^{i,*}_{j 6 + b} \sum_{a=1}^{3}Y^*_{\nu,{a c}} M_{R,{a b}}    Z_{{i 2}}^{H} +\sqrt{2} \sum_{c=1}^{3}Z^{i,*}_{j c} \sum_{b=1}^{3}Z^{i,*}_{k 6 + b} \sum_{a=1}^{3}Y^*_{\nu,{a c}} M_{R,{a b}}    Z_{{i 2}}^{H} \nonumber \\ 
 &+\sqrt{2} \sum_{c=1}^{3}Z^{i,*}_{k 6 + c} \sum_{b=1}^{3}Z^{i,*}_{j b} \sum_{a=1}^{3}M^*_{\nu,{a c}} Y_{\nu,{a b}}    Z_{{i 2}}^{H} +\sqrt{2} \sum_{c=1}^{3}Z^{i,*}_{j 6 + c} \sum_{b=1}^{3}Z^{i,*}_{k b} \sum_{a=1}^{3}M^*_{\nu,{a c}} Y_{\nu,{a b}}    Z_{{i 2}}^{H} \nonumber \\ 
 &+2 v_u \sum_{c=1}^{3}Z^{i,*}_{k c} \sum_{b=1}^{3}Z^{i,*}_{j b} \sum_{a=1}^{3}Y^*_{\nu,{a c}} Y_{\nu,{a b}}    Z_{{i 2}}^{H} +2 v_u \sum_{c=1}^{3}Z^{i,*}_{j c} \sum_{b=1}^{3}Z^{i,*}_{k b} \sum_{a=1}^{3}Y^*_{\nu,{a c}} Y_{\nu,{a b}}    Z_{{i 2}}^{H} \nonumber \\ 
 &+2 v_u \sum_{c=1}^{3}Z^{i,*}_{k 3 + c} \sum_{b=1}^{3}Z^{i,*}_{j 3 + b} \sum_{a=1}^{3}Y^*_{\nu,{c a}} Y_{\nu,{b a}}    Z_{{i 2}}^{H} \nonumber \\ 
 &+2 v_u \sum_{c=1}^{3}Z^{i,*}_{j 3 + c} \sum_{b=1}^{3}Z^{i,*}_{k 3 + b} \sum_{a=1}^{3}Y^*_{\nu,{c a}} Y_{\nu,{b a}}    Z_{{i 2}}^{H} \nonumber \\ 
 &+\Big(g_{1}^{2} + g_{2}^{2}\Big)\sum_{a=1}^{3}Z^{i,*}_{j a} Z^{i,*}_{k a}  \Big(v_d Z_{{i 1}}^{H}  - v_u Z_{{i 2}}^{H} \Big)\Big) 
\\ 
\Vhvivr_{ijk}  = \Gamma_{h_{{i}}\nu^i_{{j}}\nu^R_{{k}}}  = & \, 0 \\ 
\Vhvrvr_{ijk} =  \Gamma_{h_{{i}}\nu^R_{{j}}\nu^R_{{k}}}  = & \, -\frac{1}{4} \Big(- \sqrt{2} \mu \sum_{b=1}^{3}Z^{R,*}_{k b} \sum_{a=1}^{3}Y^*_{\nu,{a b}} Z^{R,*}_{j 3 + a}   Z_{{i 1}}^{H} - \sqrt{2} \mu \sum_{b=1}^{3}Z^{R,*}_{j b} \sum_{a=1}^{3}Y^*_{\nu,{a b}} Z^{R,*}_{k 3 + a}   Z_{{i 1}}^{H} \nonumber \\ 
 &- \sqrt{2} \mu^* \sum_{b=1}^{3}Z^{R,*}_{k b} \sum_{a=1}^{3}Z^{R,*}_{j 3 + a} Y_{\nu,{a b}}   Z_{{i 1}}^{H} - \sqrt{2} \mu^* \sum_{b=1}^{3}Z^{R,*}_{j b} \sum_{a=1}^{3}Z^{R,*}_{k 3 + a} Y_{\nu,{a b}}   Z_{{i 1}}^{H} \nonumber \\ 
 &+\sqrt{2} \sum_{b=1}^{3}Z^{R,*}_{k b} \sum_{a=1}^{3}Z^{R,*}_{j 3 + a} T^*_{\nu,{a b}}   Z_{{i 2}}^{H} +\sqrt{2} \sum_{b=1}^{3}Z^{R,*}_{j b} \sum_{a=1}^{3}Z^{R,*}_{k 3 + a} T^*_{\nu,{a b}}   Z_{{i 2}}^{H} \nonumber \\ 
 &+\sqrt{2} \sum_{b=1}^{3}Z^{R,*}_{k b} \sum_{a=1}^{3}Z^{R,*}_{j 3 + a} T_{\nu,{a b}}   Z_{{i 2}}^{H} +\sqrt{2} \sum_{b=1}^{3}Z^{R,*}_{j b} \sum_{a=1}^{3}Z^{R,*}_{k 3 + a} T_{\nu,{a b}}   Z_{{i 2}}^{H} \nonumber \\ 
 &+\sqrt{2} \sum_{c=1}^{3}Z^{R,*}_{k c} \sum_{b=1}^{3}Z^{R,*}_{j 6 + b} \sum_{a=1}^{3}Y^*_{\nu,{a c}} M_{R,{a b}}    Z_{{i 2}}^{H} +\sqrt{2} \sum_{c=1}^{3}Z^{R,*}_{j c} \sum_{b=1}^{3}Z^{R,*}_{k 6 + b} \sum_{a=1}^{3}Y^*_{\nu,{a c}} M_{R,{a b}}    Z_{{i 2}}^{H} \nonumber \\ 
 &+\sqrt{2} \sum_{c=1}^{3}Z^{R,*}_{k 6 + c} \sum_{b=1}^{3}Z^{R,*}_{j b} \sum_{a=1}^{3}M^*_{\nu,{a c}} Y_{\nu,{a b}}    Z_{{i 2}}^{H} +\sqrt{2} \sum_{c=1}^{3}Z^{R,*}_{j 6 + c} \sum_{b=1}^{3}Z^{R,*}_{k b} \sum_{a=1}^{3}M^*_{\nu,{a c}} Y_{\nu,{a b}}    Z_{{i 2}}^{H} \nonumber \\ 
 &+2 v_u \sum_{c=1}^{3}Z^{R,*}_{k c} \sum_{b=1}^{3}Z^{R,*}_{j b} \sum_{a=1}^{3}Y^*_{\nu,{a c}} Y_{\nu,{a b}}    Z_{{i 2}}^{H} +2 v_u \sum_{c=1}^{3}Z^{R,*}_{j c} \sum_{b=1}^{3}Z^{R,*}_{k b} \sum_{a=1}^{3}Y^*_{\nu,{a c}} Y_{\nu,{a b}}    Z_{{i 2}}^{H} \nonumber \\ 
 &+2 v_u \sum_{c=1}^{3}Z^{R,*}_{k 3 + c} \sum_{b=1}^{3}Z^{R,*}_{j 3 + b} \sum_{a=1}^{3}Y^*_{\nu,{c a}} Y_{\nu,{b a}}    Z_{{i 2}}^{H} \nonumber \\ 
 &+2 v_u \sum_{c=1}^{3}Z^{R,*}_{j 3 + c} \sum_{b=1}^{3}Z^{R,*}_{k 3 + b} \sum_{a=1}^{3}Y^*_{\nu,{c a}} Y_{\nu,{b a}}    Z_{{i 2}}^{H} \nonumber \\ 
 &+\Big(g_{1}^{2} + g_{2}^{2}\Big)\sum_{a=1}^{3}Z^{R,*}_{j a} Z^{R,*}_{k a}  \Big(v_d Z_{{i 1}}^{H}  - v_u Z_{{i 2}}^{H} \Big)\Big) 
\end{align} 

\subsubsection{Scalar-Vector vertices}
\begin{align} 
\Vahmwp_{ij} = \Gamma_{A^0_{{i}}H^-_{{j}}W^+_{{\mu}}}  = & \, \frac{i}{2} g_2 \Big(Z^{+,*}_{j 1} Z_{{i 1}}^{A}  + Z^{+,*}_{j 2} Z_{{i 2}}^{A} \Big) 
\\ 
\Vhhmwp_{ij} = \Gamma_{h_{{i}}H^-_{{j}}W^+_{{\mu}}}  = & \, \frac{1}{2} g_2 \Big(Z^{+,*}_{j 2} Z_{{i 2}}^{H}-Z^{+,*}_{j 1} Z_{{i 1}}^{H}  \Big) 
\\ 
\VhmhpP_{ij} = \Gamma_{H^-_{{i}}H^+_{{j}}\gamma_{{\mu}}}  = & \, -e \delta_{ij}%\, -\frac{1}{2} \Big(g_1 \cos\Theta_W   + g_2 \sin\Theta_W  \Big)\delta_{ij}
\\ 
\Vhmhpz_{ij} =\Gamma_{H^-_{{i}}H^+_{{j}}Z_{{\mu}}}  = & \, \frac{1}{2} \Big(g_1 \sin\Theta_W   - g_2 \cos\Theta_W  \Big)\delta_{ij} 
\\ 
\Veep_{ij} = \Gamma_{\tilde{e}_{{i}}\tilde{e}^*_{{j}}\gamma_{{\mu}}}  = & \, -e \delta_{ij}
\\ 
\VeeZ_{ij} = \Gamma_{\tilde{e}_{{i}}\tilde{e}^*_{{j}}Z_{{\mu}}}  = & \,  g_1 \sin\Theta_W  \sum_{a=1}^{3}Z^{E,*}_{i 3 + a} Z_{{j 3 + a}}^{E} +\frac{1}{2} \, \Big(g_1 \sin\Theta_W  - g_2 \cos\Theta_W  \Big)\sum_{a=1}^{3}Z^{E,*}_{i a} Z_{{j a}}^{E}
\\ 
\VvivrZ_{ij} = \Gamma_{\nu^i_{{i}}\nu^R_{{j}}Z_{{\mu}}}  = & \, \frac{i}{2} \Big(g_1 \sin\Theta_W   + g_2 \cos\Theta_W  \Big)\sum_{a=1}^{3}Z^{i,*}_{i a} Z^{R,*}_{j a}   
\\ 
\Vhwmwp_{i} = \Gamma_{h_{{i}}W^+_{{\sigma}}W^-_{{\mu}}}  = & \, -\frac{1}{2} g_{2}^{2} \Big(v_d Z_{{i 1}}^{H}  + v_u Z_{{i 2}}^{H} \Big) 
\\ 
\VhmwpP_{i} = \Gamma_{H^-_{{i}}W^+_{{\sigma}}\gamma_{{\mu}}}  = & \, -\frac{i}{2 \sin\Theta_W} e^2 \Big(v_d Z^{+,*}_{i 1}  - v_u Z^{+,*}_{i 2} \Big)
\\ 
\Vhmwpz_{i}  = \Gamma_{H^-_{{i}}W^+_{{\sigma}}Z_{{\mu}}}  = & \, \frac{1}{2} g_1 g_2 \Big(v_d Z^{+,*}_{i 1}  - v_u Z^{+,*}_{i 2} \Big)\sin\Theta_W   
\\ 
\end{align} 
In addition, we introduce 
\begin{eqnarray}
& \Vahpwm_{ij} = (\Vahmwp_{ij})^* \hspace{0.5cm} \Vhhpwm_{ij} = (\Vhhmwp_{ij})^* 
\hspace{0.5cm}  \VhpPwm_i = (\VhmwpP_{i})^*  \hspace{0.5cm}  \Vhpwmz_i = (\Vhmwpz_i)^*  &
\end{eqnarray}

\subsubsection{Vector vertices}
\begin{align} 
\VwpPwm = \Gamma_{W^+_{{\rho}}\gamma_{{\sigma}}W^-_{{\mu}}}  = & \,  g_2 \sin\Theta_W   
\\ 
\Vwpwmz = \Gamma_{W^+_{{\rho}}W^-_{{\sigma}}Z_{{\mu}}}  = & \, - g_2 \cos\Theta_W   
\end{align}

\section{Renormalization Group Equations}
\label{app:rges}

We give in the following the 2-loop RGEs for the considered model. For
parameters present in the MSSM we show only the difference with
respect to the MSSM RGEs. In general, the RGEs for a parameter $X$ are
defined by
\begin{equation}
\frac{d}{dt} X = \frac{1}{16 \pi^2} \beta_X^{(1)} +  \frac{1}{(16 \pi^2)} \beta_X^{(2)} 
\end{equation}
Here, $t = \log \left(Q/M\right)$, with $Q$ the renormalization scale
and $M$ a reference scale.

\subsection*{Gauge Couplings}
\begin{align} 
\Delta \beta_{g_1}^{(2)} & =  
-\frac{6}{5} g_{1}^{3} \mbox{Tr}\Big({Y_\nu  Y_{\nu}^{\dagger}}\Big) \\ 
\Delta \beta_{g_2}^{(2)} & =  
-2 g_{2}^{3} \mbox{Tr}\Big({Y_\nu  Y_{\nu}^{\dagger}}\Big) 
\end{align}

\subsection*{Gaugino Mass Parameters}
\begin{align} 
\Delta \beta_{M_1}^{(2)} & =  
-\frac{12}{5} g_{1}^{2} \Big(M_1 \mbox{Tr}\Big({Y_\nu  Y_{\nu}^{\dagger}}\Big)  - \mbox{Tr}\Big({Y_{\nu}^{\dagger}  T_\nu}\Big) \Big)\\ 
\Delta \beta_{M_2}^{(2)} & =  
4 g_{2}^{2} \Big(- M_2 \mbox{Tr}\Big({Y_\nu  Y_{\nu}^{\dagger}}\Big)  + \mbox{Tr}\Big({Y_{\nu}^{\dagger}  T_\nu}\Big)\Big)
\end{align}

\subsection*{Trilinear Superpotential Parameters}
\begin{align} 
\Delta \beta_{Y_d}^{(2)} & =  
- Y_d \mbox{Tr}\Big({Y_e  Y_{\nu}^{\dagger}  Y_\nu  Y_{e}^{\dagger}}\Big)  - {Y_d  Y_{u}^{\dagger}  Y_u} \mbox{Tr}\Big({Y_\nu  Y_{\nu}^{\dagger}}\Big) \\ 
\Delta \beta_{Y_e}^{(1)} & =  
{Y_e  Y_{\nu}^{\dagger}  Y_\nu}\\ 
\Delta \beta_{Y_e}^{(2)} & =  
-2 {Y_e  Y_{\nu}^{\dagger}  Y_\nu  Y_{e}^{\dagger}  Y_e} -2 {Y_e  Y_{\nu}^{\dagger}  Y_\nu  Y_{\nu}^{\dagger}  Y_\nu} -3 {Y_e  Y_{\nu}^{\dagger}  Y_\nu} \mbox{Tr}\Big({Y_u  Y_{u}^{\dagger}}\Big) \nonumber \\ 
 &- {Y_e  Y_{\nu}^{\dagger}  Y_\nu} \mbox{Tr}\Big({Y_\nu  Y_{\nu}^{\dagger}}\Big) - Y_e \mbox{Tr}\Big({Y_e  Y_{\nu}^{\dagger}  Y_\nu  Y_{e}^{\dagger}}\Big) \\ 
\Delta \beta_{Y_u}^{(1)} & =  
Y_u \mbox{Tr}\Big({Y_\nu  Y_{\nu}^{\dagger}}\Big) \\ 
\Delta \beta_{Y_u}^{(2)} & =  
-3 Y_u \mbox{Tr}\Big({Y_\nu  Y_{\nu}^{\dagger}  Y_\nu  Y_{\nu}^{\dagger}}\Big)  -3 {Y_u  Y_{u}^{\dagger}  Y_u} \mbox{Tr}\Big({Y_\nu  Y_{\nu}^{\dagger}}\Big)  - Y_u \mbox{Tr}\Big({Y_e  Y_{\nu}^{\dagger}  Y_\nu  Y_{e}^{\dagger}}\Big) \\ 
\beta_{Y_\nu}^{(1)} & =  
3 {Y_\nu  Y_{\nu}^{\dagger}  Y_\nu}  + Y_\nu \Big(-3 g_{2}^{2}  + 3 \mbox{Tr}\Big({Y_u  Y_{u}^{\dagger}}\Big)  -\frac{3}{5} g_{1}^{2}  + \mbox{Tr}\Big({Y_\nu  Y_{\nu}^{\dagger}}\Big)\Big) + {Y_\nu  Y_{e}^{\dagger}  Y_e}\\ 
\beta_{Y_\nu}^{(2)} & =  
+\frac{6}{5} g_{1}^{2} {Y_\nu  Y_{\nu}^{\dagger}  Y_\nu} +6 g_{2}^{2} {Y_\nu  Y_{\nu}^{\dagger}  Y_\nu} -2 {Y_\nu  Y_{e}^{\dagger}  Y_e  Y_{e}^{\dagger}  Y_e} -2 {Y_\nu  Y_{e}^{\dagger}  Y_e  Y_{\nu}^{\dagger}  Y_\nu} \nonumber \\ 
 &-4 {Y_\nu  Y_{\nu}^{\dagger}  Y_\nu  Y_{\nu}^{\dagger}  Y_\nu} +{Y_\nu  Y_{e}^{\dagger}  Y_e} \Big(-3 \mbox{Tr}\Big({Y_d  Y_{d}^{\dagger}}\Big)  + \frac{6}{5} g_{1}^{2}  - \mbox{Tr}\Big({Y_e  Y_{e}^{\dagger}}\Big) \Big)-9 {Y_\nu  Y_{\nu}^{\dagger}  Y_\nu} \mbox{Tr}\Big({Y_u  Y_{u}^{\dagger}}\Big) \nonumber \\ 
 &-3 {Y_\nu  Y_{\nu}^{\dagger}  Y_\nu} \mbox{Tr}\Big({Y_\nu  Y_{\nu}^{\dagger}}\Big) \nonumber \\ 
 &+Y_\nu \Big(\frac{207}{50} g_{1}^{4} +\frac{9}{5} g_{1}^{2} g_{2}^{2} +\frac{15}{2} g_{2}^{4} +\frac{4}{5} \Big(20 g_{3}^{2}  + g_{1}^{2}\Big)\mbox{Tr}\Big({Y_u  Y_{u}^{\dagger}}\Big) -3 \mbox{Tr}\Big({Y_d  Y_{u}^{\dagger}  Y_u  Y_{d}^{\dagger}}\Big) \nonumber \\ 
 &- \mbox{Tr}\Big({Y_e  Y_{\nu}^{\dagger}  Y_\nu  Y_{e}^{\dagger}}\Big) -9 \mbox{Tr}\Big({Y_u  Y_{u}^{\dagger}  Y_u  Y_{u}^{\dagger}}\Big) -3 \mbox{Tr}\Big({Y_\nu  Y_{\nu}^{\dagger}  Y_\nu  Y_{\nu}^{\dagger}}\Big) \Big)
\end{align} 

\subsection*{Bilinear Superpotential Parameters}
\begin{align} 
\Delta \beta_{\mu}^{(1)} & =  
\mu \mbox{Tr}\Big({Y_\nu  Y_{\nu}^{\dagger}}\Big) \\ 
\Delta \beta_{\mu}^{(2)} & =  
- \mu \Big(2 \mbox{Tr}\Big({Y_e  Y_{\nu}^{\dagger}  Y_\nu  Y_{e}^{\dagger}}\Big)  + 3 \mbox{Tr}\Big({Y_\nu  Y_{\nu}^{\dagger}  Y_\nu  Y_{\nu}^{\dagger}}\Big) \Big)\\ 
\beta_{\mu_X}^{(1)} & = \beta_{\mu_X}^{(2)} = 0\\ 
\beta_{M_R}^{(1)} & =  
2 {Y_\nu  Y_{\nu}^{\dagger}  M_R} \\ 
\beta_{M_R}^{(2)} & =  
-2 \Big({Y_\nu  Y_{e}^{\dagger}  Y_e  Y_{\nu}^{\dagger}  M_R} + {Y_\nu  Y_{\nu}^{\dagger}  Y_\nu  Y_{\nu}^{\dagger}  M_R}\Big)\nonumber \\ 
 &+{Y_\nu  Y_{\nu}^{\dagger}  M_R} \Big(-2 \mbox{Tr}\Big({Y_\nu  Y_{\nu}^{\dagger}}\Big)  + 6 g_{2}^{2}  -6 \mbox{Tr}\Big({Y_u  Y_{u}^{\dagger}}\Big)  + \frac{6}{5} g_{1}^{2} \Big)
\end{align}

\subsection*{Trilinear Soft-Breaking Parameters}
\begin{align} 
\Delta \beta_{T_d}^{(2)} & =  
-2 {Y_d  Y_{u}^{\dagger}  T_u} \mbox{Tr}\Big({Y_\nu  Y_{\nu}^{\dagger}}\Big) - {T_d  Y_{u}^{\dagger}  Y_u} \mbox{Tr}\Big({Y_\nu  Y_{\nu}^{\dagger}}\Big) -2 {Y_d  Y_{u}^{\dagger}  Y_u} \mbox{Tr}\Big({Y_{\nu}^{\dagger}  T_\nu}\Big) \nonumber \\ 
 &- T_d \mbox{Tr}\Big({Y_e  Y_{\nu}^{\dagger}  Y_\nu  Y_{e}^{\dagger}}\Big) -2 Y_d \mbox{Tr}\Big({Y_e  Y_{\nu}^{\dagger}  T_\nu  Y_{e}^{\dagger}}\Big) -2 Y_d \mbox{Tr}\Big({Y_\nu  Y_{e}^{\dagger}  T_e  Y_{\nu}^{\dagger}}\Big) \\ 
\Delta \beta_{T_e}^{(1)} & =  
2 {Y_e  Y_{\nu}^{\dagger}  T_\nu}  + {T_e  Y_{\nu}^{\dagger}  Y_\nu}\\ 
\Delta \beta_{T_e}^{(2)} & =  
-2 {Y_e  Y_{\nu}^{\dagger}  Y_\nu  Y_{e}^{\dagger}  T_e} -4 {Y_e  Y_{\nu}^{\dagger}  Y_\nu  Y_{\nu}^{\dagger}  T_\nu} -4 {Y_e  Y_{\nu}^{\dagger}  T_\nu  Y_{e}^{\dagger}  Y_e} -4 {Y_e  Y_{\nu}^{\dagger}  T_\nu  Y_{\nu}^{\dagger}  Y_\nu} \nonumber \\ 
 &-4 {T_e  Y_{\nu}^{\dagger}  Y_\nu  Y_{e}^{\dagger}  Y_e} -2 {T_e  Y_{\nu}^{\dagger}  Y_\nu  Y_{\nu}^{\dagger}  Y_\nu} -6 {Y_e  Y_{\nu}^{\dagger}  T_\nu} \mbox{Tr}\Big({Y_u  Y_{u}^{\dagger}}\Big) \nonumber \\ 
 &-3 {T_e  Y_{\nu}^{\dagger}  Y_\nu} \mbox{Tr}\Big({Y_u  Y_{u}^{\dagger}}\Big) -2 {Y_e  Y_{\nu}^{\dagger}  T_\nu} \mbox{Tr}\Big({Y_\nu  Y_{\nu}^{\dagger}}\Big) - {T_e  Y_{\nu}^{\dagger}  Y_\nu} \mbox{Tr}\Big({Y_\nu  Y_{\nu}^{\dagger}}\Big) \nonumber \\ 
 &-6 {Y_e  Y_{\nu}^{\dagger}  Y_\nu} \mbox{Tr}\Big({Y_{u}^{\dagger}  T_u}\Big) -2 {Y_e  Y_{\nu}^{\dagger}  Y_\nu} \mbox{Tr}\Big({Y_{\nu}^{\dagger}  T_\nu}\Big) - T_e \mbox{Tr}\Big({Y_e  Y_{\nu}^{\dagger}  Y_\nu  Y_{e}^{\dagger}}\Big) \nonumber \\ 
 &-2 Y_e \mbox{Tr}\Big({Y_e  Y_{\nu}^{\dagger}  T_\nu  Y_{e}^{\dagger}}\Big) -2 Y_e \mbox{Tr}\Big({Y_\nu  Y_{e}^{\dagger}  T_e  Y_{\nu}^{\dagger}}\Big) \\ 
\Delta \beta_{T_u}^{(1)} & =  
2 Y_u \mbox{Tr}\Big({Y_{\nu}^{\dagger}  T_\nu}\Big)  + T_u \mbox{Tr}\Big({Y_\nu  Y_{\nu}^{\dagger}}\Big) \\ 
\Delta \beta_{T_u}^{(2)} & =  
-4 {Y_u  Y_{u}^{\dagger}  T_u} \mbox{Tr}\Big({Y_\nu  Y_{\nu}^{\dagger}}\Big) -5 {T_u  Y_{u}^{\dagger}  Y_u} \mbox{Tr}\Big({Y_\nu  Y_{\nu}^{\dagger}}\Big) -6 {Y_u  Y_{u}^{\dagger}  Y_u} \mbox{Tr}\Big({Y_{\nu}^{\dagger}  T_\nu}\Big) \nonumber \\ 
 &- T_u \mbox{Tr}\Big({Y_e  Y_{\nu}^{\dagger}  Y_\nu  Y_{e}^{\dagger}}\Big) -2 Y_u \mbox{Tr}\Big({Y_e  Y_{\nu}^{\dagger}  T_\nu  Y_{e}^{\dagger}}\Big) -2 Y_u \mbox{Tr}\Big({Y_\nu  Y_{e}^{\dagger}  T_e  Y_{\nu}^{\dagger}}\Big) \nonumber \\ 
 &-3 T_u \mbox{Tr}\Big({Y_\nu  Y_{\nu}^{\dagger}  Y_\nu  Y_{\nu}^{\dagger}}\Big) -12 Y_u \mbox{Tr}\Big({Y_\nu  Y_{\nu}^{\dagger}  T_\nu  Y_{\nu}^{\dagger}}\Big) \\ 
\beta_{T_\nu}^{(1)} & =  
+2 {Y_\nu  Y_{e}^{\dagger}  T_e} +4 {Y_\nu  Y_{\nu}^{\dagger}  T_\nu} +{T_\nu  Y_{e}^{\dagger}  Y_e}+5 {T_\nu  Y_{\nu}^{\dagger}  Y_\nu} -\frac{3}{5} g_{1}^{2} T_\nu -3 g_{2}^{2} T_\nu \nonumber \\ 
 &+3 T_\nu \mbox{Tr}\Big({Y_u  Y_{u}^{\dagger}}\Big) +T_\nu \mbox{Tr}\Big({Y_\nu  Y_{\nu}^{\dagger}}\Big) +Y_\nu \Big(2 \mbox{Tr}\Big({Y_{\nu}^{\dagger}  T_\nu}\Big)  + 6 g_{2}^{2} M_2  + 6 \mbox{Tr}\Big({Y_{u}^{\dagger}  T_u}\Big)  + \frac{6}{5} g_{1}^{2} M_1 \Big)\\ 
\beta_{T_\nu}^{(2)} & =  
+\frac{12}{5} g_{1}^{2} {Y_\nu  Y_{e}^{\dagger}  T_e} -\frac{12}{5} g_{1}^{2} M_1 {Y_\nu  Y_{\nu}^{\dagger}  Y_\nu} -12 g_{2}^{2} M_2 {Y_\nu  Y_{\nu}^{\dagger}  Y_\nu} +\frac{6}{5} g_{1}^{2} {Y_\nu  Y_{\nu}^{\dagger}  T_\nu} \nonumber \\ 
 &+6 g_{2}^{2} {Y_\nu  Y_{\nu}^{\dagger}  T_\nu} +\frac{6}{5} g_{1}^{2} {T_\nu  Y_{e}^{\dagger}  Y_e} +\frac{12}{5} g_{1}^{2} {T_\nu  Y_{\nu}^{\dagger}  Y_\nu} +12 g_{2}^{2} {T_\nu  Y_{\nu}^{\dagger}  Y_\nu} \nonumber \\ 
 &-4 {Y_\nu  Y_{e}^{\dagger}  Y_e  Y_{e}^{\dagger}  T_e} -2 {Y_\nu  Y_{e}^{\dagger}  Y_e  Y_{\nu}^{\dagger}  T_\nu} -4 {Y_\nu  Y_{e}^{\dagger}  T_e  Y_{e}^{\dagger}  Y_e} -4 {Y_\nu  Y_{e}^{\dagger}  T_e  Y_{\nu}^{\dagger}  Y_\nu} \nonumber \\ 
 &-6 {Y_\nu  Y_{\nu}^{\dagger}  Y_\nu  Y_{\nu}^{\dagger}  T_\nu} -8 {Y_\nu  Y_{\nu}^{\dagger}  T_\nu  Y_{\nu}^{\dagger}  Y_\nu} -2 {T_\nu  Y_{e}^{\dagger}  Y_e  Y_{e}^{\dagger}  Y_e} -4 {T_\nu  Y_{e}^{\dagger}  Y_e  Y_{\nu}^{\dagger}  Y_\nu} \nonumber \\ 
 &-6 {T_\nu  Y_{\nu}^{\dagger}  Y_\nu  Y_{\nu}^{\dagger}  Y_\nu} +\frac{207}{50} g_{1}^{4} T_\nu +\frac{9}{5} g_{1}^{2} g_{2}^{2} T_\nu +\frac{15}{2} g_{2}^{4} T_\nu -6 {Y_\nu  Y_{e}^{\dagger}  T_e} \mbox{Tr}\Big({Y_d  Y_{d}^{\dagger}}\Big) \nonumber \\ 
 &-3 {T_\nu  Y_{e}^{\dagger}  Y_e} \mbox{Tr}\Big({Y_d  Y_{d}^{\dagger}}\Big) -2 {Y_\nu  Y_{e}^{\dagger}  T_e} \mbox{Tr}\Big({Y_e  Y_{e}^{\dagger}}\Big) - {T_\nu  Y_{e}^{\dagger}  Y_e} \mbox{Tr}\Big({Y_e  Y_{e}^{\dagger}}\Big) \nonumber \\ 
 &-12 {Y_\nu  Y_{\nu}^{\dagger}  T_\nu} \mbox{Tr}\Big({Y_u  Y_{u}^{\dagger}}\Big) -15 {T_\nu  Y_{\nu}^{\dagger}  Y_\nu} \mbox{Tr}\Big({Y_u  Y_{u}^{\dagger}}\Big) +\frac{4}{5} g_{1}^{2} T_\nu \mbox{Tr}\Big({Y_u  Y_{u}^{\dagger}}\Big) \nonumber \\ 
 &+16 g_{3}^{2} T_\nu \mbox{Tr}\Big({Y_u  Y_{u}^{\dagger}}\Big) -4 {Y_\nu  Y_{\nu}^{\dagger}  T_\nu} \mbox{Tr}\Big({Y_\nu  Y_{\nu}^{\dagger}}\Big) -5 {T_\nu  Y_{\nu}^{\dagger}  Y_\nu} \mbox{Tr}\Big({Y_\nu  Y_{\nu}^{\dagger}}\Big) \nonumber \\ 
 &-\frac{2}{5} {Y_\nu  Y_{e}^{\dagger}  Y_e} \Big(15 \mbox{Tr}\Big({Y_{d}^{\dagger}  T_d}\Big)  + 5 \mbox{Tr}\Big({Y_{e}^{\dagger}  T_e}\Big)  + 6 g_{1}^{2} M_1 \Big)-18 {Y_\nu  Y_{\nu}^{\dagger}  Y_\nu} \mbox{Tr}\Big({Y_{u}^{\dagger}  T_u}\Big) \nonumber \\ 
 &-6 {Y_\nu  Y_{\nu}^{\dagger}  Y_\nu} \mbox{Tr}\Big({Y_{\nu}^{\dagger}  T_\nu}\Big) -3 T_\nu \mbox{Tr}\Big({Y_d  Y_{u}^{\dagger}  Y_u  Y_{d}^{\dagger}}\Big) - T_\nu \mbox{Tr}\Big({Y_e  Y_{\nu}^{\dagger}  Y_\nu  Y_{e}^{\dagger}}\Big) \nonumber \\ 
 &-9 T_\nu \mbox{Tr}\Big({Y_u  Y_{u}^{\dagger}  Y_u  Y_{u}^{\dagger}}\Big) -3 T_\nu \mbox{Tr}\Big({Y_\nu  Y_{\nu}^{\dagger}  Y_\nu  Y_{\nu}^{\dagger}}\Big) \nonumber \\ 
 &-\frac{2}{25} Y_\nu \Big(207 g_{1}^{4} M_1 +45 g_{1}^{2} g_{2}^{2} M_1 +45 g_{1}^{2} g_{2}^{2} M_2 +375 g_{2}^{4} M_2 +20 \Big(20 g_{3}^{2} M_3  + g_{1}^{2} M_1 \Big)\mbox{Tr}\Big({Y_u  Y_{u}^{\dagger}}\Big) \nonumber \\ 
 &-20 \Big(20 g_{3}^{2}  + g_{1}^{2}\Big)\mbox{Tr}\Big({Y_{u}^{\dagger}  T_u}\Big) +75 \mbox{Tr}\Big({Y_d  Y_{u}^{\dagger}  T_u  Y_{d}^{\dagger}}\Big) +25 \mbox{Tr}\Big({Y_e  Y_{\nu}^{\dagger}  T_\nu  Y_{e}^{\dagger}}\Big) \nonumber \\ 
 &+75 \mbox{Tr}\Big({Y_u  Y_{d}^{\dagger}  T_d  Y_{u}^{\dagger}}\Big) +450 \mbox{Tr}\Big({Y_u  Y_{u}^{\dagger}  T_u  Y_{u}^{\dagger}}\Big) +25 \mbox{Tr}\Big({Y_\nu  Y_{e}^{\dagger}  T_e  Y_{\nu}^{\dagger}}\Big) +150 \mbox{Tr}\Big({Y_\nu  Y_{\nu}^{\dagger}  T_\nu  Y_{\nu}^{\dagger}}\Big) \Big)
\end{align} 

\subsection*{Bilinear Soft-Breaking Parameters}
\begin{align} 
\Delta \beta_{B_\mu}^{(1)} & =  
2 \mu \mbox{Tr}\Big({Y_{\nu}^{\dagger}  T_\nu}\Big)  + B_{\mu} \mbox{Tr}\Big({Y_\nu  Y_{\nu}^{\dagger}}\Big) \\ 
\Delta \beta_{B_\mu}^{(2)} & =  
- B_{\mu} \Big(2 \mbox{Tr}\Big({Y_e  Y_{\nu}^{\dagger}  Y_\nu  Y_{e}^{\dagger}}\Big)  + 3 \mbox{Tr}\Big({Y_\nu  Y_{\nu}^{\dagger}  Y_\nu  Y_{\nu}^{\dagger}}\Big) \Big)\nonumber \\ 
 &-4 \mu \Big(3 \mbox{Tr}\Big({Y_\nu  Y_{\nu}^{\dagger}  T_\nu  Y_{\nu}^{\dagger}}\Big)  + \mbox{Tr}\Big({Y_e  Y_{\nu}^{\dagger}  T_\nu  Y_{e}^{\dagger}}\Big) + \mbox{Tr}\Big({Y_\nu  Y_{e}^{\dagger}  T_e  Y_{\nu}^{\dagger}}\Big)\Big)\\ 
\beta_{B_{\mu_X}}^{(1)} & =  \beta_{B_{\mu_X}}^{(2)} = 0\\ 
\beta_{B_{M_R}}^{(1)} & =  
2 \Big(2 {T_\nu  Y_{\nu}^{\dagger}  M_R}  + {Y_\nu  Y_{\nu}^{\dagger}  B_{M_R}}\Big)\\ 
\beta_{B_{M_R}}^{(2)} & =  
-\frac{2}{5} \Big(-6 g_{1}^{2} {T_\nu  Y_{\nu}^{\dagger}  M_R} -30 g_{2}^{2} {T_\nu  Y_{\nu}^{\dagger}  M_R} +5 {Y_\nu  Y_{e}^{\dagger}  Y_e  Y_{\nu}^{\dagger}  B_{M_R}} +10 {Y_\nu  Y_{e}^{\dagger}  T_e  Y_{\nu}^{\dagger}  M_R} \nonumber \\ 
 &+5 {Y_\nu  Y_{\nu}^{\dagger}  Y_\nu  Y_{\nu}^{\dagger}  B_{M_R}} +10 {Y_\nu  Y_{\nu}^{\dagger}  T_\nu  Y_{\nu}^{\dagger}  M_R} +10 {T_\nu  Y_{e}^{\dagger}  Y_e  Y_{\nu}^{\dagger}  M_R} +10 {T_\nu  Y_{\nu}^{\dagger}  Y_\nu  Y_{\nu}^{\dagger}  M_R} \nonumber \\ 
 &+30 {T_\nu  Y_{\nu}^{\dagger}  M_R} \mbox{Tr}\Big({Y_u  Y_{u}^{\dagger}}\Big) +10 {T_\nu  Y_{\nu}^{\dagger}  M_R} \mbox{Tr}\Big({Y_\nu  Y_{\nu}^{\dagger}}\Big) \nonumber \\ 
 &+{Y_\nu  Y_{\nu}^{\dagger}  B_{M_R}} \Big(15 \mbox{Tr}\Big({Y_u  Y_{u}^{\dagger}}\Big)  -3 \Big(5 g_{2}^{2}  + g_{1}^{2}\Big) + 5 \mbox{Tr}\Big({Y_\nu  Y_{\nu}^{\dagger}}\Big) \Big)\nonumber \\ 
 &+2 {Y_\nu  Y_{\nu}^{\dagger}  M_R} \Big(15 g_{2}^{2} M_2  + 15 \mbox{Tr}\Big({Y_{u}^{\dagger}  T_u}\Big)  + 3 g_{1}^{2} M_1  + 5 \mbox{Tr}\Big({Y_{\nu}^{\dagger}  T_\nu}\Big) \Big)\Big)
\end{align} 

\subsection*{Soft-Breaking Scalar Masses}
The RGEs of the soft SUSY breaking masses are usually written in terms
of a set of traces (see e.g. \cite{Martin:1993zk}). In the model
considered here, only one changes with respect to the MSSM:
\begin{equation} 
\Delta \sigma_{3,1} = \frac{1}{20} \frac{1}{\sqrt{15}} g_1 \Big(-30 m_{H_u}^2 \mbox{Tr}\Big({Y_\nu  Y_{\nu}^{\dagger}}\Big) +30 \mbox{Tr}\Big({Y_\nu  m_l^{2 *}  Y_{\nu}^{\dagger}}\Big) \Big)
\end{equation} 

The resulting RGEs are:

\begin{align} 
\Delta \beta_{m_q^2}^{(2)} & =  
- \Big(2 {T_{u}^{\dagger}  T_u}  + 2 {Y_{u}^{\dagger}  m_u^2  Y_u}  + 4 m_{H_u}^2 {Y_{u}^{\dagger}  Y_u}  + {m_q^2  Y_{u}^{\dagger}  Y_u} + {Y_{u}^{\dagger}  Y_u  m_q^2}\Big)\mbox{Tr}\Big({Y_\nu  Y_{\nu}^{\dagger}}\Big) \nonumber \\ 
 &-2 \Big({T_{u}^{\dagger}  Y_u} \mbox{Tr}\Big({Y_{\nu}^{\dagger}  T_\nu}\Big) +{Y_{u}^{\dagger}  T_u} \mbox{Tr}\Big({T_\nu^*  Y_{\nu}^{T}}\Big) +{Y_{u}^{\dagger}  Y_u} \mbox{Tr}\Big({T_\nu^*  T_{\nu}^{T}}\Big) \nonumber \\ 
 &+{Y_{u}^{\dagger}  Y_u} \mbox{Tr}\Big({m_l^2  Y_{\nu}^{\dagger}  Y_\nu}\Big) +{Y_{u}^{\dagger}  Y_u} \mbox{Tr}\Big({{m_\nu^2}  Y_\nu  Y_{\nu}^{\dagger}}\Big) \Big)\\ 
\Delta \beta_{m_l^2}^{(1)} & =  
2 m_{H_u}^2 {Y_{\nu}^{\dagger}  Y_\nu}  + 2 {T_{\nu}^{\dagger}  T_\nu}  + 2 {Y_{\nu}^{\dagger}  {m_\nu^2}  Y_\nu}  + {m_l^2  Y_{\nu}^{\dagger}  Y_\nu} + {Y_{\nu}^{\dagger}  Y_\nu  m_l^2}\\ 
\Delta \beta_{m_l^2}^{(2)} & =  
-3 \Big(2 {T_{\nu}^{\dagger}  T_\nu}  + 2 {Y_{\nu}^{\dagger}  {m_\nu^2}  Y_\nu}  + 4 m_{H_u}^2 {Y_{\nu}^{\dagger}  Y_\nu}  + {m_l^2  Y_{\nu}^{\dagger}  Y_\nu} + {Y_{\nu}^{\dagger}  Y_\nu  m_l^2}\Big)\mbox{Tr}\Big({Y_u  Y_{u}^{\dagger}}\Big) \nonumber \\ 
 &- \Big(2 {T_{\nu}^{\dagger}  T_\nu}  + 2 {Y_{\nu}^{\dagger}  {m_\nu^2}  Y_\nu}  + 4 m_{H_u}^2 {Y_{\nu}^{\dagger}  Y_\nu}  + {m_l^2  Y_{\nu}^{\dagger}  Y_\nu} + {Y_{\nu}^{\dagger}  Y_\nu  m_l^2}\Big)\mbox{Tr}\Big({Y_\nu  Y_{\nu}^{\dagger}}\Big) \nonumber \\ 
 &-2 \Big(4 m_{H_u}^2 {Y_{\nu}^{\dagger}  Y_\nu  Y_{\nu}^{\dagger}  Y_\nu} +2 {Y_{\nu}^{\dagger}  Y_\nu  T_{\nu}^{\dagger}  T_\nu} +2 {Y_{\nu}^{\dagger}  T_\nu  T_{\nu}^{\dagger}  Y_\nu} +2 {T_{\nu}^{\dagger}  Y_\nu  Y_{\nu}^{\dagger}  T_\nu} \nonumber \\ 
 &+2 {T_{\nu}^{\dagger}  T_\nu  Y_{\nu}^{\dagger}  Y_\nu} +{m_l^2  Y_{\nu}^{\dagger}  Y_\nu  Y_{\nu}^{\dagger}  Y_\nu}+2 {Y_{\nu}^{\dagger}  {m_\nu^2}  Y_\nu  Y_{\nu}^{\dagger}  Y_\nu} +2 {Y_{\nu}^{\dagger}  Y_\nu  m_l^2  Y_{\nu}^{\dagger}  Y_\nu} \nonumber \\ 
 &+2 {Y_{\nu}^{\dagger}  Y_\nu  Y_{\nu}^{\dagger}  {m_\nu^2}  Y_\nu} +{Y_{\nu}^{\dagger}  Y_\nu  Y_{\nu}^{\dagger}  Y_\nu  m_l^2}+3 {T_{\nu}^{\dagger}  Y_\nu} \mbox{Tr}\Big({Y_{u}^{\dagger}  T_u}\Big) +{T_{\nu}^{\dagger}  Y_\nu} \mbox{Tr}\Big({Y_{\nu}^{\dagger}  T_\nu}\Big) \nonumber \\ 
 &+3 {Y_{\nu}^{\dagger}  T_\nu} \mbox{Tr}\Big({T_u^*  Y_{u}^{T}}\Big) +3 {Y_{\nu}^{\dagger}  Y_\nu} \mbox{Tr}\Big({T_u^*  T_{u}^{T}}\Big) +{Y_{\nu}^{\dagger}  T_\nu} \mbox{Tr}\Big({T_\nu^*  Y_{\nu}^{T}}\Big) \nonumber \\ 
 &+{Y_{\nu}^{\dagger}  Y_\nu} \mbox{Tr}\Big({T_\nu^*  T_{\nu}^{T}}\Big) +{Y_{\nu}^{\dagger}  Y_\nu} \mbox{Tr}\Big({m_l^2  Y_{\nu}^{\dagger}  Y_\nu}\Big) +3 {Y_{\nu}^{\dagger}  Y_\nu} \mbox{Tr}\Big({m_q^2  Y_{u}^{\dagger}  Y_u}\Big) \nonumber \\ 
 &+3 {Y_{\nu}^{\dagger}  Y_\nu} \mbox{Tr}\Big({m_u^2  Y_u  Y_{u}^{\dagger}}\Big) +{Y_{\nu}^{\dagger}  Y_\nu} \mbox{Tr}\Big({{m_\nu^2}  Y_\nu  Y_{\nu}^{\dagger}}\Big) \Big)\\ 
\Delta \beta_{m_{H_d}^2}^{(2)} & =  
-2 \Big(\Big(m_{H_d}^2 + m_{H_u}^2\Big)\mbox{Tr}\Big({Y_e  Y_{\nu}^{\dagger}  Y_\nu  Y_{e}^{\dagger}}\Big) +\mbox{Tr}\Big({Y_e  Y_{\nu}^{\dagger}  T_\nu  T_{e}^{\dagger}}\Big)+\mbox{Tr}\Big({Y_e  T_{\nu}^{\dagger}  T_\nu  Y_{e}^{\dagger}}\Big)+\mbox{Tr}\Big({Y_\nu  Y_{e}^{\dagger}  T_e  T_{\nu}^{\dagger}}\Big)\nonumber \\ 
 &+\mbox{Tr}\Big({Y_\nu  T_{e}^{\dagger}  T_e  Y_{\nu}^{\dagger}}\Big)+\mbox{Tr}\Big({m_e^2  Y_e  Y_{\nu}^{\dagger}  Y_\nu  Y_{e}^{\dagger}}\Big)+\mbox{Tr}\Big({m_l^2  Y_{e}^{\dagger}  Y_e  Y_{\nu}^{\dagger}  Y_\nu}\Big)+\mbox{Tr}\Big({m_l^2  Y_{\nu}^{\dagger}  Y_\nu  Y_{e}^{\dagger}  Y_e}\Big)\nonumber \\ 
 &+\mbox{Tr}\Big({{m_\nu^2}  Y_\nu  Y_{e}^{\dagger}  Y_e  Y_{\nu}^{\dagger}}\Big)\Big)\\ 
\Delta \beta_{m_{H_u}^2}^{(1)} & =  
2 \Big(m_{H_u}^2 \mbox{Tr}\Big({Y_\nu  Y_{\nu}^{\dagger}}\Big)  + \mbox{Tr}\Big({T_\nu^*  T_{\nu}^{T}}\Big) + \mbox{Tr}\Big({m_l^2  Y_{\nu}^{\dagger}  Y_\nu}\Big) + \mbox{Tr}\Big({{m_\nu^2}  Y_\nu  Y_{\nu}^{\dagger}}\Big)\Big)\\ 
\Delta \beta_{m_{H_u}^2}^{(2)} & =  
-2 \Big(\Big(m_{H_d}^2 + m_{H_u}^2\Big)\mbox{Tr}\Big({Y_e  Y_{\nu}^{\dagger}  Y_\nu  Y_{e}^{\dagger}}\Big) +\mbox{Tr}\Big({Y_e  Y_{\nu}^{\dagger}  T_\nu  T_{e}^{\dagger}}\Big)+\mbox{Tr}\Big({Y_e  T_{\nu}^{\dagger}  T_\nu  Y_{e}^{\dagger}}\Big)+\mbox{Tr}\Big({Y_\nu  Y_{e}^{\dagger}  T_e  T_{\nu}^{\dagger}}\Big)\nonumber \\ 
 &+6 m_{H_u}^2 \mbox{Tr}\Big({Y_\nu  Y_{\nu}^{\dagger}  Y_\nu  Y_{\nu}^{\dagger}}\Big) +6 \mbox{Tr}\Big({Y_\nu  Y_{\nu}^{\dagger}  T_\nu  T_{\nu}^{\dagger}}\Big) +\mbox{Tr}\Big({Y_\nu  T_{e}^{\dagger}  T_e  Y_{\nu}^{\dagger}}\Big)+6 \mbox{Tr}\Big({Y_\nu  T_{\nu}^{\dagger}  T_\nu  Y_{\nu}^{\dagger}}\Big) \nonumber \\ 
 &+\mbox{Tr}\Big({m_e^2  Y_e  Y_{\nu}^{\dagger}  Y_\nu  Y_{e}^{\dagger}}\Big)+\mbox{Tr}\Big({m_l^2  Y_{e}^{\dagger}  Y_e  Y_{\nu}^{\dagger}  Y_\nu}\Big)+\mbox{Tr}\Big({m_l^2  Y_{\nu}^{\dagger}  Y_\nu  Y_{e}^{\dagger}  Y_e}\Big)+6 \mbox{Tr}\Big({m_l^2  Y_{\nu}^{\dagger}  Y_\nu  Y_{\nu}^{\dagger}  Y_\nu}\Big) \nonumber \\ 
 &+\mbox{Tr}\Big({{m_\nu^2}  Y_\nu  Y_{e}^{\dagger}  Y_e  Y_{\nu}^{\dagger}}\Big)+6 \mbox{Tr}\Big({{m_\nu^2}  Y_\nu  Y_{\nu}^{\dagger}  Y_\nu  Y_{\nu}^{\dagger}}\Big) \Big)\\ 
\Delta \beta_{m_u^2}^{(2)} & =  
-2 \Big(\Big(2 {T_u  T_{u}^{\dagger}}  + 2 {Y_u  m_q^2  Y_{u}^{\dagger}}  + 4 m_{H_u}^2 {Y_u  Y_{u}^{\dagger}}  + {m_u^2  Y_u  Y_{u}^{\dagger}} + {Y_u  Y_{u}^{\dagger}  m_u^2}\Big)\mbox{Tr}\Big({Y_\nu  Y_{\nu}^{\dagger}}\Big) \nonumber \\ 
 &+2 \Big({Y_u  T_{u}^{\dagger}} \mbox{Tr}\Big({Y_{\nu}^{\dagger}  T_\nu}\Big) +{T_u  Y_{u}^{\dagger}} \mbox{Tr}\Big({T_\nu^*  Y_{\nu}^{T}}\Big) +{Y_u  Y_{u}^{\dagger}} \mbox{Tr}\Big({T_\nu^*  T_{\nu}^{T}}\Big) \nonumber \\ 
 &+{Y_u  Y_{u}^{\dagger}} \mbox{Tr}\Big({m_l^2  Y_{\nu}^{\dagger}  Y_\nu}\Big) +{Y_u  Y_{u}^{\dagger}} \mbox{Tr}\Big({{m_\nu^2}  Y_\nu  Y_{\nu}^{\dagger}}\Big) \Big)\Big)\\ 
\Delta \beta_{m_e^2}^{(2)} & =  
-2 \Big(2 \Big(m_{H_d}^2 + m_{H_u}^2\Big){Y_e  Y_{\nu}^{\dagger}  Y_\nu  Y_{e}^{\dagger}} +2 {Y_e  Y_{\nu}^{\dagger}  T_\nu  T_{e}^{\dagger}} +2 {Y_e  T_{\nu}^{\dagger}  T_\nu  Y_{e}^{\dagger}} +2 {T_e  Y_{\nu}^{\dagger}  Y_\nu  T_{e}^{\dagger}} \nonumber \\ 
 &+2 {T_e  T_{\nu}^{\dagger}  Y_\nu  Y_{e}^{\dagger}} +{m_e^2  Y_e  Y_{\nu}^{\dagger}  Y_\nu  Y_{e}^{\dagger}}+2 {Y_e  m_l^2  Y_{\nu}^{\dagger}  Y_\nu  Y_{e}^{\dagger}} \nonumber \\ 
 &+2 {Y_e  Y_{\nu}^{\dagger}  {m_\nu^2}  Y_\nu  Y_{e}^{\dagger}} +2 {Y_e  Y_{\nu}^{\dagger}  Y_\nu  m_l^2  Y_{e}^{\dagger}} +{Y_e  Y_{\nu}^{\dagger}  Y_\nu  Y_{e}^{\dagger}  m_e^2}\Big)\\ 
\beta_{{m_\nu^2}}^{(1)} & =  
2 \Big(2 m_{H_u}^2 {Y_\nu  Y_{\nu}^{\dagger}}  + 2 {T_\nu  T_{\nu}^{\dagger}}  + 2 {Y_\nu  m_l^2  Y_{\nu}^{\dagger}}  + {{m_\nu^2}  Y_\nu  Y_{\nu}^{\dagger}} + {Y_\nu  Y_{\nu}^{\dagger}  {m_\nu^2}}\Big)\\ 
\beta_{{m_\nu^2}}^{(2)} & =  
-\frac{2}{5} \Big(6 g_{1}^{2} M_1^* {T_\nu  Y_{\nu}^{\dagger}} +30 g_{2}^{2} M_2^* {T_\nu  Y_{\nu}^{\dagger}} -6 g_{1}^{2} {T_\nu  T_{\nu}^{\dagger}} -30 g_{2}^{2} {T_\nu  T_{\nu}^{\dagger}} \nonumber \\ 
 &-3 g_{1}^{2} {{m_\nu^2}  Y_\nu  Y_{\nu}^{\dagger}} -15 g_{2}^{2} {{m_\nu^2}  Y_\nu  Y_{\nu}^{\dagger}} -6 g_{1}^{2} {Y_\nu  m_l^2  Y_{\nu}^{\dagger}} -30 g_{2}^{2} {Y_\nu  m_l^2  Y_{\nu}^{\dagger}} \nonumber \\ 
 &-3 g_{1}^{2} {Y_\nu  Y_{\nu}^{\dagger}  {m_\nu^2}} -15 g_{2}^{2} {Y_\nu  Y_{\nu}^{\dagger}  {m_\nu^2}} +10 m_{H_d}^2 {Y_\nu  Y_{e}^{\dagger}  Y_e  Y_{\nu}^{\dagger}} \nonumber \\ 
 &+10 m_{H_u}^2 {Y_\nu  Y_{e}^{\dagger}  Y_e  Y_{\nu}^{\dagger}} +10 {Y_\nu  Y_{e}^{\dagger}  T_e  T_{\nu}^{\dagger}} +20 m_{H_u}^2 {Y_\nu  Y_{\nu}^{\dagger}  Y_\nu  Y_{\nu}^{\dagger}} +10 {Y_\nu  Y_{\nu}^{\dagger}  T_\nu  T_{\nu}^{\dagger}} \nonumber \\ 
 &+10 {Y_\nu  T_{e}^{\dagger}  T_e  Y_{\nu}^{\dagger}} +10 {Y_\nu  T_{\nu}^{\dagger}  T_\nu  Y_{\nu}^{\dagger}} +10 {T_\nu  Y_{e}^{\dagger}  Y_e  T_{\nu}^{\dagger}} +10 {T_\nu  Y_{\nu}^{\dagger}  Y_\nu  T_{\nu}^{\dagger}} \nonumber \\ 
 &+10 {T_\nu  T_{e}^{\dagger}  Y_e  Y_{\nu}^{\dagger}} +10 {T_\nu  T_{\nu}^{\dagger}  Y_\nu  Y_{\nu}^{\dagger}} +5 {{m_\nu^2}  Y_\nu  Y_{e}^{\dagger}  Y_e  Y_{\nu}^{\dagger}} +5 {{m_\nu^2}  Y_\nu  Y_{\nu}^{\dagger}  Y_\nu  Y_{\nu}^{\dagger}} \nonumber \\ 
 &+10 {Y_\nu  m_l^2  Y_{e}^{\dagger}  Y_e  Y_{\nu}^{\dagger}} +10 {Y_\nu  m_l^2  Y_{\nu}^{\dagger}  Y_\nu  Y_{\nu}^{\dagger}} +10 {Y_\nu  Y_{e}^{\dagger}  m_e^2  Y_e  Y_{\nu}^{\dagger}} \nonumber \\ 
 &+10 {Y_\nu  Y_{e}^{\dagger}  Y_e  m_l^2  Y_{\nu}^{\dagger}} +5 {Y_\nu  Y_{e}^{\dagger}  Y_e  Y_{\nu}^{\dagger}  {m_\nu^2}} +10 {Y_\nu  Y_{\nu}^{\dagger}  {m_\nu^2}  Y_\nu  Y_{\nu}^{\dagger}} +10 {Y_\nu  Y_{\nu}^{\dagger}  Y_\nu  m_l^2  Y_{\nu}^{\dagger}} \nonumber \\ 
 &+5 {Y_\nu  Y_{\nu}^{\dagger}  Y_\nu  Y_{\nu}^{\dagger}  {m_\nu^2}} +30 {T_\nu  T_{\nu}^{\dagger}} \mbox{Tr}\Big({Y_u  Y_{u}^{\dagger}}\Big) +15 {{m_\nu^2}  Y_\nu  Y_{\nu}^{\dagger}} \mbox{Tr}\Big({Y_u  Y_{u}^{\dagger}}\Big) \nonumber \\ 
 &+30 {Y_\nu  m_l^2  Y_{\nu}^{\dagger}} \mbox{Tr}\Big({Y_u  Y_{u}^{\dagger}}\Big) +15 {Y_\nu  Y_{\nu}^{\dagger}  {m_\nu^2}} \mbox{Tr}\Big({Y_u  Y_{u}^{\dagger}}\Big) +10 {T_\nu  T_{\nu}^{\dagger}} \mbox{Tr}\Big({Y_\nu  Y_{\nu}^{\dagger}}\Big) \nonumber \\ 
 &+5 {{m_\nu^2}  Y_\nu  Y_{\nu}^{\dagger}} \mbox{Tr}\Big({Y_\nu  Y_{\nu}^{\dagger}}\Big) +10 {Y_\nu  m_l^2  Y_{\nu}^{\dagger}} \mbox{Tr}\Big({Y_\nu  Y_{\nu}^{\dagger}}\Big) +5 {Y_\nu  Y_{\nu}^{\dagger}  {m_\nu^2}} \mbox{Tr}\Big({Y_\nu  Y_{\nu}^{\dagger}}\Big) \nonumber \\ 
 &+2 {Y_\nu  T_{\nu}^{\dagger}} \Big(15 g_{2}^{2} M_2  + 15 \mbox{Tr}\Big({Y_{u}^{\dagger}  T_u}\Big)  + 3 g_{1}^{2} M_1  + 5 \mbox{Tr}\Big({Y_{\nu}^{\dagger}  T_\nu}\Big) \Big)+30 {T_\nu  Y_{\nu}^{\dagger}} \mbox{Tr}\Big({T_u^*  Y_{u}^{T}}\Big) \nonumber \\ 
 &+10 {T_\nu  Y_{\nu}^{\dagger}} \mbox{Tr}\Big({T_\nu^*  Y_{\nu}^{T}}\Big) \Big)\nonumber \\ 
 &+\frac{4}{5} {Y_\nu  Y_{\nu}^{\dagger}} \Big(3 g_{1}^{2} m_{H_u}^2 +15 g_{2}^{2} m_{H_u}^2 +6 g_{1}^{2} |M_1|^2 +30 g_{2}^{2} |M_2|^2 -30 m_{H_u}^2 \mbox{Tr}\Big({Y_u  Y_{u}^{\dagger}}\Big) -10 m_{H_u}^2 \mbox{Tr}\Big({Y_\nu  Y_{\nu}^{\dagger}}\Big) \nonumber \\ 
 &-15 \mbox{Tr}\Big({T_u^*  T_{u}^{T}}\Big) -5 \mbox{Tr}\Big({T_\nu^*  T_{\nu}^{T}}\Big) -5 \mbox{Tr}\Big({m_l^2  Y_{\nu}^{\dagger}  Y_\nu}\Big) -15 \mbox{Tr}\Big({m_q^2  Y_{u}^{\dagger}  Y_u}\Big) -15 \mbox{Tr}\Big({m_u^2  Y_u  Y_{u}^{\dagger}}\Big) \nonumber \\ 
 &-5 \mbox{Tr}\Big({{m_\nu^2}  Y_\nu  Y_{\nu}^{\dagger}}\Big) \Big)\\ 
\beta_{m_X^2}^{(1)} & =  \beta_{m_X^2}^{(2)} = 0
\end{align}
\subsection*{Vacuum expectation values}
\begin{align} 
\Delta \beta_{v_d}^{(2)} & =  
v_d \mbox{Tr}\Big({Y_e  Y_{\nu}^{\dagger}  Y_\nu  Y_{e}^{\dagger}}\Big) \\ 
\Delta \beta_{v_u}^{(1)} & =  
- v_u \mbox{Tr}\Big({Y_\nu  Y_{\nu}^{\dagger}}\Big) \\ 
\Delta \beta_{v_u}^{(2)} & =  
3 v_u \mbox{Tr}\Big({Y_\nu  Y_{\nu}^{\dagger}  Y_\nu  Y_{\nu}^{\dagger}}\Big)  -\frac{3}{10} \Big(5 g_{2}^{2}  + g_{1}^{2}\Big)v_u \xi \mbox{Tr}\Big({Y_\nu  Y_{\nu}^{\dagger}}\Big)  + v_u \mbox{Tr}\Big({Y_e  Y_{\nu}^{\dagger}  Y_\nu  Y_{e}^{\dagger}}\Big) 
\end{align}

\section{Loop Integrals}
\label{app:loopintegrals}

The $B$-functions with vanishing external momenta and the arguments
$(a,b$) are given by
\begin{align}
B_0 &= 1 - \log \left(\frac{b}{Q^2}\right) + \frac{1}{b - a} \Big[ a \log \left(\frac{a}{b}\right) \Big]\,,\\
B_1 &= -\frac{1}{2} + \frac{1}{2} \log \left(\frac{b}{Q^2}\right)- \frac{1}{4 (a - b)^2} \Big[ a^2 - b^2 + 2 a^2 \log \left(\frac{b}{a}\right)\Big]\,, \\
\end{align}

The $C$-functions with vanishing external momenta and the arguments
$(a,b,c)$ read
\begin{align}
 C_0 =& -\frac{1}{(a-b) (a-c) (b-c)} \times \nonumber \\
   & \hspace{2cm} \Big[ b (c-a) \log \left(\frac{b}{a}\right)+c(a-b) \log\left(\frac{c}{a}\right) \Big]\\
 C_{00} =& \frac{1}{8(a-b) (a-c)(b-c)} \times \nonumber \\
 & \Big[(c-a) \left((a-b) (2 \log\left(\frac{a}{Q^2}\right)-3) (b-c)-2 b^2 \log\left(\frac{b}{a}\right)\right)+2 c^2
   (b-a) \log \left(\frac{c}{a}\right)\Big]\\
 C_{1} =& -\frac{1}{2 (a-b)^2 (a-c) (b-c)^2} \times \Big[ c^2 (a-b)^2 \log\left(\frac{c}{a}\right) \nonumber \\
  & +b (c-a) \left((b-a) (b-c)-(a (b-2 c)+b c) \log \left(\frac{b}{a}\right)\right) \Big]\\
 C_{2} =& -\frac{1}{2 (a-b) (a-c)^2(b-c)^2} \times \Big[a^2 (b-c)^2 \log\left(\frac{b}{a}\right) \\
  & +c (b-a) \left((a-c) (b-c)+(c (a+b)-2 a b) \log \left(\frac{b}{c}\right)\right) \Big]\\
 C_{11} =& -\frac{1}{6 (a-b)^3 (a-c) (b-c)^3} \times \nonumber \\
  &\Big[b (a-c) \big(-2 \left(a^2  \left(b^2-3 b c+3 c^2\right)+a b c (b-3 c)+b^2 c^2\right) \log
   \left(\frac{b}{a}\right) \nonumber \\
  & -(b-a ) (b-c) \left(-3 a b+5 a c+b^2-3 b c\right)\big)+2 c^3 (a-b)^3 \log \left(\frac{c}{a}\right)\Big] \\
 C_{12} =& \frac{1}{6(a-b)^2 (a-c)^2 (b-c)^3}\times \Big[(a-b) \Big((a-c) (b-c) \big(a \left(b^2+c^2\right) \nonumber \\
  & -b c (b+c)\big)+c^2 (a-b) (3 a b-c (a+2 b)) \log \left(\frac{c}{a}\right)\Big) \nonumber \\
  & +b^2 (a-c)^2 (a (b-3 c)+2 b c) \log \left(\frac{b}{a}\right) \Big] \\
 C_{22} =& \frac{1}{6 (a-b) (a-c)^3 (b-c)^3}  \times \Big[2 a^3 (b-c)^3 \log \left(\frac{b}{a}\right) \nonumber \\
  & +c (a-b) \Big(2 \left(c^2 \left(a^2+a b+b^2\right)+3 a^2 b^2-3 a b c
   (a+b)\right) \log\left(\frac{b}{c}\right) \nonumber \\
  & -(a-c) (b-c) \left(-3 c (a+b)+5 a b+c^2\right)\Big) \Big]
\end{align}
In the case of external photons, often the same combinations of
$C$-functions appear. If the arguments are $(a,b,b)$, these can
be expressed as
\begin{eqnarray}
& C_2+C_{12}+C_{22} = C_1+C_{12}+C_{11}= \frac{1}{12 b (x-1)^4}\left((x-1) (x (2 x+5)-1)-6 x^2 \log (x)\right) &\\
& C_0+C_1+C_2=  \frac{1}{2 b (x-1)^3}\left(-x^2+2 x \log (x)+1\right) &\\
& 2 C_{12}+ 2 C_{11} - C_2 = 2 C_{12}+2 C_{22} -C_1=  \frac{1}{12 b (x-1)^4}\left(6 (1-3 x) x^2 \log (x)+(x-1) (x (31 x-26)+7)\right) & \nonumber \\
&& \\
& C_1+C_2=  \frac{1}{2 b (x-1)^3}\left(2 x^2 \log (x)+(4-3 x) x-1\right) &
\end{eqnarray}
and for $(a,a,b)$ we get
\begin{eqnarray}
&C_{2}+C_{12}+C_{22}= - C_{12}=  \frac{1}{12 b (x-1)^4}\left(x ((x-6) x+3)+6 x \log (x)+2\right) &\\
&C_0+C_1+C_2 = - C_1= -\frac{1}{4 b (x-1)^3}\left(x^2-4 x+2 \log (x)+3\right) &
\end{eqnarray}
In the previous expressions we used $x=a/b$. \\

For the photonic monopole operators we define special loop functions
\begin{align}
M_{SFF}(a,b) = & \frac{((a-b) \left(16 a^2-29 a b+7 b^2\right)+6 a^2 (2 a-3 b) \log
   \left(\frac{b}{a}\right)}{36 (a-b)^4} \\
M_{FSS}(a,b) = & \frac{6 a^3 \log\left(\frac{b}{a}\right)+11 a^3-18 a^2 b+9 a b^2-2 b^3}{36
   (a-b)^4} \\ 
M_{FSV}(a,b) =& \frac{\sqrt{a} \left(2 a^3+3 a^2 b+6 a^2 b \log\left(\frac{b}{a}\right)-6
   a b^2+b^3\right)}{12 b (a-b)^4} \\
M_{FVS}(a,b) = & \frac{\sqrt{a} \left(2 a^3+3 a^2 b+6 a^2 b \log\left(\frac{b}{a}\right)-6
   a b^2+b^3\right)}{12 b (a-b)^4} \\
M_{FVV}(a,b) = & \frac{6 a^2 (a-3 b) \log \left(\frac{a}{b}\right)-(a-b) \left(5 a^2-22 a
   b+5 b^2\right)}{9 (a-b)^4}
\end{align}

The necessary box functions with the arguments $(a,b,c,d)$
read, in the limit of vanishing external momenta,
\begin{align}
 D_0 &=  -\Big[\frac{b \log \frac{b}{a} }{(b -a )(b -c )(b -d )} +
    \frac{c \log \frac{c}{a} }{(c -a )(c -b )(c -d )} \nonumber \\
     & \hspace{2cm} +\frac{d \log \frac{d}{a}}{(d -a )(d -b )(d -c )} \Big]  \\
 D_{27} &= -\frac{1}{4}\Big[ \frac{b ^2\log \frac{b}{a}}{(b -a )(b -c )(b -d )} +
    \frac{c ^2log \frac{c}{a} }{(c -a )(c -b )(c -d )}\nonumber \\
   &  \hspace{2cm} +\frac{d ^2\log \frac{d}{a}}{(d -a )(d -b )(d -c )} \Big]
\end{align}

In addition, we define
\begin{align}
I_{C_0 D_0}(a,b,c,d) = C_0(a,b,c) + d D_0 (a,b,c,d)
\end{align}

\section{Photonic penguin contributions to LFV}
\label{app:photon}

In the following appendices we present our results for the form
factors of the operators involved in our computation, done in the mass
basis. The flavor of the external fermions will be denoted with Greek
characters ($\alpha$, $\beta$, $\gamma$, $\delta$), whereas the mass
eigenstates of the particles in the loops will be denoted with Latin
characters ($a$, $b$, $c$, $d$). A sum over repeated indices will be
assumed.

\subsection{Feynman diagrams}
\parbox{0.45\linewidth}{
($a_1$)
\begin{fmffile}{Diagrams/Photon1}
\fmfframe(20,20)(20,20){
\begin{fmfgraph*}(200,90)
\fmfleft{l1,l2}
\fmfright{r1,r2}
\fmf{phantom}{r1,v1}
\fmf{phantom}{v1,r2}
\fmf{fermion}{v2,l1}
\fmf{dashes,label=$\tilde{e}_{{c}}$,tension=0.25}{v2,v3}
\fmf{dashes,label=$\tilde{e}_{{b}}$,tension=0.25}{v3,v4}
\fmf{fermion}{l2,v4}
\fmf{wiggly,tension=0.33,label=$\gamma$}{v1,v3}
\fmf{phantom,tension=0.33}{v1,v3}
\fmf{plain,tension=-0.0,label=$\tilde{\chi}^0_{{a}}$}{v2,v4}
\fmflabel{$\ell_{{\alpha}}$}{l2}
\fmflabel{$\bar{\ell}_{{\beta}}$}{l1}
\end{fmfgraph*}}
\end{fmffile}}
\hspace{1cm}
\parbox{0.45\linewidth}{
($a_2$)
\begin{fmffile}{Diagrams/Photon2}
\fmfframe(20,20)(20,20){
\begin{fmfgraph*}(200,90)
\fmfleft{l1,l2}
\fmfright{r1,r2}
\fmf{phantom}{r1,v1}
\fmf{phantom}{v1,r2}
\fmf{fermion}{v2,l1}
\fmf{plain,label=$\tilde{\chi}^-_{{c}}$,tension=0.25}{v2,v3}
\fmf{plain,label=$\tilde{\chi}^-_{{b}}$,tension=0.25}{v3,v4}
\fmf{fermion}{l2,v4}
\fmf{wiggly,tension=0.33,label=$\gamma$}{v1,v3}
\fmf{phantom,tension=0.33}{v1,v3}
\fmf{dashes,tension=-0.0,label=$\nu^i_{{a}}$}{v2,v4}
\fmflabel{$\ell_{{\alpha}}$}{l2}
\fmflabel{$\bar{\ell}_{{\beta}}$}{l1}
\end{fmfgraph*}}
\end{fmffile}}
\\
\parbox{0.45\linewidth}{
($a_3$)
\begin{fmffile}{Diagrams/Photon3}
\fmfframe(20,20)(20,20){
\begin{fmfgraph*}(200,90)
\fmfleft{l1,l2}
\fmfright{r1,r2}
\fmf{phantom}{r1,v1}
\fmf{phantom}{v1,r2}
\fmf{fermion}{v2,l1}
\fmf{plain,label=$\tilde{\chi}^-_{{c}}$,tension=0.25}{v2,v3}
\fmf{plain,label=$\tilde{\chi}^-_{{b}}$,tension=0.25}{v3,v4}
\fmf{fermion}{l2,v4}
\fmf{wiggly,tension=0.33,label=$\gamma$}{v1,v3}
\fmf{phantom,tension=0.33}{v1,v3}
\fmf{dashes,tension=-0.0,label=$\nu^R_{{a}}$}{v2,v4}
\fmflabel{$\ell_{{\alpha}}$}{l2}
\fmflabel{$\bar{\ell}_{{\beta}}$}{l1}
\end{fmfgraph*}}
\end{fmffile}}
\hspace{1cm}
\parbox{0.45\linewidth}{
($a_4$)
\begin{fmffile}{Diagrams/Photon4}
\fmfframe(20,20)(20,20){
\begin{fmfgraph*}(200,90)
\fmfleft{l1,l2}
\fmfright{r1,r2}
\fmf{phantom}{r1,v1}
\fmf{phantom}{v1,r2}
\fmf{fermion}{v2,l1}
\fmf{dashes,label=$H^-_{{c}}$,tension=0.25}{v2,v3}
\fmf{dashes,label=$H^-_{{b}}$,tension=0.25}{v3,v4}
\fmf{fermion}{l2,v4}
\fmf{wiggly,tension=0.33,label=$\gamma$}{v1,v3}
\fmf{phantom,tension=0.33}{v1,v3}
\fmf{plain,tension=-0.0,label=$\nu_{{a}}$}{v2,v4}
\fmflabel{$\ell_{{\alpha}}$}{l2}
\fmflabel{$\bar{\ell}_{{\beta}}$}{l1}
\end{fmfgraph*}}
\end{fmffile}}
\\
\parbox{0.45\linewidth}{
($a_5$)
\begin{fmffile}{Diagrams/Photon5}
\fmfframe(20,20)(20,20){
\begin{fmfgraph*}(200,90)
\fmfleft{l1,l2}
\fmfright{r1,r2}
\fmf{phantom}{r1,v1}
\fmf{phantom}{v1,r2}
\fmf{fermion}{v2,l1}
\fmf{dashes,label=$H^-_{{c}}$,tension=0.25}{v2,v3}
\fmf{wiggly,label=$W^-$,tension=0.25}{v3,v4}
\fmf{fermion}{l2,v4}
\fmf{wiggly,tension=0.33,label=$\gamma$}{v1,v3}
\fmf{phantom,tension=0.33}{v1,v3}
\fmf{plain,tension=-0.0,label=$\nu_{{a}}$}{v2,v4}
\fmflabel{$\ell_{{\alpha}}$}{l2}
\fmflabel{$\bar{\ell}_{{\beta}}$}{l1}
\end{fmfgraph*}}
\end{fmffile}}
\hspace{1cm}
\parbox{0.45\linewidth}{
($a_6$)
\begin{fmffile}{Diagrams/Photon6}
\fmfframe(20,20)(20,20){
\begin{fmfgraph*}(200,90)
\fmfleft{l1,l2}
\fmfright{r1,r2}
\fmf{phantom}{r1,v1}
\fmf{phantom}{v1,r2}
\fmf{fermion}{v2,l1}
\fmf{wiggly,label=$W^-$,tension=0.25}{v2,v3}
\fmf{dashes,label=$H^-_{{b}}$,tension=0.25}{v3,v4}
\fmf{fermion}{l2,v4}
\fmf{wiggly,tension=0.33,label=$\gamma$}{v1,v3}
\fmf{phantom,tension=0.33}{v1,v3}
\fmf{plain,tension=-0.0,label=$\nu_{{a}}$}{v2,v4}
\fmflabel{$\ell_{{\alpha}}$}{l2}
\fmflabel{$\bar{\ell}_{{\beta}}$}{l1}
\end{fmfgraph*}}
\end{fmffile}
}
\\
\parbox{0.45\linewidth}{
($a_7$)
\begin{fmffile}{Diagrams/Photon7}
\fmfframe(20,20)(20,20){
\begin{fmfgraph*}(200,90)
\fmfleft{l1,l2}
\fmfright{r1,r2}
\fmf{phantom}{r1,v1}
\fmf{phantom}{v1,r2}
\fmf{fermion}{v2,l1}
\fmf{wiggly,label=$W^-$,tension=0.25}{v2,v3}
\fmf{wiggly,label=$W^-$,tension=0.25}{v3,v4}
\fmf{fermion}{l2,v4}
\fmf{wiggly,tension=0.33,label=$\gamma$}{v1,v3}
\fmf{phantom,tension=0.33}{v1,v3}
\fmf{plain,tension=-0.0,label=$\nu_{{a}}$}{v2,v4}
\fmflabel{$\ell_{{\alpha}}$}{l2}
\fmflabel{$\bar{\ell}_{{\beta}}$}{l1}
\end{fmfgraph*}}
\end{fmffile}}

\vspace{1cm}\\
We give in the following the contribution of each diagram to the
different operators. We indicate the diagram by the corresponding
index $(a_i)$ with $i=1,\dots,7$.
\subsection{Neutralino contributions}
\begin{align} 
C_i= & C_i(m^2_{\tilde{\chi}^0_{{a}}}, m^2_{\tilde{e}_{{c}}}, m^2_{\tilde{e}_{{b}}}) \\ 
\KdiL{(a_1)}= & 2  \Veep_{c, b} (\VeneL_{\alpha, a, b} \VneeR_{a, \beta, c} (C_{12} + C_{22} + C_2) m_{\ell_\alpha} + \VeneR_{\alpha, a, b} (\VneeL_{a, \beta, c} (C_{11} \nonumber \\
& \hspace{2cm} + C_{12} + C_1) m_{\ell_\beta} - \VneeR_{a, \beta, c} (C_0 + C_1 + C_2) m_{\tilde{\chi}^0_{{a}}})) \\ 
\KmonoL{(a_1)}= &  - \VeneR_{\alpha, a, b} \VneeL_{a, \beta, c} \Veep_{c, b} \, M_{FSS}(m^2_{\tilde{\chi}^0_{{a}}}, m^2_{\tilde{e}_{{b}}})
\end{align} 

\subsection{Chargino contributions}
\begin{align} 
C_i = & C_i(m^2_{\tilde{\chi}^-_{{c}}}, m^2_{\tilde{\chi}^-_{{b}}}, m^2_{\nu^i_{{a}}}) \\ 
  \KdiL{(a_2)}= & -2  \Big( \VecviL_{\alpha, b, a} \VceviR_{c, \beta, a} (\VccpL _{b, c})^* C_{12} m_{\ell_\alpha} - \VecviR_{\alpha, b, a} ( \VceviL_{c, \beta, a} (\VccpR _{b, c})^* (C_{12} + C_{22} + C_2) m_{\ell_\beta} \nonumber \\
 & \hspace{2cm}+ \VceviR_{c, \beta, a} ( (\VccpL _{b, c})^* C_1 m_{\tilde{\chi}^-_{{b}}} - (\VccpR_{b, c})^* (C_0 + C_1 + C_2) m_{\tilde{\chi}^-_{{c}}})) \Big) \\ 
  \KmonoL{(a_2)}= &  - \VecviR_{\alpha, b, a} \VceviL_{c, \beta, a} ( \VccpR _{b, c})^* M_{SFF}(m^2_{\nu^i_{{a}}}, m^2_{\tilde{\chi}^-_{{b}}}) \\ 
\nonumber \\
C_i = & C_i(m^2_{\tilde{\chi}^-_{{c}}}, m^2_{\tilde{\chi}^-_{{b}}}, m^2_{\nu^R_{{a}}}) \\ 
  \KdiL{(a_3)}= & -2 \Big( \VecvrL_{\alpha, b, a} \VcevrR_{c, \beta, a} ( \VccpL _{b, c})^* C_{12} m_{\ell_\alpha} - \VecvrR_{\alpha, b, a} (\VcevrL_{c, \beta, a} ( \VccpR _{b, c})^* (C_{12} + C_{22} + C_2) m_{\ell_\beta} \nonumber \\
 & \hspace{2cm}+ \VcevrR_{c, \beta, a} (( \VccpL _{b, c})^* C_1 m_{\tilde{\chi}^-_{{b}}} - (\VccpR _{b, c})^* (C_0 + C_1 + C_2) m_{\tilde{\chi}^-_{{c}}})) \Big) \\ 
  \KmonoL{(a_3)}= & - \VecvrR_{\alpha, b, a} \VcevrL_{c, \beta, a} ( \VccpR _{b, c})^*  M_{SFF}(m^2_{\nu^R_{{a}}}, m^2_{\tilde{\chi}^-_{{b}}}) 
\end{align}

\subsection{$W^+$ and $H^+$ contributions}
\begin{align} 
C_i= & C_i(m^2_{\nu_{{a}}}, m^2_{H^-_{{c}}}, m^2_{H^-_{{b}}}) \\ 
  \KdiL{(a_4)}= & 2  \VhmhpP_{c, b} (\VevhmL_{\alpha, a, b}\VvehpR_{a, \beta, c} (C_{12} + C_{22} + C_2) m_{\ell_\alpha} + \VevhmR_{\alpha, a, b} (\VvehpL_{a, \beta, c} (C_{11} + C_{12} + C_1) m_{\ell_\beta} \nonumber \\
  & -\VvehpR_{a, \beta, c} (C_0 + C_1 + C_2) m_{\nu_{{a}}})) \\ 
  \KmonoL{(a_4)}= & - \VevhmR_{\alpha, a, b} \VvehpL_{a, \beta, c} \VhmhpP_{c, b} \, M_{FSS}(m^2_{\nu_{{a}}}, m^2_{H^-_{{b}}}) \label{eq:Hphoton1}\\ 
\nonumber \\
  \KdiL{(a_5)}= & 2 \,  (\VevwmL _{\alpha, a})^*\VvehpR_{a, \beta, c}\VhmwpP_{c} \, C_2(m^2_{\nu_{{a}}}, m^2_{H^-_{{c}}}, m^2_{W^-}) \\ 
  \KmonoL{(a_5)}= &  (\VevwmR _{\alpha, a})^* \VvehpL_{a, \beta, c}\VhmwpP_{c} \, M_{FVS}(m^2_{\nu_{{a}}}, m^2_{W^-}) \\ 
\nonumber \\
  \KdiL{(a_6)}= & 2  \VevhmR_{\alpha, a, b}  (\VvewpR _{a, \beta})^* \VhpPwm_{b} \, C_1(m^2_{\nu_{{a}}}, m^2_{W^-}, m^2_{H^-_{{b}}}) \\ 
  \KmonoL{(a_6)}= &  \VevhmL_{\alpha, a, b}  (\VvewpL _{a, \beta})^* \VhpPwm_{b} \, M_{FSV}(m^2_{\nu_{{a}}}, m^2_{H^-_{{b}}}) \\ 
\nonumber \\
C_i= & C_i(m^2_{\nu_{{a}}}, m^2_{W^-}, m^2_{W^-}) \\  
\KdiL{(a_7)}= & -2 \VwpPwm \Big( ( \VevwmR _{\alpha, a})^* ( \VvewpR _{a, \beta})^* (2 C_{12} - C_1 + 2 C_{22}) m_{\ell_\alpha} \nonumber \\
 & + ( \VevwmL _{\alpha, a})^* (( \VvewpL _{a, \beta})^* (2 C_{11} + 2 C_{12} - C_2) m_{\ell_\beta} + 3 ( \VvewpR _{a, \beta})^* (C_1 + C_2) m_{\nu_{{a}}}) \Big) 
 \label{eq:Wphoton2}\\ 
\KmonoL{(a_7)}= &  (\VevwmR _{\alpha, a})^*  (\VvewpR _{a, \beta})^*  \VwpPwm \, M_{FVV}(m^2_{\nu_{{a}}}, m^2_{W^-})  
\end{align} 

These coefficients are related to the ones used in the calculation of
the flavor observables by
\begin{align}
K^1_L = & \, \frac{1}{e} \sum_p \KmonoL{(a_p)}  \\
K^2_L = & - \frac{1}{2\, e\, m_{\ell_\alpha}} \sum_p \KdiL{(a_p)} 
\end{align}

\section{$Z$ and Higgs penguin contributions to LFV}
\label{app:penguins}

\subsection{Feynman diagrams}
In the following $B = Z, h_p, A^0_p$ is used. 
\subsection*{Neutralino diagrams}
\subsubsection*{Self energy corrections}
\parbox{0.45\linewidth}{
($n_1$)
\begin{fmffile}{Diagrams/PengV4LWaveNumberOfConsideredExternalStatesVZ2}
\fmfframe(20,20)(20,20){
\begin{fmfgraph*}(200,90)
\fmfleft{l2,l1}
\fmfright{r1,r2}
\fmf{fermion}{v3,l2}
\fmf{wiggly,label=$B$}{v3,v4}
\fmf{dashes}{v3,v4}
\fmf{phantom}{v4,r1}
\fmf{phantom}{v4,r2}
\fmf{phantom}{l1,v3}
\fmf{fermion}{l1,v1}
\fmf{plain,right,tension=0.2,label=$\tilde{\chi}^0_{{a}}$}{v1,v2}
\fmf{dashes,left,tension=0.2,label=$\tilde{e}_{{b}}$}{v1,v2}
\fmf{plain,label=$\bar{\ell}_{{c}}$}{v2,v3}
\fmflabel{$\ell_{{\alpha}}$}{l1}
\fmflabel{$\bar{\ell}_{{\beta}}$}{l2}
\end{fmfgraph*}}
\end{fmffile}}
\hspace{1cm} 
\parbox{0.45\linewidth}{
($n_2$)
\begin{fmffile}{Diagrams/PengV4LWaveNumberOfConsideredExternalStatesVZ10}
\fmfframe(20,20)(20,20){
\begin{fmfgraph*}(200,90)
\fmfleft{l1,l2}
\fmfright{r1,r2}
\fmf{fermion}{l2,v3}
\fmf{wiggly,label=$B$}{v3,v4}
\fmf{dashes}{v3,v4}
\fmf{phantom}{v4,r1}
\fmf{phantom}{v4,r2}
\fmf{phantom}{l1,v3}
\fmf{fermion}{v1,l1}
\fmf{dashes,right,tension=0.2,label=$\tilde{e}_{{a}}$}{v1,v2}
\fmf{plain,left,tension=0.2,label=$\tilde{\chi}^0_{{b}}$}{v1,v2}
\fmf{plain,label=$\ell_{{c}}$}{v2,v3}
\fmflabel{$\ell_{{\alpha}}$}{l2}
\fmflabel{$\bar{\ell}_{{\beta}}$}{l1}
\end{fmfgraph*}}
\end{fmffile} 
}
\subsubsection*{Vertex corrections}
\parbox{0.45\linewidth}{
($n_3$)
\begin{fmffile}{Diagrams/PengV4LPenguinNumberOfConsideredExternalStatesVZ2}
\fmfframe(20,20)(20,20){
\begin{fmfgraph*}(200,90)
\fmfleft{l1,l2}
\fmfright{r1,r2}
\fmf{phantom}{r1,v1}
\fmf{phantom}{v1,r2}
\fmf{fermion}{v2,l1}
\fmf{dashes,label=$\tilde{e}_{{c}}$,tension=0.25}{v2,v3}
\fmf{dashes,label=$\tilde{e}_{{b}}$,tension=0.25}{v3,v4}
\fmf{fermion}{l2,v4}
\fmf{wiggly,tension=0.33,label=$B$}{v1,v3}
\fmf{dashes,tension=0.33}{v1,v3}
\fmf{plain,tension=-0.0,label=$\tilde{\chi}^0_{{a}}$}{v2,v4}
\fmflabel{$\ell_{{\alpha}}$}{l2}
\fmflabel{$\bar{\ell}_{{\beta}}$}{l1}
\end{fmfgraph*}}
\end{fmffile}}
\hspace{1cm} 
\parbox{0.45\linewidth}{
($n_4$)
\begin{fmffile}{Diagrams/PengV4LPenguinNumberOfConsideredExternalStatesVZ18}
\fmfframe(20,20)(20,20){
\begin{fmfgraph*}(200,90)
\fmfleft{l1,l2}
\fmfright{r1,r2}
\fmf{phantom}{r1,v1}
\fmf{phantom}{v1,r2}
\fmf{fermion}{v2,l1}
\fmf{plain,label=$\tilde{\chi}^0_{{c}}$,tension=0.25}{v2,v3}
\fmf{plain,label=$\tilde{\chi}^0_{{b}}$,tension=0.25}{v3,v4}
\fmf{fermion}{l2,v4}
\fmf{wiggly,tension=0.33,label=$B$}{v1,v3}
\fmf{dashes,tension=0.33}{v1,v3}
\fmf{dashes,tension=-0.0,label=$\tilde{e}^*_{{a}}$}{v2,v4}
\fmflabel{$\ell_{{\alpha}}$}{l2}
\fmflabel{$\bar{\ell}_{{\beta}}$}{l1}
\end{fmfgraph*}}
\end{fmffile}
}

\subsection*{Chargino diagrams}
\subsubsection*{Self energy corrections}
\parbox{0.45\linewidth}{
($c_1$)
\begin{fmffile}{Diagrams/PengV4LWaveNumberOfConsideredExternalStatesVZ6}
\fmfframe(20,20)(20,20){
\begin{fmfgraph*}(200,90)
\fmfleft{l2,l1}
\fmfright{r1,r2}
\fmf{fermion}{v3,l2}
\fmf{wiggly,label=$B$}{v3,v4}
\fmf{dashes}{v3,v4}
\fmf{dashes}{v3,v4}
\fmf{phantom}{v4,r1}
\fmf{phantom}{v4,r2}
\fmf{phantom}{l1,v3}
\fmf{fermion}{l1,v1}
\fmf{dashes,right,tension=0.2,label=$\nu^i_{{a}}$}{v1,v2}
\fmf{plain,left,tension=0.2,label=$\tilde{\chi}^-_{{b}}$}{v1,v2}
\fmf{plain,label=$\bar{\ell}_{{c}}$}{v2,v3}
\fmflabel{$\ell_{{\alpha}}$}{l1}
\fmflabel{$\bar{\ell}_{{\beta}}$}{l2}
\end{fmfgraph*}}
\end{fmffile}}
\hspace{1cm} 
\parbox{0.45\linewidth}{
($c_2$)
\begin{fmffile}{Diagrams/PengV4LWaveNumberOfConsideredExternalStatesVZ7}
\fmfframe(20,20)(20,20){
\begin{fmfgraph*}(200,90)
\fmfleft{l2,l1}
\fmfright{r1,r2}
\fmf{fermion}{v3,l2}
\fmf{wiggly,label=$B$}{v3,v4}
\fmf{dashes}{v3,v4}
\fmf{dashes}{v3,v4}
\fmf{phantom}{v4,r1}
\fmf{phantom}{v4,r2}
\fmf{phantom}{l1,v3}
\fmf{fermion}{l1,v1}
\fmf{dashes,right,tension=0.2,label=$\nu^R_{{a}}$}{v1,v2}
\fmf{plain,left,tension=0.2,label=$\tilde{\chi}^-_{{b}}$}{v1,v2}
\fmf{plain,label=$\bar{\ell}_{{c}}$}{v2,v3}
\fmflabel{$\ell_{{\alpha}}$}{l1}
\fmflabel{$\bar{\ell}_{{\beta}}$}{l2}
\end{fmfgraph*}}
\end{fmffile}}
\\
\parbox{0.45\linewidth}{
($c_3$)
\begin{fmffile}{Diagrams/PengV4LWaveNumberOfConsideredExternalStatesVZ14}
\fmfframe(20,20)(20,20){
\begin{fmfgraph*}(200,90)
\fmfleft{l1,l2}
\fmfright{r1,r2}
\fmf{fermion}{l2,v3}
\fmf{wiggly,label=$B$}{v3,v4}
\fmf{dashes}{v3,v4}
\fmf{phantom}{v4,r1}
\fmf{phantom}{v4,r2}
\fmf{phantom}{l1,v3}
\fmf{fermion}{v1,l1}
\fmf{plain,right,tension=0.2,label=$\tilde{\chi}^-_{{a}}$}{v1,v2}
\fmf{dashes,left,tension=0.2,label=$\nu^i_{{b}}$}{v1,v2}
\fmf{plain,label=$\ell_{{c}}$}{v2,v3}
\fmflabel{$\ell_{{\alpha}}$}{l2}
\fmflabel{$\bar{\ell}_{{\beta}}$}{l1}
\end{fmfgraph*}}
\end{fmffile}}
\hspace{1cm}
\parbox{0.45\linewidth}{
($c_4$)
\begin{fmffile}{Diagrams/PengV4LWaveNumberOfConsideredExternalStatesVZ15}
\fmfframe(20,20)(20,20){
\begin{fmfgraph*}(200,90)
\fmfleft{l1,l2}
\fmfright{r1,r2}
\fmf{fermion}{l2,v3}
\fmf{wiggly,label=$B$}{v3,v4}
\fmf{dashes}{v3,v4}
\fmf{dashes}{v3,v4}
\fmf{phantom}{v4,r1}
\fmf{phantom}{v4,r2}
\fmf{phantom}{l1,v3}
\fmf{fermion}{v1,l1}
\fmf{plain,right,tension=0.2,label=$\tilde{\chi}^-_{{a}}$}{v1,v2}
\fmf{dashes,left,tension=0.2,label=$\nu^R_{{b}}$}{v1,v2}
\fmf{plain,label=$\ell_{{c}}$}{v2,v3}
\fmflabel{$\ell_{{\alpha}}$}{l2}
\fmflabel{$\bar{\ell}_{{\beta}}$}{l1}
\end{fmfgraph*}}
\end{fmffile}}

\subsubsection*{Vertex corrections}
\parbox{0.45\linewidth}{
($c_5$)
\begin{fmffile}{Diagrams/PengV4LPenguinNumberOfConsideredExternalStatesVZ8}
\fmfframe(20,20)(20,20){
\begin{fmfgraph*}(200,90)
\fmfleft{l1,l2}
\fmfright{r1,r2}
\fmf{phantom}{r1,v1}
\fmf{phantom}{v1,r2}
\fmf{fermion}{v2,l1}
\fmf{plain,label=$\tilde{\chi}^-_{{c}}$,tension=0.25}{v2,v3}
\fmf{plain,label=$\tilde{\chi}^-_{{b}}$,tension=0.25}{v3,v4}
\fmf{fermion}{l2,v4}
\fmf{wiggly,tension=0.33,label=$B$}{v1,v3}
\fmf{dashes,tension=0.33}{v1,v3}
\fmf{dashes,tension=0.33}{v1,v3}
\fmf{dashes,tension=-0.0,label=$\nu^i_{{a}}$}{v2,v4}
\fmflabel{$\ell_{{\alpha}}$}{l2}
\fmflabel{$\bar{\ell}_{{\beta}}$}{l1}
\end{fmfgraph*}}
\end{fmffile}}
\hspace{1cm}
\parbox{0.45\linewidth}{
($c_6$)
\begin{fmffile}{Diagrams/PengV4LPenguinNumberOfConsideredExternalStatesVZ9}
\fmfframe(20,20)(20,20){
\begin{fmfgraph*}(200,90)
\fmfleft{l1,l2}
\fmfright{r1,r2}
\fmf{phantom}{r1,v1}
\fmf{phantom}{v1,r2}
\fmf{fermion}{v2,l1}
\fmf{plain,label=$\tilde{\chi}^-_{{c}}$,tension=0.25}{v2,v3}
\fmf{plain,label=$\tilde{\chi}^-_{{b}}$,tension=0.25}{v3,v4}
\fmf{fermion}{l2,v4}
\fmf{wiggly,tension=0.33,label=$B$}{v1,v3}
\fmf{dashes,tension=0.33}{v1,v3}
\fmf{dashes,tension=-0.0,label=$\nu^R_{{a}}$}{v2,v4}
\fmflabel{$\ell_{{\alpha}}$}{l2}
\fmflabel{$\bar{\ell}_{{\beta}}$}{l1}
\end{fmfgraph*}}
\end{fmffile}}
\\
\parbox{0.45\linewidth}{
($c_7$)
\begin{fmffile}{Diagrams/PengV4LPenguinNumberOfConsideredExternalStatesVZ11}
\fmfframe(20,20)(20,20){
\begin{fmfgraph*}(200,90)
\fmfleft{l1,l2}
\fmfright{r1,r2}
\fmf{phantom}{r1,v1}
\fmf{phantom}{v1,r2}
\fmf{fermion}{v2,l1}
\fmf{dashes,label=$\nu^i_{{c}}$,tension=0.25}{v2,v3}
\fmf{dashes,label=$\nu^R_{{b}}$,tension=0.25}{v3,v4}
\fmf{fermion}{l2,v4}
\fmf{wiggly,tension=0.33,label=$B$}{v1,v3}
\fmf{dashes,tension=0.33}{v1,v3}
\fmf{plain,tension=-0.0,label=$\tilde{\chi}^+_{{a}}$}{v2,v4}
\fmflabel{$\ell_{{\alpha}}$}{l2}
\fmflabel{$\bar{\ell}_{{\beta}}$}{l1}
\end{fmfgraph*}}
\end{fmffile}}
\hspace{1cm}
\parbox{0.45\linewidth}{
($c_8$)
\begin{fmffile}{Diagrams/PengV4LPenguinNumberOfConsideredExternalStatesVZ12}
\fmfframe(20,20)(20,20){
\begin{fmfgraph*}(200,90)
\fmfleft{l1,l2}
\fmfright{r1,r2}
\fmf{phantom}{r1,v1}
\fmf{phantom}{v1,r2}
\fmf{fermion}{v2,l1}
\fmf{dashes,label=$\nu^R_{{c}}$,tension=0.25}{v2,v3}
\fmf{dashes,label=$\nu^i_{{b}}$,tension=0.25}{v3,v4}
\fmf{fermion}{l2,v4}
\fmf{wiggly,tension=0.33,label=$B$}{v1,v3}
\fmf{dashes,tension=0.33}{v1,v3}
\fmf{plain,tension=-0.0,label=$\tilde{\chi}^+_{{a}}$}{v2,v4}
\fmflabel{$\ell_{{\alpha}}$}{l2}
\fmflabel{$\bar{\ell}_{{\beta}}$}{l1}
\end{fmfgraph*}}
\end{fmffile}} \\
\parbox{0.45\linewidth}{
($c_9$)
\begin{fmffile}{Diagrams/PengV4LPenguinNumberOfConsideredExternalStatesVZ13}
\fmfframe(20,20)(20,20){
\begin{fmfgraph*}(200,90)
\fmfleft{l1,l2}
\fmfright{r1,r2}
\fmf{phantom}{r1,v1}
\fmf{phantom}{v1,r2}
\fmf{fermion}{v2,l1}
\fmf{dashes,label=$\nu^R_{{c}}$,tension=0.25}{v2,v3}
\fmf{dashes,label=$\nu^R_{{b}}$,tension=0.25}{v3,v4}
\fmf{fermion}{l2,v4}
\fmf{wiggly,tension=0.33,label=$B$}{v1,v3}
\fmf{dashes,tension=0.33}{v1,v3}
\fmf{plain,tension=-0.0,label=$\tilde{\chi}^+_{{a}}$}{v2,v4}
\fmflabel{$\ell_{{\alpha}}$}{l2}
\fmflabel{$\bar{\ell}_{{\beta}}$}{l1}
\end{fmfgraph*}}
\end{fmffile}}
\hspace{1cm}
\parbox{0.45\linewidth}{
($c_{10}$)
\begin{fmffile}{Diagrams/PengV4LPenguinNumberOfConsideredExternalStatesVZ14}
\fmfframe(20,20)(20,20){
\begin{fmfgraph*}(200,90)
\fmfleft{l1,l2}
\fmfright{r1,r2}
\fmf{phantom}{r1,v1}
\fmf{phantom}{v1,r2}
\fmf{fermion}{v2,l1}
\fmf{dashes,label=$\nu^i_{{c}}$,tension=0.25}{v2,v3}
\fmf{dashes,label=$\nu^i_{{b}}$,tension=0.25}{v3,v4}
\fmf{fermion}{l2,v4}
\fmf{wiggly,tension=0.33,label=$B$}{v1,v3}
\fmf{dashes,tension=0.33}{v1,v3}
\fmf{plain,tension=-0.0,label=$\tilde{\chi}^+_{{a}}$}{v2,v4}
\fmflabel{$\ell_{{\alpha}}$}{l2}
\fmflabel{$\bar{\ell}_{{\beta}}$}{l1}
\end{fmfgraph*}}
\end{fmffile}}

\subsection*{$W^+$ and $H^+$ diagrams}
\subsubsection*{Self energy corrections}
\parbox{0.45\linewidth}{
($w_1$)
\begin{fmffile}{Diagrams/PengV4LWaveNumberOfConsideredExternalStatesVZ3}
\fmfframe(20,20)(20,20){
\begin{fmfgraph*}(200,90)
\fmfleft{l2,l1}
\fmfright{r1,r2}
\fmf{fermion}{v3,l2}
\fmf{wiggly,label=$B$}{v3,v4}
\fmf{dashes}{v3,v4}
\fmf{phantom}{v4,r1}
\fmf{phantom}{v4,r2}
\fmf{phantom}{l1,v3}
\fmf{fermion}{l1,v1}
\fmf{plain,right,tension=0.2,label=$\nu_{{a}}$}{v1,v2}
\fmf{dashes,left,tension=0.2,label=$H^-_{{b}}$}{v1,v2}
\fmf{plain,label=$\bar{\ell}_{{c}}$}{v2,v3}
\fmflabel{$\ell_{{\alpha}}$}{l1}
\fmflabel{$\bar{\ell}_{{\beta}}$}{l2}
\end{fmfgraph*}}
\end{fmffile}}
\hspace{1cm} 
\parbox{0.45\linewidth}{
($w_2$)
\begin{fmffile}{Diagrams/PengV4LWaveNumberOfConsideredExternalStatesVZ4}
\fmfframe(20,20)(20,20){
\begin{fmfgraph*}(200,90)
\fmfleft{l2,l1}
\fmfright{r1,r2}
\fmf{fermion}{v3,l2}
\fmf{wiggly,label=$B$}{v3,v4}
\fmf{dashes}{v3,v4}
\fmf{phantom}{v4,r1}
\fmf{phantom}{v4,r2}
\fmf{phantom}{l1,v3}
\fmf{fermion}{l1,v1}
\fmf{plain,right,tension=0.2,label=$\nu_{{a}}$}{v1,v2}
\fmf{wiggly,left,tension=0.2,label=$W^-$}{v1,v2}
\fmf{plain,label=$\bar{\ell}_{{c}}$}{v2,v3}
\fmflabel{$\ell_{{\alpha}}$}{l1}
\fmflabel{$\bar{\ell}_{{\beta}}$}{l2}
\end{fmfgraph*}}
\end{fmffile}}
\\
\parbox{0.45\linewidth}{
($w_3$)
\begin{fmffile}{Diagrams/PengV4LWaveNumberOfConsideredExternalStatesVZ11}
\fmfframe(20,20)(20,20){
\begin{fmfgraph*}(200,90)
\fmfleft{l1,l2}
\fmfright{r1,r2}
\fmf{fermion}{l2,v3}
\fmf{wiggly,label=$B$}{v3,v4}
\fmf{dashes}{v3,v4}
\fmf{phantom}{v4,r1}
\fmf{phantom}{v4,r2}
\fmf{phantom}{l1,v3}
\fmf{fermion}{v1,l1}
\fmf{dashes,right,tension=0.2,label=$H^-_{{a}}$}{v1,v2}
\fmf{plain,left,tension=0.2,label=$\nu_{{b}}$}{v1,v2}
\fmf{plain,label=$\ell_{{c}}$}{v2,v3}
\fmflabel{$\ell_{{\alpha}}$}{l2}
\fmflabel{$\bar{\ell}_{{\beta}}$}{l1}
\end{fmfgraph*}}
\end{fmffile}}
\hspace{1cm}
\parbox{0.45\linewidth}{
($w_4$)
\begin{fmffile}{Diagrams/PengV4LWaveNumberOfConsideredExternalStatesVZ12}
\fmfframe(20,20)(20,20){
\begin{fmfgraph*}(200,90)
\fmfleft{l1,l2}
\fmfright{r1,r2}
\fmf{fermion}{l2,v3}
\fmf{wiggly,label=$B$}{v3,v4}
\fmf{dashes}{v3,v4}
\fmf{phantom}{v4,r1}
\fmf{phantom}{v4,r2}
\fmf{phantom}{l1,v3}
\fmf{fermion}{v1,l1}
\fmf{wiggly,right,tension=0.2,label=$W^-$}{v1,v2}
\fmf{plain,left,tension=0.2,label=$\nu_{{b}}$}{v1,v2}
\fmf{plain,label=$\ell_{{c}}$}{v2,v3}
\fmflabel{$\ell_{{\alpha}}$}{l2}
\fmflabel{$\bar{\ell}_{{\beta}}$}{l1}
\end{fmfgraph*}}
\end{fmffile}}

\subsubsection*{Vertex corrections}
\parbox{0.45\linewidth}{
($w_5$)
\begin{fmffile}{Diagrams/PengV4LPenguinNumberOfConsideredExternalStatesVZ3}
\fmfframe(20,20)(20,20){
\begin{fmfgraph*}(200,90)
\fmfleft{l1,l2}
\fmfright{r1,r2}
\fmf{phantom}{r1,v1}
\fmf{phantom}{v1,r2}
\fmf{fermion}{v2,l1}
\fmf{dashes,label=$H^-_{{c}}$,tension=0.25}{v2,v3}
\fmf{dashes,label=$H^-_{{b}}$,tension=0.25}{v3,v4}
\fmf{fermion}{l2,v4}
\fmf{wiggly,tension=0.33,label=$B$}{v1,v3}
\fmf{dashes,tension=0.33}{v1,v3}
\fmf{plain,tension=-0.0,label=$\nu_{{a}}$}{v2,v4}
\fmflabel{$\ell_{{\alpha}}$}{l2}
\fmflabel{$\bar{\ell}_{{\beta}}$}{l1}
\end{fmfgraph*}}
\end{fmffile}}
\hspace{1cm}
\parbox{0.45\linewidth}{
($w_6$)
\begin{fmffile}{Diagrams/PengV4LPenguinNumberOfConsideredExternalStatesVZ4}
\fmfframe(20,20)(20,20){
\begin{fmfgraph*}(200,90)
\fmfleft{l1,l2}
\fmfright{r1,r2}
\fmf{phantom}{r1,v1}
\fmf{phantom}{v1,r2}
\fmf{fermion}{v2,l1}
\fmf{dashes,label=$H^-_{{c}}$,tension=0.25}{v2,v3}
\fmf{wiggly,label=$W^-$,tension=0.25}{v3,v4}
\fmf{fermion}{l2,v4}
\fmf{wiggly,tension=0.33,label=$B$}{v1,v3}
\fmf{dashes,tension=0.33}{v1,v3}
\fmf{plain,tension=-0.0,label=$\nu_{{a}}$}{v2,v4}
\fmflabel{$\ell_{{\alpha}}$}{l2}
\fmflabel{$\bar{\ell}_{{\beta}}$}{l1}
\end{fmfgraph*}}
\end{fmffile}}
\\
\parbox{0.45\linewidth}{
($w_7$)
\begin{fmffile}{Diagrams/PengV4LPenguinNumberOfConsideredExternalStatesVZ5}
\fmfframe(20,20)(20,20){
\begin{fmfgraph*}(200,90)
\fmfleft{l1,l2}
\fmfright{r1,r2}
\fmf{phantom}{r1,v1}
\fmf{phantom}{v1,r2}
\fmf{fermion}{v2,l1}
\fmf{wiggly,label=$W^-$,tension=0.25}{v2,v3}
\fmf{dashes,label=$H^-_{{b}}$,tension=0.25}{v3,v4}
\fmf{fermion}{l2,v4}
\fmf{wiggly,tension=0.33,label=$B$}{v1,v3}
\fmf{dashes,tension=0.33}{v1,v3}
\fmf{plain,tension=-0.0,label=$\nu_{{a}}$}{v2,v4}
\fmflabel{$\ell_{{\alpha}}$}{l2}
\fmflabel{$\bar{\ell}_{{\beta}}$}{l1}
\end{fmfgraph*}}
\end{fmffile}
}
\hspace{1cm}
\parbox{0.45\linewidth}{
($w_8$)
\begin{fmffile}{Diagrams/PengV4LPenguinNumberOfConsideredExternalStatesVZ6}
\fmfframe(20,20)(20,20){
\begin{fmfgraph*}(200,90)
\fmfleft{l1,l2}
\fmfright{r1,r2}
\fmf{phantom}{r1,v1}
\fmf{phantom}{v1,r2}
\fmf{fermion}{v2,l1}
\fmf{wiggly,label=$W^-$,tension=0.25}{v2,v3}
\fmf{wiggly,label=$W^-$,tension=0.25}{v3,v4}
\fmf{fermion}{l2,v4}
\fmf{wiggly,tension=0.33,label=$B$}{v1,v3}
\fmf{dashes,tension=0.33}{v1,v3}
\fmf{plain,tension=-0.0,label=$\nu_{{a}}$}{v2,v4}
\fmflabel{$\ell_{{\alpha}}$}{l2}
\fmflabel{$\bar{\ell}_{{\beta}}$}{l1}
\end{fmfgraph*}}
\end{fmffile}}
\\
\parbox{0.45\linewidth}{
($w_9$)
\begin{fmffile}{Diagrams/PengV4LPenguinNumberOfConsideredExternalStatesVZ17}
\fmfframe(20,20)(20,20){
\begin{fmfgraph*}(200,90)
\fmfleft{l1,l2}
\fmfright{r1,r2}
\fmf{phantom}{r1,v1}
\fmf{phantom}{v1,r2}
\fmf{fermion}{v2,l1}
\fmf{plain,label=$\nu_{{c}}$,tension=0.25}{v2,v3}
\fmf{plain,label=$\nu_{{b}}$,tension=0.25}{v3,v4}
\fmf{fermion}{l2,v4}
\fmf{wiggly,tension=0.33,label=$B$}{v1,v3}
\fmf{dashes,tension=0.33}{v1,v3}
\fmf{dashes,tension=-0.0,label=$H^+_{{a}}$}{v2,v4}
\fmflabel{$\ell_{{\alpha}}$}{l2}
\fmflabel{$\bar{\ell}_{{\beta}}$}{l1}
\end{fmfgraph*}}
\end{fmffile}}
\hspace{1cm}
\parbox{0.45\linewidth}{
($w_{10}$)
\begin{fmffile}{Diagrams/PengV4LPenguinNumberOfConsideredExternalStatesVZ19}
\fmfframe(20,20)(20,20){
\begin{fmfgraph*}(200,90)
\fmfleft{l1,l2}
\fmfright{r1,r2}
\fmf{phantom}{r1,v1}
\fmf{phantom}{v1,r2}
\fmf{fermion}{v2,l1}
\fmf{plain,label=$\nu_{{c}}$,tension=0.25}{v2,v3}
\fmf{plain,label=$\nu_{{b}}$,tension=0.25}{v3,v4}
\fmf{fermion}{l2,v4}
\fmf{wiggly,tension=0.33,label=$B$}{v1,v3}
\fmf{dashes,tension=0.33}{v1,v3}
\fmf{wiggly,tension=-0.0,label=$W^+$}{v2,v4}
\fmflabel{$\ell_{{\alpha}}$}{l2}
\fmflabel{$\bar{\ell}_{{\beta}}$}{l1}
\end{fmfgraph*}}
\end{fmffile}}

\subsection{Neutralino contributions}
\subsubsection{Z-penguins}
\paragraph*{Self-energy corrections}

\begin{align} 
I_1= & B_0(m^2_{\tilde{\chi}^0_{{a}}}, m^2_{\tilde{e}_{{b}}}) \\ 
I_2= & B_1(m^2_{\tilde{\chi}^0_{{a}}}, m^2_{\tilde{e}_{{b}}}) \\ 
  \VLLZ{(n_1,n_2)}= & ( \VeezL_{\beta, c}(\VneeL_{a, \alpha, b}\VeneR_{c, a, b} I_2 m^2_{\ell_{{\alpha}}} - \VneeR_{a, \alpha, b}\VeneR_{c, a, b} I_1 m_{\ell_\alpha} m_{\tilde{\chi}^0_{{a}}} + \VneeR_{a, \alpha, b}\VeneL_{c, a, b} I_2 m_{\ell_\alpha} m_{\ell_{{c}}} \nonumber \\ & - \VneeL_{a, \alpha, b}\VeneL_{c, a, b} I_1 m_{\tilde{\chi}^0_{{a}}} m_{\ell_{{c}}}))/(m^2_{\ell_{{\alpha}}} - m^2_{\ell_{{c}}}) \hspace{1cm} + (\alpha\leftrightarrow \beta) \\ 
  \VLRZ{(n_1,n_2)}= & ( \VeezL_{\beta, c} (\VneeL_{a, \alpha, b}\VeneR_{c, a, b} I_2 m^2_{\ell_{{\alpha}}} - \VneeR_{a, \alpha, b}\VeneR_{c, a, b} I_1 m_{\ell_\alpha} m_{\tilde{\chi}^0_{{a}}} + \VneeR_{a, \alpha, b}\VeneL_{c, a, b} I_2 m_{\ell_\alpha} m_{\ell_{{c}}} \nonumber \\ & - \VneeL_{a, \alpha, b}\VeneL_{c, a, b} I_1 m_{\tilde{\chi}^0_{{a}}} m_{\ell_{{c}}}))/(m^2_{\ell_{{\alpha}}} - m^2_{\ell_{{c}}}) \hspace{1cm} + (\alpha\leftrightarrow \beta) 
\end{align}
\paragraph*{Vertex corrections} 

\begin{align} 
  \VLLZ{(n_3)}= & -2  \VneeL_{a, \alpha, b}\VeneR_{\beta, a, c} \VeeZ _{b, c} \, C_{00}(m^2_{\tilde{\chi}^0_{{a}}}, m^2_{\tilde{e}_{{c}}}, m^2_{\tilde{e}_{{b}}}) \\ 
  \VLRZ{(n_3)}= & -2  \VneeL_{a, \alpha, b}\VeneR_{\beta, a, c} \VeeZ _{b, c} \, C_{00}(m^2_{\tilde{\chi}^0_{{a}}}, m^2_{\tilde{e}_{{c}}}, m^2_{\tilde{e}_{{b}}}) \\
  \nonumber \\
I_1= & B_0(m^2_{\tilde{\chi}^0_{{b}}}, m^2_{\tilde{\chi}^0_{{c}}}) \\ 
I_2= & C_{00}(m^2_{\tilde{\chi}^0_{{c}}}, m^2_{\tilde{\chi}^0_{{b}}}, m^2_{\tilde{e}_{{a}}}) \\ 
I_3= & C_0(m^2_{\tilde{\chi}^0_{{c}}}, m^2_{\tilde{\chi}^0_{{b}}}, m^2_{\tilde{e}_{{a}}}) \\ 
  \VLLZ{(n_4)}= &  \VneeL_{b, \alpha, a}\VeneR_{\beta, c, a} (-(\VnnzL_{c, b} I_3 m_{\tilde{\chi}^0_{{b}}} m_{\tilde{\chi}^0_{{c}}}) + \VnnzR_{c, b} (I_1 - 2 I_2 + I_3 m^2_{\tilde{e}_{{a}}})) \\ 
  \VLRZ{(n_4)}= &  \VneeL_{b, \alpha, a}\VeneR_{\beta, c, a} (-(\VnnzL_{c, b} I_3 m_{\tilde{\chi}^0_{{b}}} m_{\tilde{\chi}^0_{{c}}}) + \VnnzR_{c, b} (I_1 - 2 I_2 + I_3 m^2_{\tilde{e}_{{a}}}))
\end{align} 

\subsubsection{Scalar penguins}
\paragraph{CP even scalars}
\paragraph*{Self-energy corrections}
\begin{align} 
I_1= & B_0(m^2_{\tilde{\chi}^0_{{a}}}, m^2_{\tilde{e}_{{b}}}) \\ 
I_2= & B_1(m^2_{\tilde{\chi}^0_{{a}}}, m^2_{\tilde{e}_{{b}}}) \\ 
  \SLLh{(n_1,n_2)}= & ( \VeehL_{\beta, c, p}  (-(\VneeL_{a, \alpha, b}\VeneR_{c, a, b} I_2 m^2_{\ell_{{\alpha}}}) + \VneeR_{a, \alpha, b}\VeneR_{c, a, b} I_1 m_{\ell_\alpha} m_{\tilde{\chi}^0_{{a}}} - \VneeR_{a, \alpha, b}\VeneL_{c, a, b} I_2 m_{\ell_\alpha} m_{\ell_{{c}}} \nonumber \\ &  + \VneeL_{a, \alpha, b}\VeneL_{c, a, b} I_1 m_{\tilde{\chi}^0_{{a}}} m_{\ell_{{c}}}))/(m^2_{\ell_{{\alpha}}} - m^2_{\ell_{{c}}}) \hspace{1cm} + (\alpha\leftrightarrow \beta)\\ 
  \SLRh{(n_1,n_2)}= & ( \VeehL_{\beta, c, p}  (-(\VneeL_{a, \alpha, b}\VeneR_{c, a, b} I_2 m^2_{\ell_{{\alpha}}}) + \VneeR_{a, \alpha, b}\VeneR_{c, a, b} I_1 m_{\ell_\alpha} m_{\tilde{\chi}^0_{{a}}} - \VneeR_{a, \alpha, b}\VeneL_{c, a, b} I_2 m_{\ell_\alpha} m_{\ell_{{c}}} \nonumber \\ & + \VneeL_{a, \alpha, b}\VeneL_{c, a, b} I_1 m_{\tilde{\chi}^0_{{a}}} m_{\ell_{{c}}}))/(m^2_{\ell_{{\alpha}}} - m^2_{\ell_{{c}}}) \hspace{1cm} + (\alpha\leftrightarrow \beta)
\end{align} 
 
\paragraph*{Vertex corrections}
\begin{align} 
  \SLLh{(n_3)}= & \, \VneeL_{a, \alpha, b}\VeneL_{\beta, a, c} \VeeH_{p, b, c} \, C_0(m^2_{\tilde{\chi}^0_{{a}}}, m^2_{\tilde{e}_{{c}}}, m^2_{\tilde{e}_{{b}}}) \, m_{\tilde{\chi}^0_{{a}}} \\ 
  \SLRh{(n_3)}= & \, \VneeL_{a, \alpha, b}\VeneL_{\beta, a, c} \VeeH_{p, b, c} \, C_0(m^2_{\tilde{\chi}^0_{{a}}}, m^2_{\tilde{e}_{{c}}}, m^2_{\tilde{e}_{{b}}}) \, m_{\tilde{\chi}^0_{{a}}} \\ 
  \nonumber \\  
I_1= & B_0(m^2_{\tilde{\chi}^0_{{b}}}, m^2_{\tilde{\chi}^0_{{c}}}) \\ 
I_2= & C_0(m^2_{\tilde{\chi}^0_{{c}}}, m^2_{\tilde{\chi}^0_{{b}}}, m^2_{\tilde{e}_{{a}}}) \\ 
  \SLLh{(n_4)}= &  \VneeL_{b, \alpha, a}\VeneL_{\beta, c, a}  (\VnnhL_{c, b, p} I_2 m_{\tilde{\chi}^0_{{b}}} m_{\tilde{\chi}^0_{{c}}} + \VnnhR_{c, b, p} (I_1 + I_2 m^2_{\tilde{e}_{{a}}})) \\ 
  \SLRh{(n_4)}= &  \VneeL_{b, \alpha, a}\VeneL_{\beta, c, a}  (\VnnhL_{c, b, p} I_2 m_{\tilde{\chi}^0_{{b}}} m_{\tilde{\chi}^0_{{c}}} + \VnnhR_{c, b, p} (I_1 + I_2 m^2_{\tilde{e}_{{a}}})) 
\end{align} 

\paragraph{CP odd scalars}
\paragraph*{Self-energy corrections}
\begin{align} 
I_1= & B_0(m^2_{\tilde{\chi}^0_{{a}}}, m^2_{\tilde{e}_{{b}}}) \\ 
I_2= & B_1(m^2_{\tilde{\chi}^0_{{a}}}, m^2_{\tilde{e}_{{b}}}) \\ 
  \SLLa{(n_1,n_2)}= & ( \VeeaL_{\beta, c, p}  (-(\VneeL_{a, \alpha, b}\VeneR_{c, a, b} I_2 m^2_{\ell_{{\alpha}}}) + \VneeR_{a, \alpha, b}\VeneR_{c, a, b} I_1 m_{\ell_\alpha} m_{\tilde{\chi}^0_{{a}}} - \VneeR_{a, \alpha, b}\VeneL_{c, a, b} I_2 m_{\ell_\alpha} m_{\ell_{{c}}} \nonumber \\ & + \VneeL_{a, \alpha, b}\VeneL_{c, a, b} I_1 m_{\tilde{\chi}^0_{{a}}} m_{\ell_{{c}}}))/(m^2_{\ell_{{\alpha}}} - m^2_{\ell_{{c}}}) \hspace{1cm} + (\alpha\leftrightarrow \beta)\\ 
  \SLRa{(n_1,n_2)}= & ( \VeeaL_{\beta, c, p}  (-(\VneeL_{a, \alpha, b}\VeneR_{c, a, b} I_2 m^2_{\ell_{{\alpha}}}) + \VneeR_{a, \alpha, b}\VeneR_{c, a, b} I_1 m_{\ell_\alpha} m_{\tilde{\chi}^0_{{a}}} - \VneeR_{a, \alpha, b}\VeneL_{c, a, b} I_2 m_{\ell_\alpha} m_{\ell_{{c}}} \nonumber \\ & + \VneeL_{a, \alpha, b}\VeneL_{c, a, b} I_1 m_{\tilde{\chi}^0_{{a}}} m_{\ell_{{c}}}))/(m^2_{\ell_{{\alpha}}} - m^2_{\ell_{{c}}}) \hspace{1cm} + (\alpha\leftrightarrow \beta)
\end{align} 
\paragraph*{Vertex corrections}
\begin{align} 
  \SLLa{(n_3)}= & \, \VneeL_{a, \alpha, b}\VeneL_{\beta, a, c} \Vaee_{p, b, c} \, C_0(m^2_{\tilde{\chi}^0_{{a}}}, m^2_{\tilde{e}_{{c}}}, m^2_{\tilde{e}_{{b}}}) \, m_{\tilde{\chi}^0_{{a}}} \\ 
  \SLRa{(n_3)}= & \, \VneeL_{a, \alpha, b}\VeneL_{\beta, a, c} \Vaee_{p, b, c} \, C_0(m^2_{\tilde{\chi}^0_{{a}}}, m^2_{\tilde{e}_{{c}}}, m^2_{\tilde{e}_{{b}}}) \, m_{\tilde{\chi}^0_{{a}}} \\ 
    \nonumber \\
I_1= & B_0(m^2_{\tilde{\chi}^0_{{b}}}, m^2_{\tilde{\chi}^0_{{c}}}) \\ 
I_2= & C_0(m^2_{\tilde{\chi}^0_{{c}}}, m^2_{\tilde{\chi}^0_{{b}}}, m^2_{\tilde{e}_{{a}}}) \\ 
  \SLLa{(n_4)} = & \, \VneeL_{b, \alpha, a}\VeneL_{\beta, c, a}  (\VnnaL_{c, b, p} I_2 m_{\tilde{\chi}^0_{{b}}} m_{\tilde{\chi}^0_{{c}}} + \VnnaR_{c, b, p} (I_1 + I_2 m^2_{\tilde{e}_{{a}}})) \\ 
  \SLRa{(n_4)} = & \, \VneeL_{b, \alpha, a}\VeneL_{\beta, c, a}  (\VnnaL_{c, b, p} I_2 m_{\tilde{\chi}^0_{{b}}} m_{\tilde{\chi}^0_{{c}}} + \VnnaR_{c, b, p} (I_1 + I_2 m^2_{\tilde{e}_{{a}}})) 
\end{align} 

\subsection{Chargino contributions}
\subsubsection{Z-penguins}
\paragraph*{Self-energy corrections}

\begin{align} 
I_1= & B_0(m^2_{\tilde{\chi}^-_{{b}}}, m^2_{\nu^i_{{a}}}) \\ 
I_2= & B_1(m^2_{\tilde{\chi}^-_{{b}}}, m^2_{\nu^i_{{a}}}) \\ 
  \VLLZ{(c_1,c_3)}= & ( \VeezL_{\beta, c}  (\VcevL_{b, \alpha, a} \VecviR_{c, b, a} I_2 m^2_{\ell_{{\alpha}}} - \VcevR_{b, \alpha, a} \VecviR_{c, b, a} I_1 m_{\ell_\alpha} m_{\tilde{\chi}^-_{{b}}} + \VcevR_{b, \alpha, a} \VecviL_{c, b, a} I_2 m_{\ell_\alpha} m_{\ell_{{c}}} \nonumber \\ & - \VcevL_{b, \alpha, a} \VecviL_{c, b, a} I_1 m_{\tilde{\chi}^-_{{b}}} m_{\ell_{{c}}}))/(m^2_{\ell_{{\alpha}}} - m^2_{\ell_{{c}}}) \hspace{1cm} + (\alpha\leftrightarrow \beta)\\ 
  \VLRZ{(c_1,c_3)}= & ( \VeezL_{\beta, c}  (\VcevL_{b, \alpha, a} \VecviR_{c, b, a} I_2 m^2_{\ell_{{\alpha}}} - \VcevR_{b, \alpha, a} \VecviR_{c, b, a} I_1 m_{\ell_\alpha} m_{\tilde{\chi}^-_{{b}}} + \VcevR_{b, \alpha, a} \VecviL_{c, b, a} I_2 m_{\ell_\alpha} m_{\ell_{{c}}} \nonumber \\ & - \VcevL_{b, \alpha, a} \VecviL_{c, b, a} I_1 m_{\tilde{\chi}^-_{{b}}} m_{\ell_{{c}}}))/(m^2_{\ell_{{\alpha}}} - m^2_{\ell_{{c}}}) \hspace{1cm} + (\alpha\leftrightarrow \beta)
\end{align} 
\begin{align} 
I_1= & B_0(m^2_{\tilde{\chi}^-_{{b}}}, m^2_{\nu^R_{{a}}}) \\ 
I_2= & B_1(m^2_{\tilde{\chi}^-_{{b}}}, m^2_{\nu^R_{{a}}}) \\ 
  \VLLZ{(c_2,c_4)}= & ( \VeezL_{\beta, c}  (\VcevrL_{b, \alpha, a} \VecvrR_{c, b, a} I_2 m^2_{\ell_{{\alpha}}} - \VcevrR_{b, \alpha, a} \VecvrR_{c, b, a} I_1 m_{\ell_\alpha} m_{\tilde{\chi}^-_{{b}}} + \VcevrR_{b, \alpha, a} \VecvrL_{c, b, a} I_2 m_{\ell_\alpha} m_{\ell_{{c}}} \nonumber \\ & - \VcevrL_{b, \alpha, a} \VecvrL_{c, b, a} I_1 m_{\tilde{\chi}^-_{{b}}} m_{\ell_{{c}}}))/(m^2_{\ell_{{\alpha}}} - m^2_{\ell_{{c}}}) \hspace{1cm} + (\alpha\leftrightarrow \beta)\\ 
  \VLRZ{(c_2,c_4)}= & ( \VeezL_{\beta, c}  (\VcevrL_{b, \alpha, a} \VecvrR_{c, b, a} I_2 m^2_{\ell_{{\alpha}}} - \VcevrR_{b, \alpha, a} \VecvrR_{c, b, a} I_1 m_{\ell_\alpha} m_{\tilde{\chi}^-_{{b}}} + \VcevrR_{b, \alpha, a} \VecvrL_{c, b, a} I_2 m_{\ell_\alpha} m_{\ell_{{c}}} \nonumber \\ & - \VcevrL_{b, \alpha, a} \VecvrL_{c, b, a} I_1 m_{\tilde{\chi}^-_{{b}}} m_{\ell_{{c}}}))/(m^2_{\ell_{{\alpha}}} - m^2_{\ell_{{c}}}) \hspace{1cm} + (\alpha\leftrightarrow \beta)
\end{align} 
\paragraph*{Vertex corrections}

\begin{align} 
I_1= & B_0(m^2_{\tilde{\chi}^-_{{b}}}, m^2_{\tilde{\chi}^-_{{c}}}) \\ 
I_2= & C_{00}(m^2_{\tilde{\chi}^-_{{c}}}, m^2_{\tilde{\chi}^-_{{b}}}, m^2_{\nu^i_{{a}}}) \\ 
I_3= & C_0(m^2_{\tilde{\chi}^-_{{c}}}, m^2_{\tilde{\chi}^-_{{b}}}, m^2_{\nu^i_{{a}}}) \\ 
  \VLLZ{(c_5)}= &  \VcevL_{b, \alpha, a} \VecviR_{\beta, c, a}  (-(\VcczL_{c, b} I_3 m_{\tilde{\chi}^-_{{b}}} m_{\tilde{\chi}^-_{{c}}}) + \VcczR_{c, b} (I_1 - 2 I_2 + I_3 m^2_{\nu^i_{{a}}})) \\ 
  \VLRZ{(c_5)}= &  \VcevL_{b, \alpha, a} \VecviR_{\beta, c, a}  (-(\VcczL_{c, b} I_3 m_{\tilde{\chi}^-_{{b}}} m_{\tilde{\chi}^-_{{c}}}) + \VcczR_{c, b} (I_1 - 2 I_2 + I_3 m^2_{\nu^i_{{a}}})) \\ 
  \nonumber \\  
I_1= & B_0(m^2_{\tilde{\chi}^-_{{b}}}, m^2_{\tilde{\chi}^-_{{c}}}) \\ 
I_2= & C_{00}(m^2_{\tilde{\chi}^-_{{c}}}, m^2_{\tilde{\chi}^-_{{b}}}, m^2_{\nu^R_{{a}}}) \\ 
I_3= & C_0(m^2_{\tilde{\chi}^-_{{c}}}, m^2_{\tilde{\chi}^-_{{b}}}, m^2_{\nu^R_{{a}}}) \\ 
  \VLLZ{(c_6)}= &  \VcevrL_{b, \alpha, a} \VecvrR_{\beta, c, a}  (-(\VcczL_{c, b} I_3 m_{\tilde{\chi}^-_{{b}}} m_{\tilde{\chi}^-_{{c}}}) + \VcczR_{c, b} (I_1 - 2 I_2 + I_3 m^2_{\nu^R_{{a}}})) \\ 
  \VLRZ{(c_6)}= &  \VcevrL_{b, \alpha, a} \VecvrR_{\beta, c, a}  (-(\VcczL_{c, b} I_3 m_{\tilde{\chi}^-_{{b}}} m_{\tilde{\chi}^-_{{c}}}) + \VcczR_{c, b} (I_1 - 2 I_2 + I_3 m^2_{\nu^R_{{a}}})) \\ 
  \nonumber \\  
  \VLLZ{(c_7)}= & -2  \VcevrL_{a, \alpha, b} \VecviR_{\beta, a, c} \VvivrZ _{c, b} \,  C_{00}(m^2_{\tilde{\chi}^-_{{a}}}, m^2_{\nu^i_{{c}}}, m^2_{\nu^R_{{b}}}) \\ 
  \VLRZ{(c_7)}= & -2  \VcevrL_{a, \alpha, b} \VecviR_{\beta, a, c} \VvivrZ _{c, b} \,  C_{00}(m^2_{\tilde{\chi}^-_{{a}}}, m^2_{\nu^i_{{c}}}, m^2_{\nu^R_{{b}}}) \\ 
    \nonumber \\
  \VLLZ{(c_8)}= & \, 2  \VcevL_{a, \alpha, b} \VecvrR_{\beta, a, c} \VvivrZ_{b, c} \, C_{00}(m^2_{\tilde{\chi}^-_{{a}}}, m^2_{\nu^R_{{c}}}, m^2_{\nu^i_{{b}}}) \\ 
  \VLRZ{(c_8)}= & \, 2  \VcevL_{a, \alpha, b} \VecvrR_{\beta, a, c} \VvivrZ_{b, c} \, C_{00}(m^2_{\tilde{\chi}^-_{{a}}}, m^2_{\nu^R_{{c}}}, m^2_{\nu^i_{{b}}})
\end{align}

\subsubsection{Scalar penguins}
\paragraph{CP even scalars}
\paragraph*{Self-energy corrections}
\begin{align} 
I_1= & B_0(m^2_{\tilde{\chi}^-_{{b}}}, m^2_{\nu^i_{{a}}}) \\ 
I_2= & B_1(m^2_{\tilde{\chi}^-_{{b}}}, m^2_{\nu^i_{{a}}}) \\ 
  \SLLh{(c_1,c_3)}= & ( \VeehL_{\beta, c, p}  (-(\VcevL_{b, \alpha, a} \VecviR_{c, b, a} I_2 m^2_{\ell_{{\alpha}}}) + \VcevR_{b, \alpha, a} \VecviR_{c, b, a} I_1 m_{\ell_\alpha} m_{\tilde{\chi}^-_{{b}}} - \VcevR_{b, \alpha, a} \VecviL_{c, b, a} I_2 m_{\ell_\alpha} m_{\ell_{{c}}} \nonumber \\ & + \VcevL_{b, \alpha, a} \VecviL_{c, b, a} I_1 m_{\tilde{\chi}^-_{{b}}} m_{\ell_{{c}}}))/(m^2_{\ell_{{\alpha}}} - m^2_{\ell_{{c}}}) \hspace{1cm} + (\alpha\leftrightarrow \beta)\\ 
  \SLRh{(c_1,c_3)}= & ( \VeehL_{\beta, c, p}  (-(\VcevL_{b, \alpha, a} \VecviR_{c, b, a} I_2 m^2_{\ell_{{\alpha}}}) + \VcevR_{b, \alpha, a} \VecviR_{c, b, a} I_1 m_{\ell_\alpha} m_{\tilde{\chi}^-_{{b}}} - \VcevR_{b, \alpha, a} \VecviL_{c, b, a} I_2 m_{\ell_\alpha} m_{\ell_{{c}}} \nonumber \\ & + \VcevL_{b, \alpha, a} \VecviL_{c, b, a} I_1 m_{\tilde{\chi}^-_{{b}}} m_{\ell_{{c}}}))/(m^2_{\ell_{{\alpha}}} - m^2_{\ell_{{c}}}) \hspace{1cm} + (\alpha\leftrightarrow \beta)
  \nonumber \\  
I_1= & B_0(m^2_{\tilde{\chi}^-_{{b}}}, m^2_{\nu^R_{{a}}}) \\ 
I_2= & B_1(m^2_{\tilde{\chi}^-_{{b}}}, m^2_{\nu^R_{{a}}}) \\ 
  \SLLh{(c_2,c_4)}= & ( \VeehL_{\beta, c, p}  (-(\VcevrL_{b, \alpha, a} \VecvrR_{c, b, a} I_2 m^2_{\ell_{{\alpha}}}) + \VcevrR_{b, \alpha, a} \VecvrR_{c, b, a} I_1 m_{\ell_\alpha} m_{\tilde{\chi}^-_{{b}}} - \VcevrR_{b, \alpha, a} \VecvrL_{c, b, a} I_2 m_{\ell_\alpha} m_{\ell_{{c}}} \nonumber \\ &  + \VcevrL_{b, \alpha, a} \VecvrL_{c, b, a} I_1 m_{\tilde{\chi}^-_{{b}}} m_{\ell_{{c}}}))/(m^2_{\ell_{{\alpha}}} - m^2_{\ell_{{c}}}) \hspace{1cm} + (\alpha\leftrightarrow \beta)\\ 
  \SLRh{(c_2,c_4)}= & ( \VeehL_{\beta, c, p}  (-(\VcevrL_{b, \alpha, a} \VecvrR_{c, b, a} I_2 m^2_{\ell_{{\alpha}}}) + \VcevrR_{b, \alpha, a} \VecvrR_{c, b, a} I_1 m_{\ell_\alpha} m_{\tilde{\chi}^-_{{b}}} - \VcevrR_{b, \alpha, a} \VecvrL_{c, b, a} I_2 m_{\ell_\alpha} m_{\ell_{{c}}} \nonumber \\ & + \VcevrL_{b, \alpha, a} \VecvrL_{c, b, a} I_1 m_{\tilde{\chi}^-_{{b}}} m_{\ell_{{c}}}))/(m^2_{\ell_{{\alpha}}} - m^2_{\ell_{{c}}}) \hspace{1cm} + (\alpha\leftrightarrow \beta) 
\end{align} 
\paragraph*{Vertex corrections}
\begin{align} 
I_1= & B_0(m^2_{\tilde{\chi}^-_{{b}}}, m^2_{\tilde{\chi}^-_{{c}}}) \\ 
I_2= & C_0(m^2_{\tilde{\chi}^-_{{c}}}, m^2_{\tilde{\chi}^-_{{b}}}, m^2_{\nu^i_{{a}}}) \\ 
  \SLLh{(c_5)}= & \, \VcevL_{b, \alpha, a} \VecviL_{\beta, c, a}  (\VcchL_{c, b, p} I_2 m_{\tilde{\chi}^-_{{b}}} m_{\tilde{\chi}^-_{{c}}} + \VcchR_{c, b, p} (I_1 + I_2 m^2_{\nu^i_{{a}}})) \\ 
  \SLRh{(c_5)}= & \, \VcevL_{b, \alpha, a} \VecviL_{\beta, c, a}  (\VcchL_{c, b, p} I_2 m_{\tilde{\chi}^-_{{b}}} m_{\tilde{\chi}^-_{{c}}} + \VcchR_{c, b, p} (I_1 + I_2 m^2_{\nu^i_{{a}}})) \\ 
  \nonumber \\
I_1= & B_0(m^2_{\tilde{\chi}^-_{{b}}}, m^2_{\tilde{\chi}^-_{{c}}}) \\ 
I_2= & C_0(m^2_{\tilde{\chi}^-_{{c}}}, m^2_{\tilde{\chi}^-_{{b}}}, m^2_{\nu^R_{{a}}}) \\ 
  \SLLh{(c_6)}= & \, \VcevrL_{b, \alpha, a} \VecvrL_{\beta, c, a}  (\VcchL_{c, b, p} I_2 m_{\tilde{\chi}^-_{{b}}} m_{\tilde{\chi}^-_{{c}}} + \VcchR_{c, b, p} (I_1 + I_2 m^2_{\nu^R_{{a}}})) \\ 
  \SLRh{(c_6)}= & \, \VcevrL_{b, \alpha, a} \VecvrL_{\beta, c, a}  (\VcchL_{c, b, p} I_2 m_{\tilde{\chi}^-_{{b}}} m_{\tilde{\chi}^-_{{c}}} + \VcchR_{c, b, p} (I_1 + I_2 m^2_{\nu^R_{{a}}})) \\ 
  \nonumber \\
  \SLLh{(c_{10})} = & \, \VcevL_{a, \alpha, b} \VecviL_{\beta, a, c}\Vhvivi_{p, c, b}  C_0(m^2_{\tilde{\chi}^-_{{a}}}, m^2_{\nu^i_{{c}}}, m^2_{\nu^i_{{b}}}) \, m_{\tilde{\chi}^-_{{a}}} \\ 
  \SLRh{(c_{10})} = & \, \VcevL_{a, \alpha, b} \VecviL_{\beta, a, c}\Vhvivi_{p, c, b}  C_0(m^2_{\tilde{\chi}^-_{{a}}}, m^2_{\nu^i_{{c}}}, m^2_{\nu^i_{{b}}}) \, m_{\tilde{\chi}^-_{{a}}} \\ 
  \nonumber \\
  \SLLh{(c_7)}= & \, \VcevrL_{a, \alpha, b} \VecviL_{\beta, a, c} \Vhvivr_{p, c, b}  C_0(m^2_{\tilde{\chi}^-_{{a}}}, m^2_{\nu^i_{{c}}}, m^2_{\nu^R_{{b}}}) \, m_{\tilde{\chi}^-_{{a}}} \\ 
  \SLRh{(c_7)}= & \, \VcevrL_{a, \alpha, b} \VecviL_{\beta, a, c} \Vhvivr_{p, c, b}  C_0(m^2_{\tilde{\chi}^-_{{a}}}, m^2_{\nu^i_{{c}}}, m^2_{\nu^R_{{b}}}) \, m_{\tilde{\chi}^-_{{a}}} \\ 
  \nonumber \\
  \SLLh{(c_8)}= & \, \VcevL_{a, \alpha, b} \VecvrL_{\beta, a, c} \Vhvivr_{p, b, c}  C_0(m^2_{\tilde{\chi}^-_{{a}}}, m^2_{\nu^R_{{c}}}, m^2_{\nu^i_{{b}}}) \, m_{\tilde{\chi}^-_{{a}}} \\ 
  \SLRh{(c_8)}= & \, \VcevL_{a, \alpha, b} \VecvrL_{\beta, a, c} \Vhvivr_{p, b, c}  C_0(m^2_{\tilde{\chi}^-_{{a}}}, m^2_{\nu^R_{{c}}}, m^2_{\nu^i_{{b}}}) \, m_{\tilde{\chi}^-_{{a}}} \\ 
  \nonumber \\
  \SLLh{(c_9)}= & \, \VcevrL_{a, \alpha, b} \VecvrL_{\beta, a, c}\Vhvrvr_{p, c, b}  C_0(m^2_{\tilde{\chi}^-_{{a}}}, m^2_{\nu^R_{{c}}}, m^2_{\nu^R_{{b}}}) \, m_{\tilde{\chi}^-_{{a}}} \\ 
  \SLRh{(c_9)}= & \, \VcevrL_{a, \alpha, b} \VecvrL_{\beta, a, c}\Vhvrvr_{p, c, b}  C_0(m^2_{\tilde{\chi}^-_{{a}}}, m^2_{\nu^R_{{c}}}, m^2_{\nu^R_{{b}}}) \, m_{\tilde{\chi}^-_{{a}}} \\ 
\end{align}

\paragraph{CP odd scalars}
\paragraph*{Self-energy corrections}
\begin{align} 
I_1= & B_0(m^2_{\tilde{\chi}^-_{{b}}}, m^2_{\nu^i_{{a}}}) \\ 
I_2= & B_1(m^2_{\tilde{\chi}^-_{{b}}}, m^2_{\nu^i_{{a}}}) \\ 
  \SLLa{(c_1,c_3)}= & ( \VeeaL_{\beta, c, p}  (-(\VcevL_{b, \alpha, a} \VecviR_{c, b, a} I_2 m^2_{\ell_{{\alpha}}}) + \VcevR_{b, \alpha, a} \VecviR_{c, b, a} I_1 m_{\ell_\alpha} m_{\tilde{\chi}^-_{{b}}} - \VcevR_{b, \alpha, a} \VecviL_{c, b, a} I_2 m_{\ell_\alpha} m_{\ell_{{c}}} \nonumber \\ & + \VcevL_{b, \alpha, a} \VecviL_{c, b, a} I_1 m_{\tilde{\chi}^-_{{b}}} m_{\ell_{{c}}}))/(m^2_{\ell_{{\alpha}}} - m^2_{\ell_{{c}}}) \hspace{1cm} + (\alpha\leftrightarrow \beta)\\ 
  \SLRa{(c_1,c_3)}= & ( \VeeaL_{\beta, c, p}  (-(\VcevL_{b, \alpha, a} \VecviR_{c, b, a} I_2 m^2_{\ell_{{\alpha}}}) + \VcevR_{b, \alpha, a} \VecviR_{c, b, a} I_1 m_{\ell_\alpha} m_{\tilde{\chi}^-_{{b}}} - \VcevR_{b, \alpha, a} \VecviL_{c, b, a} I_2 m_{\ell_\alpha} m_{\ell_{{c}}} \nonumber \\ & + \VcevL_{b, \alpha, a} \VecviL_{c, b, a} I_1 m_{\tilde{\chi}^-_{{b}}} m_{\ell_{{c}}}))/(m^2_{\ell_{{\alpha}}} - m^2_{\ell_{{c}}}) \hspace{1cm} + (\alpha\leftrightarrow \beta) 
  \nonumber \\
I_1= & B_0(m^2_{\tilde{\chi}^-_{{b}}}, m^2_{\nu^R_{{a}}}) \\ 
I_2= & B_1(m^2_{\tilde{\chi}^-_{{b}}}, m^2_{\nu^R_{{a}}}) \\ 
  \SLLa{(c_2,c_4)}= & ( \VeeaL_{\beta, c, p}  (-(\VcevrL_{b, \alpha, a} \VecvrR_{c, b, a} I_2 m^2_{\ell_{{\alpha}}}) + \VcevrR_{b, \alpha, a} \VecvrR_{c, b, a} I_1 m_{\ell_\alpha} m_{\tilde{\chi}^-_{{b}}} - \VcevrR_{b, \alpha, a} \VecvrL_{c, b, a} I_2 m_{\ell_\alpha} m_{\ell_{{c}}} \nonumber \\ & + \VcevrL_{b, \alpha, a} \VecvrL_{c, b, a} I_1 m_{\tilde{\chi}^-_{{b}}} m_{\ell_{{c}}}))/(m^2_{\ell_{{\alpha}}} - m^2_{\ell_{{c}}}) \hspace{1cm} + (\alpha\leftrightarrow \beta)\\ 
  \SLRa{(c_2,c_4)}= & ( \VeeaL_{\beta, c, p}  (-(\VcevrL_{b, \alpha, a} \VecvrR_{c, b, a} I_2 m^2_{\ell_{{\alpha}}}) + \VcevrR_{b, \alpha, a} \VecvrR_{c, b, a} I_1 m_{\ell_\alpha} m_{\tilde{\chi}^-_{{b}}} - \VcevrR_{b, \alpha, a} \VecvrL_{c, b, a} I_2 m_{\ell_\alpha} m_{\ell_{{c}}} \nonumber \\ & + \VcevrL_{b, \alpha, a} \VecvrL_{c, b, a} I_1 m_{\tilde{\chi}^-_{{b}}} m_{\ell_{{c}}}))/(m^2_{\ell_{{\alpha}}} - m^2_{\ell_{{c}}}) \hspace{1cm} + (\alpha\leftrightarrow \beta)
\end{align} 

\paragraph*{Vertex corrections}
\begin{align} 
I_1= & B_0(m^2_{\tilde{\chi}^-_{{b}}}, m^2_{\tilde{\chi}^-_{{c}}}) \\ 
I_2= & C_0(m^2_{\tilde{\chi}^-_{{c}}}, m^2_{\tilde{\chi}^-_{{b}}}, m^2_{\nu^i_{{a}}}) \\ 
  \SLLa{(c_5)}= & \, \VcevL_{b, \alpha, a} \VecviL_{\beta, c, a}  (\VccaL_{c, b, p} I_2 m_{\tilde{\chi}^-_{{b}}} m_{\tilde{\chi}^-_{{c}}} + \VccaR_{c, b, p} (I_1 + I_2 m^2_{\nu^i_{{a}}})) \\ 
  \SLRa{(c_5)}= & \, \VcevL_{b, \alpha, a} \VecviL_{\beta, c, a}  (\VccaL_{c, b, p} I_2 m_{\tilde{\chi}^-_{{b}}} m_{\tilde{\chi}^-_{{c}}} + \VccaR_{c, b, p} (I_1 + I_2 m^2_{\nu^i_{{a}}})) \\ 
  \nonumber \\
I_1= & B_0(m^2_{\tilde{\chi}^-_{{b}}}, m^2_{\tilde{\chi}^-_{{c}}}) \\ 
I_2= & C_0(m^2_{\tilde{\chi}^-_{{c}}}, m^2_{\tilde{\chi}^-_{{b}}}, m^2_{\nu^R_{{a}}}) \\ 
  \SLLa{(c_6)}= & \, \VcevrL_{b, \alpha, a} \VecvrL_{\beta, c, a}  (\VccaL_{c, b, p} I_2 m_{\tilde{\chi}^-_{{b}}} m_{\tilde{\chi}^-_{{c}}} + \VccaR_{c, b, p} (I_1 + I_2 m^2_{\nu^R_{{a}}})) \\ 
  \SLRa{(c_6)}= & \, \VcevrL_{b, \alpha, a} \VecvrL_{\beta, c, a}  (\VccaL_{c, b, p} I_2 m_{\tilde{\chi}^-_{{b}}} m_{\tilde{\chi}^-_{{c}}} + \VccaR_{c, b, p} (I_1 + I_2 m^2_{\nu^R_{{a}}})) \\ 
  \nonumber \\
  \SLLa{(c_{10})}= & \, \VcevL_{a, \alpha, b} \VecviL_{\beta, a, c} \Vavivi_{p, c, b}  C_0(m^2_{\tilde{\chi}^-_{{a}}}, m^2_{\nu^i_{{c}}}, m^2_{\nu^i_{{b}}}) \, m_{\tilde{\chi}^-_{{a}}} \\ 
  \SLRa{(c_{10})}= & \, \VcevL_{a, \alpha, b} \VecviL_{\beta, a, c} \Vavivi_{p, c, b}  C_0(m^2_{\tilde{\chi}^-_{{a}}}, m^2_{\nu^i_{{c}}}, m^2_{\nu^i_{{b}}}) \, m_{\tilde{\chi}^-_{{a}}} \\ 
  \nonumber \\
  \SLLa{(c_7)}= & \, \VcevrL_{a, \alpha, b} \VecviL_{\beta, a, c} \Vavivr_{p, c, b}  C_0(m^2_{\tilde{\chi}^-_{{a}}}, m^2_{\nu^i_{{c}}}, m^2_{\nu^R_{{b}}}) \, m_{\tilde{\chi}^-_{{a}}} \\ 
  \SLRa{(c_7)}= & \, \VcevrL_{a, \alpha, b} \VecviL_{\beta, a, c} \Vavivr_{p, c, b}  C_0(m^2_{\tilde{\chi}^-_{{a}}}, m^2_{\nu^i_{{c}}}, m^2_{\nu^R_{{b}}}) \, m_{\tilde{\chi}^-_{{a}}} \\ 
  \nonumber \\
  \SLLa{(c_8)}= & \, \VcevL_{a, \alpha, b} \VecvrL_{\beta, a, c} \Vavivr_{p, b, c}  C_0(m^2_{\tilde{\chi}^-_{{a}}}, m^2_{\nu^R_{{c}}}, m^2_{\nu^i_{{b}}}) \, m_{\tilde{\chi}^-_{{a}}} \\ 
  \SLRa{(c_8)}= & \, \VcevL_{a, \alpha, b} \VecvrL_{\beta, a, c} \Vavivr_{p, b, c}  C_0(m^2_{\tilde{\chi}^-_{{a}}}, m^2_{\nu^R_{{c}}}, m^2_{\nu^i_{{b}}}) \, m_{\tilde{\chi}^-_{{a}}} \\ 
  \nonumber \\
  \SLLa{(c_9)}= & \, \VcevrL_{a, \alpha, b} \VecvrL_{\beta, a, c} \Vavrvr_{p, c, b}   C_0(m^2_{\tilde{\chi}^-_{{a}}}, m^2_{\nu^R_{{c}}}, m^2_{\nu^R_{{b}}}) \, m_{\tilde{\chi}^-_{{a}}} \\ 
  \SLRa{(c_9)}= & \, \VcevrL_{a, \alpha, b} \VecvrL_{\beta, a, c} \Vavrvr_{p, c, b}   C_0(m^2_{\tilde{\chi}^-_{{a}}}, m^2_{\nu^R_{{c}}}, m^2_{\nu^R_{{b}}}) \, m_{\tilde{\chi}^-_{{a}}} 
\end{align}

\subsection{$W^+$ and $H^+$ contributions}
\subsubsection{Z-penguins}
\paragraph*{Self-energy corrections}

\begin{align} 
I_1= & B_0(m^2_{\nu_{{a}}}, m^2_{H^-_{{b}}}) \\ 
I_2= & B_1(m^2_{\nu_{{a}}}, m^2_{H^-_{{b}}}) \\ 
  \VLLZ{(w_1,w_3)}= & ( \VeezL_{\beta, c} (\VvehpL_{a, \alpha, b} \VevhmR_{c, a, b} I_2 m^2_{\ell_{{\alpha}}} - \VvehpR_{a, \alpha, b} \VevhmR_{c, a, b} I_1 m_{\ell_\alpha} m_{\nu_{{a}}} + \VvehpR_{a, \alpha, b} \VevhmL_{c, a, b} I_2 m_{\ell_\alpha} m_{\ell_{{c}}} \nonumber \\ & - \VvehpL_{a, \alpha, b} \VevhmL_{c, a, b} I_1 m_{\nu_{{a}}} m_{\ell_{{c}}}))/(m^2_{\ell_{{\alpha}}} - m^2_{\ell_{{c}}}) \hspace{1cm} + (\alpha\leftrightarrow \beta)\\ 
  \VLRZ{(w_1,w_3)}= & ( \VeezL_{\beta, c} (\VvehpL_{a, \alpha, b} \VevhmR_{c, a, b} I_2 m^2_{\ell_{{\alpha}}} - \VvehpR_{a, \alpha, b} \VevhmR_{c, a, b} I_1 m_{\ell_\alpha} m_{\nu_{{a}}} + \VvehpR_{a, \alpha, b} \VevhmL_{c, a, b} I_2 m_{\ell_\alpha} m_{\ell_{{c}}} \nonumber \\ & - \VvehpL_{a, \alpha, b} \VevhmL_{c, a, b} I_1 m_{\nu_{{a}}} m_{\ell_{{c}}}))/(m^2_{\ell_{{\alpha}}} - m^2_{\ell_{{c}}}) \hspace{1cm} + (\alpha\leftrightarrow \beta)
  \nonumber \\
I_1= & B_0(m^2_{\nu_{{a}}}, m^2_{W^-}) \\ 
I_2= & B_1(m^2_{\nu_{{a}}}, m^2_{W^-}) \\ 
  \VLLZ{(w_2,w_4)}= & ( \VeezL_{\beta, c}  (\VvewpR_{a, i} m_{\ell_\alpha} (-2 \VevwmL_{c, a} (1 - 2 I_1) m_{\nu_{{a}}} + \VevwmR_{c, a} (1 + 2 I_2) m_{\ell_{{c}}}) \nonumber \\ & + \VvewpL_{a, i} (\VevwmL_{c, a} (1 + 2 I_2) m^2_{\ell_{{\alpha}}} - 2 \VevwmR_{c, a} (1 - 2 I_1) m_{\nu_{{a}}} m_{\ell_{{c}}})))/(m^2_{\ell_{{\alpha}}} - m^2_{\ell_{{c}}}) \hspace{1cm} + (\alpha\leftrightarrow \beta)\\ 
  \VLRZ{(w_2,w_4)}= & ( \VeezL_{\beta, c}  (\VvewpR_{a, i} m_{\ell_\alpha} (-2 \VevwmL_{c, a} (1 - 2 I_1) m_{\nu_{{a}}} + \VevwmR_{c, a} (1 + 2 I_2) m_{\ell_{{c}}}) \nonumber \\ & + \VvewpL_{a, i} (\VevwmL_{c, a} (1 + 2 I_2) m^2_{\ell_{{\alpha}}} - 2 \VevwmR_{c, a} (1 - 2 I_1) m_{\nu_{{a}}} m_{\ell_{{c}}})))/(m^2_{\ell_{{\alpha}}} - m^2_{\ell_{{c}}}) \hspace{1cm} + (\alpha\leftrightarrow \beta)
\end{align} 
\paragraph*{Vertex corrections}
\begin{align} 
  \VLLZ{(w_5)}= & -2  \VvehpL_{a, \alpha, b} \VevhmR_{\beta, a, c} \Vhmhpz_{b, c} \, C_{00}(m^2_{\nu_{{a}}}, m^2_{H^-_{{c}}}, m^2_{H^-_{{b}}}) \\ 
  \VLRZ{(w_5)}= & -2  \VvehpL_{a, \alpha, b} \VevhmR_{\beta, a, c} \Vhmhpz_{b, c} \,  C_{00}(m^2_{\nu_{{a}}}, m^2_{H^-_{{c}}}, m^2_{H^-_{{b}}}) \\ 
  \nonumber \\
  \VLLZ{(w_6)}= & \, \VvewpL_{a, i} \VevhmR_{\beta, a, c} \Vhpwmz_{c} \, C_0(m^2_{\nu_{{a}}}, m^2_{H^-_{{c}}}, m^2_{W^-}) \, m_{\nu_{{a}}} \\ 
  \VLRZ{(w_6)}= & \, \VvewpL_{a, i} \VevhmR_{\beta, a, c} \Vhpwmz_{c} \, C_0(m^2_{\nu_{{a}}}, m^2_{H^-_{{c}}}, m^2_{W^-}) \, m_{\nu_{{a}}} \\ 
  \nonumber \\
  \VLLZ{(w_7)}= & \, \VvehpL_{a, \alpha, b} \VevwmL_{\beta, a} \Vhmwpz_{b} \, C_0(m^2_{\nu_{{a}}}, m^2_{W^-}, m^2_{H^-_{{b}}}) \, m_{\nu_{{a}}} \\ 
  \VLRZ{(w_7)}= & \, \VvehpL_{a, \alpha, b} \VevwmL_{\beta, a} \Vhmwpz_{b} \, C_0(m^2_{\nu_{{a}}}, m^2_{W^-}, m^2_{H^-_{{b}}}) \, m_{\nu_{{a}}} \\ 
  \nonumber \\
I_1= & B_0(m^2_{W^-}, m^2_{W^-}) \\ 
I_2= & C_{00}(m^2_{\nu_{{a}}}, m^2_{W^-}, m^2_{W^-}) \\ 
I_3= & C_0(m^2_{\nu_{{a}}}, m^2_{W^-}, m^2_{W^-}) \\ 
  \VLLZ{(w_8)}= & - \VvewpL_{a, i} \VevwmL_{\beta, a} \Vwpwmz  (-1 + 2 (I_1 + 2 I_2 + I_3 m^2_{\nu_{{a}}})) \\ 
  \VLRZ{(w_8)}= & - \VvewpL_{a, i} \VevwmL_{\beta, a} \Vwpwmz  (-1 + 2 (I_1 + 2 I_2 + I_3 m^2_{\nu_{{a}}})) \\ 
  \nonumber \\
I_1= & B_0(m^2_{\nu_{{b}}}, m^2_{\nu_{{c}}}) \\ 
I_2= & C_{00}(m^2_{\nu_{{c}}}, m^2_{\nu_{{b}}}, m^2_{H^-_{{a}}}) \\ 
I_3= & C_0(m^2_{\nu_{{c}}}, m^2_{\nu_{{b}}}, m^2_{H^-_{{a}}}) \\ 
  \VLLZ{(w_9)}= & \, \VvehpL_{b, \alpha, a} \VevhmR_{\beta, c, a}  (-(\VvvzL_{c, b} I_3 m_{\nu_{{b}}} m_{\nu_{{c}}}) + \VvvzR_{c, b} (I_1 - 2 I_2 + I_3 m^2_{H^-_{{a}}})) \\ 
  \VLRZ{(w_9)}= & \, \VvehpL_{b, \alpha, a} \VevhmR_{\beta, c, a}  (-(\VvvzL_{c, b} I_3 m_{\nu_{{b}}} m_{\nu_{{c}}}) + \VvvzR_{c, b} (I_1 - 2 I_2 + I_3 m^2_{H^-_{{a}}})) \\ 
  \nonumber \\
I_1= & B_0(m^2_{\nu_{{b}}}, m^2_{\nu_{{c}}}) \\ 
I_2= & C_{00}(m^2_{\nu_{{c}}}, m^2_{\nu_{{b}}}, m^2_{W^-}) \\ 
I_3= & C_0(m^2_{\nu_{{c}}}, m^2_{\nu_{{b}}}, m^2_{W^-}) \\ 
  \VLLZ{(w_{10})}= & -( \VvewpL_{b, i} \VevwmL_{\beta, c}  (2 \VvvzR_{c, b} I_3 m_{\nu_{{b}}} m_{\nu_{{c}}} + \VvvzL_{c, b} (1 - 2 (I_1 - 2 I_2 + I_3 m^2_{W^-})))) \\ 
  \VLRZ{(w_{10})}= & -( \VvewpL_{b, i} \VevwmL_{\beta, c}  (2 \VvvzR_{c, b} I_3 m_{\nu_{{b}}} m_{\nu_{{c}}} + \VvvzL_{c, b} (1 - 2 (I_1 - 2 I_2 + I_3 m^2_{W^-})))) 
\end{align} 

\subsubsection{Scalar penguins}
\paragraph{CP even scalars}
\paragraph*{Self-energy corrections}
\begin{align} 
I_1= & B_0(m^2_{\nu_{{a}}}, m^2_{H^-_{{b}}}) \\ 
I_2= & B_1(m^2_{\nu_{{a}}}, m^2_{H^-_{{b}}}) \\ 
  \SLLh{(w_1,w_3)}= & ( \VeehL_{\beta, c, p}  (-(\VvehpL_{a, \alpha, b} \VevhmR_{c, a, b} I_2 m^2_{\ell_{{\alpha}}}) + \VvehpR_{a, \alpha, b} \VevhmR_{c, a, b} I_1 m_{\ell_\alpha} m_{\nu_{{a}}} - \VvehpR_{a, \alpha, b} \VevhmL_{c, a, b} I_2 m_{\ell_\alpha} m_{\ell_{{c}}} \nonumber \\ & + \VvehpL_{a, \alpha, b} \VevhmL_{c, a, b} I_1 m_{\nu_{{a}}} m_{\ell_{{c}}}))/(m^2_{\ell_{{\alpha}}} - m^2_{\ell_{{c}}}) \hspace{1cm} + (\alpha\leftrightarrow \beta) \\ 
  \SLRh{(w_1,w_3)}= & ( \VeehL_{\beta, c, p}  (-(\VvehpL_{a, \alpha, b} \VevhmR_{c, a, b} I_2 m^2_{\ell_{{\alpha}}}) + \VvehpR_{a, \alpha, b} \VevhmR_{c, a, b} I_1 m_{\ell_\alpha} m_{\nu_{{a}}} - \VvehpR_{a, \alpha, b} \VevhmL_{c, a, b} I_2 m_{\ell_\alpha} m_{\ell_{{c}}} \nonumber \\ & + \VvehpL_{a, \alpha, b} \VevhmL_{c, a, b} I_1 m_{\nu_{{a}}} m_{\ell_{{c}}}))/(m^2_{\ell_{{\alpha}}} - m^2_{\ell_{{c}}}) \hspace{1cm} + (\alpha\leftrightarrow \beta)
  \nonumber \\
I_1= & B_0(m^2_{\nu_{{a}}}, m^2_{W^-}) \\ 
I_2= & B_1(m^2_{\nu_{{a}}}, m^2_{W^-}) \\ 
  \SLLh{(w_2,w_4)}= & -(( \VeehL_{\beta, c, p}  (\VvewpR_{a, i} m_{\ell_\alpha} (-2 \VevwmL_{c, a} (1 - 2 I_1) m_{\nu_{{a}}} + \VevwmR_{c, a} (1 + 2 I_2) m_{\ell_{{c}}}) + \VvewpL_{a, i} (\VevwmL_{c, a} (1 + 2 I_2) m^2_{\ell_{{\alpha}}} \nonumber \\ & - 2 \VevwmR_{c, a} (1 - 2 I_1) m_{\nu_{{a}}} m_{\ell_{{c}}})))/(m^2_{\ell_{{\alpha}}} - m^2_{\ell_{{c}}})) \hspace{1cm} + (\alpha\leftrightarrow \beta)\\ 
  \SLRh{(w_2,w_4)}= & -(( \VeehL_{\beta, c, p}  (\VvewpR_{a, i} m_{\ell_\alpha} (-2 \VevwmL_{c, a} (1 - 2 I_1) m_{\nu_{{a}}} + \VevwmR_{c, a} (1 + 2 I_2) m_{\ell_{{c}}}) + \VvewpL_{a, i} (\VevwmL_{c, a} (1 + 2 I_2) m^2_{\ell_{{\alpha}}} \nonumber \\ & - 2 \VevwmR_{c, a} (1 - 2 I_1) m_{\nu_{{a}}} m_{\ell_{{c}}})))/(m^2_{\ell_{{\alpha}}} - m^2_{\ell_{{c}}})) \hspace{1cm} + (\alpha\leftrightarrow \beta)
\end{align} 
\paragraph*{Vertex corrections}
\begin{align} 
  \SLLh{(w_5)}= & \, \VvehpL_{a, \alpha, b} \VevhmL_{\beta, a, c} \Vhhmhp_{p, b, c} \, C_0(m^2_{\nu_{{a}}}, m^2_{H^-_{{c}}}, m^2_{H^-_{{b}}}) \, m_{\nu_{{a}}} \\ 
  \SLRh{(w_5)}= & \, \VvehpL_{a, \alpha, b} \VevhmL_{\beta, a, c} \Vhhmhp_{p, b, c} \, C_0(m^2_{\nu_{{a}}}, m^2_{H^-_{{c}}}, m^2_{H^-_{{b}}}) \, m_{\nu_{{a}}} \\ 
  \nonumber \\
I_1= & B_0(m^2_{H^-_{{c}}}, m^2_{W^-}) \\ 
I_2= & C_0(m^2_{\nu_{{a}}}, m^2_{H^-_{{c}}}, m^2_{W^-}) \\ 
  \SLLh{(w_6)}= & - \VvewpL_{a, i} \VevhmL_{\beta, a, c} \Vhhpwm_{p, c} \, (I_1 + I_2 m^2_{\nu_{{a}}}) \\ 
  \SLRh{(w_6)}= & - \VvewpL_{a, i} \VevhmL_{\beta, a, c} \Vhhpwm_{p, c} \, (I_1 + I_2 m^2_{\nu_{{a}}}) \\ 
  \nonumber \\
I_1= & B_0(m^2_{H^-_{{b}}}, m^2_{W^-}) \\ 
I_2= & C_0(m^2_{\nu_{{a}}}, m^2_{W^-}, m^2_{H^-_{{b}}}) \\ 
  \SLLh{(w_7)}= & \, \VvehpL_{a, \alpha, b} \VevwmR_{\beta, a} \Vhhmwp_{p, b} \, (I_1 + I_2 m^2_{\nu_{{a}}}) \\ 
  \SLRh{(w_7)}= & \, \VvehpL_{a, \alpha, b} \VevwmR_{\beta, a} \Vhhmwp_{p, b} \, (I_1 + I_2 m^2_{\nu_{{a}}}) \\ 
  \nonumber \\
  \SLLh{(w_8)}= & \, 4 \, \VvewpL_{a, i} \VevwmR_{\beta, a} \Vhwmwp_{p} \, C_0(m^2_{\nu_{{a}}}, m^2_{W^-}, m^2_{W^-}) \, m_{\nu_{{a}}} \\ 
  \SLRh{(w_8)}= & \, 4 \, \VvewpL_{a, i} \VevwmR_{\beta, a} \Vhwmwp_{p} \, C_0(m^2_{\nu_{{a}}}, m^2_{W^-}, m^2_{W^-}) \, m_{\nu_{{a}}} \\ 
  \nonumber \\
I_1= & B_0(m^2_{\nu_{{b}}}, m^2_{\nu_{{c}}}) \\ 
I_2= & C_0(m^2_{\nu_{{c}}}, m^2_{\nu_{{b}}}, m^2_{H^-_{{a}}}) \\ 
  \SLLh{(w_9)}= & \, \VvehpL_{b, \alpha, a} \VevhmL_{\beta, c, a}  (\VvvhL_{c, b, p} I_2 m_{\nu_{{b}}} m_{\nu_{{c}}} + \VvvhR_{c, b, p} (I_1 + I_2 m^2_{H^-_{{a}}})) \\ 
  \SLRh{(w_9)}= & \, \VvehpL_{b, \alpha, a} \VevhmL_{\beta, c, a}  (\VvvhL_{c, b, p} I_2 m_{\nu_{{b}}} m_{\nu_{{c}}} + \VvvhR_{c, b, p} (I_1 + I_2 m^2_{H^-_{{a}}})) \\ 
  \nonumber \\
I_1= & B_0(m^2_{\nu_{{b}}}, m^2_{\nu_{{c}}}) \\ 
I_2= & C_0(m^2_{\nu_{{c}}}, m^2_{\nu_{{b}}}, m^2_{W^-}) \\ 
  \SLLh{(w_{10})}= & \, 2 \, \VvewpL_{b, i} \VevwmR_{\beta, c}  \, (-2 \VvvhR_{c, b, p} I_2 m_{\nu_{{b}}} m_{\nu_{{c}}} + \VvvhL_{c, b, p} (1 - 2 (I_1 + I_2 m^2_{W^-}))) \\ 
  \SLRh{(w_{10})}= & \, 2 \, \VvewpL_{b, i} \VevwmR_{\beta, c}  \, (-2 \VvvhR_{c, b, p} I_2 m_{\nu_{{b}}} m_{\nu_{{c}}} + \VvvhL_{c, b, p} (1 - 2 (I_1 + I_2 m^2_{W^-}))) \\ 
\end{align} 

\paragraph{CP odd scalars}
\paragraph*{Self-energy corrections}
\begin{align} 
I_1= & B_0(m^2_{\nu_{{a}}}, m^2_{H^-_{{b}}}) \\ 
I_2= & B_1(m^2_{\nu_{{a}}}, m^2_{H^-_{{b}}}) \\ 
  \SLLa{(w_1,w_3)}= & ( \VeeaL_{\beta, c, p}  (-(\VvehpL_{a, \alpha, b} \VevhmR_{c, a, b} I_2 m^2_{\ell_{{\alpha}}}) + \VvehpR_{a, \alpha, b} \VevhmR_{c, a, b} I_1 m_{\ell_\alpha} m_{\nu_{{a}}} - \VvehpR_{a, \alpha, b} \VevhmL_{c, a, b} I_2 m_{\ell_\alpha} m_{\ell_{{c}}} \nonumber \\ & + \VvehpL_{a, \alpha, b} \VevhmL_{c, a, b} I_1 m_{\nu_{{a}}} m_{\ell_{{c}}}))/(m^2_{\ell_{{\alpha}}} - m^2_{\ell_{{c}}}) \hspace{1cm} + (\alpha\leftrightarrow \beta)\\ 
  \SLRa{(w_1,w_3)}= & ( \VeeaL_{\beta, c, p}  (-(\VvehpL_{a, \alpha, b} \VevhmR_{c, a, b} I_2 m^2_{\ell_{{\alpha}}}) + \VvehpR_{a, \alpha, b} \VevhmR_{c, a, b} I_1 m_{\ell_\alpha} m_{\nu_{{a}}} - \VvehpR_{a, \alpha, b} \VevhmL_{c, a, b} I_2 m_{\ell_\alpha} m_{\ell_{{c}}} \nonumber \\ & + \VvehpL_{a, \alpha, b} \VevhmL_{c, a, b} I_1 m_{\nu_{{a}}} m_{\ell_{{c}}}))/(m^2_{\ell_{{\alpha}}} - m^2_{\ell_{{c}}}) \hspace{1cm} + (\alpha\leftrightarrow \beta) 
  \nonumber \\
I_1= & B_0(m^2_{\nu_{{a}}}, m^2_{W^-}) \\ 
I_2= & B_1(m^2_{\nu_{{a}}}, m^2_{W^-}) \\ 
  \SLLa{(w_2,w_4)}= & -( \VeeaL_{\beta, c, p}  (\VvewpR_{a, i} m_{\ell_\alpha} (-2 \VevwmL_{c, a} (1 - 2 I_1) m_{\nu_{{a}}} + \VevwmR_{c, a} (1 + 2 I_2) m_{\ell_{{c}}}) + \VvewpL_{a, i} (\VevwmL_{c, a} (1 + 2 I_2) m^2_{\ell_{{\alpha}}} \nonumber \\ & - 2 \VevwmR_{c, a} (1 - 2 I_1) m_{\nu_{{a}}} m_{\ell_{{c}}})))/(m^2_{\ell_{{\alpha}}} - m^2_{\ell_{{c}}}) \hspace{1cm} + (\alpha\leftrightarrow \beta)\\ 
  \SLRa{(w_2,w_4)}= & -( \VeeaL_{\beta, c, p}  (\VvewpR_{a, i} m_{\ell_\alpha} (-2 \VevwmL_{c, a} (1 - 2 I_1) m_{\nu_{{a}}} + \VevwmR_{c, a} (1 + 2 I_2) m_{\ell_{{c}}}) + \VvewpL_{a, i} (\VevwmL_{c, a} (1 + 2 I_2) m^2_{\ell_{{\alpha}}} \nonumber \\ & - 2 \VevwmR_{c, a} (1 - 2 I_1) m_{\nu_{{a}}} m_{\ell_{{c}}})))/(m^2_{\ell_{{\alpha}}} - m^2_{\ell_{{c}}}) \hspace{1cm} + (\alpha\leftrightarrow \beta) 
\end{align}

\paragraph*{Vertex corrections}
\begin{align} 
  \SLLa{(w_5)}= & \, \VvehpL_{a, \alpha, b} \VevhmL_{\beta, a, c} \Vahmhp_{p, b, c}  \, C_0(m^2_{\nu_{{a}}}, m^2_{H^-_{{c}}}, m^2_{H^-_{{b}}})  \, m_{\nu_{{a}}} \\ 
  \SLRa{(w_5)}= & \, \VvehpL_{a, \alpha, b} \VevhmL_{\beta, a, c} \Vahmhp_{p, b, c}  \, C_0(m^2_{\nu_{{a}}}, m^2_{H^-_{{c}}}, m^2_{H^-_{{b}}})  \, m_{\nu_{{a}}} \\ 
  \nonumber \\
I_1= & B_0(m^2_{H^-_{{c}}}, m^2_{W^-}) \\ 
I_2= & C_0(m^2_{\nu_{{a}}}, m^2_{H^-_{{c}}}, m^2_{W^-}) \\ 
  \SLLa{(w_6)}= & - \VvewpL_{a, i} \VevhmL_{\beta, a, c} \Vahpwm_{p, c} \, (I_1 + I_2 m^2_{\nu_{{a}}}) \\ 
  \SLRa{(w_6)}= & - \VvewpL_{a, i} \VevhmL_{\beta, a, c} \Vahpwm_{p, c} \, (I_1 + I_2 m^2_{\nu_{{a}}}) \\ 
  \nonumber \\
I_1= & B_0(m^2_{H^-_{{b}}}, m^2_{W^-}) \\ 
I_2= & C_0(m^2_{\nu_{{a}}}, m^2_{W^-}, m^2_{H^-_{{b}}}) \\ 
  \SLLa{(w_7)}= & \, \VvehpL_{a, \alpha, b} \VevwmR_{\beta, a} \Vahmwp_{p, b} \, (I_1 + I_2 m^2_{\nu_{{a}}}) \\ 
  \SLRa{(w_7)}= & \, \VvehpL_{a, \alpha, b} \VevwmR_{\beta, a} \Vahmwp_{p, b} \, (I_1 + I_2 m^2_{\nu_{{a}}}) \\ 
  \nonumber \\
I_1= & B_0(m^2_{\nu_{{b}}}, m^2_{\nu_{{c}}}) \\ 
I_2= & C_0(m^2_{\nu_{{c}}}, m^2_{\nu_{{b}}}, m^2_{H^-_{{a}}}) \\ 
  \SLLa{(w_9)}= & \, \VvehpL_{b, \alpha, a} \VevhmL_{\beta, c, a} \, (\VvvaL_{c, b, p} I_2 m_{\nu_{{b}}} m_{\nu_{{c}}} + \VvvaR_{c, b, p} (I_1 + I_2 m^2_{H^-_{{a}}})) \\ 
  \SLRa{(w_9)}= & \, \VvehpL_{b, \alpha, a} \VevhmL_{\beta, c, a} \, (\VvvaL_{c, b, p} I_2 m_{\nu_{{b}}} m_{\nu_{{c}}} + \VvvaR_{c, b, p} (I_1 + I_2 m^2_{H^-_{{a}}})) \\ 
  \nonumber \\
I_1= & B_0(m^2_{\nu_{{b}}}, m^2_{\nu_{{c}}}) \\ 
I_2= & C_0(m^2_{\nu_{{c}}}, m^2_{\nu_{{b}}}, m^2_{W^-}) \\ 
  \SLLa{(w_{10})}= & \, 2 \, \VvewpL_{b, i} \VevwmR_{\beta, c} \, (-2 \VvvaR_{c, b, p} I_2 m_{\nu_{{b}}} m_{\nu_{{c}}} + \VvvaL_{c, b, p} (1 - 2 (I_1 + I_2 m^2_{W^-}))) \\ 
  \SLRa{(w_{10})}= & \, 2 \, \VvewpL_{b, i} \VevwmR_{\beta, c} \, (-2 \VvvaR_{c, b, p} I_2 m_{\nu_{{b}}} m_{\nu_{{c}}} + \VvvaL_{c, b, p} (1 - 2 (I_1 + I_2 m^2_{W^-}))) \\ 
\end{align}

\section{Box contributions to LFV}
\label{app:boxes}

\subsection{Four lepton boxes}
\subsubsection{Feynman diagrams}
\subsection*{Neutralino diagrams}

\parbox{0.45\linewidth}{
($n^l_1$)
\begin{fmffile}{Diagrams/Box4LNumberOfConsideredExternalStatesBox23} 
\fmfframe(20,20)(20,20){ 
\begin{fmfgraph*}(120,80) 
\fmftop{t1,t2}
\fmfbottom{b1,b2}
\fmf{fermion}{t1,v1}
\fmf{plain,label=$\tilde{\chi}^0_{{a}}$,tension=0.5}{v1,v2}
\fmf{fermion}{v2,t2}
\fmf{fermion}{v3,b1}
\fmf{plain,label=$\tilde{\chi}^0_{{c}}$,tension=0.5}{v3,v4}
\fmf{fermion}{b2,v4}
\fmf{dashes,label=$\tilde{e}^*_{{d}}$,tension=0.1}{v1,v3}
\fmf{dashes,label=$\tilde{e}^*_{{b}}$,tension=0.1}{v2,v4}
\fmflabel{$\ell_{{\alpha}}$}{t1}
\fmflabel{$\bar{\ell}_{{\beta}}$}{t2}
\fmflabel{$\ell_{{\gamma}}$}{b2}
\fmflabel{$\bar{\ell}_{{\delta}}$}{b1}
\end{fmfgraph*}}
\end{fmffile}}
\hspace{1cm}
\parbox{0.45\linewidth}{
($n^l_2$)
\begin{fmffile}{Diagrams/Box4LNumberOfConsideredExternalStatesBox48}
\fmfframe(20,20)(20,20){
\begin{fmfgraph*}(120,80)
\fmftop{t1,t2}
\fmfbottom{b1,b2}
\fmf{fermion}{t1,v1}
\fmf{plain,label=$\tilde{\chi}^0_{{a}}$,tension=0.5}{v1,v2}
\fmf{phantom}{v2,t2}
\fmf{fermion}{v3,b1}
\fmf{plain,label=$\tilde{\chi}^0_{{c}}$,tension=0.5}{v3,v4}
\fmf{phantom}{v4,b2}
\fmf{dashes,label=$\tilde{e}^*_{{d}}$,tension=0.1}{v1,v3}
\fmf{dashes,label=$\tilde{e}_{{b}}$,tension=0.1}{v2,v4}
\fmffreeze
\fmf{fermion}{v4,t2}
\fmf{fermion}{b2,v2}
\fmflabel{$\ell_{{\alpha}}$}{t1}
\fmflabel{$\ell_{{\gamma}}$}{b2}
\fmflabel{$\bar{\ell}_{{\beta}}$}{t2}
\fmflabel{$\bar{\ell}_{{\delta}}$}{b1}
\end{fmfgraph*}}
\end{fmffile}
}

\subsection*{Chargino diagrams}

\parbox{0.45\linewidth}{
($c^l_1$)
\begin{fmffile}{Diagrams/Box4LNumberOfConsideredExternalStatesBox9} 
\fmfframe(20,20)(20,20){ 
\begin{fmfgraph*}(120,80) 
\fmftop{t1,t2}
\fmfbottom{b1,b2}
\fmf{fermion}{t1,v1}
\fmf{plain,label=$\tilde{\chi}^-_{{a}}$,tension=0.5}{v1,v2}
\fmf{fermion}{v2,t2}
\fmf{fermion}{v3,b1}
\fmf{plain,label=$\tilde{\chi}^-_{{c}}$,tension=0.5}{v3,v4}
\fmf{fermion}{b2,v4}
\fmf{dashes,label=$\nu^i_{{d}}$,tension=0.1}{v1,v3}
\fmf{dashes,label=$\nu^i_{{b}}$,tension=0.1}{v2,v4}
\fmflabel{$\ell_{{\alpha}}$}{t1}
\fmflabel{$\bar{\ell}_{{\beta}}$}{t2}
\fmflabel{$\ell_{{\gamma}}$}{b2}
\fmflabel{$\bar{\ell}_{{\delta}}$}{b1}
\end{fmfgraph*}}
\end{fmffile}}
\hspace{1cm}
\parbox{0.45\linewidth}{
($c^l_2$)
\begin{fmffile}{Diagrams/Box4LNumberOfConsideredExternalStatesBox10} 
\fmfframe(20,20)(20,20){ 
\begin{fmfgraph*}(120,80) 
\fmftop{t1,t2}
\fmfbottom{b1,b2}
\fmf{fermion}{t1,v1}
\fmf{plain,label=$\tilde{\chi}^-_{{a}}$,tension=0.5}{v1,v2}
\fmf{fermion}{v2,t2}
\fmf{fermion}{v3,b1}
\fmf{plain,label=$\tilde{\chi}^-_{{c}}$,tension=0.5}{v3,v4}
\fmf{fermion}{b2,v4}
\fmf{dashes,label=$\nu^i_{{d}}$,tension=0.1}{v1,v3}
\fmf{dashes,label=$\nu^R_{{b}}$,tension=0.1}{v2,v4}
\fmflabel{$\ell_{{\alpha}}$}{t1}
\fmflabel{$\bar{\ell}_{{\beta}}$}{t2}
\fmflabel{$\ell_{{\gamma}}$}{b2}
\fmflabel{$\bar{\ell}_{{\delta}}$}{b1}
\end{fmfgraph*}}
\end{fmffile}}
\\
\parbox{0.45\linewidth}{
($c^l_3$)
\begin{fmffile}{Diagrams/Box4LNumberOfConsideredExternalStatesBox11} 
\fmfframe(20,20)(20,20){ 
\begin{fmfgraph*}(120,80) 
\fmftop{t1,t2}
\fmfbottom{b1,b2}
\fmf{fermion}{t1,v1}
\fmf{plain,label=$\tilde{\chi}^-_{{a}}$,tension=0.5}{v1,v2}
\fmf{fermion}{v2,t2}
\fmf{fermion}{v3,b1}
\fmf{plain,label=$\tilde{\chi}^-_{{c}}$,tension=0.5}{v3,v4}
\fmf{fermion}{b2,v4}
\fmf{dashes,label=$\nu^R_{{d}}$,tension=0.1}{v1,v3}
\fmf{dashes,label=$\nu^i_{{b}}$,tension=0.1}{v2,v4}
\fmflabel{$\ell_{{\alpha}}$}{t1}
\fmflabel{$\bar{\ell}_{{\beta}}$}{t2}
\fmflabel{$\ell_{{\gamma}}$}{b2}
\fmflabel{$\bar{\ell}_{{\delta}}$}{b1}
\end{fmfgraph*}}
\end{fmffile}}
\hspace{1cm}
\parbox{0.45\linewidth}{
($c^l_4$)
\begin{fmffile}{Diagrams/Box4LNumberOfConsideredExternalStatesBox12} 
\fmfframe(20,20)(20,20){ 
\begin{fmfgraph*}(120,80) 
\fmftop{t1,t2}
\fmfbottom{b1,b2}
\fmf{fermion}{t1,v1}
\fmf{plain,label=$\tilde{\chi}^-_{{a}}$,tension=0.5}{v1,v2}
\fmf{fermion}{v2,t2}
\fmf{fermion}{v3,b1}
\fmf{plain,label=$\tilde{\chi}^-_{{c}}$,tension=0.5}{v3,v4}
\fmf{fermion}{b2,v4}
\fmf{dashes,label=$\nu^R_{{d}}$,tension=0.1}{v1,v3}
\fmf{dashes,label=$\nu^R_{{b}}$,tension=0.1}{v2,v4}
\fmflabel{$\ell_{{\alpha}}$}{t1}
\fmflabel{$\bar{\ell}_{{\beta}}$}{t2}
\fmflabel{$\ell_{{\gamma}}$}{b2}
\fmflabel{$\bar{\ell}_{{\delta}}$}{b1}
\end{fmfgraph*}}
\end{fmffile}}
\\
\parbox{0.45\linewidth}{
($c^l_5$)
\begin{fmffile}{Diagrams/Box4LNumberOfConsideredExternalStatesBox34}
\fmfframe(20,20)(20,20){
\begin{fmfgraph*}(120,80)
\fmftop{t1,t2}
\fmfbottom{b1,b2}
\fmf{fermion}{t1,v1}
\fmf{plain,label=$\tilde{\chi}^-_{{a}}$,tension=0.5}{v1,v2}
\fmf{fermion}{v2,t2}
\fmf{phantom}{b1,v3}
\fmf{plain,label=$\tilde{\chi}^+_{{c}}$,tension=0.5}{v3,v4}
\fmf{phantom}{v4,b2}
\fmf{dashes,label=$\nu^i_{{d}}$,tension=0.1}{v1,v3}
\fmf{dashes,label=$\nu^i_{{b}}$,tension=0.1}{v2,v4}
\fmffreeze 
\fmf{fermion}{v4,b1}
\fmf{fermion}{b2,v3}
\fmflabel{$\ell_{{\alpha}}$}{t1}
\fmflabel{$\bar{\ell}_{{\beta}}$}{t2}
\fmflabel{$\bar{\ell}_{{\delta}}$}{b1}
\fmflabel{$\ell_{{\gamma}}$}{b2}
\end{fmfgraph*}}
\end{fmffile}}
\hspace{1cm}
\parbox{0.45\linewidth}{
($c^l_6$)
\begin{fmffile}{Diagrams/Box4LNumberOfConsideredExternalStatesBox35}
\fmfframe(20,20)(20,20){
\begin{fmfgraph*}(120,80)
\fmftop{t1,t2}
\fmfbottom{b1,b2}
\fmf{fermion}{t1,v1}
\fmf{plain,label=$\tilde{\chi}^-_{{a}}$,tension=0.5}{v1,v2}
\fmf{fermion}{v2,t2}
\fmf{phantom}{b1,v3}
\fmf{plain,label=$\tilde{\chi}^+_{{c}}$,tension=0.5}{v3,v4}
\fmf{phantom}{v4,b2}
\fmf{dashes,label=$\nu^i_{{d}}$,tension=0.1}{v1,v3}
\fmf{dashes,label=$\nu^R_{{b}}$,tension=0.1}{v2,v4}
\fmffreeze 
\fmf{fermion}{v4,b1}
\fmf{fermion}{b2,v3}
\fmflabel{$\ell_{{\alpha}}$}{t1}
\fmflabel{$\bar{\ell}_{{\beta}}$}{t2}
\fmflabel{$\bar{\ell}_{{\delta}}$}{b1}
\fmflabel{$\ell_{{\gamma}}$}{b2}
\end{fmfgraph*}}
\end{fmffile}}
\\
\parbox{0.45\linewidth}{
($c^l_7$)
\begin{fmffile}{Diagrams/Box4LNumberOfConsideredExternalStatesBox36}
\fmfframe(20,20)(20,20){
\begin{fmfgraph*}(120,80)
\fmftop{t1,t2}
\fmfbottom{b1,b2}
\fmf{fermion}{t1,v1}
\fmf{plain,label=$\tilde{\chi}^-_{{a}}$,tension=0.5}{v1,v2}
\fmf{fermion}{v2,t2}
\fmf{phantom}{b1,v3}
\fmf{plain,label=$\tilde{\chi}^+_{{c}}$,tension=0.5}{v3,v4}
\fmf{phantom}{v4,b2}
\fmf{dashes,label=$\nu^R_{{d}}$,tension=0.1}{v1,v3}
\fmf{dashes,label=$\nu^i_{{b}}$,tension=0.1}{v2,v4}
\fmffreeze 
\fmf{fermion}{v4,b1}
\fmf{fermion}{b2,v3}
\fmflabel{$\ell_{{\alpha}}$}{t1}
\fmflabel{$\bar{\ell}_{{\beta}}$}{t2}
\fmflabel{$\bar{\ell}_{{\delta}}$}{b1}
\fmflabel{$\ell_{{\gamma}}$}{b2}
\end{fmfgraph*}}
\end{fmffile}}
\hspace{1cm}
\parbox{0.45\linewidth}{
($c^l_8$)
\begin{fmffile}{Diagrams/Box4LNumberOfConsideredExternalStatesBox37}
\fmfframe(20,20)(20,20){
\begin{fmfgraph*}(120,80)
\fmftop{t1,t2}
\fmfbottom{b1,b2}
\fmf{fermion}{t1,v1}
\fmf{plain,label=$\tilde{\chi}^-_{{a}}$,tension=0.5}{v1,v2}
\fmf{fermion}{v2,t2}
\fmf{phantom}{b1,v3}
\fmf{plain,label=$\tilde{\chi}^+_{{c}}$,tension=0.5}{v3,v4}
\fmf{phantom}{v4,b2}
\fmf{dashes,label=$\nu^R_{{d}}$,tension=0.1}{v1,v3}
\fmf{dashes,label=$\nu^R_{{b}}$,tension=0.1}{v2,v4}
\fmffreeze 
\fmf{fermion}{v4,b1}
\fmf{fermion}{b2,v3}
\fmflabel{$\ell_{{\alpha}}$}{t1}
\fmflabel{$\bar{\ell}_{{\beta}}$}{t2}
\fmflabel{$\bar{\ell}_{{\delta}}$}{b1}
\fmflabel{$\ell_{{\gamma}}$}{b2}
\end{fmfgraph*}}
\end{fmffile}
}

\subsection*{$W^+$ and $H^+$ diagrams}
\parbox{0.45\linewidth}{
($w^l_1$)
\begin{fmffile}{Diagrams/Box4LNumberOfConsideredExternalStatesBox21} 
\fmfframe(20,20)(20,20){ 
\begin{fmfgraph*}(120,80) 
\fmftop{t1,t2}
\fmfbottom{b1,b2}
\fmf{fermion}{t1,v1}
\fmf{plain,label=$\nu_{{a}}$,tension=0.5}{v1,v2}
\fmf{fermion}{v2,t2}
\fmf{fermion}{v3,b1}
\fmf{plain,label=$\nu_{{c}}$,tension=0.5}{v3,v4}
\fmf{fermion}{b2,v4}
\fmf{dashes,label=$H^+_{{d}}$,tension=0.1}{v1,v3}
\fmf{dashes,label=$H^+_{{b}}$,tension=0.1}{v2,v4}
\fmflabel{$\ell_{{\alpha}}$}{t1}
\fmflabel{$\bar{\ell}_{{\beta}}$}{t2}
\fmflabel{$\ell_{{\gamma}}$}{b2}
\fmflabel{$\bar{\ell}_{{\delta}}$}{b1}
\end{fmfgraph*}}
\end{fmffile}}
\hspace{1cm}
\parbox{0.45\linewidth}{
($w^l_2$)
\begin{fmffile}{Diagrams/Box4LNumberOfConsideredExternalStatesBox22} 
\fmfframe(20,20)(20,20){ 
\begin{fmfgraph*}(120,80) 
\fmftop{t1,t2}
\fmfbottom{b1,b2}
\fmf{fermion}{t1,v1}
\fmf{plain,label=$\nu_{{a}}$,tension=0.5}{v1,v2}
\fmf{fermion}{v2,t2}
\fmf{fermion}{v3,b1}
\fmf{plain,label=$\nu_{{c}}$,tension=0.5}{v3,v4}
\fmf{fermion}{b2,v4}
\fmf{dashes,label=$H^+_{{d}}$,tension=0.1}{v1,v3}
\fmf{wiggly,label=$W^+$,tension=0.1}{v2,v4}
\fmflabel{$\ell_{{\alpha}}$}{t1}
\fmflabel{$\bar{\ell}_{{\beta}}$}{t2}
\fmflabel{$\ell_{{\gamma}}$}{b2}
\fmflabel{$\bar{\ell}_{{\delta}}$}{b1}
\end{fmfgraph*}}
\end{fmffile}}
\\
\parbox{0.45\linewidth}{
($w^l_3$)
\begin{fmffile}{Diagrams/Box4LNumberOfConsideredExternalStatesBox24} 
\fmfframe(20,20)(20,20){ 
\begin{fmfgraph*}(120,80) 
\fmftop{t1,t2}
\fmfbottom{b1,b2}
\fmf{fermion}{t1,v1}
\fmf{plain,label=$\nu_{{a}}$,tension=0.5}{v1,v2}
\fmf{fermion}{v2,t2}
\fmf{fermion}{v3,b1}
\fmf{plain,label=$\nu_{{c}}$,tension=0.5}{v3,v4}
\fmf{fermion}{b2,v4}
\fmf{wiggly,label=$W^+$,tension=0.1}{v1,v3}
\fmf{dashes,label=$H^+_{{b}}$,tension=0.1}{v2,v4}
\fmflabel{$\ell_{{\alpha}}$}{t1}
\fmflabel{$\bar{\ell}_{{\beta}}$}{t2}
\fmflabel{$\ell_{{\gamma}}$}{b2}
\fmflabel{$\bar{\ell}_{{\delta}}$}{b1}
\end{fmfgraph*}}
\end{fmffile}}
\hspace{1cm}
\parbox{0.45\linewidth}{
($w^l_4$)
\begin{fmffile}{Diagrams/Box4LNumberOfConsideredExternalStatesBox25} 
\fmfframe(20,20)(20,20){ 
\begin{fmfgraph*}(120,80) 
\fmftop{t1,t2}
\fmfbottom{b1,b2}
\fmf{fermion}{t1,v1}
\fmf{plain,label=$\nu_{{a}}$,tension=0.5}{v1,v2}
\fmf{fermion}{v2,t2}
\fmf{fermion}{v3,b1}
\fmf{plain,label=$\nu_{{c}}$,tension=0.5}{v3,v4}
\fmf{fermion}{b2,v4}
\fmf{wiggly,label=$W^+$,tension=0.1}{v1,v3}
\fmf{wiggly,label=$W^+$,tension=0.1}{v2,v4}
\fmflabel{$\ell_{{\alpha}}$}{t1}
\fmflabel{$\bar{\ell}_{{\beta}}$}{t2}
\fmflabel{$\ell_{{\gamma}}$}{b2}
\fmflabel{$\bar{\ell}_{{\delta}}$}{b1}
\end{fmfgraph*}}
\end{fmffile}}
\\
\parbox{0.45\linewidth}{
($w^l_5$)
\begin{fmffile}{Diagrams/Box4LNumberOfConsideredExternalStatesBox46}
\fmfframe(20,20)(20,20){
\begin{fmfgraph*}(120,80)
\fmftop{t1,t2}
\fmfbottom{b1,b2}
\fmf{fermion}{t1,v1}
\fmf{plain,label=$\nu_{{a}}$,tension=0.5}{v1,v2}
\fmf{fermion}{v3,b1}
\fmf{phantom}{v2,t2}
\fmf{plain,label=$\nu_{{c}}$,tension=0.5}{v3,v4}
\fmf{phantom}{v4,b2}
\fmf{dashes,label=$H^+_{{d}}$,tension=0.1}{v1,v3}
\fmf{dashes,label=$H^-_{{b}}$,tension=0.1}{v2,v4}
\fmffreeze
\fmf{fermion}{v4,t2}
\fmf{fermion}{b2,v2}
\fmflabel{$\ell_{{\alpha}}$}{t1}
\fmflabel{$\ell_{{\gamma}}$}{b2}
\fmflabel{$\bar{\ell}_{{\beta}}$}{t2}
\fmflabel{$\bar{\ell}_{{\delta}}$}{b1}
\end{fmfgraph*}}
\end{fmffile}}
\hspace{1cm}
\parbox{0.45\linewidth}{
($w^l_6$)
\begin{fmffile}{Diagrams/Box4LNumberOfConsideredExternalStatesBox47}
\fmfframe(20,20)(20,20){
\begin{fmfgraph*}(120,80)
\fmftop{t1,t2}
\fmfbottom{b1,b2}
\fmf{fermion}{t1,v1}
\fmf{plain,label=$\nu_{{a}}$,tension=0.5}{v1,v2}
\fmf{fermion}{v3,b1}
\fmf{phantom}{v2,t2}
\fmf{plain,label=$\nu_{{c}}$,tension=0.5}{v3,v4}
\fmf{phantom}{v4,b2}
\fmf{dashes,label=$H^+_{{d}}$,tension=0.1}{v1,v3}
\fmf{wiggly,label=$W^-$,tension=0.1}{v2,v4}
\fmffreeze
\fmf{fermion}{v4,t2}
\fmf{fermion}{b2,v2}
\fmflabel{$\ell_{{\alpha}}$}{t1}
\fmflabel{$\ell_{{\gamma}}$}{b2}
\fmflabel{$\bar{\ell}_{{\beta}}$}{t2}
\fmflabel{$\bar{\ell}_{{\delta}}$}{b1}
\end{fmfgraph*}}
\end{fmffile} }
\\
\parbox{0.45\linewidth}{
($w^l_7$)
\begin{fmffile}{Diagrams/Box4LNumberOfConsideredExternalStatesBox49}
\fmfframe(20,20)(20,20){
\begin{fmfgraph*}(120,80)
\fmftop{t1,t2}
\fmfbottom{b1,b2}
\fmf{fermion}{t1,v1}
\fmf{plain,label=$\nu_{{a}}$,tension=0.5}{v1,v2}
\fmf{fermion}{v3,b1}
\fmf{phantom}{v2,t2}
\fmf{plain,label=$\nu_{{c}}$,tension=0.5}{v3,v4}
\fmf{phantom}{v4,b2}
\fmf{wiggly,label=$W^+$,tension=0.1}{v1,v3}
\fmf{dashes,label=$H^-_{{b}}$,tension=0.1}{v2,v4}
\fmffreeze
\fmf{fermion}{v4,t2}
\fmf{fermion}{b2,v2}
\fmflabel{$\ell_{{\alpha}}$}{t1}
\fmflabel{$\ell_{{\gamma}}$}{b2}
\fmflabel{$\bar{\ell}_{{\beta}}$}{t2}
\fmflabel{$\bar{\ell}_{{\delta}}$}{b1}
\end{fmfgraph*}}
\end{fmffile} }
\hspace{1cm}
\parbox{0.45\linewidth}{
($w^l_8$)
\begin{fmffile}{Diagrams/Box4LNumberOfConsideredExternalStatesBox50}
\fmfframe(20,20)(20,20){
\begin{fmfgraph*}(120,80)
\fmftop{t1,t2}
\fmfbottom{b1,b2}
\fmf{fermion}{t1,v1}
\fmf{plain,label=$\nu_{{a}}$,tension=0.5}{v1,v2}
\fmf{phantom}{v2,t2}
\fmf{fermion}{v3,b1}
\fmf{plain,label=$\nu_{{c}}$,tension=0.5}{v3,v4}
\fmf{phantom}{v4,b2}
\fmf{wiggly,label=$W^+$,tension=0.1}{v1,v3}
\fmf{wiggly,label=$W^-$,tension=0.1}{v2,v4}
\fmffreeze
\fmf{fermion}{v4,t2}
\fmf{fermion}{b2,v2}
\fmflabel{$\ell_{{\alpha}}$}{t1}
\fmflabel{$\ell_{{\gamma}}$}{b2}
\fmflabel{$\bar{\ell}_{{\beta}}$}{t2}
\fmflabel{$\bar{\ell}_{{\delta}}$}{b1}
\end{fmfgraph*}}
\end{fmffile}
}

\subsubsection{Neutralino contributions}
\begin{align} 
  \SLLb{(n^l_1)} = & - \VneeL_{a, \alpha, d}\VeneL_{\beta, a, b} \VneeL_{c, \gamma, b}\VeneL_{\delta, c, d}  m_{\tilde{\chi}^0_{{a}}} m_{\tilde{\chi}^0_{{c}}} D_0(m^2_{\tilde{\chi}^0_{{a}}}, m^2_{\tilde{\chi}^0_{{c}}}, m^2_{\tilde{e}_{{d}}}, m^2_{\tilde{e}_{{b}}}) \\ 
  \SLRb{(n^l_1)} = & - \VneeL_{a, \alpha, d}\VeneL_{\beta, a, b} \VneeR_{c, \gamma, b}\VeneR_{\delta, c, d}  m_{\tilde{\chi}^0_{{a}}} m_{\tilde{\chi}^0_{{c}}} D_0(m^2_{\tilde{\chi}^0_{{a}}}, m^2_{\tilde{\chi}^0_{{c}}}, m^2_{\tilde{e}_{{d}}}, m^2_{\tilde{e}_{{b}}}) \\ 
  \VLLb{(n^l_1)} = & - \VneeL_{a, \alpha, d}\VeneR_{\beta, a, b} \VneeL_{c, \gamma, b}\VeneR_{\delta, c, d} D_{27}(m^2_{\tilde{\chi}^0_{{a}}}, m^2_{\tilde{\chi}^0_{{c}}}, m^2_{\tilde{e}_{{d}}}, m^2_{\tilde{e}_{{b}}}) \\ 
  \VLRb{(n^l_1)} = & - \VneeL_{a, \alpha, d}\VeneR_{\beta, a, b} \VneeR_{c, \gamma, b}\VeneL_{\delta, c, d} D_{27}(m^2_{\tilde{\chi}^0_{{a}}}, m^2_{\tilde{\chi}^0_{{c}}}, m^2_{\tilde{e}_{{d}}}, m^2_{\tilde{e}_{{b}}}) \\ 
  \nonumber \\  
  \SLLb{(n^l_2)} = & \, \frac{1}{2} \VneeL_{a, \alpha, d} \VneeL_{a, \gamma, b}\VeneL_{\beta, c, b}\VeneL_{\delta, c, d} m_{\tilde{\chi}^0_{{a}}} m_{\tilde{\chi}^0_{{c}}} D_0(m^2_{\tilde{\chi}^0_{{a}}}, m^2_{\tilde{\chi}^0_{{c}}}, m^2_{\tilde{e}_{{d}}}, m^2_{\tilde{e}_{{b}}}) \\ 
  \SLRb{(n^l_2)} = & -2  \VneeL_{a, \alpha, d} \VneeR_{a, \gamma, b}\VeneL_{\beta, c, b}\VeneR_{\delta, c, d} D_{27}(m^2_{\tilde{\chi}^0_{{a}}}, m^2_{\tilde{\chi}^0_{{c}}}, m^2_{\tilde{e}_{{d}}}, m^2_{\tilde{e}_{{b}}}) \\ 
  \VLLb{(n^l_2)} = & -\frac{1}{2} \VneeL_{a, \alpha, d} \VneeL_{a, \gamma, b}\VeneR_{\beta, c, b}\VeneR_{\delta, c, d} m_{\tilde{\chi}^0_{{a}}} m_{\tilde{\chi}^0_{{c}}} D_0(m^2_{\tilde{\chi}^0_{{a}}}, m^2_{\tilde{\chi}^0_{{c}}}, m^2_{\tilde{e}_{{d}}}, m^2_{\tilde{e}_{{b}}}) \\ 
  \VLRb{(n^l_2)} = & - \VneeL_{a, \alpha, d} \VneeR_{a, \gamma, b}\VeneR_{\beta, c, b}\VeneL_{\delta, c, d} \, D_{27}(m^2_{\tilde{\chi}^0_{{a}}}, m^2_{\tilde{\chi}^0_{{c}}}, m^2_{\tilde{e}_{{d}}}, m^2_{\tilde{e}_{{b}}}) \\ 
  \TLLb{(n^l_2)} = & \, \frac{1}{8} \VneeL_{a, \alpha, d} \VneeL_{a, \gamma, b}\VeneL_{\beta, c, b}\VeneL_{\delta, c, d} m_{\tilde{\chi}^0_{{a}}} m_{\tilde{\chi}^0_{{c}}} D_0(m^2_{\tilde{\chi}^0_{{a}}}, m^2_{\tilde{\chi}^0_{{c}}}, m^2_{\tilde{e}_{{d}}}, m^2_{\tilde{e}_{{b}}})
\end{align} 

\subsubsection{Chargino contributions}
\begin{align} 
  \SLLb{(c^l_1)} = & - \VcevL_{a, \alpha, d} \VecviL_{\beta, a, b} \VcevL_{c, \gamma, b} \VecviL_{\delta, c, d} m_{\tilde{\chi}^-_{{a}}} m_{\tilde{\chi}^-_{{c}}} D_0(m^2_{\tilde{\chi}^-_{{a}}}, m^2_{\tilde{\chi}^-_{{c}}}, m^2_{\nu^i_{{d}}}, m^2_{\nu^i_{{b}}}) \\ 
  \SLRb{(c^l_1)} = & - \VcevL_{a, \alpha, d} \VecviL_{\beta, a, b} \VcevR_{c, \gamma, b} \VecviR_{\delta, c, d} m_{\tilde{\chi}^-_{{a}}} m_{\tilde{\chi}^-_{{c}}} D_0(m^2_{\tilde{\chi}^-_{{a}}}, m^2_{\tilde{\chi}^-_{{c}}}, m^2_{\nu^i_{{d}}}, m^2_{\nu^i_{{b}}}) \\ 
  \VLLb{(c^l_1)} = & - \VcevL_{a, \alpha, d} \VecviR_{\beta, a, b} \VcevL_{c, \gamma, b} \VecviR_{\delta, c, d} D_{27}(m^2_{\tilde{\chi}^-_{{a}}}, m^2_{\tilde{\chi}^-_{{c}}}, m^2_{\nu^i_{{d}}}, m^2_{\nu^i_{{b}}}) \\ 
  \VLRb{(c^l_1)} = & - \VcevL_{a, \alpha, d} \VecviR_{\beta, a, b} \VcevR_{c, \gamma, b} \VecviL_{\delta, c, d} D_{27}(m^2_{\tilde{\chi}^-_{{a}}}, m^2_{\tilde{\chi}^-_{{c}}}, m^2_{\nu^i_{{d}}}, m^2_{\nu^i_{{b}}}) \\ 
  \nonumber \\ 
  \SLLb{(c^l_2)} = & - \VcevL_{a, \alpha, d} \VecvrL_{\beta, a, b} \VcevrL_{c, \gamma, b} \VecviL_{\delta, c, d} m_{\tilde{\chi}^-_{{a}}}  m_{\tilde{\chi}^-_{{c}}} \, D_0(m^2_{\tilde{\chi}^-_{{a}}}, m^2_{\tilde{\chi}^-_{{c}}}, m^2_{\nu^i_{{d}}}, m^2_{\nu^R_{{b}}}) \\ 
  \SLRb{(c^l_2)} = & - \VcevL_{a, \alpha, d} \VecvrL_{\beta, a, b} \VcevrR_{c, \gamma, b} \VecviR_{\delta, c, d} m_{\tilde{\chi}^-_{{a}}}  m_{\tilde{\chi}^-_{{c}}} \, D_0(m^2_{\tilde{\chi}^-_{{a}}}, m^2_{\tilde{\chi}^-_{{c}}}, m^2_{\nu^i_{{d}}}, m^2_{\nu^R_{{b}}}) \\ 
  \VLLb{(c^l_2)} = & - \VcevL_{a, \alpha, d} \VecvrR_{\beta, a, b} \VcevrL_{c, \gamma, b} \VecviR_{\delta, c, d} \, D_{27}(m^2_{\tilde{\chi}^-_{{a}}}, m^2_{\tilde{\chi}^-_{{c}}}, m^2_{\nu^i_{{d}}}, m^2_{\nu^R_{{b}}}) \\ 
  \VLRb{(c^l_2)} = & - \VcevL_{a, \alpha, d} \VecvrR_{\beta, a, b} \VcevrR_{c, \gamma, b} \VecviL_{\delta, c, d} \, D_{27}(m^2_{\tilde{\chi}^-_{{a}}}, m^2_{\tilde{\chi}^-_{{c}}}, m^2_{\nu^i_{{d}}}, m^2_{\nu^R_{{b}}}) \\ 
  \nonumber \\
  \SLLb{(c^l_3)} = & - \VcevrL_{a, \alpha, d} \VecviL_{\beta, a, b} \VcevL_{c, \gamma, b} \VecvrL_{\delta, c, d} m_{\tilde{\chi}^-_{{a}}} m_{\tilde{\chi}^-_{{c}}} \, D_0(m^2_{\tilde{\chi}^-_{{a}}}, m^2_{\tilde{\chi}^-_{{c}}}, m^2_{\nu^R_{{d}}}, m^2_{\nu^i_{{b}}}) \\ 
  \SLRb{(c^l_3)} = & - \VcevrL_{a, \alpha, d} \VecviL_{\beta, a, b} \VcevR_{c, \gamma, b} \VecvrR_{\delta, c, d} m_{\tilde{\chi}^-_{{a}}} m_{\tilde{\chi}^-_{{c}}} \, D_0(m^2_{\tilde{\chi}^-_{{a}}}, m^2_{\tilde{\chi}^-_{{c}}}, m^2_{\nu^R_{{d}}}, m^2_{\nu^i_{{b}}}) \\ 
  \VLLb{(c^l_3)} = & - \VcevrL_{a, \alpha, d} \VecviR_{\beta, a, b} \VcevL_{c, \gamma, b} \VecvrR_{\delta, c, d} \, D_{27}(m^2_{\tilde{\chi}^-_{{a}}}, m^2_{\tilde{\chi}^-_{{c}}}, m^2_{\nu^R_{{d}}}, m^2_{\nu^i_{{b}}}) \\ 
  \VLRb{(c^l_3)} = & - \VcevrL_{a, \alpha, d} \VecviR_{\beta, a, b} \VcevR_{c, \gamma, b} \VecvrL_{\delta, c, d} \, D_{27}(m^2_{\tilde{\chi}^-_{{a}}}, m^2_{\tilde{\chi}^-_{{c}}}, m^2_{\nu^R_{{d}}}, m^2_{\nu^i_{{b}}}) \\ 
  \nonumber \\
  \SLLb{(c^l_4)} = & - \VcevrL_{a, \alpha, d} \VecvrL_{\beta, a, b} \VcevrL_{c, \gamma, b} \VecvrL_{\delta, c, d} m_{\tilde{\chi}^-_{{a}}} m_{\tilde{\chi}^-_{{c}}} \, D_0(m^2_{\tilde{\chi}^-_{{a}}}, m^2_{\tilde{\chi}^-_{{c}}}, m^2_{\nu^R_{{d}}}, m^2_{\nu^R_{{b}}}) \\ 
  \SLRb{(c^l_4)} = & - \VcevrL_{a, \alpha, d} \VecvrL_{\beta, a, b} \VcevrR_{c, \gamma, b} \VecvrR_{\delta, c, d} m_{\tilde{\chi}^-_{{a}}} m_{\tilde{\chi}^-_{{c}}} \, D_0(m^2_{\tilde{\chi}^-_{{a}}}, m^2_{\tilde{\chi}^-_{{c}}}, m^2_{\nu^R_{{d}}}, m^2_{\nu^R_{{b}}}) \\ 
  \VLLb{(c^l_4)} = & - \VcevrL_{a, \alpha, d} \VecvrR_{\beta, a, b} \VcevrL_{c, \gamma, b} \VecvrR_{\delta, c, d} \, D_{27}(m^2_{\tilde{\chi}^-_{{a}}}, m^2_{\tilde{\chi}^-_{{c}}}, m^2_{\nu^R_{{d}}}, m^2_{\nu^R_{{b}}}) \\ 
  \VLRb{(c^l_4)} = & - \VcevrL_{a, \alpha, d} \VecvrR_{\beta, a, b} \VcevrR_{c, \gamma, b} \VecvrL_{\delta, c, d} \, D_{27}(m^2_{\tilde{\chi}^-_{{a}}}, m^2_{\tilde{\chi}^-_{{c}}}, m^2_{\nu^R_{{d}}}, m^2_{\nu^R_{{b}}}) \\ 
  \nonumber \\
  \SLLb{(c^l_5)} = & - \VcevL_{a, \alpha, d} \VecviL_{\beta, a, b} \VecviL_{\delta, c, b} \VcevL_{c, \gamma, d} m_{\tilde{\chi}^-_{{a}}} m_{\tilde{\chi}^-_{{c}}} \, D_0(m^2_{\tilde{\chi}^-_{{a}}}, m^2_{\tilde{\chi}^-_{{c}}}, m^2_{\nu^i_{{d}}}, m^2_{\nu^i_{{b}}}) \\ 
  \SLRb{(c^l_5)} = & - \VcevL_{a, \alpha, d} \VecviL_{\beta, a, b} \VecviR_{\delta, c, b} \VcevR_{c, \gamma, d} m_{\tilde{\chi}^-_{{a}}} m_{\tilde{\chi}^-_{{c}}} \, D_0(m^2_{\tilde{\chi}^-_{{a}}}, m^2_{\tilde{\chi}^-_{{c}}}, m^2_{\nu^i_{{d}}}, m^2_{\nu^i_{{b}}}) \\ 
  \VLLb{(c^l_5)} = &  \, \VcevL_{a, \alpha, d} \VecviR_{\beta, a, b} \VecviR_{\delta, c, b} \VcevL_{c, \gamma, d} \, D_{27}(m^2_{\tilde{\chi}^-_{{a}}}, m^2_{\tilde{\chi}^-_{{c}}}, m^2_{\nu^i_{{d}}}, m^2_{\nu^i_{{b}}}) \\ 
  \VLRb{(c^l_5)} = &  \, \VcevL_{a, \alpha, d} \VecviR_{\beta, a, b} \VecviL_{\delta, c, b} \VcevR_{c, \gamma, d} \, D_{27}(m^2_{\tilde{\chi}^-_{{a}}}, m^2_{\tilde{\chi}^-_{{c}}}, m^2_{\nu^i_{{d}}}, m^2_{\nu^i_{{b}}}) \\ 
  \nonumber \\
  \SLLb{(c^l_6)} = & - \VcevL_{a, \alpha, d} \VecvrL_{\beta, a, b} \VecvrL_{\delta, c, b} \VcevL_{c, \gamma, d} m_{\tilde{\chi}^-_{{a}}} m_{\tilde{\chi}^-_{{c}}} \, D_0(m^2_{\tilde{\chi}^-_{{a}}}, m^2_{\tilde{\chi}^-_{{c}}}, m^2_{\nu^i_{{d}}}, m^2_{\nu^R_{{b}}}) \\ 
  \SLRb{(c^l_6)} = & - \VcevL_{a, \alpha, d} \VecvrL_{\beta, a, b} \VecvrR_{\delta, c, b} \VcevR_{c, \gamma, d} m_{\tilde{\chi}^-_{{a}}} m_{\tilde{\chi}^-_{{c}}} \, D_0(m^2_{\tilde{\chi}^-_{{a}}}, m^2_{\tilde{\chi}^-_{{c}}}, m^2_{\nu^i_{{d}}}, m^2_{\nu^R_{{b}}}) \\ 
  \VLLb{(c^l_6)} = & \, \VcevL_{a, \alpha, d} \VecvrR_{\beta, a, b} \VecvrR_{\delta, c, b} \VcevL_{c, \gamma, d} \, D_{27}(m^2_{\tilde{\chi}^-_{{a}}}, m^2_{\tilde{\chi}^-_{{c}}}, m^2_{\nu^i_{{d}}}, m^2_{\nu^R_{{b}}}) \\ 
  \VLRb{(c^l_6)} = & \, \VcevL_{a, \alpha, d} \VecvrR_{\beta, a, b} \VecvrL_{\delta, c, b} \VcevR_{c, \gamma, d} \, D_{27}(m^2_{\tilde{\chi}^-_{{a}}}, m^2_{\tilde{\chi}^-_{{c}}}, m^2_{\nu^i_{{d}}}, m^2_{\nu^R_{{b}}}) \\ 
  \nonumber \\
  \SLLb{(c^l_7)} = & - \VcevrL_{a, \alpha, d} \VecviL_{\beta, a, b} \VecviL_{\delta, c, b} \VcevrL_{c, \gamma, d} m_{\tilde{\chi}^-_{{a}}}, m_{\tilde{\chi}^-_{{c}}} \, D_0(m^2_{\tilde{\chi}^-_{{a}}}, m^2_{\tilde{\chi}^-_{{c}}}, m^2_{\nu^R_{{d}}}, m^2_{\nu^i_{{b}}}) \\ 
  \SLRb{(c^l_7)} = & - \VcevrL_{a, \alpha, d} \VecviL_{\beta, a, b} \VecviR_{\delta, c, b} \VcevrR_{c, \gamma, d} m_{\tilde{\chi}^-_{{a}}}, m_{\tilde{\chi}^-_{{c}}} \, D_0(m^2_{\tilde{\chi}^-_{{a}}}, m^2_{\tilde{\chi}^-_{{c}}}, m^2_{\nu^R_{{d}}}, m^2_{\nu^i_{{b}}}) \\ 
  \VLLb{(c^l_7)} = & \, \VcevrL_{a, \alpha, d} \VecviR_{\beta, a, b} \VecviR_{\delta, c, b} \VcevrL_{c, \gamma, d} \, D_{27}(m^2_{\tilde{\chi}^-_{{a}}}, m^2_{\tilde{\chi}^-_{{c}}}, m^2_{\nu^R_{{d}}}, m^2_{\nu^i_{{b}}}) \\ 
  \VLRb{(c^l_7)} = & \, \VcevrL_{a, \alpha, d} \VecviR_{\beta, a, b} \VecviL_{\delta, c, b} \VcevrR_{c, \gamma, d} \, D_{27}(m^2_{\tilde{\chi}^-_{{a}}}, m^2_{\tilde{\chi}^-_{{c}}}, m^2_{\nu^R_{{d}}}, m^2_{\nu^i_{{b}}}) \\ 
  \nonumber \\
  \SLLb{(c^l_8)} = & - \VcevrL_{a, \alpha, d} \VecvrL_{\beta, a, b} \VecvrL_{\delta, c, b} \VcevrL_{c, \gamma, d} m_{\tilde{\chi}^-_{{a}}} m_{\tilde{\chi}^-_{{c}}} \, D_0(m^2_{\tilde{\chi}^-_{{a}}}, m^2_{\tilde{\chi}^-_{{c}}}, m^2_{\nu^R_{{d}}}, m^2_{\nu^R_{{b}}}) \\ 
  \SLRb{(c^l_8)} = & - \VcevrL_{a, \alpha, d} \VecvrL_{\beta, a, b} \VecvrR_{\delta, c, b} \VcevrR_{c, \gamma, d} m_{\tilde{\chi}^-_{{a}}} m_{\tilde{\chi}^-_{{c}}} \, D_0(m^2_{\tilde{\chi}^-_{{a}}}, m^2_{\tilde{\chi}^-_{{c}}}, m^2_{\nu^R_{{d}}}, m^2_{\nu^R_{{b}}}) \\ 
  \VLLb{(c^l_8)} = & \, \VcevrL_{a, \alpha, d} \VecvrR_{\beta, a, b} \VecvrR_{\delta, c, b} \VcevrL_{c, \gamma, d} \, D_{27}(m^2_{\tilde{\chi}^-_{{a}}}, m^2_{\tilde{\chi}^-_{{c}}}, m^2_{\nu^R_{{d}}}, m^2_{\nu^R_{{b}}}) \\ 
  \VLRb{(c^l_8)} = & \, \VcevrL_{a, \alpha, d} \VecvrR_{\beta, a, b} \VecvrL_{\delta, c, b} \VcevrR_{c, \gamma, d} \, D_{27}(m^2_{\tilde{\chi}^-_{{a}}}, m^2_{\tilde{\chi}^-_{{c}}}, m^2_{\nu^R_{{d}}}, m^2_{\nu^R_{{b}}}) 
\end{align} 

\subsubsection{$W^+$ and $H^+$ contributions}
\begin{align} 
  \SLLb{(w^l_1)} = & - \VvehpL_{a, \alpha, d} \VevhmL_{\beta, a, b} \VvehpL_{c, \gamma, b} \VevhmL_{\delta, c, d} m_{\nu_{{a}}} m_{\nu_{{c}}} \, D_0(m^2_{\nu_{{a}}}, m^2_{\nu_{{c}}}, m^2_{H^-_{{d}}}, m^2_{H^-_{{b}}}) \\ 
  \SLRb{(w^l_1)} = & - \VvehpL_{a, \alpha, d} \VevhmL_{\beta, a, b} \VvehpR_{c, \gamma, b} \VevhmR_{\delta, c, d} m_{\nu_{{a}}} m_{\nu_{{c}}} \, D_0(m^2_{\nu_{{a}}}, m^2_{\nu_{{c}}}, m^2_{H^-_{{d}}}, m^2_{H^-_{{b}}}) \\ 
  \VLLb{(w^l_1)} = & - \VvehpL_{a, \alpha, d} \VevhmR_{\beta, a, b} \VvehpL_{c, \gamma, b} \VevhmR_{\delta, c, d} \, D_{27}(m^2_{\nu_{{a}}}, m^2_{\nu_{{c}}}, m^2_{H^-_{{d}}}, m^2_{H^-_{{b}}}) \\ 
  \VLRb{(w^l_1)} = & - \VvehpL_{a, \alpha, d} \VevhmR_{\beta, a, b} \VvehpR_{c, \gamma, b} \VevhmL_{\delta, c, d} \, D_{27}(m^2_{\nu_{{a}}}, m^2_{\nu_{{c}}}, m^2_{H^-_{{d}}}, m^2_{H^-_{{b}}}) \\ 
  \nonumber \\
  \SLLb{(w^l_2)} = & \, 2 \, \VvehpL_{a, \alpha, d} \VevwmR_{\beta, a} \VvewpL_{c, \gamma} \VevhmL_{\delta, c, d} (I_{C_0 D_0}(m^2_{\nu_{{c}}}, m^2_{W^-},  m^2_{H^-_{{d}}}, m^2_{\nu_{{a}}}) - 2  D_{27}(m^2_{\nu_{{a}}}, m^2_{\nu_{{c}}}, m^2_{W^-},  m^2_{H^-_{{d}}})) \\ 
  \SLRb{(w^l_2)} = & \, 2 \, \VvehpL_{a, \alpha, d} \VevwmR_{\beta, a} \VvewpR_{c, \gamma} \VevhmR_{\delta, c, d} (I_{C_0 D_0}(m^2_{\nu_{{c}}}, m^2_{W^-},  m^2_{H^-_{{d}}}, m^2_{\nu_{{a}}}) - 2  D_{27}(m^2_{\nu_{{a}}}, m^2_{\nu_{{c}}}, m^2_{W^-},  m^2_{H^-_{{d}}})) \\ 
  \VLLb{(w^l_2)} = & \, \VvehpL_{a, \alpha, d} \VevwmL_{\beta, a} \VvewpL_{c, \gamma} \VevhmR_{\delta, c, d} m_{\nu_{{a}}} m_{\nu_{{c}}} D_0(m^2_{\nu_{{a}}}, m^2_{\nu_{{c}}}, m^2_{W^-},  m^2_{H^-_{{d}}}) \\ 
  \VLRb{(w^l_2)} = & \, \VvehpL_{a, \alpha, d} \VevwmL_{\beta, a} \VvewpR_{c, \gamma} \VevhmL_{\delta, c, d} m_{\nu_{{a}}} m_{\nu_{{c}}} D_0(m^2_{\nu_{{a}}}, m^2_{\nu_{{c}}}, m^2_{W^-},  m^2_{H^-_{{d}}}) \\ 
  \nonumber \\
  \SLLb{(w^l_3)} = & \, 2 \, \VvewpL_{a, i} \VevhmL_{\beta, a, b} \VvehpL_{c, \gamma, b} \VevwmR_{\delta, c} (I_{C_0 D_0}(m^2_{\nu_{{c}}},  m^2_{H^-_{{b}}}, m^2_{W^-}, m^2_{\nu_{{a}}}) - 2 D_{27}(m^2_{\nu_{{a}}}, m^2_{\nu_{{c}}}, m^2_{H^-_{{b}}}, m^2_{W^-})) \\ 
  \SLRb{(w^l_3)} = & \, 2 \, \VvewpL_{a, i} \VevhmL_{\beta, a, b} \VvehpR_{c, \gamma, b} \VevwmL_{\delta, c} (I_{C_0 D_0}(m^2_{\nu_{{c}}},  m^2_{H^-_{{b}}}, m^2_{W^-}, m^2_{\nu_{{a}}}) - 2 D_{27}(m^2_{\nu_{{a}}}, m^2_{\nu_{{c}}}, m^2_{H^-_{{b}}}, m^2_{W^-})) \\ 
  \VLLb{(w^l_3)} = & \, \VvewpL_{a, i} \VevhmR_{\beta, a, b} \VvehpL_{c, \gamma, b} \VevwmL_{\delta, c} m_{\nu_{{a}}}  m_{\nu_{{c}}} D_0(m^2_{\nu_{{a}}}, m^2_{\nu_{{c}}}, m^2_{H^-_{{b}}}, m^2_{W^-}) \\ 
  \VLRb{(w^l_3)} = & \, \VvewpL_{a, i} \VevhmR_{\beta, a, b} \VvehpR_{c, \gamma, b} \VevwmR_{\delta, c} m_{\nu_{{a}}}  m_{\nu_{{c}}} D_0(m^2_{\nu_{{a}}}, m^2_{\nu_{{c}}}, m^2_{H^-_{{b}}}, m^2_{W^-}) \\ 
  \TLLb{(w^l_3)} = & - \VvewpL_{a, i} \VevhmL_{\beta, a, b} \VvehpL_{c, \gamma, b} \VevwmR_{\delta, c} \, D_{27}(m^2_{\nu_{{a}}}, m^2_{\nu_{{c}}}, m^2_{H^-_{{b}}}, m^2_{W^-}) \\ 
  \nonumber \\
  \SLLb{(w^l_4)} = & -4  \VvewpL_{a, i} \VevwmR_{\beta, a} \VvewpL_{c, \gamma} \VevwmR_{\delta, c} m_{\nu_{{a}}} m_{\nu_{{c}}} D_0(m^2_{\nu_{{a}}}, m^2_{\nu_{{c}}}, m^2_{W^-}, m^2_{W^-}) \\ 
  \SLRb{(w^l_4)} = & -4  \VvewpL_{a, i} \VevwmR_{\beta, a} \VvewpR_{c, \gamma} \VevwmL_{\delta, c} m_{\nu_{{a}}} m_{\nu_{{c}}} D_0(m^2_{\nu_{{a}}}, m^2_{\nu_{{c}}}, m^2_{W^-}, m^2_{W^-}) \\ 
  \VLLb{(w^l_4)} = & -4  \VvewpL_{a, i} \VevwmL_{\beta, a} \VvewpL_{c, \gamma} \VevwmL_{\delta, c} (I_{C_0 D_0}(m^2_{\nu_{{c}}}, m^2_{W^-}, m^2_{W^-}, m^2_{\nu_{{a}}}) - 3 D_{27}(m^2_{\nu_{{a}}}, m^2_{\nu_{{c}}}, m^2_{W^-}, m^2_{W^-})) \\ 
  \VLRb{(w^l_4)} = & -4  \VvewpL_{a, i} \VevwmL_{\beta, a} \VvewpR_{c, \gamma} \VevwmR_{\delta, c} I_{C_0 D_0}(m^2_{\nu_{{c}}}, m^2_{W^-}, m^2_{W^-}, m^2_{\nu_{{a}}}) \\ 
  \TLLb{(w^l_4)} = &  \, \VvewpL_{a, i} \VevwmR_{\beta, a} \VvewpL_{c, \gamma} \VevwmR_{\delta, c} m_{\nu_{{a}}} m_{\nu_{{c}}} D_0(m^2_{\nu_{{a}}}, m^2_{\nu_{{c}}}, m^2_{W^-}, m^2_{W^-}) \\ 
  \nonumber \\
    \SLLb{(w^l_5)} = & \, \frac{1}{2} \VvehpL_{a, \alpha, d} \VevhmL_{\delta, c, d} \VvehpL_{a, \gamma, b} \VevhmL_{\beta, c, b} m_{\nu_{{a}}} m_{\nu_{{c}}} D_0(m^2_{\nu_{{a}}}, m^2_{\nu_{{c}}}, m^2_{H^-_{{d}}}, m^2_{H^-_{{b}}}) \\ 
    \SLRb{(w^l_5)} = & -2  \VvehpL_{a, \alpha, d} \VevhmR_{\delta, c, d} \VvehpR_{a, \gamma, b} \VevhmL_{\beta, c, b} D_{27}(m^2_{\nu_{{a}}}, m^2_{\nu_{{c}}}, m^2_{H^-_{{d}}}, m^2_{H^-_{{b}}})  \\ 
    \VLLb{(w^l_5)} = & -\frac{1}{2} \VvehpL_{a, \alpha, d} \VevhmR_{\delta, c, d} \VvehpL_{a, \gamma, b} \VevhmR_{\beta, c, b} m_{\nu_{{a}}} m_{\nu_{{c}}} D_0(m^2_{\nu_{{a}}}, m^2_{\nu_{{c}}}, m^2_{H^-_{{d}}}, m^2_{H^-_{{b}}}) \\ 
    \VLRb{(w^l_5)} = & - \VvehpL_{a, \alpha, d} \VevhmL_{\delta, c, d} \VvehpR_{a, \gamma, b} \VevhmR_{\beta, c, b} D_{27}(m^2_{\nu_{{a}}}, m^2_{\nu_{{c}}}, m^2_{H^-_{{d}}}, m^2_{H^-_{{b}}})  \\ 
    \TLLb{(w^l_5)} = & \, \frac{1}{8} \VvehpL_{a, \alpha, d} \VevhmL_{\delta, c, d} \VvehpL_{a, \gamma, b} \VevhmL_{\beta, c, b} m_{\nu_{{a}}} m_{\nu_{{c}}} D_0(m^2_{\nu_{{a}}}, m^2_{\nu_{{c}}}, m^2_{H^-_{{d}}}, m^2_{H^-_{{b}}}) \\ 
    \SLLb{(w^l_6)} = & - \VvehpL_{a, \alpha, d} \VevhmL_{\delta, c, d} \VvewpL_{a, \gamma} \VevwmR_{\beta, c} ( I_{C_0 D_0}(m^2_{\nu_{{c}}}, m^2_{W^-},  m^2_{H^-_{{d}}}, m^2_{\nu_{{a}}})  - 8 D_{27}(m^2_{\nu_{{a}}}, m^2_{\nu_{{c}}}, m^2_{W^-},  m^2_{H^-_{{d}}})) \\ 
    \SLRb{(w^l_6)} = & \, 2 \, \VvehpL_{a, \alpha, d} \VevhmR_{\delta, c, d} \VvewpR_{a, \gamma} \VevwmR_{\beta, c} m_{\nu_{{a}}} m_{\nu_{{c}}} D_0(m^2_{\nu_{{a}}}, m^2_{\nu_{{c}}}, m^2_{W^-},  m^2_{H^-_{{d}}}) \\ 
    \VLLb{(w^l_6)} = & \, \VvehpL_{a, \alpha, d} \VevhmR_{\delta, c, d} \VvewpL_{a, \gamma} \VevwmL_{\beta, c} ( I_{C_0 D_0}(m^2_{\nu_{{c}}}, m^2_{W^-},  m^2_{H^-_{{d}}}, m^2_{\nu_{{a}}})  - 2 D_{27}(m^2_{\nu_{{a}}}, m^2_{\nu_{{c}}}, m^2_{W^-},  m^2_{H^-_{{d}}})) \\ 
    \VLRb{(w^l_6)}l = & \, \VvehpL_{a, \alpha, d} \VevhmL_{\delta, c, d} \VvewpR_{a, \gamma} \VevwmL_{\beta, c} m_{\nu_{{a}}} m_{\nu_{{c}}} D_0(m^2_{\nu_{{a}}}, m^2_{\nu_{{c}}}, m^2_{W^-},  m^2_{H^-_{{d}}}) \\ 
    \TLLb{(w^l_6)} = & - \frac{1}{4} \VvehpL_{a, \alpha, d} \VevhmL_{\delta, c, d} \VvewpL_{a, \gamma} \VevwmR_{\beta, c}  I_{C_0 D_0}(m^2_{\nu_{{c}}}, m^2_{W^-},  m^2_{H^-_{{d}}}, m^2_{\nu_{{a}}})  \\ 
 \SLLb{(w^l_7)} = & -\VvewpL_{a, i} \VevwmR_{\delta, c} \VvehpL_{a, \gamma, b} \VevhmL_{\beta, c, b} ( I_{C_0 D_0}(m^2_{\nu_{{c}}}, m^2_{H^-_{{b}}}, m^2_{W^-}, m^2_{\nu_{{a}}}) - 8 D_{27}(m^2_{\nu_{{a}}}, m^2_{\nu_{{c}}}, m^2_{H^-_{{b}}}, m^2_{W^-})) \\ 
  \SLRb{(w^l_7)} = & \, 2 \, \VvewpR_{a, i} \VevwmR_{\delta, c} \VvehpL_{a, \gamma, b} \VevhmR_{\beta, c, b} m_{\nu_{{a}}} m_{\nu_{{c}}} D_0(m^2_{\nu_{{a}}}, m^2_{\nu_{{c}}}, m^2_{H^-_{{b}}}, m^2_{W^-})  \\ 
  \VLLb{(w^l_7)} = & \, \VvewpL_{a, i} \VevwmL_{\delta, c} \VvehpL_{a, \gamma, b} \VevhmR_{\beta, c, b}  ( I_{C_0 D_0}(m^2_{\nu_{{c}}}, m^2_{H^-_{{b}}}, m^2_{W^-}, m^2_{\nu_{{a}}}) - 2 D_{27}(m^2_{\nu_{{a}}}, m^2_{\nu_{{c}}}, m^2_{H^-_{{b}}}, m^2_{W^-})) \\ 
  \VLRb{(w^l_7)} = & \, \VvewpL_{a, i} \VevwmR_{\delta, c} \VvehpR_{a, \gamma, b} \VevhmR_{\beta, c, b} m_{\nu_{{a}}} m_{\nu_{{c}}} D_0(m^2_{\nu_{{a}}}, m^2_{\nu_{{c}}}, m^2_{H^-_{{b}}}, m^2_{W^-})  \\ 
  \TLLb{(w^l_7)} = & -\frac{1}{4} \VvewpL_{a, i} \VevwmR_{\delta, c} \VvehpL_{a, \gamma, b} \VevhmL_{\beta, c, b}  I_{C_0 D_0}(m^2_{\nu_{{c}}}, m^2_{H^-_{{b}}}, m^2_{W^-}, m^2_{\nu_{{a}}}) \\ 
  \nonumber \\
  \SLLb{(w^l_8)} = & -4   \VvewpL_{a, i} \VevwmR_{\delta, c} \VvewpL_{a, \gamma} \VevwmR_{\beta, c}  m_{\nu_{{a}}} m_{\nu_{{c}}} D_0(m^2_{\nu_{{a}}}, m^2_{\nu_{{c}}}, m^2_{W^-}, m^2_{W^-}) \\ 
  \SLRb{(w^l_8)} = & -8   \VvewpL_{a, i} \VevwmL_{\delta, c} \VvewpR_{a, \gamma} \VevwmR_{\beta, c} (I_{C_0 D_0}(m^2_{\nu_{{c}}}, m^2_{W^-}, m^2_{W^-}, m^2_{\nu_{{a}}}) - 3 D_{27}(m^2_{\nu_{{a}}}, m^2_{\nu_{{c}}}, m^2_{W^-}, m^2_{W^-})) \\ 
  \VLLb{(w^l_8)} = & -2   \VvewpL_{a, i} \VevwmL_{\delta, c} \VvewpL_{a, \gamma} \VevwmL_{\beta, c} m_{\nu_{{a}}} m_{\nu_{{c}}} D_0(m^2_{\nu_{{a}}}, m^2_{\nu_{{c}}}, m^2_{W^-}, m^2_{W^-}) \\ 
  \VLRb{(w^l_8)} = & -4  \VvewpL_{a, i} \VevwmR_{\delta, c} \VvewpR_{a, \gamma} \VevwmL_{\beta, c} I_{C_0 D_0}(m^2_{\nu_{{c}}}, m^2_{W^-}, m^2_{W^-}, m^2_{\nu_{{a}}}) \\ 
  \TLLb{(w^l_8)} = & \, \VvewpL_{a, i} \VevwmR_{\delta, c} \VvewpL_{a, \gamma} \VevwmR_{\beta, c}  m_{\nu_{{a}}} m_{\nu_{{c}}} D_0(m^2_{\nu_{{a}}}, m^2_{\nu_{{c}}}, m^2_{W^-}, m^2_{W^-}) 
\end{align} 

\subsection{Additional boxes for $\ell_\alpha^- \to \ell_\beta^- \ell_\gamma^+ \ell_\gamma^-$}
In the case of $\ell_\alpha^- \to \ell_\beta^- \ell_\gamma^+
\ell_\gamma^-$ it is necessary to calculate the crossed diagrams with
exchanged indices $\beta \leftrightarrow \gamma$ explicitly.
\subsubsection{Crossed neutralino contributions}
\begin{align} 
  \SLLb{(n^l_{1'})}= & \, \frac{1}{2} \VneeL_{d, \alpha, a} \VeneL_{\beta, b, a} \VneeL_{b, \gamma, c} \VeneL_{\delta, d, c}  m_{\tilde{\chi}^0_{{b}}}  m_{\tilde{\chi}^0_{{d}}} D_0(m^2_{\tilde{\chi}^0_{{d}}}, m^2_{\tilde{\chi}^0_{{b}}}, m^2_{\tilde{e}_{{a}}}, m^2_{\tilde{e}_{{c}}}) \\ 
  \SLRb{(n^l_{1'})}= & \, 2 \, \VneeL_{d, \alpha, a} \VeneL_{\beta, b, a} \VneeR_{b, \gamma, c} \VeneR_{\delta, d, c} D_{27}(m^2_{\tilde{\chi}^0_{{d}}}, m^2_{\tilde{\chi}^0_{{b}}}, m^2_{\tilde{e}_{{a}}}, m^2_{\tilde{e}_{{c}}}) \\ 
  \VLLb{(n^l_{1'})}= & - \VneeL_{d, \alpha, a} \VeneR_{\beta, b, a} \VneeL_{b, \gamma, c} \VeneR_{\delta, d, c} D_{27}(m^2_{\tilde{\chi}^0_{{d}}}, m^2_{\tilde{\chi}^0_{{b}}}, m^2_{\tilde{e}_{{a}}}, m^2_{\tilde{e}_{{c}}}) \\ 
  \VLRb{(n^l_{1'})}= & \, \frac{1}{2} \VneeL_{d, \alpha, a} \VeneR_{\beta, b, a} \VneeR_{b, \gamma, c} \VeneL_{\delta, d, c}  m_{\tilde{\chi}^0_{{b}}}  m_{\tilde{\chi}^0_{{d}}} D_0(m^2_{\tilde{\chi}^0_{{d}}}, m^2_{\tilde{\chi}^0_{{b}}}, m^2_{\tilde{e}_{{a}}}, m^2_{\tilde{e}_{{c}}}) \\ 
  \TLLb{(n^l_{1'})}= & - \frac{1}{8} \VneeL_{d, \alpha, a} \VeneL_{\beta, b, a} \VneeL_{b, \gamma, c} \VeneL_{\delta, d, c} m_{\tilde{\chi}^0_{{b}}}  m_{\tilde{\chi}^0_{{d}}} D_0(m^2_{\tilde{\chi}^0_{{d}}}, m^2_{\tilde{\chi}^0_{{b}}}, m^2_{\tilde{e}_{{a}}}, m^2_{\tilde{e}_{{c}}}) \\ 
\nonumber\\
  \SLLb{(n^l_{2'})}= & \, \frac{1}{2} \VneeL_{d, \alpha, a} \VeneL_{\beta, b, a} \VeneL_{\delta, b, c} \VneeL_{d, \gamma, c}  m_{\tilde{\chi}^0_{{b}}}  m_{\tilde{\chi}^0_{{d}}} D_0(m^2_{\tilde{\chi}^0_{{b}}}, m^2_{\tilde{\chi}^0_{{d}}}, m^2_{\tilde{e}_{{a}}}, m^2_{\tilde{e}_{{c}}}) \\ 
  \SLRb{(n^l_{2'})}= & \, 2 \, \VneeL_{d, \alpha, a} \VeneL_{\beta, b, a} \VeneR_{\delta, b, c} \VneeR_{d, \gamma, c} D_{27}(m^2_{\tilde{\chi}^0_{{b}}}, m^2_{\tilde{\chi}^0_{{d}}}, m^2_{\tilde{e}_{{a}}}, m^2_{\tilde{e}_{{c}}}) \\ 
  \VLLb{(n^l_{2'})}= & -\frac{1}{2} \VneeL_{d, \alpha, a} \VeneR_{\beta, b, a} \VeneR_{\delta, b, c} \VneeL_{d, \gamma, c}  m_{\tilde{\chi}^0_{{b}}}  m_{\tilde{\chi}^0_{{d}}} D_0(m^2_{\tilde{\chi}^0_{{b}}}, m^2_{\tilde{\chi}^0_{{d}}}, m^2_{\tilde{e}_{{a}}}, m^2_{\tilde{e}_{{c}}}) \\ 
  \VLRb{(n^l_{2'})}= & \, \VneeL_{d, \alpha, a} \VeneR_{\beta, b, a} \VeneL_{\delta, b, c} \VneeR_{d, \gamma, c}   D_{27}(m^2_{\tilde{\chi}^0_{{b}}}, m^2_{\tilde{\chi}^0_{{d}}}, m^2_{\tilde{e}_{{a}}}, m^2_{\tilde{e}_{{c}}}) \\ 
  \TLLb{(n^l_{2'})}= & \, \frac{1}{8} \VneeL_{d, \alpha, a} \VeneL_{\beta, b, a} \VeneL_{\delta, b, c} \VneeL_{d, \gamma, c}  m_{\tilde{\chi}^0_{{b}}}  m_{\tilde{\chi}^0_{{d}}} D_0(m^2_{\tilde{\chi}^0_{{b}}}, m^2_{\tilde{\chi}^0_{{d}}}, m^2_{\tilde{e}_{{a}}}, m^2_{\tilde{e}_{{c}}}) \\ 
\end{align} 

\subsubsection{Crossed chargino contributions}
\begin{align} 
  \SLLb{(c^l_{1'})}= & \, \frac{1}{2} \VceviL_{d, \alpha, a}\VecviL_{\delta, d, c}\VceviL_{b, \gamma, a}\VecviL_{\beta, b, c} m_{\tilde{\chi}^-_{{b}}}  m_{\tilde{\chi}^-_{{d}}} D_0(m^2_{\tilde{\chi}^-_{{d}}}, m^2_{\tilde{\chi}^-_{{b}}}, m^2_{\nu^i_{{a}}}, m^2_{\nu^i_{{c}}}) \\ 
  \SLRb{(c^l_{1'})}= & -2 \VceviL_{d, \alpha, a}\VecviR_{\delta, d, c}\VceviR_{b, \gamma, a}\VecviL_{\beta, b, c} D_{27}(m^2_{\tilde{\chi}^-_{{d}}}, m^2_{\tilde{\chi}^-_{{b}}}, m^2_{\nu^i_{{a}}}, m^2_{\nu^i_{{c}}}) \\ 
  \VLLb{(c^l_{1'})}= & \, \VceviL_{d, \alpha, a}\VecviR_{\delta, d, c}\VceviL_{b, \gamma, a}\VecviR_{\beta, b, c} D_{27}(m^2_{\tilde{\chi}^-_{{d}}}, m^2_{\tilde{\chi}^-_{{b}}}, m^2_{\nu^i_{{a}}}, m^2_{\nu^i_{{c}}})) \\ 
  \VLRb{(c^l_{1'})}= & \, \frac{1}{2} \VceviL_{d, \alpha, a}\VecviL_{\delta, d, c}\VceviR_{b, \gamma, a}\VecviR_{\beta, b, c} m_{\tilde{\chi}^-_{{b}}}  m_{\tilde{\chi}^-_{{d}}} D_0(m^2_{\tilde{\chi}^-_{{d}}}, m^2_{\tilde{\chi}^-_{{b}}}, m^2_{\nu^i_{{a}}}, m^2_{\nu^i_{{c}}}) \\ 
  \TLLb{(c^l_{1'})}= & - \frac{1}{8} \VceviL_{d, \alpha, a}\VecviL_{\delta, d, c}\VceviL_{b, \gamma, a}\VecviL_{\beta, b, c}  m_{\tilde{\chi}^-_{{d}}} D_0(m^2_{\tilde{\chi}^-_{{d}}}, m^2_{\tilde{\chi}^-_{{b}}}, m^2_{\nu^i_{{a}}}, m^2_{\nu^i_{{c}}}) \\ 
\nonumber\\
  \SLLb{(c^l_{2'})}= & \, \frac{1}{2} \VceviL_{d, \alpha, a}\VecviL_{\beta, b, a}\VcevrL_{b, \gamma, c}\VecvrL_{\delta, d, c} m_{\tilde{\chi}^-_{{b}}}  m_{\tilde{\chi}^-_{{d}}} D_0(m^2_{\tilde{\chi}^-_{{d}}}, m^2_{\tilde{\chi}^-_{{b}}}, m^2_{\nu^i_{{a}}}, m^2_{\nu^R_{{c}}}) \\ 
  \SLRb{(c^l_{2'})}= & \, 2 \, \VceviL_{d, \alpha, a}\VecviL_{\beta, b, a}\VcevrR_{b, \gamma, c}\VecvrR_{\delta, d, c} D_{27}(m^2_{\tilde{\chi}^-_{{d}}}, m^2_{\tilde{\chi}^-_{{b}}}, m^2_{\nu^i_{{a}}}, m^2_{\nu^R_{{c}}}) \\ 
  \VLLb{(c^l_{2'})}= & -\VceviL_{d, \alpha, a}\VecviR_{\beta, b, a}\VcevrL_{b, \gamma, c}\VecvrR_{\delta, d, c} D_{27}(m^2_{\tilde{\chi}^-_{{d}}}, m^2_{\tilde{\chi}^-_{{b}}}, m^2_{\nu^i_{{a}}}, m^2_{\nu^R_{{c}}}) \\ 
  \VLRb{(c^l_{2'})}= & \,\frac{1}{2} \VceviL_{d, \alpha, a}\VecviR_{\beta, b, a}\VcevrR_{b, \gamma, c}\VecvrL_{\delta, d, c} m_{\tilde{\chi}^-_{{b}}}  m_{\tilde{\chi}^-_{{d}}} D_0(m^2_{\tilde{\chi}^-_{{d}}}, m^2_{\tilde{\chi}^-_{{b}}}, m^2_{\nu^i_{{a}}}, m^2_{\nu^R_{{c}}}) \\ 
  \TLLb{(c^l_{2'})}= & - \frac{1}{8} \VceviL_{d, \alpha, a}\VecviL_{\beta, b, a}\VcevrL_{b, \gamma, c} m_{\tilde{\chi}^-_{{b}}}  m_{\tilde{\chi}^-_{{d}}} D_0(m^2_{\tilde{\chi}^-_{{d}}}, m^2_{\tilde{\chi}^-_{{b}}}, m^2_{\nu^i_{{a}}}, m^2_{\nu^R_{{c}}}) \\ 
\nonumber\\
  \SLLb{(c^l_{3'})}= & \, \frac{1}{2} \VcevrL_{d, \alpha, a}\VecvrL_{\beta, b, a}\VceviL_{b, \gamma, c}\VecviL_{\delta, d, c} m_{\tilde{\chi}^-_{{b}}}  m_{\tilde{\chi}^-_{{d}}} D_0(m^2_{\tilde{\chi}^-_{{d}}}, m^2_{\tilde{\chi}^-_{{b}}}, m^2_{\nu^R_{{a}}}, m^2_{\nu^i_{{c}}}) \\ 
  \SLRb{(c^l_{3'})}= & \, 2 \, \VcevrL_{d, \alpha, a}\VecvrL_{\beta, b, a}\VceviR_{b, \gamma, c}\VecviR_{\delta, d, c} D_{27}(m^2_{\tilde{\chi}^-_{{d}}}, m^2_{\tilde{\chi}^-_{{b}}}, m^2_{\nu^R_{{a}}}, m^2_{\nu^i_{{c}}}) \\ 
  \VLLb{(c^l_{3'})}= & -\VcevrL_{d, \alpha, a}\VecvrR_{\beta, b, a}\VceviL_{b, \gamma, c}\VecviR_{\delta, d, c} D_{27}(m^2_{\tilde{\chi}^-_{{d}}}, m^2_{\tilde{\chi}^-_{{b}}}, m^2_{\nu^R_{{a}}}, m^2_{\nu^i_{{c}}}) \\ 
  \VLRb{(c^l_{3'})}= & \, \frac{1}{2} \VcevrL_{d, \alpha, a}\VecvrR_{\beta, b, a}\VceviR_{b, \gamma, c}\VecviL_{\delta, d, c} m_{\tilde{\chi}^-_{{b}}}  m_{\tilde{\chi}^-_{{d}}} D_0(m^2_{\tilde{\chi}^-_{{d}}}, m^2_{\tilde{\chi}^-_{{b}}}, m^2_{\nu^R_{{a}}}, m^2_{\nu^i_{{c}}}) \\ 
  \TLLb{(c^l_{3'})}= & - \frac{1}{8} \VcevrL_{d, \alpha, a}\VecvrL_{\beta, b, a}\VceviL_{b, \gamma, c}\VecviL_{\delta, d, c} m_{\tilde{\chi}^-_{{b}}}  m_{\tilde{\chi}^-_{{d}}} D_0(m^2_{\tilde{\chi}^-_{{d}}}, m^2_{\tilde{\chi}^-_{{b}}}, m^2_{\nu^R_{{a}}}, m^2_{\nu^i_{{c}}}) \\ 
\nonumber\\
  \SLLb{(c^l_{4'})}= & \, \frac{1}{2} \VcevrL_{d, \alpha, a}\VecvrL_{\beta, b, a}\VcevrL_{b, \gamma, c}\VecvrL_{\delta, d, c}   m_{\tilde{\chi}^-_{{b}}}  m_{\tilde{\chi}^-_{{d}}} D_0(m^2_{\tilde{\chi}^-_{{d}}}, m^2_{\tilde{\chi}^-_{{b}}}, m^2_{\nu^R_{{a}}}, m^2_{\nu^R_{{c}}}) \\ 
  \SLRb{(c^l_{4'})}= & \, 2 \, \VcevrL_{d, \alpha, a}\VecvrL_{\beta, b, a}\VcevrR_{b, \gamma, c}\VecvrR_{\delta, d, c}  D_{27}(m^2_{\tilde{\chi}^-_{{d}}}, m^2_{\tilde{\chi}^-_{{b}}}, m^2_{\nu^R_{{a}}}, m^2_{\nu^R_{{c}}}) \\ 
  \VLLb{(c^l_{4'})}= & -\VcevrL_{d, \alpha, a}\VecvrR_{\beta, b, a}\VcevrL_{b, \gamma, c}\VecvrR_{\delta, d, c}  D_{27}(m^2_{\tilde{\chi}^-_{{d}}}, m^2_{\tilde{\chi}^-_{{b}}}, m^2_{\nu^R_{{a}}}, m^2_{\nu^R_{{c}}}) \\ 
  \VLRb{(c^l_{4'})}= & \, \frac{1}{2}\VcevrL_{d, \alpha, a}\VecvrR_{\beta, b, a}\VcevrR_{b, \gamma, c}\VecvrL_{\delta, d, c}   m_{\tilde{\chi}^-_{{b}}}  m_{\tilde{\chi}^-_{{d}}} D_0(m^2_{\tilde{\chi}^-_{{d}}}, m^2_{\tilde{\chi}^-_{{b}}}, m^2_{\nu^R_{{a}}}, m^2_{\nu^R_{{c}}}) \\ 
  \TLLb{(c^l_{4'})}= & -\frac{1}{8}\VcevrL_{d, \alpha, a}\VecvrL_{\beta, b, a}\VcevrL_{b, \gamma, c}\VecvrL_{\delta, d, c}  m_{\tilde{\chi}^-_{{b}}}  m_{\tilde{\chi}^-_{{d}}} D_0(m^2_{\tilde{\chi}^-_{{d}}}, m^2_{\tilde{\chi}^-_{{b}}}, m^2_{\nu^R_{{a}}}, m^2_{\nu^R_{{c}}}) \\ 
\nonumber\\
   \SLLb{(c^l_{5'})}= & \, \frac{1}{2}\VceviL_{d, \alpha, a}\VecviL_{\delta, d, c}\VceviL_{b, \gamma, a}\VecviL_{\beta, b, c}   m_{\tilde{\chi}^-_{{b}}}  m_{\tilde{\chi}^-_{{d}}} D_0(m^2_{\tilde{\chi}^-_{{d}}}, m^2_{\tilde{\chi}^-_{{b}}}, m^2_{\nu^i_{{a}}}, m^2_{\nu^i_{{c}}}) \\ 
  \SLRb{(c^l_{5'})}= & -2 \VceviL_{d, \alpha, a}\VecviR_{\delta, d, c}\VceviR_{b, \gamma, a}\VecviL_{\beta, b, c} D_{27}(m^2_{\tilde{\chi}^-_{{d}}}, m^2_{\tilde{\chi}^-_{{b}}}, m^2_{\nu^i_{{a}}}, m^2_{\nu^i_{{c}}}) \\ 
  \VLLb{(c^l_{5'})}= & \, \VceviL_{d, \alpha, a}\VecviR_{\delta, d, c}\VceviL_{b, \gamma, a}\VecviR_{\beta, b, c}    D_{27}(m^2_{\tilde{\chi}^-_{{d}}}, m^2_{\tilde{\chi}^-_{{b}}}, m^2_{\nu^i_{{a}}}, m^2_{\nu^i_{{c}}}) \\ 
  \VLRb{(c^l_{5'})}= & \, \frac{1}{2}\VceviL_{d, \alpha, a}\VecviL_{\delta, d, c}\VceviR_{b, \gamma, a}\VecviR_{\beta, b, c}   m_{\tilde{\chi}^-_{{b}}}  m_{\tilde{\chi}^-_{{d}}} D_0(m^2_{\tilde{\chi}^-_{{d}}}, m^2_{\tilde{\chi}^-_{{b}}}, m^2_{\nu^i_{{a}}}, m^2_{\nu^i_{{c}}}) \\ 
  \TLLb{(c^l_{5'})}= & -\frac{1}{8}\VceviL_{d, \alpha, a}\VecviL_{\delta, d, c}\VceviL_{b, \gamma, a}\VecviL_{\beta, b, c}  m_{\tilde{\chi}^-_{{b}}}  m_{\tilde{\chi}^-_{{d}}} D_0(m^2_{\tilde{\chi}^-_{{d}}}, m^2_{\tilde{\chi}^-_{{b}}}, m^2_{\nu^i_{{a}}}, m^2_{\nu^i_{{c}}}) \\ 
\nonumber\\
   \SLLb{(c^l_{6'})}= & \, \frac{1}{2}\VceviL_{d, \alpha, a}\VecvrL_{\delta, d, c}\VceviL_{b, \gamma, a}\VecvrL_{\beta, b, c}    m_{\tilde{\chi}^-_{{b}}}  m_{\tilde{\chi}^-_{{d}}} D_0(m^2_{\tilde{\chi}^-_{{d}}}, m^2_{\tilde{\chi}^-_{{b}}}, m^2_{\nu^i_{{a}}}, m^2_{\nu^R_{{c}}}) \\ 
  \SLRb{(c^l_{6'})}= & -2 \VceviL_{d, \alpha, a}\VecvrR_{\delta, d, c}\VceviR_{b, \gamma, a}\VecvrL_{\beta, b, c}  D_{27}(m^2_{\tilde{\chi}^-_{{d}}}, m^2_{\tilde{\chi}^-_{{b}}}, m^2_{\nu^i_{{a}}}, m^2_{\nu^R_{{c}}}) \\ 
  \VLLb{(c^l_{6'})}= & \, \VceviL_{d, \alpha, a}\VecvrR_{\delta, d, c}\VceviL_{b, \gamma, a}\VecvrR_{\beta, b, c}     D_{27}(m^2_{\tilde{\chi}^-_{{d}}}, m^2_{\tilde{\chi}^-_{{b}}}, m^2_{\nu^i_{{a}}}, m^2_{\nu^R_{{c}}}) \\ 
  \VLRb{(c^l_{6'})}= & \, \frac{1}{2}\VceviL_{d, \alpha, a}\VecvrL_{\delta, d, c}\VceviR_{b, \gamma, a}\VecvrR_{\beta, b, c}    m_{\tilde{\chi}^-_{{b}}}  m_{\tilde{\chi}^-_{{d}}} D_0(m^2_{\tilde{\chi}^-_{{d}}}, m^2_{\tilde{\chi}^-_{{b}}}, m^2_{\nu^i_{{a}}}, m^2_{\nu^R_{{c}}}) \\ 
  \TLLb{(c^l_{6'})}= & -\frac{1}{8}\VceviL_{d, \alpha, a}\VecvrL_{\delta, d, c}\VceviL_{b, \gamma, a}\VecvrL_{\beta, b, c}   m_{\tilde{\chi}^-_{{b}}}  m_{\tilde{\chi}^-_{{d}}} D_0(m^2_{\tilde{\chi}^-_{{d}}}, m^2_{\tilde{\chi}^-_{{b}}}, m^2_{\nu^i_{{a}}}, m^2_{\nu^R_{{c}}}) \\ 
\nonumber\\
  \SLLb{(c^l_{7'})}= & \, \frac{1}{2}\VcevrL_{d, \alpha, a}\VecviL_{\delta, d, c}\VcevrL_{b, \gamma, a}\VecviL_{\beta, b, c}   m_{\tilde{\chi}^-_{{b}}}  m_{\tilde{\chi}^-_{{d}}} D_0(m^2_{\tilde{\chi}^-_{{d}}}, m^2_{\tilde{\chi}^-_{{b}}}, m^2_{\nu^R_{{a}}}, m^2_{\nu^i_{{c}}}) \\ 
  \SLRb{(c^l_{7'})}= & -2 \VcevrL_{d, \alpha, a}\VecviR_{\delta, d, c}\VcevrR_{b, \gamma, a}\VecviL_{\beta, b, c} D_{27}(m^2_{\tilde{\chi}^-_{{d}}}, m^2_{\tilde{\chi}^-_{{b}}}, m^2_{\nu^R_{{a}}}, m^2_{\nu^i_{{c}}}) \\ 
  \VLLb{(c^l_{7'})}= & \, \VcevrL_{d, \alpha, a}\VecviR_{\delta, d, c}\VcevrL_{b, \gamma, a}\VecviR_{\beta, b, c}    D_{27}(m^2_{\tilde{\chi}^-_{{d}}}, m^2_{\tilde{\chi}^-_{{b}}}, m^2_{\nu^R_{{a}}}, m^2_{\nu^i_{{c}}}) \\ 
  \VLRb{(c^l_{7'})}= & \, \frac{1}{2}\VcevrL_{d, \alpha, a}\VecviL_{\delta, d, c}\VcevrR_{b, \gamma, a}\VecviR_{\beta, b, c}   m_{\tilde{\chi}^-_{{b}}}  m_{\tilde{\chi}^-_{{d}}} D_0(m^2_{\tilde{\chi}^-_{{d}}}, m^2_{\tilde{\chi}^-_{{b}}}, m^2_{\nu^R_{{a}}}, m^2_{\nu^i_{{c}}}) \\ 
  \TLLb{(c^l_{7'})}= & -\frac{1}{8}\VcevrL_{d, \alpha, a}\VecviL_{\delta, d, c}\VcevrL_{b, \gamma, a}\VecviL_{\beta, b, c}  m_{\tilde{\chi}^-_{{b}}}  m_{\tilde{\chi}^-_{{d}}} D_0(m^2_{\tilde{\chi}^-_{{d}}}, m^2_{\tilde{\chi}^-_{{b}}}, m^2_{\nu^R_{{a}}}, m^2_{\nu^i_{{c}}}) \\ 
\nonumber \\  
  \SLLb{(c^l_{8'})}= & \, \frac{1}{2}\VcevrL_{d, \alpha, a}\VecvrL_{\delta, d, c}\VcevrL_{b, \gamma, a}\VecvrL_{\beta, b, c}   m_{\tilde{\chi}^-_{{b}}}  m_{\tilde{\chi}^-_{{d}}} D_0(m^2_{\tilde{\chi}^-_{{d}}}, m^2_{\tilde{\chi}^-_{{b}}}, m^2_{\nu^R_{{a}}}, m^2_{\nu^R_{{c}}}) \\ 
  \SLRb{(c^l_{8'})}= & -2 \VcevrL_{d, \alpha, a}\VecvrR_{\delta, d, c}\VcevrR_{b, \gamma, a}\VecvrL_{\beta, b, c} D_{27}(m^2_{\tilde{\chi}^-_{{d}}}, m^2_{\tilde{\chi}^-_{{b}}}, m^2_{\nu^R_{{a}}}, m^2_{\nu^R_{{c}}}) \\ 
  \VLLb{(c^l_{8'})}= & \, \VcevrL_{d, \alpha, a}\VecvrR_{\delta, d, c}\VcevrL_{b, \gamma, a}\VecvrR_{\beta, b, c}    D_{27}(m^2_{\tilde{\chi}^-_{{d}}}, m^2_{\tilde{\chi}^-_{{b}}}, m^2_{\nu^R_{{a}}}, m^2_{\nu^R_{{c}}}) \\ 
  \VLRb{(c^l_{8'})}= & \, \frac{1}{2}\VcevrL_{d, \alpha, a}\VecvrL_{\delta, d, c}\VcevrR_{b, \gamma, a}\VecvrR_{\beta, b, c}    m_{\tilde{\chi}^-_{{b}}}  m_{\tilde{\chi}^-_{{d}}} D_0(m^2_{\tilde{\chi}^-_{{d}}}, m^2_{\tilde{\chi}^-_{{b}}}, m^2_{\nu^R_{{a}}}, m^2_{\nu^R_{{c}}}) \\ 
  \TLLb{(c^l_{8'})}= & -\frac{1}{8}\VcevrL_{d, \alpha, a}\VecvrL_{\delta, d, c}\VcevrL_{b, \gamma, a}\VecvrL_{\beta, b, c}  m_{\tilde{\chi}^-_{{b}}}  m_{\tilde{\chi}^-_{{d}}} D_0(m^2_{\tilde{\chi}^-_{{d}}}, m^2_{\tilde{\chi}^-_{{b}}}, m^2_{\nu^R_{{a}}}, m^2_{\nu^R_{{c}}}) 
\end{align} 

\subsubsection{Crossed $W^+$ and $H^+$ contributions}
\begin{align} 
 \SLLb{(w^l_{1'})}= & \, \frac{1}{2}\VvehpL_{d, \alpha, a} \VevhmL_{\beta, b, a}\VvehpL_{b, \gamma, c} \VevhmL_{\delta, d, c}  m_{\nu_{{b}}}  m_{\nu_{{d}}} D_0(m^2_{\nu_{{d}}}, m^2_{\nu_{{b}}}, m^2_{H^-_{{a}}}, m^2_{H^-_{{c}}}) \\ 
  \SLRb{(w^l_{1'})}= & \, 2 \, \VvehpL_{d, \alpha, a} \VevhmL_{\beta, b, a}\VvehpR_{b, \gamma, c} \VevhmR_{\delta, d, c} D_{27}(m^2_{\nu_{{d}}}, m^2_{\nu_{{b}}}, m^2_{H^-_{{a}}}, m^2_{H^-_{{c}}}) \\ 
  \VLLb{(w^l_{1'})}= & -\VvehpL_{d, \alpha, a} \VevhmR_{\beta, b, a}\VvehpL_{b, \gamma, c} \VevhmR_{\delta, d, c} D_{27}(m^2_{\nu_{{d}}}, m^2_{\nu_{{b}}}, m^2_{H^-_{{a}}}, m^2_{H^-_{{c}}}) \\ 
  \VLRb{(w^l_{1'})}= & \, \frac{1}{2}\VvehpL_{d, \alpha, a} \VevhmR_{\beta, b, a}\VvehpR_{b, \gamma, c} \VevhmL_{\delta, d, c}  m_{\nu_{{b}}}  m_{\nu_{{d}}} D_0(m^2_{\nu_{{d}}}, m^2_{\nu_{{b}}}, m^2_{H^-_{{a}}}, m^2_{H^-_{{c}}}) \\ 
  \TLLb{(w^l_{1'})}= & -\frac{1}{8}\VvehpL_{d, \alpha, a} \VevhmL_{\beta, b, a}\VvehpL_{b, \gamma, c} \VevhmL_{\delta, d, c} m_{\nu_{{b}}}  m_{\nu_{{d}}} D_0(m^2_{\nu_{{d}}}, m^2_{\nu_{{b}}}, m^2_{H^-_{{a}}}, m^2_{H^-_{{c}}}) \\ 
\nonumber \\
I_1 = & I_{C_0 D_0}(m^2_{\nu_{{d}}}, m^2_{\nu_{{b}}}, m^2_{W^-}, m^2_{H^-_{{a}}}) \\ 
I_2 = & D_{27}(m^2_{\nu_{{d}}}, m^2_{\nu_{{b}}}, m^2_{H^-_{{a}}}, m^2_{W^-}) \\ 
I_3 = & m_{\nu_{{b}}}  m_{\nu_{{d}}} D_0(m^2_{\nu_{{d}}}, m^2_{\nu_{{b}}}, m^2_{H^-_{{a}}}, m^2_{W^-}) \\ 
  \SLLb{(w^l_{2'})}= & \, \frac{1}{4} (-3\VvehpR_{d, \alpha, a} \VevhmR_{\beta, b, a} \VvewpR_{b, \gamma} \VevwmL_{\delta, d} I_2 -\VvehpL_{d, \alpha, a} \VevhmL_{\beta, b, a} \VvewpL_{b, \gamma} \VevwmR_{\delta, d} (4 I_1 + 13 I_2)) \\ 
  \SLRb{(w^l_{2'})}= & \, \frac{3}{4} (-(\VvehpR_{d, \alpha, a} \VevhmR_{\beta, b, a} \VvewpR_{b, \gamma} \VevwmL_{\delta, d}) +\VvehpL_{d, \alpha, a} \VevhmL_{\beta, b, a} \VvewpL_{b, \gamma} \VevwmR_{\delta, d}) I_2 - 2\VvehpL_{d, \alpha, a} \VevhmL_{\beta, b, a} \VvewpR_{b, \gamma} \VevwmL_{\delta, d} I_3 \\ 
  \VLLb{(w^l_{2'})}= & \, \VvehpL_{d, \alpha, a} \VevhmR_{\beta, b, a} \VvewpL_{b, \gamma} \VevwmL_{\delta, d} I_3 \\ 
  \VLRb{(w^l_{2'})}= & -\VvehpL_{d, \alpha, a} \VevhmR_{\beta, b, a} \VvewpR_{b, \gamma} \VevwmR_{\delta, d} (I_1 - 2 I_2) \\ 
  \TLLb{(w^l_{2'})}= &  \, \frac{1}{16} (\VvehpL_{d, \alpha, a} \VevhmL_{\beta, b, a} \VvewpL_{b, \gamma} \VevwmR_{\delta, d} (4 I_1 - 21 I_2) + 5\VvehpR_{d, \alpha, a} \VevhmR_{\beta, b, a} \VvewpR_{b, \gamma} \VevwmL_{\delta, d} I_2) \\ 
\nonumber\\
I_1 = & I_{C_0 D_0}(m^2_{\nu_{{d}}}, m^2_{\nu_{{b}}}, m^2_{H^-_{{c}}}, m^2_{W^-}) \\ 
I_2 = & D_{27}(m^2_{\nu_{{d}}}, m^2_{\nu_{{b}}}, m^2_{W^-}, m^2_{H^-_{{c}}}) \\ 
I_3 = & m_{\nu_{{b}}}  m_{\nu_{{d}}} D_0(m^2_{\nu_{{d}}}, m^2_{\nu_{{b}}}, m^2_{W^-}, m^2_{H^-_{{c}}}) \\ 
  \SLLb{(w^l_{3'})}= & \, \frac{1}{4} (-3 \VvewpR_{d, i} \VevwmL_{\beta, b}\VvehpR_{b, \gamma, c} \VevhmR_{\delta, d, c} I_2 - \VvewpL_{d, i} \VevwmR_{\beta, b}\VvehpL_{b, \gamma, c} \VevhmL_{\delta, d, c} (4 I_1 + 13 I_2)) \\ 
  \SLRb{(w^l_{3'})}= & \, \frac{3}{4} (\VvewpL_{d, i} \VevwmR_{\beta, b}\VvehpL_{b, \gamma, c} \VevhmL_{\delta, d, c} - \VvewpR_{d, i} \VevwmL_{\beta, b}\VvehpR_{b, \gamma, c} \VevhmR_{\delta, d, c}) I_2 - 2 \VvewpL_{d, i} \VevwmR_{\beta, b}\VvehpR_{b, \gamma, c} \VevhmR_{\delta, d, c} I_3 \\ 
  \VLLb{(w^l_{3'})}= & \, \VvewpL_{d, i} \VevwmL_{\beta, b}\VvehpL_{b, \gamma, c} \VevhmR_{\delta, d, c}     I_3 \\ 
  \VLRb{(w^l_{3'})}= & - \VvewpL_{d, i} \VevwmL_{\beta, b}\VvehpR_{b, \gamma, c} \VevhmL_{\delta, d, c}  (I_1 - 2 I_2) \\ 
  \TLLb{(w^l_{3'})}= & \, \frac{1}{16} (\VvewpL_{d, i} \VevwmR_{\beta, b}\VvehpL_{b, \gamma, c} \VevhmL_{\delta, d, c}  (4 I_1 - 21 I_2) + 5 \VvewpR_{d, i} \VevwmL_{\beta, b} \VvehpR_{b, \gamma, c} \VevhmR_{\delta, d, c} I_2) \\ 
\nonumber\\
  \SLLb{(w^l_{4'})}= & \, 8 \, \VvewpL_{d, i} \VevwmR_{\beta, b} \VvewpL_{b, \gamma} \VevwmR_{\delta, d}  m_{\nu_{{b}}}  m_{\nu_{{d}}} D_0(m^2_{\nu_{{d}}}, m^2_{\nu_{{b}}}, m^2_{W^-}, m^2_{W^-}) \\ 
  \SLRb{(w^l_{4'})}= & \, 8 \, \VvewpL_{d, i} \VevwmR_{\beta, b} \VvewpR_{b, \gamma} \VevwmL_{\delta, d}  I_{C_0 D_0}(m^2_{\nu_{{d}}}, m^2_{\nu_{{b}}}, m^2_{W^-}, m^2_{W^-}) \\ 
  \VLLb{(w^l_{4'})}= &-4  \VvewpL_{d, i} \VevwmL_{\beta, b} \VvewpL_{b, \gamma} \VevwmL_{\delta, d} (I_{C_0 D_0}(m^2_{\nu_{{d}}}, m^2_{\nu_{{b}}}, m^2_{W^-}, m^2_{W^-}) - 3 D_{27}(m^2_{\nu_{{d}}}, m^2_{\nu_{{b}}}, m^2_{W^-}, m^2_{W^-})) \\ 
  \VLRb{(w^l_{4'})}= & \, 2 \, \VvewpL_{d, i} \VevwmL_{\beta, b} \VvewpR_{b, \gamma} \VevwmR_{\delta, d}  m_{\nu_{{b}}}  m_{\nu_{{d}}} D_0(m^2_{\nu_{{d}}}, m^2_{\nu_{{b}}}, m^2_{W^-}, m^2_{W^-}) \\ 
\nonumber \\
  \SLLb{(w^l_{5'})}= & \, \frac{1}{2} \VvehpL_{d, \alpha, a} \VevhmL_{\beta, b, a} \VevhmL_{\delta, b, c} \VvehpL_{d, \gamma, c}  m_{\nu_{{b}}}  m_{\nu_{{d}}} D_0(m^2_{\nu_{{b}}}, m^2_{\nu_{{d}}}, m^2_{H^-_{{a}}}, m^2_{H^-_{{c}}}) \\ 
  \SLRb{(w^l_{5'})}= & \, 2 \, \VvehpL_{d, \alpha, a} \VevhmL_{\beta, b, a} \VevhmR_{\delta, b, c} \VvehpR_{d, \gamma, c} D_{27}(m^2_{\nu_{{b}}}, m^2_{\nu_{{d}}}, m^2_{H^-_{{a}}}, m^2_{H^-_{{c}}}) \\ 
  \VLLb{(w^l_{5'})}= &- \frac{1}{2} \VvehpL_{d, \alpha, a} \VevhmR_{\beta, b, a} \VevhmR_{\delta, b, c} \VvehpL_{d, \gamma, c} m_{\nu_{{b}}}  m_{\nu_{{d}}} D_0(m^2_{\nu_{{b}}}, m^2_{\nu_{{d}}}, m^2_{H^-_{{a}}}, m^2_{H^-_{{c}}}) \\ 
  \VLRb{(w^l_{5'})}= & \, \VvehpL_{d, \alpha, a} \VevhmR_{\beta, b, a} \VevhmL_{\delta, b, c} \VvehpR_{d, \gamma, c}   D_{27}(m^2_{\nu_{{b}}}, m^2_{\nu_{{d}}}, m^2_{H^-_{{a}}}, m^2_{H^-_{{c}}}) \\ 
  \TLLb{(w^l_{5'})}= & \, \frac{1}{8} \VvehpL_{d, \alpha, a} \VevhmL_{\beta, b, a} \VevhmL_{\delta, b, c} \VvehpL_{d, \gamma, c}  m_{\nu_{{b}}}  m_{\nu_{{d}}} D_0(m^2_{\nu_{{b}}}, m^2_{\nu_{{d}}}, m^2_{H^-_{{a}}}, m^2_{H^-_{{c}}}) \\ 
\nonumber\\
  \SLLb{(w^l_{6'})}= & -8 \VvehpL_{d, \alpha, a} \VevhmL_{\beta, b, a} \VevwmR_{\delta, b} \VvewpL_{d, \gamma} D_{27}(m^2_{\nu_{{b}}}, m^2_{\nu_{{d}}}, m^2_{H^-_{{a}}}, m^2_{W^-}) \\ 
  \SLRb{(w^l_{6'})}= & -2 \VvehpL_{d, \alpha, a} \VevhmL_{\beta, b, a} \VevwmL_{\delta, b} \VvewpR_{d, \gamma} m_{\nu_{{b}}}  m_{\nu_{{d}}} D_0(m^2_{\nu_{{b}}}, m^2_{\nu_{{d}}}, m^2_{H^-_{{a}}}, m^2_{W^-}) \\ 
  \VLLb{(w^l_{6'})}= & \, 2 \, \VvehpL_{d, \alpha, a} \VevhmR_{\beta, b, a} \VevwmL_{\delta, b} \VvewpL_{d, \gamma}  D_{27}(m^2_{\nu_{{b}}}, m^2_{\nu_{{d}}}, m^2_{H^-_{{a}}}, m^2_{W^-}) \\ 
  \VLRb{(w^l_{6'})}= & -\VvehpL_{d, \alpha, a} \VevhmR_{\beta, b, a} \VevwmR_{\delta, b} \VvewpR_{d, \gamma}  m_{\nu_{{b}}}  m_{\nu_{{d}}} D_0(m^2_{\nu_{{b}}}, m^2_{\nu_{{d}}}, m^2_{H^-_{{a}}}, m^2_{W^-}) \\ 
\nonumber\\
  \SLLb{(w^l_{7'})}= &  -8  \VvewpL_{d, i} \VevwmR_{\beta, b} \VevhmL_{\delta, b, c}\VvehpL_{d, \gamma, c} D_{27}(m^2_{\nu_{{b}}}, m^2_{\nu_{{d}}}, m^2_{W^-}, m^2_{H^-_{{c}}}) \\ 
  \SLRb{(w^l_{7'})}= &  -2  \VvewpL_{d, i} \VevwmR_{\beta, b} \VevhmR_{\delta, b, c}\VvehpR_{d, \gamma, c} m_{\nu_{{b}}}  m_{\nu_{{d}}} D_0(m^2_{\nu_{{b}}}, m^2_{\nu_{{d}}}, m^2_{W^-}, m^2_{H^-_{{c}}}) \\ 
  \VLLb{(w^l_{7'})}= &  \, 2 \, \VvewpL_{d, i} \VevwmL_{\beta, b} \VevhmR_{\delta, b, c}\VvehpL_{d, \gamma, c}  D_{27}(m^2_{\nu_{{b}}}, m^2_{\nu_{{d}}}, m^2_{W^-}, m^2_{H^-_{{c}}}) \\ 
  \VLRb{(w^l_{7'})}= &  - \VvewpL_{d, i} \VevwmL_{\beta, b} \VevhmL_{\delta, b, c}\VvehpR_{d, \gamma, c}  m_{\nu_{{b}}}  m_{\nu_{{d}}} D_0(m^2_{\nu_{{b}}}, m^2_{\nu_{{d}}}, m^2_{W^-}, m^2_{H^-_{{c}}}) \\ 
\nonumber\\
  \SLLb{(w^l_{8'})}= & \, 8 \, \VvewpL_{d, i} \VevwmR_{\beta, b} \VevwmR_{\delta, b} \VvewpL_{d, \gamma}  m_{\nu_{{b}}}  m_{\nu_{{d}}} D_0(m^2_{\nu_{{b}}}, m^2_{\nu_{{d}}}, m^2_{W^-}, m^2_{W^-}) \\ 
  \SLRb{(w^l_{8'})}= & \, 32 \, \VvewpL_{d, i} \VevwmR_{\beta, b} \VevwmL_{\delta, b} \VvewpR_{d, \gamma} D_{27}(m^2_{\nu_{{b}}}, m^2_{\nu_{{d}}}, m^2_{W^-}, m^2_{W^-}) \\ 
  \VLLb{(w^l_{8'})}= & -2  \VvewpL_{d, i} \VevwmL_{\beta, b} \VevwmL_{\delta, b} \VvewpL_{d, \gamma} m_{\nu_{{b}}}  m_{\nu_{{d}}} D_0(m^2_{\nu_{{b}}}, m^2_{\nu_{{d}}}, m^2_{W^-}, m^2_{W^-}) \\ 
  \VLRb{(w^l_{8'})}= & \, 4 \, \VvewpL_{d, i} \VevwmL_{\beta, b} \VevwmR_{\delta, b} \VvewpR_{d, \gamma}  D_{27}(m^2_{\nu_{{b}}}, m^2_{\nu_{{d}}}, m^2_{W^-}, m^2_{W^-})  
\end{align} 

\subsection{Two-Lepton -- Two-Quark boxes}
\subsubsection{Feynman diagrams}
\subsection*{Neutralino diagrams}
\parbox{0.45\linewidth}{
($n^d_1$)
\begin{fmffile}{Diagrams/Box2L2dNumberOfConsideredExternalStatesBox5} 
\fmfframe(20,20)(20,20){ 
\begin{fmfgraph*}(120,80) 
\fmftop{t1,t2}
\fmfbottom{b1,b2}
\fmf{fermion}{t1,v1}
\fmf{dashes,label=$\tilde{e}_{{a}}$,tension=0.5}{v1,v2}
\fmf{fermion}{v2,t2}
\fmf{fermion}{v3,b1}
\fmf{dashes,label=$\tilde{d}_{{c}}$,tension=0.5}{v3,v4}
\fmf{fermion}{b2,v4}
\fmf{plain,label=$\tilde{\chi}^0_{{d}}$,tension=0.1}{v1,v3}
\fmf{plain,label=$\tilde{\chi}^0_{{b}}$,tension=0.1}{v2,v4}
\fmflabel{$\ell_{{\alpha}}$}{t1}
\fmflabel{$\bar{\ell}_{{\beta}}$}{t2}
\fmflabel{$d_{{\gamma}}$}{b2}
\fmflabel{$\bar{d}_{{\delta}}$}{b1}
\end{fmfgraph*}}
\end{fmffile}}
\hspace{1cm}
\parbox{0.45\linewidth}{
($n^d_2$)
\begin{fmffile}{Diagrams/Box2L2dNumberOfConsideredExternalStatesBox28}
\fmfframe(20,20)(20,20){
\begin{fmfgraph*}(120,80)
\fmftop{t1,t2}
\fmfbottom{b1,b2}
\fmf{fermion}{t1,v1}
\fmf{dashes,label=$\tilde{e}_{{a}}$,tension=0.5}{v1,v2}
\fmf{fermion}{v2,t2}
\fmf{phantom}{b1,v3}
\fmf{dashes,label=$\tilde{d}^*_{{c}}$,tension=0.5}{v3,v4}
\fmf{phantom}{v4,b2}
\fmf{plain,label=$\tilde{\chi}^0_{{d}}$,tension=0.1}{v1,v3}
\fmf{plain,label=$\tilde{\chi}^0_{{b}}$,tension=0.1}{v2,v4}
\fmffreeze 
\fmf{fermion}{v4,b1}
\fmf{fermion}{b2,v3}
\fmflabel{$\ell_{{\alpha}}$}{t1}
\fmflabel{$\bar{\ell}_{{\beta}}$}{t2}
\fmflabel{$\bar{d}_{{\delta}}$}{b1}
\fmflabel{$d_{{\gamma}}$}{b2}
\end{fmfgraph*}}
\end{fmffile}} \\
\parbox{0.45\linewidth}{
($n^u_1$)
\begin{fmffile}{Diagrams/Box2L2uN2}
\fmfframe(20,20)(20,20){
\begin{fmfgraph*}(120,80)
\fmftop{t1,t2}
\fmfbottom{b1,b2}
\fmf{fermion}{t1,v1}
\fmf{dashes,label=$\tilde{e}_{{a}}$,tension=0.5}{v1,v2}
\fmf{fermion}{v2,t2}
\fmf{phantom}{b1,v3}
\fmf{dashes,label=$\tilde{u}^*_{{c}}$,tension=0.5}{v3,v4}
\fmf{phantom}{v4,b2}
\fmf{plain,label=$\tilde{\chi}^0_{{d}}$,tension=0.1}{v1,v3}
\fmf{plain,label=$\tilde{\chi}^0_{{b}}$,tension=0.1}{v2,v4}
\fmffreeze 
\fmf{fermion}{v4,b1}
\fmf{fermion}{b2,v3}
\fmflabel{$\ell_{{\alpha}}$}{t1}
\fmflabel{$\bar{\ell}_{{\beta}}$}{t2}
\fmflabel{$\bar{u}_{{\delta}}$}{b1}
\fmflabel{$u_{{\gamma}}$}{b2}
\end{fmfgraph*}}
\end{fmffile}}
\hspace{1cm}
\parbox{0.45\linewidth}{
($n^u_2$)
\begin{fmffile}{Diagrams/Box2L2uN1} 
\fmfframe(20,20)(20,20){ 
\begin{fmfgraph*}(120,80) 
\fmftop{t1,t2}
\fmfbottom{b1,b2}
\fmf{fermion}{t1,v1}
\fmf{dashes,label=$\tilde{e}_{{a}}$,tension=0.5}{v1,v2}
\fmf{fermion}{v2,t2}
\fmf{fermion}{v3,b1}
\fmf{dashes,label=$\tilde{u}_{{c}}$,tension=0.5}{v3,v4}
\fmf{fermion}{b2,v4}
\fmf{plain,label=$\tilde{\chi}^0_{{d}}$,tension=0.1}{v1,v3}
\fmf{plain,label=$\tilde{\chi}^0_{{b}}$,tension=0.1}{v2,v4}
\fmflabel{$\ell_{{\alpha}}$}{t1}
\fmflabel{$\bar{\ell}_{{\beta}}$}{t2}
\fmflabel{$u_{{\gamma}}$}{b2}
\fmflabel{$\bar{u}_{{\delta}}$}{b1}
\end{fmfgraph*}}
\end{fmffile}}

\subsection*{Chargino diagrams}
\parbox{0.45\linewidth}{
($c^d_1$)
\begin{fmffile}{Diagrams/Box2L2dNumberOfConsideredExternalStatesBox18} 
\fmfframe(20,20)(20,20){ 
\begin{fmfgraph*}(120,80) 
\fmftop{t1,t2}
\fmfbottom{b1,b2}
\fmf{fermion}{t1,v1}
\fmf{dashes,label=$\nu^i_{{a}}$,tension=0.5}{v1,v2}
\fmf{fermion}{v2,t2}
\fmf{fermion}{v3,b1}
\fmf{dashes,label=$\tilde{u}_{{c}}$,tension=0.5}{v3,v4}
\fmf{fermion}{b2,v4}
\fmf{plain,label=$\tilde{\chi}^+_{{d}}$,tension=0.1}{v1,v3}
\fmf{plain,label=$\tilde{\chi}^+_{{b}}$,tension=0.1}{v2,v4}
\fmflabel{$\ell_{{\alpha}}$}{t1}
\fmflabel{$\bar{\ell}_{{\beta}}$}{t2}
\fmflabel{$d_{{\gamma}}$}{b2}
\fmflabel{$\bar{d}_{{\delta}}$}{b1}
\end{fmfgraph*}}
\end{fmffile}}
\hspace{1cm}
\parbox{0.45\linewidth}{
($c^d_2$)
\begin{fmffile}{Diagrams/Box2L2dNumberOfConsideredExternalStatesBox19} 
\fmfframe(20,20)(20,20){ 
\begin{fmfgraph*}(120,80) 
\fmftop{t1,t2}
\fmfbottom{b1,b2}
\fmf{fermion}{t1,v1}
\fmf{dashes,label=$\nu^R_{{a}}$,tension=0.5}{v1,v2}
\fmf{fermion}{v2,t2}
\fmf{fermion}{v3,b1}
\fmf{dashes,label=$\tilde{u}_{{c}}$,tension=0.5}{v3,v4}
\fmf{fermion}{b2,v4}
\fmf{plain,label=$\tilde{\chi}^+_{{d}}$,tension=0.1}{v1,v3}
\fmf{plain,label=$\tilde{\chi}^+_{{b}}$,tension=0.1}{v2,v4}
\fmflabel{$\ell_{{\alpha}}$}{t1}
\fmflabel{$\bar{\ell}_{{\beta}}$}{t2}
\fmflabel{$d_{{\gamma}}$}{b2}
\fmflabel{$\bar{d}_{{\delta}}$}{b1}
\end{fmfgraph*}}
\end{fmffile}} \\
\parbox{0.45\linewidth}{
($c^u_1$)
\begin{fmffile}{Diagrams/Box2L2uC1} 
\fmfframe(20,20)(20,20){ 
\begin{fmfgraph*}(120,80) 
\fmftop{t1,t2}
\fmfbottom{b1,b2}
\fmf{fermion}{t1,v1}
\fmf{dashes,label=$\nu^i_{{a}}$,tension=0.5}{v1,v2}
\fmf{fermion}{v2,t2}
\fmf{phantom}{v3,b1}
\fmf{dashes,label=$\tilde{d}_{{c}}$,tension=0.5}{v3,v4}
\fmf{phantom}{b2,v4}
\fmf{plain,label=$\tilde{\chi}^+_{{d}}$,tension=0.1}{v1,v3}
\fmf{plain,label=$\tilde{\chi}^+_{{b}}$,tension=0.1}{v2,v4}
\fmffreeze 
\fmf{fermion}{b1,v4}
\fmf{fermion}{b2,v3}
\fmflabel{$\ell_{{\alpha}}$}{t1}
\fmflabel{$\bar{\ell}_{{\beta}}$}{t2}
\fmflabel{$u_{{\gamma}}$}{b2}
\fmflabel{$\bar{u}_{{\delta}}$}{b1}
\end{fmfgraph*}}
\end{fmffile}}
\hspace{1cm}
\parbox{0.45\linewidth}{
($c^u_2$)
\begin{fmffile}{Diagrams/Box2L2uC2} 
\fmfframe(20,20)(20,20){ 
\begin{fmfgraph*}(120,80) 
\fmftop{t1,t2}
\fmfbottom{b1,b2}
\fmf{fermion}{t1,v1}
\fmf{dashes,label=$\nu^R_{{a}}$,tension=0.5}{v1,v2}
\fmf{fermion}{v2,t2}
\fmf{phantom}{b1,v3}
\fmf{dashes,label=$\tilde{d}_{{c}}$,tension=0.5}{v3,v4}
\fmf{phantom}{v4,b2}
\fmf{plain,label=$\tilde{\chi}^+_{{d}}$,tension=0.1}{v1,v3}
\fmf{plain,label=$\tilde{\chi}^+_{{b}}$,tension=0.1}{v2,v4}
\fmffreeze 
\fmf{fermion}{b1,v4}
\fmf{fermion}{b2,v3}
\fmflabel{$\ell_{{\alpha}}$}{t1}
\fmflabel{$\bar{\ell}_{{\beta}}$}{t2}
\fmflabel{$u_{{\gamma}}$}{b2}
\fmflabel{$\bar{u}_{{\delta}}$}{b1}
\end{fmfgraph*}}
\end{fmffile}}

\subsection*{$W^+$ and $H^+$ diagrams}
\parbox{0.45\linewidth}{
($w^d_1$)
\begin{fmffile}{Diagrams/Box2L2dNumberOfConsideredExternalStatesBox20} 
\fmfframe(20,20)(20,20){ 
\begin{fmfgraph*}(120,80) 
\fmftop{t1,t2}
\fmfbottom{b1,b2}
\fmf{fermion}{t1,v1}
\fmf{plain,label=$\nu_{{a}}$,tension=0.5}{v1,v2}
\fmf{fermion}{v2,t2}
\fmf{fermion}{v3,b1}
\fmf{plain,label=$u_{{c}}$,tension=0.5}{v3,v4}
\fmf{fermion}{b2,v4}
\fmf{dashes,label=$H^+_{{d}}$,tension=0.1}{v1,v3}
\fmf{dashes,label=$H^+_{{b}}$,tension=0.1}{v2,v4}
\fmflabel{$\ell_{{\alpha}}$}{t1}
\fmflabel{$\bar{\ell}_{{\beta}}$}{t2}
\fmflabel{$d_{{\gamma}}$}{b2}
\fmflabel{$\bar{d}_{{\delta}}$}{b1}
\end{fmfgraph*}}
\end{fmffile}}
\hspace{1cm}
\parbox{0.45\linewidth}{
($w^d_2$)
\begin{fmffile}{Diagrams/Box2L2dNumberOfConsideredExternalStatesBox21} 
\fmfframe(20,20)(20,20){ 
\begin{fmfgraph*}(120,80) 
\fmftop{t1,t2}
\fmfbottom{b1,b2}
\fmf{fermion}{t1,v1}
\fmf{plain,label=$\nu_{{a}}$,tension=0.5}{v1,v2}
\fmf{fermion}{v2,t2}
\fmf{fermion}{v3,b1}
\fmf{plain,label=$u_{{c}}$,tension=0.5}{v3,v4}
\fmf{fermion}{b2,v4}
\fmf{dashes,label=$H^+_{{d}}$,tension=0.1}{v1,v3}
\fmf{wiggly,label=$W^+$,tension=0.1}{v2,v4}
\fmflabel{$\ell_{{\alpha}}$}{t1}
\fmflabel{$\bar{\ell}_{{\beta}}$}{t2}
\fmflabel{$d_{{\gamma}}$}{b2}
\fmflabel{$\bar{d}_{{\delta}}$}{b1}
\end{fmfgraph*}}
\end{fmffile}}
\\
\parbox{0.45\linewidth}{
($w^d_3$)
\begin{fmffile}{Diagrams/Box2L2dNumberOfConsideredExternalStatesBox22} 
\fmfframe(20,20)(20,20){ 
\begin{fmfgraph*}(120,80) 
\fmftop{t1,t2}
\fmfbottom{b1,b2}
\fmf{fermion}{t1,v1}
\fmf{plain,label=$\nu_{{a}}$,tension=0.5}{v1,v2}
\fmf{fermion}{v2,t2}
\fmf{fermion}{v3,b1}
\fmf{plain,label=$u_{{c}}$,tension=0.5}{v3,v4}
\fmf{fermion}{b2,v4}
\fmf{wiggly,label=$W^+$,tension=0.1}{v1,v3}
\fmf{dashes,label=$H^+_{{b}}$,tension=0.1}{v2,v4}
\fmflabel{$\ell_{{\alpha}}$}{t1}
\fmflabel{$\bar{\ell}_{{\beta}}$}{t2}
\fmflabel{$d_{{\gamma}}$}{b2}
\fmflabel{$\bar{d}_{{\delta}}$}{b1}
\end{fmfgraph*}}
\end{fmffile}
}
\hspace{1cm}
\parbox{0.45\linewidth}{
($w^d_4$)
\begin{fmffile}{Diagrams/Box2L2dNumberOfConsideredExternalStatesBox23} 
\fmfframe(20,20)(20,20){ 
\begin{fmfgraph*}(120,80) 
\fmftop{t1,t2}
\fmfbottom{b1,b2}
\fmf{fermion}{t1,v1}
\fmf{plain,label=$\nu_{{a}}$,tension=0.5}{v1,v2}
\fmf{fermion}{v2,t2}
\fmf{fermion}{v3,b1}
\fmf{plain,label=$u_{{c}}$,tension=0.5}{v3,v4}
\fmf{fermion}{b2,v4}
\fmf{wiggly,label=$W^+$,tension=0.1}{v1,v3}
\fmf{wiggly,label=$W^+$,tension=0.1}{v2,v4}
\fmflabel{$\ell_{{\alpha}}$}{t1}
\fmflabel{$\bar{\ell}_{{\beta}}$}{t2}
\fmflabel{$d_{{\gamma}}$}{b2}
\fmflabel{$\bar{d}_{{\delta}}$}{b1}
\end{fmfgraph*}}
\end{fmffile}
}\\
\parbox{0.45\linewidth}{
($w^u_1$)
\begin{fmffile}{Diagrams/Box2L2uW1} 
\fmfframe(20,20)(20,20){ 
\begin{fmfgraph*}(120,80) 
\fmftop{t1,t2}
\fmfbottom{b1,b2}
\fmf{fermion}{t1,v1}
\fmf{plain,label=$\nu_{{a}}$,tension=0.5}{v1,v2}
\fmf{fermion}{v2,t2}
\fmf{phantom}{v3,b1}
\fmf{plain,label=$d_{{c}}$,tension=0.5}{v3,v4}
\fmf{phantom}{b2,v4}
\fmf{dashes,label=$H^+_{{d}}$,tension=0.1}{v1,v3}
\fmf{dashes,label=$H^+_{{b}}$,tension=0.1}{v2,v4}
\fmffreeze 
\fmf{fermion}{b1,v4}
\fmf{fermion}{b2,v3}
\fmflabel{$\ell_{{\alpha}}$}{t1}
\fmflabel{$\bar{\ell}_{{\beta}}$}{t2}
\fmflabel{$u_{{\gamma}}$}{b2}
\fmflabel{$\bar{u}_{{\delta}}$}{b1}
\end{fmfgraph*}}
\end{fmffile}}
\hspace{1cm}
\parbox{0.45\linewidth}{
($w^u_2$)
\begin{fmffile}{Diagrams/Box2L2uW2} 
\fmfframe(20,20)(20,20){ 
\begin{fmfgraph*}(120,80) 
\fmftop{t1,t2}
\fmfbottom{b1,b2}
\fmf{fermion}{t1,v1}
\fmf{plain,label=$\nu_{{a}}$,tension=0.5}{v1,v2}
\fmf{fermion}{v2,t2}
\fmf{phantom}{v3,b1}
\fmf{plain,label=$d_{{c}}$,tension=0.5}{v3,v4}
\fmf{phantom}{b2,v4}
\fmf{dashes,label=$H^+_{{d}}$,tension=0.1}{v1,v3}
\fmf{wiggly,label=$W^+$,tension=0.1}{v2,v4}
\fmffreeze 
\fmf{fermion}{b1,v4}
\fmf{fermion}{b2,v3}
\fmflabel{$\ell_{{\alpha}}$}{t1}
\fmflabel{$\bar{\ell}_{{\beta}}$}{t2}
\fmflabel{$u_{{\gamma}}$}{b2}
\fmflabel{$\bar{u}_{{\delta}}$}{b1}
\end{fmfgraph*}}
\end{fmffile}}
\\
\parbox{0.45\linewidth}{
($w^u_3$)
\begin{fmffile}{Diagrams/Box2L2uW3} 
\fmfframe(20,20)(20,20){ 
\begin{fmfgraph*}(120,80) 
\fmftop{t1,t2}
\fmfbottom{b1,b2}
\fmf{fermion}{t1,v1}
\fmf{plain,label=$\nu_{{a}}$,tension=0.5}{v1,v2}
\fmf{fermion}{v2,t2}
\fmf{phantom}{v3,b1}
\fmf{plain,label=$d_{{c}}$,tension=0.5}{v3,v4}
\fmf{phantom}{b2,v4}
\fmf{wiggly,label=$W^+$,tension=0.1}{v1,v3}
\fmf{dashes,label=$H^+_{{b}}$,tension=0.1}{v2,v4}
\fmffreeze 
\fmf{fermion}{b1,v4}
\fmf{fermion}{b2,v3}
\fmflabel{$\ell_{{\alpha}}$}{t1}
\fmflabel{$\bar{\ell}_{{\beta}}$}{t2}
\fmflabel{$u_{{\gamma}}$}{b2}
\fmflabel{$\bar{u}_{{\delta}}$}{b1}
\end{fmfgraph*}}
\end{fmffile}
}
\hspace{1cm}
\parbox{0.45\linewidth}{
($w^u_4$)
\begin{fmffile}{Diagrams/Box2L2uW4} 
\fmfframe(20,20)(20,20){ 
\begin{fmfgraph*}(120,80) 
\fmftop{t1,t2}
\fmfbottom{b1,b2}
\fmf{fermion}{t1,v1}
\fmf{plain,label=$\nu_{{a}}$,tension=0.5}{v1,v2}
\fmf{fermion}{v2,t2}
\fmf{phantom}{v3,b1}
\fmf{plain,label=$d_{{c}}$,tension=0.5}{v3,v4}
\fmf{phantom}{b2,v4}
\fmf{wiggly,label=$W^+$,tension=0.1}{v1,v3}
\fmf{wiggly,label=$W^+$,tension=0.1}{v2,v4}
\fmffreeze 
\fmf{fermion}{b1,v4}
\fmf{fermion}{b2,v3}
\fmflabel{$\ell_{{\alpha}}$}{t1}
\fmflabel{$\bar{\ell}_{{\beta}}$}{t2}
\fmflabel{$u_{{\gamma}}$}{b2}
\fmflabel{$\bar{u}_{{\delta}}$}{b1}
\end{fmfgraph*}}
\end{fmffile}
}

\subsubsection{Down quarks}
\paragraph{Neutralino}
\begin{align} 
  \SLLb{(n^d_1)}= & \, \frac{1}{2} \VneeL_{d, \alpha, a} \VeneL_{\beta, b, a} \VnddL_{b, \gamma, c} \VdndL_{\delta, d, c} m_{\tilde{\chi}^0_{{b}}} m_{\tilde{\chi}^0_{{d}}} D_0(m^2_{\tilde{\chi}^0_{{d}}}, m^2_{\tilde{\chi}^0_{{b}}}, m^2_{\tilde{e}_{{a}}}, m^2_{\tilde{d}_{{c}}}) \\ 
  \SLRb{(n^d_1)}= & \, 2 \, \VneeL_{d, \alpha, a} \VeneL_{\beta, b, a} \VnddR_{b, \gamma, c} \VdndR_{\delta, d, c} D_{27}(m^2_{\tilde{\chi}^0_{{d}}}, m^2_{\tilde{\chi}^0_{{b}}}, m^2_{\tilde{e}_{{a}}}, m^2_{\tilde{d}_{{c}}}) \\ 
  \VLLb{(n^d_1)}= & - \VneeL_{d, \alpha, a} \VeneR_{\beta, b, a} \VnddL_{b, \gamma, c} \VdndR_{\delta, d, c} D_{27}(m^2_{\tilde{\chi}^0_{{d}}}, m^2_{\tilde{\chi}^0_{{b}}}, m^2_{\tilde{e}_{{a}}}, m^2_{\tilde{d}_{{c}}}) \\ 
  \VLRb{(n^d_1)}= & \, \frac{1}{2} \VneeL_{d, \alpha, a} \VeneR_{\beta, b, a} \VnddR_{b, \gamma, c} \VdndL_{\delta, d, c} m_{\tilde{\chi}^0_{{b}}} m_{\tilde{\chi}^0_{{d}}} D_0(m^2_{\tilde{\chi}^0_{{d}}}, m^2_{\tilde{\chi}^0_{{b}}}, m^2_{\tilde{e}_{{a}}}, m^2_{\tilde{d}_{{c}}}) \\ 
  \TLLb{(n^d_1)}= & -\frac{1}{8} \VneeL_{d, \alpha, a} \VeneL_{\beta, b, a} \VnddL_{b, \gamma, c} \VdndL_{\delta, d, c} m_{\tilde{\chi}^0_{{b}}} m_{\tilde{\chi}^0_{{d}}} D_0(m^2_{\tilde{\chi}^0_{{d}}}, m^2_{\tilde{\chi}^0_{{b}}}, m^2_{\tilde{e}_{{a}}}, m^2_{\tilde{d}_{{c}}}) \\ 
\nonumber \\ 
  \SLLb{(n^d_2)}= & \, \frac{1}{2} \VneeL_{d, \alpha, a} \VeneL_{\beta, b, a} \VdndL_{\delta, b, c} \VnddL_{d, \gamma, c} m_{\tilde{\chi}^0_{{b}}} m_{\tilde{\chi}^0_{{d}}} D_0(m^2_{\tilde{\chi}^0_{{b}}}, m^2_{\tilde{\chi}^0_{{d}}}, m^2_{\tilde{e}_{{a}}}, m^2_{\tilde{d}_{{c}}}) \\ 
  \SLRb{(n^d_2)}= & \, 2 \, \VneeL_{d, \alpha, a} \VeneL_{\beta, b, a} \VdndR_{\delta, b, c} \VnddR_{d, \gamma, c} D_{27}(m^2_{\tilde{\chi}^0_{{b}}}, m^2_{\tilde{\chi}^0_{{d}}}, m^2_{\tilde{e}_{{a}}}, m^2_{\tilde{d}_{{c}}}) \\ 
  \VLLb{(n^d_2)}= & -\frac{1}{2} \VneeL_{d, \alpha, a} \VeneR_{\beta, b, a} \VdndR_{\delta, b, c} \VnddL_{d, \gamma, c} m_{\tilde{\chi}^0_{{b}}} m_{\tilde{\chi}^0_{{d}}} D_0(m^2_{\tilde{\chi}^0_{{b}}}, m^2_{\tilde{\chi}^0_{{d}}}, m^2_{\tilde{e}_{{a}}}, m^2_{\tilde{d}_{{c}}}) \\ 
  \VLRb{(n^d_2)}= & \, \VneeL_{d, \alpha, a} \VeneR_{\beta, b, a} \VdndL_{\delta, b, c} \VnddR_{d, \gamma, c} D_{27}(m^2_{\tilde{\chi}^0_{{b}}}, m^2_{\tilde{\chi}^0_{{d}}}, m^2_{\tilde{e}_{{a}}}, m^2_{\tilde{d}_{{c}}}) \\ 
  \TLLb{(n^d_2)}= & \, \frac{1}{8} \VneeL_{d, \alpha, a} \VeneL_{\beta, b, a} \VdndL_{\delta, b, c} \VnddL_{d, \gamma, c} m_{\tilde{\chi}^0_{{b}}} m_{\tilde{\chi}^0_{{d}}} D_0(m^2_{\tilde{\chi}^0_{{b}}}, m^2_{\tilde{\chi}^0_{{d}}}, m^2_{\tilde{e}_{{a}}}, m^2_{\tilde{d}_{{c}}}) 
\end{align} 

\paragraph{Chargino}
\begin{align} 
  \SLLb{(c^d_1)}= & \, \frac{1}{2} \VceviL_{d, \alpha, a} \VecviL_{\beta, b, a} \VcduL_{b, \gamma, c} \VdcuL_{\delta, d, c} m_{\tilde{\chi}^-_{{b}}} m_{\tilde{\chi}^-_{{d}}} D_0(m^2_{\tilde{\chi}^-_{{d}}}, m^2_{\tilde{\chi}^-_{{b}}}, m^2_{\nu^i_{{a}}}, m^2_{\tilde{u}_{{c}}}) \\ 
  \SLRb{(c^d_1)}= & \, 2 \, \VceviL_{d, \alpha, a} \VecviL_{\beta, b, a} \VcduR_{b, \gamma, c} \VdcuR_{\delta, d, c} D_{27}(m^2_{\tilde{\chi}^-_{{d}}}, m^2_{\tilde{\chi}^-_{{b}}}, m^2_{\nu^i_{{a}}}, m^2_{\tilde{u}_{{c}}}) \\ 
  \VLLb{(c^d_1)}= & - \VceviL_{d, \alpha, a} \VecviR_{\beta, b, a} \VcduL_{b, \gamma, c} \VdcuR_{\delta, d, c} D_{27}(m^2_{\tilde{\chi}^-_{{d}}}, m^2_{\tilde{\chi}^-_{{b}}}, m^2_{\nu^i_{{a}}}, m^2_{\tilde{u}_{{c}}}) \\ 
  \VLRb{(c^d_1)}= & \, \frac{1}{2} \VceviL_{d, \alpha, a} \VecviR_{\beta, b, a} \VcduR_{b, \gamma, c} \VdcuL_{\delta, d, c} m_{\tilde{\chi}^-_{{b}}} m_{\tilde{\chi}^-_{{d}}} D_0(m^2_{\tilde{\chi}^-_{{d}}}, m^2_{\tilde{\chi}^-_{{b}}}, m^2_{\nu^i_{{a}}}, m^2_{\tilde{u}_{{c}}}) \\ 
  \TLLb{(c^d_1)}= & -\frac{1}{8} \VceviL_{d, \alpha, a} \VecviL_{\beta, b, a} \VcduL_{b, \gamma, c} \VdcuL_{\delta, d, c} m_{\tilde{\chi}^-_{{b}}} m_{\tilde{\chi}^-_{{d}}} D_0(m^2_{\tilde{\chi}^-_{{d}}}, m^2_{\tilde{\chi}^-_{{b}}}, m^2_{\nu^i_{{a}}}, m^2_{\tilde{u}_{{c}}}) \\ 
\nonumber \\ 
  \SLLb{(c^d_2)}= & \, \frac{1}{2} \VcevrL_{d, \alpha, a} \VecvrL_{\beta, b, a} \VcduL_{b, \gamma, c} \VdcuL_{\delta, d, c} m_{\tilde{\chi}^-_{{b}}} m_{\tilde{\chi}^-_{{d}}} D_0(m^2_{\tilde{\chi}^-_{{d}}}, m^2_{\tilde{\chi}^-_{{b}}}, m^2_{\nu^R_{{a}}}, m^2_{\tilde{u}_{{c}}}) \\ 
  \SLRb{(c^d_2)}= & \, 2 \, \VcevrL_{d, \alpha, a} \VecvrL_{\beta, b, a} \VcduR_{b, \gamma, c} \VdcuR_{\delta, d, c} D_{27}(m^2_{\tilde{\chi}^-_{{d}}}, m^2_{\tilde{\chi}^-_{{b}}}, m^2_{\nu^R_{{a}}}, m^2_{\tilde{u}_{{c}}}) \\ 
  \VLLb{(c^d_2)}= & - \VcevrL_{d, \alpha, a} \VecvrR_{\beta, b, a} \VcduL_{b, \gamma, c} \VdcuR_{\delta, d, c} D_{27}(m^2_{\tilde{\chi}^-_{{d}}}, m^2_{\tilde{\chi}^-_{{b}}}, m^2_{\nu^R_{{a}}}, m^2_{\tilde{u}_{{c}}}) \\ 
  \VLRb{(c^d_2)}= & \, \frac{1}{2} \VcevrL_{d, \alpha, a} \VecvrR_{\beta, b, a} \VcduR_{b, \gamma, c} \VdcuL_{\delta, d, c} m_{\tilde{\chi}^-_{{b}}} m_{\tilde{\chi}^-_{{d}}} D_0(m^2_{\tilde{\chi}^-_{{d}}}, m^2_{\tilde{\chi}^-_{{b}}}, m^2_{\nu^R_{{a}}}, m^2_{\tilde{u}_{{c}}}) \\ 
  \TLLb{(c^d_2)}= & -\frac{1}{8} \VcevrL_{d, \alpha, a} \VecvrL_{\beta, b, a} \VcduL_{b, \gamma, c} \VdcuL_{\delta, d, c} m_{\tilde{\chi}^-_{{b}}} m_{\tilde{\chi}^-_{{d}}} D_0(m^2_{\tilde{\chi}^-_{{d}}}, m^2_{\tilde{\chi}^-_{{b}}}, m^2_{\nu^R_{{a}}}, m^2_{\tilde{u}_{{c}}})  
\end{align} 

\paragraph{$W^+$ and $H^+$}
\begin{align} 
  \SLLb{(w^d_1)}= & - \VvehpL_{a, \alpha, d} \VevhmL_{\beta, a, b} \VudhpL_{c, \gamma, b} \VduhmL_{\delta, c, d} m_{\nu_{{a}}} m_{u_{{c}}} \, D_0(m^2_{\nu_{{a}}}, m^2_{u_{{c}}}, m^2_{H^-_{{d}}}, m^2_{H^-_{{b}}}) \\ 
  \SLRb{(w^d_1)}= & - \VvehpL_{a, \alpha, d} \VevhmL_{\beta, a, b} \VudhpR_{c, \gamma, b} \VduhmR_{\delta, c, d} m_{\nu_{{a}}} m_{u_{{c}}} \, D_0(m^2_{\nu_{{a}}}, m^2_{u_{{c}}}, m^2_{H^-_{{d}}}, m^2_{H^-_{{b}}}) \\ 
  \VLLb{(w^d_1)}= & - \VvehpL_{a, \alpha, d} \VevhmR_{\beta, a, b} \VudhpL_{c, \gamma, b} \VduhmR_{\delta, c, d} \, D_{27}(m^2_{\nu_{{a}}}, m^2_{u_{{c}}}, m^2_{H^-_{{d}}}, m^2_{H^-_{{b}}}) \\ 
  \VLRb{(w^d_1)}= & - \VvehpL_{a, \alpha, d} \VevhmR_{\beta, a, b} \VudhpR_{c, \gamma, b} \VduhmL_{\delta, c, d} \, D_{27}(m^2_{\nu_{{a}}}, m^2_{u_{{c}}}, m^2_{H^-_{{d}}}, m^2_{H^-_{{b}}}) \\ 
\nonumber \\
  \SLLb{(w^d_2)}= & \, 2 \, \VvehpL_{a, \alpha, d} \VevwmR_{\beta, a} \VudwpL_{c, \gamma} \VduhmL_{\delta, c, d} (I_{C_0 D_0}(m^2_{u_{{c}}}, m^2_{W^-}, m^2_{H^-_{{d}}}, m^2_{\nu_{{a}}}) - 2 D_{27}(m^2_{\nu_{{a}}}, m^2_{u_{{c}}}, m^2_{W^-}, m^2_{H^-_{{d}}})) \\ 
  \SLRb{(w^d_2)}= & \, 2  \, \VvehpL_{a, \alpha, d} \VevwmR_{\beta, a} \VudwpR_{c, \gamma} \VduhmR_{\delta, c, d} (I_{C_0 D_0}(m^2_{u_{{c}}}, m^2_{W^-}, m^2_{H^-_{{d}}}, m^2_{\nu_{{a}}}) - 2 D_{27}(m^2_{\nu_{{a}}}, m^2_{u_{{c}}}, m^2_{W^-}, m^2_{H^-_{{d}}})) \\ 
  \VLLb{(w^d_2)}= &  \VvehpL_{a, \alpha, d} \VevwmL_{\beta, a} \VudwpL_{c, \gamma} \VduhmR_{\delta, c, d} m_{\nu_{{a}}} m_{u_{{c}}} D_0(m^2_{\nu_{{a}}}, m^2_{u_{{c}}}, m^2_{W^-}, m^2_{H^-_{{d}}}) \\ 
  \VLRb{(w^d_2)}= &  \VvehpL_{a, \alpha, d} \VevwmL_{\beta, a} \VudwpR_{c, \gamma} \VduhmL_{\delta, c, d} m_{\nu_{{a}}} m_{u_{{c}}} D_0(m^2_{\nu_{{a}}}, m^2_{u_{{c}}}, m^2_{W^-}, m^2_{H^-_{{d}}}) \\ 
  \TLLb{(w^d_2)}= & - \VvehpL_{a, \alpha, d} \VevwmR_{\beta, a} \VudwpL_{c, \gamma} \VduhmL_{\delta, c, d} D_{27}(m^2_{\nu_{{a}}}, m^2_{u_{{c}}}, m^2_{W^-}, m^2_{H^-_{{d}}}) \\ 
\nonumber \\
  \SLLb{(w^d_3)}= & \, 2 \, \VvewpL_{a, i} \VevhmL_{\beta, a, b} \VudhpL_{c, \gamma, b} \VduwmR_{\delta, c} ( I_{C_0 D_0}(m^2_{u_{{c}}}, m^2_{H^-_{{b}}}, m^2_{W^-}, m^2_{\nu_{{a}}}) - 2 D_{27}(m^2_{\nu_{{a}}}, m^2_{u_{{c}}}, mS22, m^2_{W^-})) \\ 
  \SLRb{(w^d_3)}= & \, 2 \, \VvewpL_{a, i} \VevhmL_{\beta, a, b} \VudhpR_{c, \gamma, b} \VduwmL_{\delta, c} ( I_{C_0 D_0}(m^2_{u_{{c}}}, m^2_{H^-_{{b}}}, m^2_{W^-}, m^2_{\nu_{{a}}}) - 2 D_{27}(m^2_{\nu_{{a}}}, m^2_{u_{{c}}}, mS22, m^2_{W^-})) \\ 
  \VLLb{(w^d_3)}= & \, \VvewpL_{a, i} \VevhmR_{\beta, a, b} \VudhpL_{c, \gamma, b} \VduwmL_{\delta, c} m_{\nu_{{a}}} m_{u_{{c}}} D_0(m^2_{\nu_{{a}}}, m^2_{u_{{c}}}, m^2_{H^-_{{b}}}, m^2_{W^-}) \\ 
  \VLRb{(w^d_3)}= &  \, \VvewpL_{a, i} \VevhmR_{\beta, a, b} \VudhpR_{c, \gamma, b} \VduwmR_{\delta, c} m_{\nu_{{a}}} m_{u_{{c}}} D_0(m^2_{\nu_{{a}}}, m^2_{u_{{c}}}, m^2_{H^-_{{b}}}, m^2_{W^-}) \\ 
  \TLLb{(w^d_3)}d= & - \VvewpL_{a, i} \VevhmL_{\beta, a, b} \VudhpL_{c, \gamma, b} \VduwmR_{\delta, c} D_{27}(m^2_{\nu_{{a}}}, m^2_{u_{{c}}}, m^2_{H^-_{{b}}}, m^2_{W^-}) \\ 
\nonumber \\
  \SLLb{(w^d_4)}= & -4  \VvewpL_{a, i} \VevwmR_{\beta, a} \VudwpL_{c, \gamma} \VduwmR_{\delta, c} m_{\nu_{{a}}} m_{u_{{c}}} D_0(m^2_{\nu_{{a}}}, m^2_{u_{{c}}}, m^2_{W^-}, m^2_{W^-}) \\ 
  \SLRb{(w^d_4)}= & -4  \VvewpL_{a, i} \VevwmR_{\beta, a} \VudwpR_{c, \gamma} \VduwmL_{\delta, c} m_{\nu_{{a}}} m_{u_{{c}}} D_0(m^2_{\nu_{{a}}}, m^2_{u_{{c}}}, m^2_{W^-}, m^2_{W^-}) \\ 
  \VLLb{(w^d_4)}= & -4  \VvewpL_{a, i} \VevwmL_{\beta, a} \VudwpL_{c, \gamma} \VduwmL_{\delta, c} (I_{C_0 D_0}(m^2_{u_{{c}}}, m^2_{W^-}, m^2_{W^-}, m^2_{\nu_{{a}}}) - 3 D_{27}(m^2_{\nu_{{a}}}, m^2_{u_{{c}}}, m^2_{W^-}, m^2_{W^-})) \\ 
  \VLRb{(w^d_4)}= & -4  \VvewpL_{a, i} \VevwmL_{\beta, a} \VudwpR_{c, \gamma} \VduwmR_{\delta, c} I_{C_0 D_0}(m^2_{u_{{c}}}, m^2_{W^-}, m^2_{W^-}, m^2_{\nu_{{a}}}) \\ 
  \TLLb{(w^d_4)}d= & \, \VvewpL_{a, i} \VevwmR_{\beta, a} \VudwpL_{c, \gamma} \VduwmR_{\delta, c} m_{\nu_{{a}}} m_{u_{{c}}} D_0(m^2_{\nu_{{a}}}, m^2_{u_{{c}}}, m^2_{W^-}, m^2_{W^-}) 
\end{align}

\subsubsection{Up quarks}
\paragraph{Neutralino}
\begin{align} 
  \SLLb{(n^u_1)}= & \, \frac{1}{2} \VneeL_{d, \alpha, a} \VeneL_{\beta, b, a} \VnuuL_{b, \gamma, c} \VunuL_{\delta, d, c} D_{27}(m^2_{\tilde{\chi}^0_{{d}}}, m^2_{\tilde{\chi}^0_{{b}}}, m^2_{\tilde{e}_{{a}}}, m^2_{\tilde{u}_{{c}}}) \\ 
  \SLRb{(n^u_1)}= & \, 2 \, \VneeL_{d, \alpha, a} \VeneL_{\beta, b, a} \VnuuR_{b, \gamma, c} \VunuR_{\delta, d, c} D_{27}(m^2_{\tilde{\chi}^0_{{d}}}, m^2_{\tilde{\chi}^0_{{b}}}, m^2_{\tilde{e}_{{a}}}, m^2_{\tilde{u}_{{c}}}) \\ 
  \VLLb{(n^u_1)}= & - \VneeL_{d, \alpha, a} \VeneR_{\beta, b, a} \VnuuL_{b, \gamma, c} \VunuR_{\delta, d, c} D_{27}(m^2_{\tilde{\chi}^0_{{d}}}, m^2_{\tilde{\chi}^0_{{b}}}, m^2_{\tilde{e}_{{a}}}, m^2_{\tilde{u}_{{c}}}) \\ 
  \VLRb{(n^u_1)}= & \, \frac{1}{2} \VneeL_{d, \alpha, a} \VeneR_{\beta, b, a} \VnuuR_{b, \gamma, c} \VunuL_{\delta, d, c} D_{27}(m^2_{\tilde{\chi}^0_{{d}}}, m^2_{\tilde{\chi}^0_{{b}}}, m^2_{\tilde{e}_{{a}}}, m^2_{\tilde{u}_{{c}}}) \\ 
\nonumber \\
  \SLLb{(n^u_2)}= & \, \frac{1}{2} \VneeL_{d, \alpha, a} \VeneL_{\beta, b, a} \VunuL_{\delta, b, c} \VnuuL_{d, \gamma, c} m_{\tilde{\chi}^0_{{b}}} m_{\tilde{\chi}^0_{{d}}} D_0(m^2_{\tilde{\chi}^0_{{b}}}, m^2_{\tilde{\chi}^0_{{d}}}, m^2_{\tilde{e}_{{a}}}, m^2_{\tilde{u}_{{c}}}) \\ 
  \SLRb{(n^u_2)}= & \, 2 \, \VneeL_{d, \alpha, a} \VeneL_{\beta, b, a} \VunuR_{\delta, b, c} \VnuuR_{d, \gamma, c} D_{27}(m^2_{\tilde{\chi}^0_{{b}}}, m^2_{\tilde{\chi}^0_{{d}}}, m^2_{\tilde{e}_{{a}}}, m^2_{\tilde{u}_{{c}}}) \\ 
  \VLLb{(n^u_2)}= & -\frac{1}{2} \VneeL_{d, \alpha, a} \VeneR_{\beta, b, a} \VunuR_{\delta, b, c} \VnuuL_{d, \gamma, c} m_{\tilde{\chi}^0_{{b}}} m_{\tilde{\chi}^0_{{d}}} D_0(m^2_{\tilde{\chi}^0_{{b}}}, m^2_{\tilde{\chi}^0_{{d}}}, m^2_{\tilde{e}_{{a}}}, m^2_{\tilde{u}_{{c}}}) \\ 
  \VLRb{(n^u_2)}= & \, \VneeL_{d, \alpha, a} \VeneR_{\beta, b, a} \VunuL_{\delta, b, c} \VnuuR_{d, \gamma, c} D_{27}(m^2_{\tilde{\chi}^0_{{b}}}, m^2_{\tilde{\chi}^0_{{d}}}, m^2_{\tilde{e}_{{a}}}, m^2_{\tilde{u}_{{c}}}) \\ 
  \TLLb{(n^u_2)}= & \, \frac{1}{8} \VneeL_{d, \alpha, a} \VeneL_{\beta, b, a} \VunuL_{\delta, b, c} \VnuuL_{d, \gamma, c} m_{\tilde{\chi}^0_{{b}}} m_{\tilde{\chi}^0_{{d}}} D_0(m^2_{\tilde{\chi}^0_{{b}}}, m^2_{\tilde{\chi}^0_{{d}}}, m^2_{\tilde{e}_{{a}}}, m^2_{\tilde{u}_{{c}}}) 
\end{align}

\paragraph{Chargino}

\begin{align} 
  \SLLb{(c^u_1)}= & \, \frac{1}{2} \VceviL_{d, \alpha, a} \VecviL_{\beta, b, a} \VcudL_{b, \delta, c} \VcudLc_{d, \gamma, c} m_{\tilde{\chi}^-_{{b}}} m_{\tilde{\chi}^-_{{d}}} D_0(m^2_{\tilde{\chi}^-_{{b}}}, m^2_{\tilde{\chi}^-_{{d}}}, m^2_{\nu^i_{{a}}}, m^2_{\tilde{d}_{{c}}}) \\ 
  \SLRb{(c^u_1)}= & \, 2 \, \VceviL_{d, \alpha, a} \VecviL_{\beta, b, a} \VcudR_{b, \delta, c} \VcudRc_{d, \gamma, c} D_{27}(m^2_{\tilde{\chi}^-_{{b}}}, m^2_{\tilde{\chi}^-_{{d}}}, m^2_{\nu^i_{{a}}}, m^2_{\tilde{d}_{{c}}}) \\ 
  \VLLb{(c^u_1)}= & -\frac{1}{2} \VceviL_{d, \alpha, a} \VecviR_{\beta, b, a} \VcudR_{b, \delta, c} \VcudLc_{d, \gamma, c} m_{\tilde{\chi}^-_{{b}}} m_{\tilde{\chi}^-_{{d}}} D_0(m^2_{\tilde{\chi}^-_{{b}}}, m^2_{\tilde{\chi}^-_{{d}}}, m^2_{\nu^i_{{a}}}, m^2_{\tilde{d}_{{c}}}) \\ 
  \VLRb{(c^u_1)}= & \, \VceviL_{d, \alpha, a} \VecviR_{\beta, b, a} \VcudL_{b, \delta, c} \VcudRc_{d, \gamma, c} D_{27}(m^2_{\tilde{\chi}^-_{{b}}}, m^2_{\tilde{\chi}^-_{{d}}}, m^2_{\nu^i_{{a}}}, m^2_{\tilde{d}_{{c}}}) \\ 
  \TLLb{(c^u_1)}= & \, \frac{1}{8} \VceviL_{d, \alpha, a} \VecviL_{\beta, b, a} \VcudL_{b, \delta, c} \VcudLc_{d, \gamma, c} m_{\tilde{\chi}^-_{{b}}} m_{\tilde{\chi}^-_{{d}}} D_0(m^2_{\tilde{\chi}^-_{{b}}}, m^2_{\tilde{\chi}^-_{{d}}}, m^2_{\nu^i_{{a}}}, m^2_{\tilde{d}_{{c}}}) \\ 
\nonumber \\ 
  \SLLb{(c^u_2)}= & \, \frac{1}{2} \VcevrL_{d, \alpha, a} \VecvrL_{\beta, b, a} \VcudL_{b, \delta, c} \VcudLc_{d, \gamma, c}  m_{\tilde{\chi}^-_{{b}}}  m_{\tilde{\chi}^-_{{d}}} D_0(m^2_{\tilde{\chi}^-_{{b}}}, m^2_{\tilde{\chi}^-_{{d}}}, m^2_{\nu^R_{{a}}}, m^2_{\tilde{d}_{{c}}}) \\ 
  \SLRb{(c^u_2)}= & \, 2 \, \VcevrL_{d, \alpha, a} \VecvrL_{\beta, b, a} \VcudR_{b, \delta, c} \VcudRc_{d, \gamma, c} D_{27}(m^2_{\tilde{\chi}^-_{{b}}}, m^2_{\tilde{\chi}^-_{{d}}}, m^2_{\nu^R_{{a}}}, m^2_{\tilde{d}_{{c}}}) \\ 
  \VLLb{(c^u_2)}= & -\frac{1}{2} \VcevrL_{d, \alpha, a} \VecvrR_{\beta, b, a} \VcudR_{b, \delta, c} \VcudLc_{d, \gamma, c}  m_{\tilde{\chi}^-_{{b}}}  m_{\tilde{\chi}^-_{{d}}} D_0(m^2_{\tilde{\chi}^-_{{b}}}, m^2_{\tilde{\chi}^-_{{d}}}, m^2_{\nu^R_{{a}}}, m^2_{\tilde{d}_{{c}}}) \\ 
  \VLRb{(c^u_2)}= & \, \VcevrL_{d, \alpha, a} \VecvrR_{\beta, b, a} \VcudL_{b, \delta, c} \VcudRc_{d, \gamma, c} D_{27}(m^2_{\tilde{\chi}^-_{{b}}}, m^2_{\tilde{\chi}^-_{{d}}}, m^2_{\nu^R_{{a}}}, m^2_{\tilde{d}_{{c}}}) \\ 
  \TLLb{(c^u_2)}= & \, \frac{1}{8} \VcevrL_{d, \alpha, a} \VecvrL_{\beta, b, a} \VcudL_{b, \delta, c} \VcudLc_{d, \gamma, c}  m_{\tilde{\chi}^-_{{b}}}  m_{\tilde{\chi}^-_{{d}}} D_0(m^2_{\tilde{\chi}^-_{{b}}}, m^2_{\tilde{\chi}^-_{{d}}}, m^2_{\nu^R_{{a}}}, m^2_{\tilde{d}_{{c}}}) 
\end{align} 

\paragraph{$W^+$ and $H^+$}
\begin{align} 
  \SLLb{(w^u_1)}= & - \VvehpL_{a, \alpha, d} \VevhmL_{\beta, a, b} \VudhpL_{\delta, c, b} \VduhmL_{c, \gamma, d} m_{\nu_{{a}}} m_{d_{{c}}} D_0(m^2_{\nu_{{a}}}, m^2_{d_{{c}}}, m^2_{H^-_{{d}}}, m^2_{H^-_{{b}}}) \\ 
  \SLRb{(w^u_1)}= & - \VvehpL_{a, \alpha, d} \VevhmL_{\beta, a, b} \VudhpR_{\delta, c, b} \VduhmR_{c, \gamma, d} m_{\nu_{{a}}} m_{d_{{c}}} D_0(m^2_{\nu_{{a}}}, m^2_{d_{{c}}}, m^2_{H^-_{{d}}}, m^2_{H^-_{{b}}}) \\ 
  \VLLb{(w^u_1)}= & \,  \VvehpL_{a, \alpha, d} \VevhmR_{\beta, a, b} \VudhpR_{\delta, c, b} \VduhmL_{c, \gamma, d} D_{27}(m^2_{\nu_{{a}}}, m^2_{d_{{c}}}, m^2_{H^-_{{d}}}, m^2_{H^-_{{b}}}) \\ 
  \VLRb{(w^u_1)}= & \,  \VvehpL_{a, \alpha, d} \VevhmR_{\beta, a, b} \VudhpL_{\delta, c, b} \VduhmR_{c, \gamma, d} D_{27}(m^2_{\nu_{{a}}}, m^2_{d_{{c}}}, m^2_{H^-_{{d}}}, m^2_{H^-_{{b}}}) \\ 
\nonumber \\
  \SLLb{(w^u_2)}= & -4  \VvehpL_{a, \alpha, d} \VevwmR_{\beta, a} \VudwpR_{\delta, c} \VduhmL_{c, \gamma, d} D_{27}(m^2_{\nu_{{a}}}, m^2_{d_{{c}}}, m^2_{W^-}, m^2_{H^-_{{d}}}) \\ 
  \SLRb{(w^u_2)}= & -4  \VvehpL_{a, \alpha, d} \VevwmR_{\beta, a} \VudwpL_{\delta, c} \VduhmR_{c, \gamma, d} D_{27}(m^2_{\nu_{{a}}}, m^2_{d_{{c}}}, m^2_{W^-}, m^2_{H^-_{{d}}}) \\ 
  \VLLb{(w^u_2)}= & \, \VvehpL_{a, \alpha, d} \VevwmL_{\beta, a} \VudwpL_{\delta, c} \VduhmL_{c, \gamma, d} m_{\nu_{{a}}} m_{d_{{c}}} D_0(m^2_{\nu_{{a}}}, m^2_{d_{{c}}}, m^2_{W^-}, m^2_{H^-_{{d}}}) \\ 
  \VLRb{(w^u_2)}= & \, \VvehpL_{a, \alpha, d} \VevwmL_{\beta, a} \VudwpR_{\delta, c} \VduhmR_{c, \gamma, d} m_{\nu_{{a}}} m_{d_{{c}}} D_0(m^2_{\nu_{{a}}}, m^2_{d_{{c}}}, m^2_{W^-}, m^2_{H^-_{{d}}}) \\ 
  \TLLb{(w^u_2)}= & - \VvehpL_{a, \alpha, d} \VevwmR_{\beta, a} \VudwpR_{\delta, c} \VduhmL_{c, \gamma, d}  D_{27}(m^2_{\nu_{{a}}}, m^2_{d_{{c}}}, m^2_{W^-}, m^2_{H^-_{{d}}}) \\ 
\nonumber \\ 
  \SLLb{(w^u_3)}= & -4  \VvewpL_{a, i} \VevhmL_{\beta, a, b} \VudhpL_{\delta, c, b} \VduwmL_{c, \gamma} D_{27}(m^2_{\nu_{{a}}}, m^2_{d_{{c}}}, m^2_{H^-_{{b}}}, m^2_{W^-})\\ 
  \SLRb{(w^u_3)}= & -4  \VvewpL_{a, i} \VevhmL_{\beta, a, b} \VudhpR_{\delta, c, b} \VduwmR_{c, \gamma} D_{27}(m^2_{\nu_{{a}}}, m^2_{d_{{c}}}, m^2_{H^-_{{b}}}, m^2_{W^-}) \\ 
  \VLLb{(w^u_3)}= & \, \VvewpL_{a, i} \VevhmR_{\beta, a, b} \VudhpR_{\delta, c, b} \VduwmL_{c, \gamma} m_{\nu_{{a}}} m_{d_{{c}}} D_0(m^2_{\nu_{{a}}}, m^2_{d_{{c}}}, m^2_{H^-_{{b}}}, m^2_{W^-}) \\ 
  \VLRb{(w^u_3)}= & \, \VvewpL_{a, i} \VevhmR_{\beta, a, b} \VudhpL_{\delta, c, b} \VduwmR_{c, \gamma} m_{\nu_{{a}}} m_{d_{{c}}} D_0(m^2_{\nu_{{a}}}, m^2_{d_{{c}}}, m^2_{H^-_{{b}}}, m^2_{W^-}) \\ 
  \TLLb{(w^u_3)}= & - \VvewpL_{a, i} \VevhmL_{\beta, a, b} \VudhpL_{\delta, c, b} \VduwmL_{c, \gamma} D_{27}(m^2_{\nu_{{a}}}, m^2_{d_{{c}}}, m^2_{H^-_{{b}}}, m^2_{W^-}) \\ 
\nonumber \\
  \SLLb{(w^u_4)}= & -4  \VvewpL_{a, i} \VevwmR_{\beta, a} \VudwpR_{\delta, c} \VduwmL_{c, \gamma}  m_{\nu_{{a}}} m_{d_{{c}}} D_0(m^2_{\nu_{{a}}}, m^2_{d_{{c}}}, m^2_{W^-}, m^2_{W^-}) \\ 
  \SLRb{(w^u_4)}= & -4  \VvewpL_{a, i} \VevwmR_{\beta, a} \VudwpL_{\delta, c} \VduwmR_{c, \gamma}  m_{\nu_{{a}}} m_{d_{{c}}} D_0(m^2_{\nu_{{a}}}, m^2_{d_{{c}}}, m^2_{W^-}, m^2_{W^-}) \\ 
  \VLLb{(w^u_4)}= & \, 16 \, \VvewpL_{a, i} \VevwmL_{\beta, a} \VudwpL_{\delta, c} \VduwmL_{c, \gamma} D_{27}(m^2_{\nu_{{a}}}, m^2_{d_{{c}}}, m^2_{W^-}, m^2_{W^-}) \\ 
  \VLRb{(w^u_4)}= & \, 4 \, \VvewpL_{a, i} \VevwmL_{\beta, a} \VudwpR_{\delta, c} \VduwmR_{c, \gamma} D_{27}(m^2_{\nu_{{a}}}, m^2_{d_{{c}}}, m^2_{W^-}, m^2_{W^-}) \\ 
  \TLLb{(w^u_4)}= & - \VvewpL_{a, i} \VevwmR_{\beta, a} \VudwpR_{\delta, c} \VduwmL_{c, \gamma}  m_{\nu_{{a}}} m_{d_{{c}}} D_0(m^2_{\nu_{{a}}}, m^2_{d_{{c}}}, m^2_{W^-}, m^2_{W^-}) 
\end{align}

\section{Form factors of the 4-fermion operators}
\label{app:operators}

We define the sum over all penguin diagrams as  
\begin{align}
 \VLLZ{sum} = & \, \frac{1}{16\pi^2} \left( \sum_a  \VLLZ{(n_a)} + \sum_a  \VLLZ{(c_a)} +\sum_a  \VLLZ{(w_a)} \right) \\
 \VLRZ{sum} = & \, \frac{1}{16\pi^2} \left(\sum_a  \VLRZ{(n_a)} + \sum_a  \VLRZ{(c_a)} +\sum_a  \VLRZ{(w_a)} \right)\\
 \SLLh{sum} = & \, \frac{1}{16\pi^2} \left(\sum_a  \SLLh{(n_a)} + \sum_a  \SLLh{(c_a)} +\sum_a  \SLLh{(w_a)} \right)\\
 \SLRh{sum} = & \, \frac{1}{16\pi^2} \left(\sum_a  \SLRh{(n_a)} + \sum_a  \SLRh{(c_a)} +\sum_a  \SLRh{(w_a)} \right)\\
 \SLLa{sum} = & \, \frac{1}{16\pi^2} \left(\sum_a  \SLLa{(n_a)} + \sum_a  \SLLa{(c_a)} +\sum_a  \SLLa{(w_a)} \right)\\
 \SLRa{sum} = & \, \frac{1}{16\pi^2} \left(\sum_a  \SLRa{(n_a)} + \sum_a  \SLRa{(c_a)} +\sum_a  \SLRa{(w_a)} \right)
\end{align}
and the sum over all boxes as
\begin{align}
\XLLb{(sum^x)} = & \, \frac{1}{16\pi^2} \left(\sum_a \XLLb{(n^x_a)} +\sum_a \XLLb{(c^x_a)} + \sum_a \XLLb{(w^x_a)} \right)\\
\XLRb{(sum^x)} = & \, \frac{1}{16\pi^2} \left(\sum_a \XLRb{(n^x_a)} +\sum_a \XLRb{(c^x_a)} + \sum_a \XLRb{(w^x_a)} \right)
\end{align}
with $X=V,S,T$ and $x=\ell,d,u$.  With these, we can finally obtain the
form factors of the 4-lepton operators as follows:
\begin{align}
 A^V_{LL} = & \, \VLLZ{sum} \VeezL_{\gamma, \delta} \frac{1}{m^2_{Z}} + \VLLb{(sum^\ell)} \\
 A^V_{LR} = & \, \VLRZ{sum} \VeezR_{\gamma, \delta} \frac{1}{m^2_{Z}} + \VLRb{(sum^\ell)}\\
 A^S_{LL} = & \, \sum_p \SLLh{sum} \VeehL_{\gamma, \delta, p} \frac{1}{m^2_{h_{{p}}}} + \sum_p \SLLa{sum} \VeeaL_{\gamma, \delta, p} \frac{1}{m^2_{A^0_{{p}}}} + \SLLb{(sum^\ell)} \\
 A^S_{LR} = &  \sum_p \SLRh{sum} \VeehR_{\gamma, \delta, p} \frac{1}{m^2_{h_{{p}}}} + \sum_p \SLRa{sum} \VeeaR_{\gamma, \delta, p} \frac{1}{m^2_{A^0_{{p}}}} + \SLRb{(sum^\ell)}\\
 A^T_{LL} = & \, \TLLb{(sum^\ell)}\\
 A^T_{LR} = & \, \TLRb{(sum^\ell)} \\
\nonumber \\
 B^V_{LL} = & \, \VLLZ{sum} \VddzL_{\gamma, \delta} \frac{1}{m^2_{Z}} + \VLLb{(sum^\ell)} \\
 B^V_{LR} = & \, \VLRZ{sum} \VddzR_{\gamma, \delta} \frac{1}{m^2_{Z}} + \VLRb{(sum^\ell)}\\
 B^S_{LL} = & \, \sum_p \SLLh{sum} \VddhL_{\gamma, \delta, p} \frac{1}{m^2_{h_{{p}}}} + \sum_p \SLLa{sum} \VddaL_{\gamma, \delta, p} \frac{1}{m^2_{A^0_{{p}}}} + \SLLb{(sum^d)} \\
 B^S_{LR} = &  \sum_p \SLRh{sum} \VddhR_{\gamma, \delta, p} \frac{1}{m^2_{h_{{p}}}} + \sum_p \SLRa{sum} \VddaR_{\gamma, \delta, p} \frac{1}{m^2_{A^0_{{p}}}} + \SLRb{(sum^d)}\\
 B^T_{LL} = & \, \TLLb{(sum^d)}\\
 B^T_{LR} = & \, \TLRb{(sum^d)} \\
\nonumber \\
 C^V_{LL} = & \, \VLLZ{sum} \VuuzL_{\gamma, \delta} \frac{1}{m^2_{Z}} + \VLLb{(sum^u)} \\
 C^V_{LR} = & \, \VLRZ{sum} \VuuzR_{\gamma, \delta} \frac{1}{m^2_{Z}} + \VLRb{(sum^u)}\\
 C^S_{LL} = & \, \sum_p \SLLh{sum} \VuuhL_{\gamma, \delta, p} \frac{1}{m^2_{h_{{p}}}} + \sum_p \SLLa{sum} \VuuaL_{\gamma, \delta, p} \frac{1}{m^2_{A^0_{{p}}}} + \SLLb{(sum^u)} \\
 C^S_{LR} = &  \sum_p \SLRh{sum} \VuuhR_{\gamma, \delta, p} \frac{1}{m^2_{h_{{p}}}} + \sum_p \SLRa{sum} \VuuaR_{\gamma, \delta, p} \frac{1}{m^2_{A^0_{{p}}}} + \SLRb{(sum^u)}\\
 C^T_{LL} = & \, \TLLb{(sum^u)}\\
 C^T_{LR} = & \, \TLRb{(sum^u)}
\end{align}
and the other chiralities are given by $X^W_{RL} =X^W_{LR}
(R\leftrightarrow L)$ and $X^W_{RR} =X^W_{LL} (R\leftrightarrow L)$
($X=A,B,C$; $W=S,T,V$).

\end{appendix}

\bibliographystyle{h-physrev5}

\end{document}